\documentclass[a4paper,11pt]{article}
\usepackage{jheppub} % use jheppub.sty
\usepackage{amsfonts,mathrsfs,amsthm,textcomp,dsfont,esint,array,multirow, mathtools,bbm,slashed,xcolor,tikz}
\usepackage[mathscr]{euscript}
\usepackage[all,2cell]{xy}

\usepackage[hyphens]{url} 
\definecolor{linkc}{rgb}{0,0,1}
\usepackage[pagebackref=true,colorlinks=true,linkcolor=linkc,citecolor=linkc,urlcolor=linkc,linktoc=all,pdfusetitle=true]{hyperref}
\renewcommand*{\backref}[1]{}
\renewcommand*{\backrefalt}[4]{{%
		\ifcase #1 %Not cited.%
		\else [Cited: #2.]%
		\fi	}}

\allowdisplaybreaks
\def\IC{\mathbb{C}}
\def\IH{\mathbb{H}}\def\IN{\mathbb{N}}\def\IP{\mathbb{CP}}
\def\IR{\mathbb{R}}\def\IZ{\mathbb{Z}}

\def\CA{{\cal A}}\def\CB{{\cal B}}\def\CD{{\cal D}}
\def\CE{{\cal E}}\def\CF{{\cal F}}\def\CG{{\cal G}}
\def\CH{{\cal H}}\def\CI{{\cal I}}
\def\CK{{\cal K}}\def\CL{{\cal L}}\def\CM{{\cal M}}
\def\CN{{\cal N}}\def\CO{{\cal O}}
\def\CQ{{\cal Q}}\def\CR{{\cal R}}

\def\CX{{\cal X}}\def\CY{{\cal Y}}
\def\CZ{{\cal Z}}\def\CC{{\cal C}}

\def\e{\epsilon}

\def\t{\widetilde}

\def\half{\frac{1}{2}}

\def\tr{{\rm tr}}

\newcommand{\be}{\begin{equation}}
\newcommand{\ee}{\end{equation}}
\newcommand{\bea}{\begin{equation} \begin{aligned}} \newcommand{\eea}{\end{aligned} \end{equation}}
\newcommand{\ba}{\begin{array}}

\newcommand{\sgn}{\mathrm{sgn}}
\newcommand{\smashlastub}[1]{%
  \sbox0{\let\smash\relax$#1$}
  \vphantom{\usebox0}
  \sbox2{$#1$}
  \raisebox{\dimexpr(\ht0-\ht2)}{\usebox2}
}

\newcommand{\BC}{\mathbb{C}}
\newcommand{\BH}{\mathbb{H}}
\newcommand{\BT}{\mathbb{T}}

\newcommand{\eps}{\epsilon}

\newcommand{\bfb}{{\boldsymbol b}}

\newcommand{\bfn}{{\boldsymbol n}} 
 
\newcommand{\bfx}{{\boldsymbol x}}

\newcommand{\bfnu}{{\boldsymbol \nu}}

\newcommand{\bfk}{{\boldsymbol k}}
\newcommand{\bfz}{{\boldsymbol z}}
\newcommand{\bfmu}{{\boldsymbol \mu}}

\newcommand{\cR}{\mathcal{R}}
\newcommand{\Tr}{\text{Tr}}
\newcommand{\ttu}{\mathtt{u}}

\newcommand{\fa}{\mathfrak{a}}

\title{Path Integral Derivations Of K-Theoretic Donaldson Invariants}

\author[a]{Heeyeon Kim,}
\author[b,c]{Jan Manschot,}
\author[d]{Gregory W. Moore,}
\author[d]{Runkai Tao,}
\author[e]{Xinyu Zhang}
\affiliation[a]{Department of Physics, Korea Advanced Institute of Science and Technology \\ 291 Daehak-ro, Yuseong-gu, Daejeon 34141, Republic of Korea}
\affiliation[b]{School of Mathematics, Trinity College, Dublin 2, Ireland}
\affiliation[c]{Hamilton Mathematical Institute, Trinity College, Dublin 2, Ireland}
\affiliation[d]{NHETC and Department of Physics and Astronomy, Rutgers University, 126 Frelinghuysen Rd., Piscataway NJ 08855, USA}
\affiliation[e]{Zhejiang Institute of Modern Physics, School of Physics, Zhejiang University, Hangzhou, Zhejiang, 310058, China}
\emailAdd{heeyeon.kim@kaist.ac.kr}
\emailAdd{jan.manschot@tcd.ie}
\emailAdd{gwmoore@physics.rutgers.edu}
\emailAdd{runkai.tao@physics.rutgers.edu}
\emailAdd{xinyu.zhang@zju.edu.cn}

\abstract{
We consider 5d $\mathcal{N}=1$ SU(2) super Yang-Mills theory on $X\times S^1$, with $X$ a closed smooth four-manifold. A partial topological twisting along $X$ renders the theory formally independent of the metric on $X$. The theory depends on the spin structure and the circumference $R$ of $S^1$. The coefficients of the $R$-expansion of the partition function are Witten indices, which are identified with $L^2$-indices of Dirac operators on moduli spaces of instantons. The partition function encodes BPS indices for instanton particles on a spatial manifold $X$, and these indices are special cases of K-theoretic Donaldson invariants. When the 't Hooft flux of the gauge theory is nonzero and $X$ is not spin, the 5d theory can be anomalous, but this anomaly can be canceled by coupling to a line bundle with connection for the global $U(1)$ ``instanton number symmetry''. For $b_2^+(X)>0$ we can derive the partition function from integration over the Coulomb branch of the effective 4d low-energy theory. When $X$ is toric we can also use equivariant localization with respect to the $\mathbb{C}^* \times \mathbb{C}^*$ symmetry. The two methods lead to the same results for the wall-crossing formula. We also determine path integrals for four-manifolds with $b_2^+(X)>1$. Our results agree with those for algebraic surfaces by G\"ottsche, Kool, Nakajima, Yoshioka, and Williams, but apply to a larger class of manifolds. When the circumference of the circle is tuned to special values, the path integral is associated with the 5d superconformal $E_1$ theory. Topological invariants in this case involve generalizations of Seiberg-Witten invariants.
}

\begin{document}

\maketitle
\pagestyle{empty}

\clearpage

\pagestyle{plain}
\setcounter{page}{1} 
\pagenumbering{arabic} 

%%%%%%%%%Body of the paper%%%%%%%%%

\section{A Brief Summary Of This Paper}\label{sec:BriefSummary}

This paper is a contribution to a long and fruitful dialogue between mathematicians and physicists. The dialogue concerns the relation between quantum field theory and the differential topology of four-dimensional (4d) manifolds. A small sample of papers in this subject is \cite{Belavin:1975fg, Witten:1988ze, Donaldson90, Witten:1994cg, Vafa:1994tf, Witten:1995gf, Nekrasov:1996cz, Moore:1997pc, LoNeSha, Marino:1998bm, Marino:1998rg, Ne, Dorey:2002ik, Nakajima:2003uh, Nakajima:2005fg, Gottsche:2006tn, Korpas:2019ava, Korpas:2019cwg, Manschot:2019pog, Manschot:2021qqe, Manschot:2023rdh, Moore:2024vsd}.
The physics of instantons, BPS equations, and (topologically twisted) supersymmetric gauge theories has led to the discovery of new four-manifold invariants, while the mathematical study of the invariants and relevant moduli spaces has stimulated many developments in physics. This article considers ``K-theoretic'' generalizations of Donaldson invariants. Such generalizations are related to the indices of Dirac operators on moduli spaces of instantons. 

In this paper $X$ denotes a compact, smooth four-manifold without boundary. For simplicity, we will take $X$ to be simply connected. We also restrict our attention to those $X$ that admit an almost complex structure. This is equivalent to the restriction that $b_2^+:=b_2^+(X)$ is odd.  
The physical theory which computes the K-theoretic Donaldson invariants is five-dimensional (5d) $\CN=1$ super Yang-Mills theory (SYM) formulated on $X\times S^1$, as proposed by Nekrasov some time ago
\cite{Nekrasov:1996cz}.  
The 5d theory on $X\times S^1$ can be partially topologically twisted, essentially extending the standard Donaldson-Witten twist \cite{Witten:1988ze} from four dimensions. 
A more precise description of the twisting is given in Sec.  \ref{sec:TopTwist}. 
The partially twisted field theory is not a 5d topological field theory. It depends on the metric and spin structure on $S^1$. 
We will focus on the computation of the partition function as a function of a dimensionless measure of the circumference of the circle, denoted $\CR$, and defined in Eq. \eqref{eq:CR=RLambda} below. The partition function is a power series in $\CR$ with coefficients given by the $L^2$-index of a suitable Dirac operator on the moduli space of $\mathrm{SU}(2)$ instantons on $X$. See Eqs. \eqref{eq:Zee-QM-1} and \eqref{eq:WittenIndex} below. 
We will demonstrate that, under proper identifications (spelled out in Sec.  \ref{sec:ThePartitionFunction}), the partition function nicely reproduces the results of G\"ottsche, Kool, Nakajima, Yoshioka and others on generating functions for holomorphic Euler characteristics of moduli spaces of sheaves \cite{Gottsche:2006bm, Gottsche:2019vbi, Gottsche:2020ass}. 
Whereas these previous results apply to the case that $X$ is a projective algebraic surface, our results should extend to the broader class of four-manifolds described above.

An important aspect of the 5d gauge theory is the $\mathrm{U}(1)$ global symmetry whose current is the Pontryagin density \cite{Seiberg:1996bd}. We will denote this $\mathrm{U}(1)$ group by $\mathrm{U}(1)^{(I)}$. See Eq. \eqref{eq:InstCurrent} below. The electric coupling of a background $\mathrm{U}(1)^{(I)}$ gauge field  to the $\mathrm{U}(1)^{(I)}$ current leads to a term in the action of the form 
\be
\label{eq:mixedCSintro}
S_{\rm mixed}=\frac{i}{8\pi^2}\int_{X\times S^1} A^{(I)}\wedge {\rm Tr}\left( F \wedge F \right)+\cdots,
\ee 
where $A^{(I)}$ is a background connection on a line bundle $L^{(I)}$ with structure group $\mathrm{U}(1)^{(I)}$ 
and $F$ is the field strength of the  dynamical $\mathrm{SU}(2)$ gauge field. When both $A^{(I)}$ and $A$ are topologically nontrivial this term is best defined using methods of differential cohomology. It is also related to the mixed Chern-Simons terms discussed in \cite{Losev:1995cr, LoNeSha, Baulieu:1997nj}. The supersymmetric completion of this term leads to the entire action of the 5d SYM coupled to a $\mathrm{U}(1)^{(I)}$ background vector multiplet. See Eq. \eqref{mixed CS action} below. The partition function depends on the line bundle $L^{(I)}$ with connection $A^{(I)}$ only through $\bfn_I:=c_1(L^{(I)})$ modulo torsion. Moreover, since all fields in the 5d vector multiplet are in the adjoint representation of the $\mathrm{SU}(2)$ gauge group, we can formulate the path integral when the gauge bundle $P\to X$ has structure group $\mathrm{SO}(3)$ and does not lift to an $\mathrm{SU}(2)$-bundle. The obstruction to lifting is the second Stiefel-Whitney class $w_2(P)$, often referred to as an 't Hooft flux. We generally denote an integral lift of the 't Hooft flux by $2\bfmu$. Of course, the lift is only defined modulo $2$ and our results are independent of the choice of lift. 
Altogether, we will study the partition function as a function not only of $\CR$ but also as a function of $\bfmu$ and $\bfn_I$.  

As in the case of standard Donaldson invariants, when $b_2^+=1$ the theory is not quite topological on $X$ but does depend on the metric through a choice of period point $J\in H^2(X,\IR)$ with $*J=J$ and $J^2=1$ (together with a choice of root of this equation). 
Thus, finally, the central object of study in this paper is the partition function of the partially topologically twisted theory on $X\times S^1$, denoted $Z^J_{\bfmu,\bfn_I}(\CR)$. 
The dependence on $J$ is piecewise-linear, and we derive a wall-crossing formula for $Z^J_{\bfmu,\bfn_I}(\CR)$ in Eq. \eqref{eq:Physical-WC} below and present an alternative derivation in Eq. \eqref{eq:WCLocalization} below. As we will explain in more detail below, when $\CR^4=1$ special features arise and the wall-crossing behavior changes. 

We present very explicit formulae for $Z^J_{\bfmu,\bfn_I}(\CR)$ in Sec.  \ref{Sec:DEvalU} for manifolds with $b_2^+=1$. For $b_2^+>1$, the $J$-dependence of the partition function disappears and the full function can be expressed in terms of Seiberg-Witten (SW) invariants.  
We give very general results for such manifolds in Eqs. \eqref{ZSWjmun} and \eqref{fullZSW} below. In cases where we can compare, our results agree with \cite{Gottsche:2006bm, Gottsche:2019vbi, Gottsche:2020ass}.\footnote{One exception is for instanton charge $k=1$ and $X=\mathbb{CP}^2$. This difference is discussed in some detail at the end of Sec.  \ref{sec:evaluate}.}
 
The 5d theory under study can be mathematically inconsistent due to global anomalies. An easy way to see that such anomalies can arise is to consider the reduction of the 5d theory on $X\times S^1$ to a supersymmetric quantum mechanics (SQM) on $S^1$, whose target space is the moduli space of instantons on $X$. This reduction is described in Sec.  \ref{subsec:redu2SQM} below. It is well known that SQM is anomalous, and hence mathematically inconsistent, if the target space is odd-dimensional, or if it is even-dimensional but not spin. In the latter case, the path integral over fermions gives a Pfaffian whose sign cannot be consistently defined.
(A general formula for the anomaly in any SQM is given in \cite[Proposition 5.8]{Freed:2014iua}.) 
In our case, the moduli space of instantons is not spin when $B(w_2(P), w_2(X)) \not= 0$. Here $w_2(X)$ is the second Stiefel-Whitney class of the tangent bundle $TX$, and $B$ is the intersection form on cohomology. 
A general formula for the first and second Stiefel-Whitney classes of moduli spaces of instantons for gauge theory with compact Lie group on a general four-manifold has recently been derived by D. Freed, M. Hopkins and the third named author \cite{FHMW}, and our assertion above is a corollary of that general result.
\footnote{GM thanks E. Witten for several very important discussions at the beginning of that project.}
A general study of global anomalies in 5d SYM remains to be done. A general formula for anomalies of theories with fermions can be found in \cite[Conjecture  9.70]{Freed:2016rqq}, but the equation needs some unpacking to be useful to physicists. 

It turns out that the anomaly in the effective SQM can be canceled by coupling the 5d SYM to a suitable background line bundle with connection $L^{(I)}$. When the 't Hooft flux $w_2(P)$ is non-vanishing, the exponentiated term \eqref{eq:mixedCSintro} can be anomalous. (This was independently observed in \cite{BenettiGenolini:2020doj, Apruzzi:2021nmk}.) 
On the other hand, in the reduction along $X$ via collective coordinates to an SQM the term \eqref{eq:mixedCSintro} induces a ``line bundle'' $\mathcal{L}^{(I)}$ on the moduli space of instantons. See Sec.  \ref{subsec:ChernSimonsLineBundle} below. The scare quotes we have just used alert us to the fact that when Eq. \eqref{eq:mixedCSintro} has a global anomaly, the line bundle $\mathcal{L}^{(I)}$ is not quite defined --- only its square is well defined. Indeed, when $X$ admits an almost complex structure, the moduli space of instantons is Spin$^c$ and we propose that the product $S^+\otimes \CL^{(I)}$ is well defined, even though the factors are not separately well defined. Here $S^+$ is the spin bundle on the moduli space of instantons. We argue that the anomaly cancellation condition is
\be
\label{AnomCanCond}
B\left(w_2(P) , c_1\!\left(L^{(I)}\right) + w_2(X)\right)=0 \mod 2.
\ee
When Eq. \eqref{AnomCanCond} is satisfied, the SQM is anomaly-free. It is the reduction of the theory to SQM with target space the moduli space of instantons on $X$ that leads to the interpretation \eqref{eq:Zee-QM-1} and \eqref{eq:WittenIndex} of $Z^J_{\bfmu,\bfn_I}(\CR)$ as a generating function of $L^2$-indices of Dirac operators on the moduli space of instantons. 

There is an interesting interplay of the anomalies we have just discussed with a one-form symmetry $\mathbb{Z}_2^{(1)}$
\footnote{This is also known to an older generation as the center symmetry of the Polyakov loop.}
of the 5d theory on $X\times S^1$. This symmetry flips the sign of all Wilson line defects in the fundamental representation of $\mathrm{SU}(2)$. The order parameter of the theory on $\IR^4 \times S^1$ is the vacuum expectation value (vev), denoted $U$, of a supersymmetric Wilson line \cite{Nekrasov:1996cz}. See Eqs. \eqref{DefWF} and \eqref{eq:Define-U} below. The low-energy effective theory (LEET) has a mixed anomaly between the one-form symmetry and the $\mathrm{U}(1)^{(I)}$ symmetry. This is expressed in terms of the behavior of the measure of the $U$-plane integral under the one-form symmetry given in Eq. \eqref{OmegT} below. 

We approach the explicit evaluation of the path integral $Z^J_{\bfmu,\bfn_I}(\CR)$ in two very different ways. The second method is the subject of Sec.  \ref{Sec:SusyLoc} and will be discussed in this introductory section below. The first method uses an integral over the Coulomb branch of the 4d Kaluza-Klein (KK) reduced theory, following the general framework developed for 4d theories in \cite{Moore:1997pc, Marino:1997gj, Marino:1998bm, Marino:1998rg, Manschot:2019pog, Manschot:2021qqe, Aspman:2022sfj, Aspman:2023ate}. When $b_2^+=1$, the path integral of the partially twisted theory can be derived from the study of the ``Coulomb branch integral'' or, more specifically, ``$U$-plane integral''. This integral is taken over the zero modes on the Coulomb branch of the 4d LEET on $\IR^4$, with a measure computable from the low-energy effective action (LEEA) of the theory.  However, some aspects of this effective 4d theory present novel and nontrivial difficulties not encountered in previous studies of twisted 4d theories: 
\begin{enumerate}
\item First, the SW geometry is much more complicated. We review what is known about the SW solution in Sec.  \ref{sec:5dR4S1} through \ref{SecFundDom} below. One salient feature is that the SW differential, as usually presented, is multi-valued, and one must work equivariantly on a cyclic covering of the SW curve. In fact, in our view, the analog of special K\"ahler geometry for 5d KK reduced supersymmetric gauge theories has \emph{not} been properly explained in the literature. 
\item Second, the consideration of multi-valued couplings introduces numerous order-of-limits issues. Two key limits, which do \emph{not} commute, are the weak-coupling limit and the limit in which the theory becomes effectively 4d or 5d. We will return to this issue later. 
\item Third, a very important upshot of Sec.  \ref{subsubsec:MoularParametrization} is that the Coulomb branch is a nontrivial double-cover of a modular curve for $\Gamma^0(4)$. This is explained in some detail in Sec.  \ref{subsubsec:MoularParametrization} and Sec.  \ref{SecFundDom} below. There is a strong analogy with the modular parametrization of the Coulomb branch of $\CN=2$ $\mathrm{SU}(2)$ theories with fundamental flavors, as described in \cite{Aspman:2021vhs}. In our case, the deck transformation of the double-covering map is precisely the action of the one-form symmetry: $U\to - U$. 
\end{enumerate}
  
Sec.  \ref{Sec:Uplane} gives an explicit description of the $U$-plane integral. The key formula is Eq. \eqref{UplaneintDef}  (and Eq. \eqref{DefUplaneIntnIK} when the KK flux is turned on). The various factors in the measure are defined in Eqs. \eqref{nuR}, \eqref{Ctauv}, \eqref{DefPsi} and \eqref{vsmallR}. The integration region $\CF_R$ is described in Sec.  \ref{SecFundDom}. More conceptually, the integral is over the $U$-plane, which is a double-cover of a modular curve for $\Gamma^0(4)$. Thus, the path integral can be expressed as a sum of two contributions exchanged by the $\mathbb{Z}_2$ action. The magnitude of the two contributions is identical, and if Eq.  (\ref{AnomCanCond}) is not satisfied, the path integral vanishes since the two contributions cancel. This is a manifestation of the mixed anomaly between the one-form symmetry and the $\mathrm{U}(1)^{(I)}$ symmetry. The vanishing is very similar in spirit to Witten's original description of the global anomaly in $\mathrm{SU}(2)$ gauge theories in four dimensions based on $\pi_4(\mathrm{SU}(2))=\mathbb{Z}_2$ \cite{witten19822}.

After determining the domain of integration, the next task is to demonstrate that the measure is well defined on this domain. This involves a nontrivial computation and is checked in Sec.  \ref{IntMono} based on the monodromy behavior of the relevant couplings derived in Sec.  \ref{Monos}. 

Once a well-defined measure is established, we proceed to evaluate the integral. Although the $U$-plane integral appears to be hopelessly divergent at first glance, it can nevertheless be assigned a meaningful mathematical definition. Because of the order-of-limits issues mentioned previously, one can in fact give it two very different definitions. The situation is quite reminiscent of other subtleties in the compactification of supersymmetric theories on $S^1$ that have appeared in the past \cite{Seiberg:1996nz, Marino:1998eg, Aharony:2013dha, Hwang:2017nop, Hwang:2018riu, Jia:2021ikh, Closset:2022vjj}. The 4d limit ($\CR\to 0$) of the theory can be obtained by taking a scaling limit of physical quantities around the region $U=2$ or its image under the one-form symmetry, $U=-2$. Expressing this scaling limit in the modular parametrization leads to the definition \eqref{PhiIntDef}. In practice, this requires that various factors in the $U$-plane measure should be first expanded as an expansion in series in $\CR$ around $\CR=0$, and then the coefficients of the integrand in this expansion should be evaluated by integrating over $\tau$, using regularization techniques familiar from previous work on Coulomb branch integrals \cite{Korpas:2019ava}. A second definition, in which this order of operations is reversed, yields results inconsistent with the mathematical findings of \cite{Gottsche:2006bm, Gottsche:2019vbi, Gottsche:2020ass}. This crucial point is discussed further in Sec.  \ref{subsubsec:WhichCorrect} below, and deserves a deeper understanding.

From the explicit $U$-plane measure we can study wall-crossing as a function of the period point $J$. The variation of the $U$-plane integral with respect to $J$ can be written as a total derivative in a natural way using the theory of mock modular and mock Jacobi forms \cite{Korpas:2019ava,Manschot:2021qqe}. Wall-crossing receives contributions from both weak- and strong-coupling regimes. The weak-coupling regime leads to results that agree with those in \cite{Gottsche:2006bm, Gottsche:2019vbi, Gottsche:2020ass}, whereas the strong-coupling wall-crossing, which occurs at walls associated with Spin$^c$ structures, is canceled by another contribution to the path integral associated with the LEET valid in a scaling region around the strong-coupling singularities. This contribution is proportional to SW invariants. In fact, the couplings in this LEET can most easily be deduced from the requirement that they cancel the strong-coupling wall-crossing of the $U$-plane integral. This method was first used in \cite{Moore:1997pc}. These couplings in the LEET in the scaling region around the strong-coupling singularities are closely related to universal functions appearing in the work of \cite{Gottsche:2019vbi}. We adopt the method of \cite{Moore:1997pc} to determine the partition function for manifolds with $b_2^+>1$ in Sec.  \ref{sec:SWConts}. 

The 5d SYM is not a ultraviolet (UV) complete theory. However, there are UV completions that involve 5d superconformal fixed points \cite{Seiberg:1996bd,Intriligator:1997pq}. One such theory, the $E_1$ fixed point, can be accessed from the 4d KK expressions by viewing the partition function not as a series in $\CR$ around $\CR=0$, but as a well-defined function of $\CR$ and then studying the limit $\CR^4 \to 1$. A partial justification of this claim follows from Eq. \eqref{CR4}, which shows that $\CR$ should be a phase. The specific choice of phase is determined by our explicit results. In Sec.  \ref{Sec:E1Uplane}, we examine the behavior of the $U$-plane integral in this limit and find some intriguing results. First, there is another issue of order-of-limits: the limit of the $U$-plane integral as $\CR^4 \to 1$ is not the same as taking the limit of the integrand and then doing the integral. Second, the wall-crossing behavior at the strong-coupling singularities involves generalizations of the SW invariants, and the wall-crossing walls do \emph{not} correspond to Spin$^c$ structures. Although these issues present rich avenues for deeper investigation, their full exploration lies beyond the scope of this paper. 

As mentioned previously, in Sec.  \ref{Sec:SusyLoc} we approach the study of $Z^J_{\bfmu,\bfn_I}(\CR)$ from a second, entirely different, point of view. Here we restrict our attention to toric surfaces. The $\IC^* \times \IC^*$ action allows us to introduce a refined partition function which is a function of the equivariant parameters $\epsilon_1$, $\epsilon_2$ \cite{Moore:1997dj, Lossev:1997bz, Moore:1998et, Ne}, and of a lift of the background flux $\bfn_I$ to equivariant cohomology $\mathfrak{p}^{(I)}$. See Eqs. \eqref{eq:Zee-QM-Equiv} and \eqref{eq:nfrakp}. This function serves as a generating function for character-valued indices of Dirac-like operators, as in Eq. \eqref{eq:dChar}. 
The extra symmetry allows one to localize the path integral directly to an integral whose integrand is related to the Nekrasov partition function. One may attempt to localize directly to solutions of the BPS equations. This involves a reduction to the abelian theory with BPS equations described in Sec.  \ref{subsubsec:BPS configurations}. The resulting expression is extremely delicate. It turns out that localization directly to the BPS locus is too singular. Just as the $U$-plane integral is \emph{not} an integral over the exact solutions of the BPS equations and one must localize to a slightly larger space of fields by incorporating non-BPS zero modes, in Sec.  \ref{Sec:SusyLoc} we introduce a zero mode, denoted $h$, for the auxiliary field. See Eqs. \eqref{eq:DOO-auxfield} and \eqref{def: non-bps-h} below. 
We claim that the partition function can be written as a finite-dimensional integral \eqref{partition ftn zero}. Similar expressions have been successfully applied in lower-dimensional analogues of twisted indices preserving the one-dimensional (1d) $\CN=(0,2)$ superalgebra, as seen, for example, in \cite{Hori:2014tda,Benini:2013nda,Benini:2015noa,Closset:2015rna}. 
The finite-dimensional integral \eqref{partition ftn zero} is conceptually similar to the $U$-plane integral. It equates the entire path integral with a finite-dimensional integral over the Coulomb branch, with an integrand derived from the LEET. However, Eq. \eqref{partition ftn zero}   differs substantially from the $U$-plane integral in several important respects. The integrand involves a function defined in Eq. \eqref{eq:g-full-defined}, obtained by integrating out nonzero modes. When restricted to the BPS locus, it can be expressed as a product of Nekrasov partition functions, as in Eq. \eqref{ga}. As with the $U$-plane integral, the integral requires careful treatment due to the singularities of the integrand, which are discussed in Sec.  \ref{subsubsec:g-function-singularities-contours}. The expression can be further reduced to a sum of contour integrals, as in Eq. \eqref{zero mode contour}.  

Using the localization formula \eqref{partition ftn zero}, we rederive the wall-crossing expressions given in Eq. \eqref{equivariant wall crossing}. In the non-equivariant limit, we recover the wall-crossing formula derived from the $U$-plane integral, as demonstrated in Sec.  \ref{sec: loc to U}. 

In Sec.  \ref{sec: puzzle}, we describe a significant puzzle that arises from our analysis. Focusing on the case of $X= \IP^2$ for simplicity and starting from Eq. \eqref{partition ftn zero}, we find that if we define the integrals in Sec.  \ref{subsubsec:g-function-singularities-contours} in a way that would appear natural, we are led to a formula for the partition function, Eq. \eqref{JK contour}, which cannot be correct. For example, it is inconsistent with the expected general expression \eqref{eq:Zee-QM-Equiv}. After some nontrivial manipulations, including nontrivial identities for the Nekrasov partition,  described in App. \ref{app:NekPartProps}, we show that Eq. \eqref{JK contour} is equivalent to Eq. \eqref{eq:SumAllBundles}, which can be given an interpretation as a sum over stable, semi-stable, and unstable bundles. If we restrict to the contributions from semi-stable bundles, which amounts to a restriction on contours computing the residues of the instanton partition function, we arrive at the closely related and similar expression, Eq. \eqref{ZesP2}. The expression \eqref{ZesP2} is very likely to be correct, since it passes several consistency checks. Among other things, Eq. \eqref{ZesP2} yields the correct non-equivariant limit.
In Sec.  \ref{sec: puzzle}, we list the known weaknesses in the chain of reasoning leading to Eq. \eqref{JK contour}. We note that expressions very similar to Eq. \eqref{ZesP2} have appeared in numerous places in the literature, but in our view, such expressions have not been properly derived from the path integral viewpoint. 

In Sec.  \ref{sec:equivariantKtheoreticP2}, we evaluate Eq. \eqref{ZesP2} explicitly to produce Eq. \eqref{ZesnI}, which can be expanded numerically. Some sample expansions to low orders are given in Tables \ref{P2eqKhalf} and \ref{P2eqK0}. These are new results and it would be nice to test them using different methods. 

Besides a background flux for the $\mathrm{U}(1)^{(I)}$ symmetry, one can also consider a non-vanishing background flux for the $\mathrm{U}(1)$ symmetry associated with translations along the $S^1$ factor in $X\times S^1$. We denote this $\mathrm{U}(1)$ group by $\mathrm{U}(1)^{(K)}$. It has long been known \cite{Gross:1983hb}
that this is equivalent to considering the partition function on a nontrivial circle bundle over $X$. Some aspects of the topological twisting in this setting, together with certain holomorphic objects entering the measure (and their equivariant extension on toric manifolds), have been discussed in \cite{Closset:2022vjj}. We believe that the methods presented in our paper could be used to give complete and explicit results for the partition function in this generalized case. While we leave a detailed evaluation to future work, we comment on the generalization to non-vanishing KK flux at several points in this article. In particular, the formula for the measure of the $U$-plane integral is given in Eq. \eqref{DefUplaneIntnIK} below. We perform a nontrivial check confirming that this measure is single-valued on the $U$-plane in Sec.  \ref{IntMono}. 

Our results have some bearing on the important but poorly understood issue of the quantum status of ``instanton particles''. Such particles exist classically: A Yang-Mills instanton in four dimensions explicitly represents some kind of solitonic particle in five dimensions. The quantum status of these particles is less clear because standard collective coordinate quantization associates them with $L^2$-waveforms on the moduli space of instantons on $\mathbb{R}^4$, but such waveforms do not exist. Nevertheless, the existence of quantum instanton particles on Minkowski space $\mathbb{M}^{1,4}$ is crucial to a number of aspects of string/M-theory duality. For discussions, see, for example 
\cite{Seiberg:1996bd, Aharony:1997an, Douglas:2010iu, Papageorgakis:2014foa}. The expressions \eqref{eq:Zee-QM-1} and \eqref{eq:WittenIndex} below show that our partition functions are meaningful BPS state counting functions when the spatial manifold $\IR^4$ is compactified to $X$. We may take this as evidence that the relevant quantum particles do exist, and our counting functions might conceivably provide some useful information about them. 

We have already pointed out several important open issues in our summary above. 
Sketches of other potentially fruitful directions for continuing this research are given in Sec.  \ref{Sec:ConOUt}, among which the most challenging direction, but also the direction with the greatest potential impact, is the generalization of our considerations to six-dimensional (6d) supersymmetric field theories. 
 
A number of appendices supplement the text. 
App. \ref{app:conventions} spells out some basic field-theory conventions. 
App. \ref{App:MForms} summarizes our definitions of some of the basic modular objects that we use. 
App. \ref{app4dN=2} recalls some basic formulae from the 4d SW geometry. 
App. \ref{app:ExactU} recalls some exact formulae for expansions of the Coulomb branch parameter $U$ as functions of $\CR$ and the special K\"ahler coordinate.
App. \ref{app:Polylogs} summarizes some basic facts about polylogarithms that are important to our discussion of monodromy, SW geometry, and properties of the Nekrasov partition function. 
App. \ref{ResultsGNY} addresses an important point: Our wall-crossing formula and explicit evaluation of the partition function for $\IP^2$ are not obviously consistent with the results of \cite{Gottsche:2006bm, Gottsche:2019vbi, Gottsche:2020ass}. Some nontrivial manipulations with modular functions are needed to establish the equivalence. Details of these manipulations can be found in this appendix.
App. \ref{app:ExpCusps} gives detailed expansions of a crucial coupling near strong-coupling cusps. These are needed in the derivation of the wall-crossing formulae and hence in the determination of the partition functions for manifolds with $b_2^+>1$. 
App. \ref{app:MonopoleEquations} discusses some aspects of the generalized SW equations that appear in our discussion of the $E_1$ theory. 
App. \ref{sec:Toric Geometry} and App. \ref{app:Loc Hilbert Schemes} recall facts and identities important in the derivation of the refined partition function on toric manifolds. In particular, App.  \ref{app:Loc Hilbert Schemes} derives the result that the holomorphic part of the integrand of Eq. \eqref{partition ftn zero} can be expressed as a product of Nekrasov partition functions. 
Moreover, App. \ref{app:NekPartProps} establishes some new and remarkable properties of the 5d Nekrasov partition function which turn out to be essential to the discussion of Sec.  \ref{Sec:SusyLoc} and the puzzle discussed in Sec.  \ref{sec: puzzle}.   
App. \ref{app:SQMinOmega} demonstrates how the reduction to SQM is modified by the 
$\Omega$-deformation and justifies the important result \eqref{eq:Zee-QM-Equiv} and  \eqref{eq:dChar}.
Finally, we recall the definition of the Donaldson $\mu$-map, denoted $\mu_D$ in this paper, in App. \ref{app:DonaldsonMuMap}.

%%%%%%%%%%%%%%%%%%%%%%%%%%%%%%%%%%%%%%%%%%%%%%

\acknowledgments
We thank Cyril Closset, Dan Freed, Elias Furrer, Pietro Genolini, Lothar G\"ottsche, Mike Hopkins, Martijn Kool, Nikita Nekrasov, Du Pei, Samson Shatashvili, Nathan Seiberg, Yuji Tachikawa, Edward Witten, and Xingyang Yu for valuable comments and discussions. We particularly thank Cyril Closset and Elias Furrer for detailed comments on an earlier draft of the paper. 
The work of H.K. is supported by the National Research Foundation of Korea (NRF) grant NRF2023R1A2C1004965 and RS-2024-00405629, and also by POSCO Science Fellowship of POSCO TJPark Foundation. 
The research of J.M. was supported in part by Laureate Award 15175 “Modularity in Quantum Field Theory and Gravity” of the Irish Research Council, and the Ambrose Monell Foundation. J.M. thanks the Institute for Advanced Study for hospitality during part of this project. 
The work of G.M. was supported by the US Department of Energy under grant DE-SC0010008. G.M. thanks the Institute for Advanced Study for hospitality while much of this manuscript was written. In particular, G.M. was supported by the IBM Einstein Fellow Fund.   
The research of X.Z. was supported by National Natural Science Foundation of China under Grant No. 12475073, and by National Science Foundation of China under Grant No. 12347103.

\bigskip
\centerline{\emph{No part of this paper was written by AI.}}

%%%%%%%%%%%%%%%%%%%%%%%%%%%%%%%%%%%%%%%%%%%%%%

\section{The Untwisted Theory}\label{sec:UntwistedTheory}

In this section, we collect various basic aspects of 5d $\CN=1$ SYM defined on a smooth, oriented, spin five-manifold $X_5$. After reviewing the 5d $\CN=1$ superalgebra and the supersymmetric action on $X_5$, we discuss the LEET on $X\times S^1$, with $X$ a compact, smooth, oriented four-manifold. Here an orientation on $X$ determines one on $X \times S^1$.  We will initially assume $X$ to be spin, later extending the analysis to non-spin cases. 

\subsection{Five-dimensional $\CN=1$ Super Yang-Mills Theory On A Smooth Oriented Riemannian Five-Manifold $X_5$}

We begin with the theory defined on flat space $\mathbb{R}^{1,4}$. 
The field content of 5d $\CN=1$ SYM comprises a vector multiplet $V$, which consists of a gauge field $A_m$, a real scalar $\sigma$, symplectic Majorana spinors $\lambda_A$, and a bosonic auxiliary field $D_{AB}=D_{BA}$, 
\be
V = \left( A_m, \sigma, \lambda_A, D_{AB}\right) , \label{eq:vectormultiplet}
\ee
where $m,n=1,\cdots, 5$ are spacetime indices, $A,B=1,2$ are $\mathrm{SU}(2)_\mathrm{R}$ indices. All fields transform in the adjoint representation of the gauge group $G$, and transform under $\mathrm{SO}(5)\times \mathrm{SU}(2)_\mathrm{R}$ as
\be
(\mathbf{5},\mathbf{1}), \quad (\mathbf{1},\mathbf{1}), \quad
(\mathbf{4},\mathbf{2}), \quad (\mathbf{1},\mathbf{3}).
\ee

After Wick rotation to Euclidean signature, the action of the theory on a smooth, Riemannian, spin five-manifold $X_5$ is
\be\label{YM action}
S_{\text{SYM}}[V] = \int_{X_5} \mathrm{d}^5 x\,  \sqrt{g}\CL_{\text{SYM}},
\ee
where $g={\rm det}(g_{mn})$, and 
\be\label{eq:5d-FlatSpace} 
\CL_{\text{SYM}} =  \frac{1}{g^2_{5d}} \text{tr}\left(\frac{1}{2} F_{mn} F^{mn}  + D^m\sigma D_m\sigma +\frac{1}{2} D^{AB}D_{AB} +  i\lambda^A \Gamma^m D_m \lambda_A +i\lambda^A \left[\sigma,\lambda_A \right] \right).
\ee
Here the gauge covariant derivative $D_m$ and the field strength $F_{mn}$ are defined to be
\be
D_m = \nabla_m -i A_m, \quad F_{mn} = \partial_m A_n - \partial_n A_m -i \left[A_m, A_n \right].
\ee
The covariant derivative acting on $\lambda^A$ includes the spin connection term, and after performing a partial topological twist, will also incorporate a coupling to the background $\mathrm{SU}(2)_\mathrm{R}$ gauge field.  
For $G=\mathrm{SU}(N)$, $\text{tr}$ denotes the trace in the $N$ dimensional defining representation. To ensure the convergence of the Euclidean path integral 
\be
\int[\mathrm{d}V]\, e^{-S_\text{SYM}[V]},
\ee
the following reality conditions on the bosonic fields are imposed,
\be
A^\dagger_m = A_m, \quad \sigma^\dagger = \sigma, \quad \left({D^{AB}}\right)^\dagger = -\e^{AA'}\e^{BB'}D_{A'B'}.
\ee

On $X_5=\mathbb{R}^5$, the theory has off-shell Poincar\'e supersymmetry:
\footnote{The difference between our formulas and those in the literature (after a suitable change of variables) is due to the difference between commuting versus anti-commuting spinor $\xi_A$. }
\begin{equation}\label{eq:5dSUSY-TMNS}
\begin{aligned}\delta A_m & =i \xi_A \Gamma_m \lambda^A, \\
\delta \sigma & =-\xi_A \lambda^A, \\
\delta \lambda_A & =\frac{1}{2} \Gamma^{m n} \xi_A F_{m n}+i \Gamma^m \xi_A D_m \sigma-i D_{A B} \xi^B, \\
\delta D^{A B} & =\xi^A \Gamma^m D_m \lambda^B+\xi^B \Gamma^m D_m \lambda^A+\xi^A\left[\sigma, \lambda^B\right]+\xi^B\left[\sigma, \lambda^A\right].
\end{aligned}
\end{equation}
Here we define the supersymmetry variation as $\delta = -i \xi_A \mathbf{Q}^A$, where the supercharges $\mathbf{Q}^A$ $(A=1,2)$ are two 5d Dirac spinors. For a general five-manifold $X_5$, this Poincar\'e supersymmetry is broken. 

As pointed out by Seiberg \cite{Seiberg:1996bd}, the theory also possesses a global $\mathrm{U}(1)^{(I)}$ symmetry associated with the 4-form current
\be\label{eq:InstCurrent}
j^{(I)}= \frac{1}{8\pi^2}\text{tr}\left(F\wedge F\right).
\ee
For $G=\mathrm{SU}(N)$, this current is normalized so that it represents an integral cohomology class, the second Chern class of the $\mathrm{SU}(N)$-bundle. On Minkowski spacetime $\IR^{1,4}$, time-independent Yang-Mills instantons define solitonic particles, at least classically, which carry nontrivial $\mathrm{U}(1)^{(I)}$ charges. 
\footnote{\label{foot:quantissues} 
Semiclassical quantization of these solitons involves unresolved issues. See \cite{Douglas:2010iu, Papageorgakis:2014foa} for discussions. }
Fields in the vector multiplet are neutral under this symmetry.

It is often useful to promote couplings to background superfields \cite{Seiberg:1993vc}. Accordingly, we introduce a background vector multiplet for the $\mathrm{U}(1)^{(I)}$ symmetry,
\be\label{eq:U(1)-I-Supermultiplet}
V^{(I)} = \left(A_m^{(I)},\sigma^{(I)},\lambda_A^{(I)}, D_{AB}^{(I)}\right),
\ee
where $A^{(I)}$ is the gauge connection of a principal $\mathrm{U}(1)$-bundle $L^{(I)}$ over $X_5$. Crucially, the torsion-free part of the first Chern class of $L^{(I)}$, 
\be
\bfn_I := \left[\frac{F^{(I)}}{2\pi}\right] = \overline{c_1(L^{(I)})}  \in H^2(X_5,\mathbb{Z})/\mathrm{Tors},
\ee
will play an important role in subsequent computations.  
To keep the notation from becoming too heavy, we may sometimes drop the subscript $I$ from $\bfn_I$. 

For an oriented five-manifold $X_5$, we can introduce an action coupling the gauge group $G$ and the global symmetry group $\mathrm{U}(1)^{(I)}$,
\begin{align}
\label{mixed CS action}
    S_{\text{mixed}} &= \frac{i}{8\pi^2}\int_{X_5} A^{(I)}\wedge \text{tr}\left( F\wedge F\right) \nonumber  \\ 
    &\quad +\frac{1}{8\pi^2}\int_{X_5}  \mathrm{d}^5x\, \sqrt{g}\,\text{tr}\bigg[\frac{1}{2} {\lambda^{(I)A}} \Gamma^{mn} \lambda_A
    F_{mn} -\frac{1}{4} \lambda^A \Gamma^{mn} \lambda_A F^{(I)}_{mn} +\frac{i}{2} \lambda^A \lambda^B D^{(I)}_{AB} +i {\lambda^{(I)A}} \lambda^B  D_{AB} \nonumber \\
    &\quad + \sigma^{(I)}\bigg( \half F^{mn} F_{mn} +D^m \sigma D_m \sigma + \half D^{AB} D_{AB} +i\lambda^A \Gamma^m D_m \lambda_A + i \lambda^A  \left[\sigma,\lambda_A \right] \bigg)\nonumber \\
    & \quad + \sigma \left( F^{(I)}_{mn} F^{mn} + 2 D^m \sigma^{(I)} D_m \sigma + D^{(I)AB} D_{AB} +i \lambda^{(I)A} \Gamma^m D_m \lambda_A +i\lambda^A \Gamma^m D_m \lambda^{(I)}_A \right) \bigg].
\end{align}
For $X_5=\mathbb{R}^5$, this is invariant under the supersymmetry transformations
\eqref{eq:5dSUSY-TMNS} applied to both the dynamical and the background vector multiplets. 
Hence, Eq. \eqref{mixed CS action} gives the supersymmetric completion of the mixed Chern-Simons term in the first line.
We recover Eq. \eqref{eq:5d-FlatSpace} from Eq. \eqref{mixed CS action} by setting all fields in the background $\mathrm{U}(1)^{(I)}$ vector multiplet \eqref{eq:U(1)-I-Supermultiplet} to zero except for $\sigma^{(I)}$,
\be\label{eq:sigmaI-5dcoupling}
\sigma^{(I)} = -\frac{8\pi^2}{ g_{5d}^2}.
\ee

It is important to note that the mixed Chern-Simons term remains well defined even when both the background gauge connection $A^{(I)}$ and the dynamical gauge connection $A$ are topologically nontrivial. The connection $A^{(I)}$ defines an element in the differential cohomology group $\check H^2(X_5)$, while the connection $A$ defines a 3d Chern-Simons form which corresponds to an element in the differential cohomology group $\check H^4(X_5)$. Through the usual bilinear pairing of the cup product and pushforward along $X_5$, these produce a well-defined map $\check H^2(X_5) \times \check H^4(X_5) \rightarrow \check H^1(\text{pt}) \cong \IR/\IZ$. 
For our chosen normalizations, the exponentiated action $e^{-S}$ is unambiguous when the gauge bundle is an $\mathrm{SU}(N)$-bundle. However, for a $\mathrm{PSU}(N)$-bundle with nontrivial 't Hooft flux, it acquires a root-of-unity ambiguity.
The specific case of $\mathrm{PSU}(2)\cong \mathrm{SO}(3)$ and $X_5 = X_4 \times S^1$ is analyzed in detail in the derivation of Eq. \eqref{nI condition} below. In addition, we expect a global anomaly in the fermion determinant. These issues are examined in the context of collective coordinate reduction to the moduli space of instantons in Sec.  \ref{subsec:ChernSimonsLineBundle} and Sec.  \ref{subsec:AnomalyCancellation}. 

The 5d supersymmetric Chern-Simons action involving only the gauge group $G$ \cite{kim20125}, 
\bea\label{eq:ChernSimonsAction}
S^{\text{CS}} &= \frac{-i\kappa_{\text{CS}}}{24\pi^2} \int_{X_5} \text{tr} \left[ A \wedge F \wedge F + \frac{i}{2} A \wedge A \wedge A \wedge F - \frac{1}{10} A \wedge A \wedge A \wedge A \wedge A\right] \\
& \quad + \frac{\kappa_{\text{CS}}}{2\pi^2} \int_{X_5} \mathrm{d}^5x\,\sqrt{g}\, \text{tr} \left[\sigma \left(\frac12 F_{mn} F^{mn}  + D^m\sigma D_m\sigma +\frac12 D^{AB}D_{AB} +  i\lambda^A \Gamma^m D_m \lambda_A +i\lambda^A [\sigma,\lambda_A] \right)\right]\\
& \quad + \frac{\kappa_{\text{CS}}}{8\pi^2} \int_{X_5} \mathrm{d}^5x\,\sqrt{g} \text{tr} \left[ \lambda^A \Gamma^{mn} \lambda_A F_{mn}-2i \lambda^A \lambda^B D_{AB}\right],
\eea
is independently supersymmetric under the transformations \eqref{eq:5dSUSY-TMNS} for $X_5 = \mathbb{R}^5$. The coupling $\kappa_{\text{CS}}$ is a free parameter, subject to quantization constraints. In this paper, we will put $\kappa_{\text{CS}}=0$.

Since 5d gauge theories are non-renormalizable, they must be interpreted as effective field theories with a UV cutoff. Indeed, they emerge as low-energy effective field theories of nontrivial strongly-coupled UV fixed points of the RG flow where $g^2_{5d}\to \infty$ \cite{Seiberg:1996bd}. 
For rank-one gauge groups, although Eq.  \eqref{eq:ChernSimonsAction} vanishes identically, there are two UV fixed points: the $E_1$ and the $\tilde E_1$ theories \cite{Intriligator:1997pq,Douglas:1996xp}. They are distinguished by a topological term that significantly affects both the instanton partition function \cite{Tachikawa:2004ur,Bergman:2013ala} and the SW curve \cite{Gottsche:2006bm}.
\footnote{The topological term corresponds to an invariant for $\Omega_5^{\mathrm{Spin}}(B\mathrm{SO}(3))$ and is realized as an exponentiated $\eta$-invariant --- the partition function of an invertible TQFT known as Dai-Freed theory
\cite{Dai:1994kq, Monnier:2019ytc, Witten:2019bou}. 
It provides the bulk theory that renders $\mathrm{SU}(2)$ gauge theory with a single doublet non-anomalous. Similarly, for higher-rank groups, the ``Chern-Simons terms'' we introduce should be understood as $\eta$-invariants. The authors thank D. Freed, N. Seiberg, and Y. Tachikawa for very useful remarks clarifying these topological terms. }
 
In this work, we focus on the $E_1$ theory, which is the UV completion of the 5d $\mathrm{SU}(2)$ gauge theory with trivial discrete theta angle. Although it should be possible to extend our $U$-plane computations to the $\tilde{E}_1$ theory, the distinguishing invertible theory is a topological invariant of spin manifolds. 
Thus, the Coulomb branch integral for the $\tilde E_1$ theory would presumably require $X_5$ to be spin manifolds. 
Verification of this constraint would be valuable but falls outside the scope of this paper.

Furthermore, we restrict ourselves to the case $X_5 = X \times S^1$, where $X$ is a compact, oriented, smooth Riemannian four-manifold without boundary, and $X_5$ is equipped with a product metric. 
\footnote{The generalization to circle bundles over $X$ can be considered by turning on a KK flux.}
In this setup, we turn on not only a nonzero $\sigma^{(I)}$ but also a nontrivial holonomy for the background connection $A^{(I)}$ along $S^1$,
\be\label{eq:thetadef}
\theta = \oint_{S^1} A^{(I)}.
\ee
Upon dimensional reduction along $S^1$, $\theta$ becomes the 4d $\theta$-angle, coupling to the instanton charge via 
\be\label{eq:theta-couple}
\exp\left(-\frac{i \theta}{8\pi^2} \int_X  \text{tr} F\wedge F \right),
\ee
where $F$ is here the field strength of the gauge field in the 4d effective theory. 

In Sec.  \ref{sec:TopTwist}, we define a partial topological twist for $X_5 = X \times S^1$.  
Of particular interest are situations where the background $\mathrm{U}(1)^{(I)}$-bundle is the pullback of a nontrivial principal $\mathrm{U}(1)$-bundle over $X$.

\subsection{Five-dimensional $\CN=1$ $\mathrm{SU}(2)$ Super Yang-Mills Theory On $\mathbb{R}^4\times S^1$}
\label{sec:5dR4S1}

We now consider 5d $E_1$ theory (perturbed by a relevant operator) on $\mathbb{R}^4\times S^1$, where the circumference of $S^1$ is $R$.
At long distances, this is described by 5d $\CN=1$ SYM. We specialize to $G=\mathrm{SU}(2)$. At distances large compared to the radius $R$ the theory is described by an effective 4d $\CN=2$ $\mathrm{SU}(2)$ theory on $\mathbb{R}^4$.  For the purposes of this paper, the exact solution of the LEET \cite{Nekrasov:1996cz} is particularly important. It takes into account the infinite tower of KK states of the 5d SYM, instanton particles, and other nonperturbative effects. 

\subsubsection{Moduli Space Of Vacua}

In conventional 4d $\CN=2$ SYM, the space of classical vacua consists of gauge-inequivalent configurations with trivial gauge connections and a constant adjoint scalar field $\phi$ satisfying $[\phi,\bar\phi]=0$. Up to gauge transformation, we can take
\begin{equation}
\phi = \left(\begin{array}{cc} a & \\ & -a \end{array}\right),
\end{equation}
with Weyl group action $a\to -a$. 
Thus, the classical moduli space of vacua is parameterized by the complex coordinate $u_{\mathrm{cl}} = a^2$. 
Quantum mechanically, the moduli space is parametrized by the vev 
\be\label{4du}
u=\frac{1}{2} \left\langle \tr \phi^2 \right\rangle,
\ee
and the LEET is a 4d $\CN=2$ $\mathrm{U}(1)$ gauge theory with the effective prepotential determined by the SW geometry \cite{Seiberg:1994rs}.

When we consider 5d SYM on $\IR^4 \times S^1$ at length scales much larger than $g^2_{5d}$ and the circumference $R$ of the circle, the low-energy physics is described by an effective 4d $\CN=2$ field theory. 
In this LEET the classical vacua correspond to trivial gauge connections $A_{\mu}$ $(\mu=1,\cdots, 4)$, and constant fields $\sigma$ and $A_5$ that can be simultaneously diagonalized by a gauge transformation,
\be \label{asigmaa5}
\sigma + i a_5 = \left(\begin{array}{cc} a & \\ & -a \end{array}\right),
\ee 
where $a_5$ is the holonomy of the 5d gauge field around the circle:
\be\label{asigmaa5-2}
a_5 := \frac{1}{R}\oint_{S^1} A_5 \mathrm{d}x^5.
\ee
It is useful to define the dimensionless variable
\be 
\mathfrak{a}:=Ra.
\ee
A crucial difference between the LEET from 5d SYM on a circle and conventional 4d $\CN=2$ SYM is the existence of extra residual discrete gauge transformations along $S^1$, which act on $\mathfrak{a}$ via shifts,
\be
\mathfrak{a}\rightarrow \mathfrak{a} + 2\pi i n,\quad n\in\mathbb{Z}.
\ee
Therefore, single-valued functions on the classical moduli space of vacua must depend on $e^{\mathfrak{a}}$ and also be invariant under the $\mathbb{Z}_2$ action of the Weyl group, $\mathfrak{a}\to -\mathfrak{a}$. 

To parametrize the quantum moduli space of vacua in the 5d KK theory, we introduce a supersymmetric Wilson loop wrapping $S^1$ in the fundamental representation,
\be
\label{DefWF}
W_F(p) := \text{tr}_F\,\text{P}\! \exp \left[\oint_{p\times S^1} \left(\sigma + i A_5 \right)\mathrm{d}x^5\right],
\ee
where $p\in \IR^4$. This is invariant under the supersymmetry \eqref{eq:Qbar-nonabelian} which survives the partial topological twist described below. 
The gauge-invariant order parameter is then
\be\label{eq:Define-U}
U:= \left\langle W_F(p) \right\rangle,
\ee
and is independent of $p$ by supersymmetry. 

The order parameter $U$ is a function of $\mathfrak{a}$ and $\CR^4$, where $\CR^4$ is a dimensionless parameter combining $g_{5d}$ and $\theta$,
\begin{equation} \label{CR4}
\mathcal{R}^4 := \exp\left(-\frac{8\pi^2 R}{g_{5d}^2} + i \theta \right).
\end{equation}
Notice that $\vert \CR\vert \leq 1$, with the 5d $E_1$ theory giving $\vert \CR\vert=1$.
Crucially, \eqref{CR4} does not specify $\CR$, i.e., the fourth root of the right-hand side. This ambiguity is related to the phase indeterminacy of $\Lambda$ in 4d $\CN=2$ $\mathrm{SU}(2)$ SYM, where $\Lambda^4$ is the 4d instanton counting parameter.
We will show later that changing the phase of $\Lambda$ by a fourth root of one changes the partition function by an overall phase. There are exact results in the literature, recalled in App. \ref{app:ExactU}, for $U$ as a function of $\mathfrak{a}$ and $\CR$. In an expansion around $\CR=0$ the first few terms are: 
\bea\label{eq:Exact-formula-U}
U(\fa,\CR)  &= e^{\fa}+ e^{-\fa} + \CR^4 \frac{e^{\fa}+e^{-\fa}}{(e^{\fa}-e^{-\fa})^2} +\CR^8 \frac{5(e^{\fa}+e^{-\fa})}{(e^{\fa}-e^{-\fa})^6} \\ 
& \quad + \CR^{12} \frac{(e^{\fa}+e^{-\fa})(7e^{-2\fa}+ 58 + 7 e^{2\fa})}{(e^{\fa}-e^{-\fa})^{10}} + \CO(\CR^{16}).
\eea
 
Similarly to (\ref{asigmaa5}), we define a local parameter $a^{(I)}$ for the background field of the global $\mathrm{U}(1)^{(I)}$ symmetry,
\be 
a^{(I)} :=\sigma^{(I)}+\frac{i}{R} \oint_{S^1} A^{(I)}_m \mathrm{d}x^m,
\ee 
which is related to $\CR^4$ via
\be 
\CR^4 = \exp\left[ R a^{(I)}\right].
\ee

We note that the 4d effective theory also possesses a distinguished $\mathrm{U}(1)^{(K)}$ KK symmetry. We denote the associated background vector multiplet by
\be
V^{(K)} = \left(A_\mu^{(K)},\phi^{(K)},\lambda_A^{(K)}, D_{AB}^{(K)}\right).
\ee
The vev of $\phi^{(K)}$ is denoted by $a^{(K)}$, which is identified with the KK momentum
\be
a^{(K)} = \frac{2\pi i}{R}.
\ee

\subsubsection{Low-Energy Effective Prepotential}

Due to 4d $\CN=2$ supersymmetry, the LEET on $\mathbb{R}^4$ is governed by a prepotential $\CF$. It can be computed by summing contributions from all KK modes and instantons
\cite{Nekrasov:1996cz, Lawrence:1997jr, Nakajima:2005fg,Gottsche:2006bm}, yielding a function of the local Coulomb branch coordinate $a$, the circumference $R$, and the dynamically generated scale $\Lambda$ of the 4d effective theory,
\footnote{Note that our convention for the prepotential $\CF$ follows  \cite{Nakajima:2003pg, Nakajima:2005fg}, but is different from the more commonly used convention in the physics literature.}
\be \label{prepotential}
\CF(a, R,\Lambda)= \CF^{\rm pert}(a, R,\Lambda) + 
\CF^{\rm inst}(a, R, \Lambda).
\ee
The perturbative part is given by 
\be\label{prepotential2}
\CF^{\rm pert}(a, R,\Lambda) = -4 a^2 \log(R\Lambda)+ \frac{2\left(\zeta(3)-\mathrm{Li}_3 (e^{-2 Ra}) \right)}{R^2}   - \frac{2\pi^2}{3R}a + 2\pi ia^2 +\frac{4 R }{3} a^3.
\ee
In terms of the dimensionless variable $\mathfrak{a}$, Eq. \eqref{prepotential2} is expressed as 
\be
R^2 \CF^{\rm pert}(\fa, R\Lambda) = -4 \fa^2 \log(R\Lambda)+ 2\left(\zeta(3)-\mathrm{Li}_3 (e^{-2\fa}) \right)   - \frac{2\pi^2}{3} \fa + 2\pi i \fa^2 +\frac{4 }{3}\fa^3.
\ee
Here $\Lambda$ serves as a UV cut-off of the one-loop determinants, and the trilogarithm $\text{Li}_3$ arises from the sum of the KK modes. 
Notice that $\text{Li}_3(e^{-2\fa})$ is single-valued when ${\rm Re}(\fa)>0$ but exhibits monodromy upon analytic continuation.
For more properties of the polylogarithm, see App. \ref{app:Polylogs}. 

The instanton contribution has the structure
\be\label{eq:StructureInstantonExp}
\mathcal{F}^{\text {inst }}(a, R,\Lambda)=\frac{\sinh ^2(Ra)}{R^2} \sum_{n=1}^{\infty} \mathcal{F}_n\left(\sinh ^2(Ra)\right)\left(\frac{\mathcal{R}}{\sinh (Ra)}\right)^{4 n},
\ee
or equivalently in dimensionless form
\be
R^2 \mathcal{F}^{\text {inst }}(\fa, \CR)= \sum_{n=1}^{\infty} \frac{\mathcal{F}_n\left(\sinh ^2(\mathfrak{a})\right)}{\sinh^{4n-2} (\mathfrak{a})}\mathcal{R}^{4 n}.
\ee
Here $\mathcal{F}_n(x) \in \mathbb{Q}\left[x\right]$ are polynomials with rational coefficients, and the first few terms are 
\be
\mathcal{F}_1(x) = -\frac{1}{2},\quad \mathcal{F}_2(x) = -\frac{5+x}{64},\quad \cdots
\ee
In fact, $\CR$ is a dimensionless circumference scale of the compactification, and is expressed in terms of $R$ and $\Lambda$ by a simple relation
\be\label{eq:CR=RLambda}
\CR = R\Lambda. 
\ee
The correctness of this relation will be tested by considering various limits. Note that the phase of $\CR^4$ is determined by Eq. \eqref{CR4}. The choice of a fourth root of $\CR^4$ is ambiguous by a fourth root of unity. Eq. \eqref{eq:CR=RLambda} gives an interpretation to that fourth root as determining the phase of the 4d scale $\Lambda$. 

\subsubsection{Four-dimensional And Five-dimensional limits}\label{4d5dlimits}

Let us analyze how the 5d KK theory reduces to familiar theories through distinct limits.

We start with the standard 4d limit:
\be 
\label{4dlimit}
R\to 0, \text{ while holding } a \text{ and } \Lambda \text{ fixed}.
\ee 
In dimensionless terms, this corresponds to $\CR \to 0$ and $\mathfrak{a}\to 0$, with the ratio $\mathfrak{a}/\CR = a/\Lambda$ fixed. Physically, since $\vert aR\vert \ll 1$, the $W$-boson mass ($\sim \left(\vert a\vert\right)$) becomes much smaller than the KK scale ($\sim R^{-1}$). Thus, the KK modes decouple, leaving the conventional 4d $\CN=2$ $\mathrm{SU}(2)$ SYM parametrized by $\Lambda$ and $a$. Given the relation \eqref{eq:CR=RLambda}, the standard low-energy effective prepotentials of the 4d theory is recovered, 
\be\label{eq:4dPrepotential}
\CF(a, R,\Lambda) \to 4a^2 \left(\log\left(\frac{2a}{\Lambda}\right)-\frac{3}{2} \right) - \frac{\Lambda^4}{2a^2} - \frac{5\Lambda^8}{64a^6} + \cdots.
\ee
Here we used the small $\fa$ expansion of $\text{Li}_3(e^{-2\fa})$,
\begin{equation}
    \mathrm{Li}_3\left(e^{-2\fa}\right)= \zeta(3) - \frac{\pi^2 \fa}{3} - 2 \mathfrak{a}^2\log(\mathfrak{a}) +\left(3-2\log 2 \right)\mathfrak{a}^2 + \mathcal{O}\left(\mathfrak{a}^3 \right),
\end{equation}
which holds in the principal branch of both logarithm and polylogarithm functions near $\mathfrak{a} = 0$. 

Similarly for the order parameter, we expand the exponential in the definition of $W_F$ and take the limit \eqref{4dlimit}. The 4d gauge-invariant order parameter $u$ defined in Eq. \eqref{4du} can be obtained from $U$ via
\be\label{eq:4dLimit-WF}
\lim_{R\to 0}\frac{U-2}{R^2} = \frac{1}{2} \left\langle \tr_F \phi^2 \right\rangle = u,
\ee
where  
\be
\phi = \frac{1}{R}\oint_{S^1} \left(\sigma + i A_5 \right) \mathrm{d}x^5.
\ee
The standard SW solution for $u$ can also be obtained from Eq. \eqref{eq:Exact-formula-U} in this limit,
\be\label{eq:4dLimit-(U-2)}
\frac{U-2}{R^2} \rightarrow  a^2 \left[ 1+ \frac{\Lambda^4}{2a^4}   +  \CO \left(  \frac{\Lambda^8}{a^8}  \right)  \right].
\ee

An important symmetry throughout this paper is the electric one-form symmetry $\mathbb{Z}_2^{(1)}$ of the $\mathrm{SU}(2)$ gauge field. This multiplies the holonomies by a nontrivial $\{\pm 1\}$-valued character of the fundamental group of spacetime. 
\footnote{In general for $G$-gauge theory on any spacetime $X$ with compact gauge group $G$, the gauge equivalence class of a gauge field is completely determined by its holonomy function on 
based loops at some point $x_0\in X$.   
More precisely, assume that $X$ is connected. After choosing a basepoint $x_0 \in X$ and a trivialization of the fiber of $P\to X$ over $x_0$, the holonomy function $h_{A,x_0}: \Omega_{x_0}X \to G$, defined by $\gamma \mapsto {\rm P}\!\exp\oint_{\gamma} A$, determines $A$ up to gauge equivalence by gauge transformations $g$ with $g(x_0)=1$. Here $\gamma\in \Omega_{x_0}X$ is a based loop in $X$ at $x_0$.   
Now, given a homomorphism $\chi: \pi_1(X,x_0) \to Z(G)$, where $Z(G)$ is the center of $G$, we may lift $\chi$ uniquely to a function $\tilde\chi: \Omega_{x_0}X \to Z(G)$. We now consider the function from $\Omega_{x_0}X$ to $G$ defined by $\gamma \mapsto h_{A,x_0}(\gamma)\tilde \chi(\gamma)$. Since $G$ is compact, this will be the holonomy function of a new gauge field $A'$, unique up to gauge equivalence, 
i.e. $h_{A',x_0}(\gamma) = h_{A,x_0}(\gamma)\tilde\chi(\gamma)$ for some $A'$. The transformation of gauge equivalence classes $[A]\to [A']$ is what is meant by the ``shift by a flat $Z(G)$-valued connection''. }
Denoting the symmetry by ${\bf T}$, the action on the Coulomb branch parameter is the involution: 
\be
\label{eq:T1form}
{\bf T}: U \mapsto -U.
\ee 
In terms of the parameter $\mathfrak{a}$ this corresponds to the shift $\mathfrak{a} \to \mathfrak{a} +  i \pi $. 

Due to $\mathbb{Z}_2^{(1)}$, there is an alternative 4d limit. To this end, we define $\tilde a$ and $\tilde{\mathfrak{a}}$ as
\be\label{eq:4dlimit-two}
R \tilde{a} = Ra + i \pi,\qquad \tilde{\mathfrak{a}}=\mathfrak{a}+i\pi,
\ee
and take $R\to 0$ while keeping $\tilde a$ and $\Lambda$ fixed. The fact that only even powers of $\sinh(Ra)$ appear in $\CF^{\rm inst}$ leads to 
\be
\CF^{\rm inst}(a, R, \Lambda)=\CF^{\rm inst}(\tilde{a}, R, \Lambda),
\ee
and in the limit specified below \eqref{eq:4dlimit-two}, we obtain exactly the same 4d instanton prepotential but with $a \to \tilde{a}$. When $\mathrm{Re}(Ra)>0$, the perturbative part becomes 
\begin{align}\label{eq:5dPrepotential:U=-2}
     \mathcal{F}^{\text{pert}}\!\left(\tilde{a}-\frac{i\pi}{R}\mid R,\Lambda \right)&=-4 \tilde{a}^2 \log (R \Lambda)+\frac{2\left(\zeta(3)-\text{Li}_3\left(e^{-2 R \tilde{a}}\right)\right)}{R^2}-\frac{2 \pi^2 \tilde{a}}{3 R}-2 \pi i \tilde{a}^2+\frac{4 R \tilde{a}^3}{3} \nonumber \\ & \quad + \frac{4\pi^2}{R^2}\log(R\Lambda) + \frac{8\pi i \tilde{a} \log(R\Lambda)}{R}.
\end{align}
Despite two divergent terms in the second line as $R\to 0$ with fixed $\Lambda, \tilde{a}$, the physical quantities such as the couplings \eqref{vandxi} do admit a smooth 4d limit. The order parameter (\ref{eq:Exact-formula-U}) also behaves consistently for $R\to 0$,
\be
\label{eq:4dLimit-U2}
\frac{U(\tilde{\mathfrak{a}}-i\pi,\CR)+2}{R^2} \rightarrow  -\tilde{a}^2 \left[ 1+  \frac{\Lambda^4}{2\tilde{a}^4} +  \CO \left( \frac{\Lambda^8}{\tilde{a}^8}  \right)  \right].
\ee

To recover the 5d physics, we take the limit
\be\label{5dlimit}
R\to \infty, \text{ while holding } g^2_{5d} \text{ fixed,}
\ee
which implies $\CR\to 0$ and $\Lambda \to 0$. Since the holonomy of $A_5$ is bounded, we know from \eqref{asigmaa5} and \eqref{asigmaa5-2} that $a$ becomes a real scalar $\sigma$. For $\sigma>0$ so that $\mathrm{Re}(Ra) \to +\infty$, instantons are exponentially suppressed,
\footnote{To justify this rigorously without imposing any condition on the relative size of $\mathfrak{a}$ and $\log \CR$, one should prove that the degree of $\CF_n(x)$ is less than $2n-1$, which is indeed the case for lower order terms. If this is true for all $n$, then it is easy to see that $\CF^{\rm inst}$ vanishes in the limit \eqref{5dlimit}. Unfortunately, it is not easy to determine the degree of $\CF_n$ from the localization expression for the instanton partition function. }
\be
\CF^{\rm inst} \to 0.
\ee
Substituting \eqref{CR4} into the perturbative prepotential and taking the limit we find that 
\be 
\CF^{\rm pert} \to R \left( \frac{8\pi^2}{g^2_{5d}} \sigma^2 + \frac{4}{3} \sigma^3 \right),
\ee
matching the expected 5d prepotential \cite{Seiberg:1996bd}. Note that the relative factor of $R$ between the 4d and 5d prepotentials comes from the relative superspace volumes in 4d and 5d.

\subsection{Seiberg-Witten Geometry For Five-dimensional $\mathrm{SU}(2)$ Super Yang-Mills Theory On $\IR^4 \times S^1$}

Similarly to the renowned case of 4d $\mathrm{SU}(2)$ SYM \cite{Seiberg:1994rs}, the prepotential for the LEET of 5d $\mathrm{SU}(2)$ SYM on $\mathbb{R}^4 \times S^1$ defined in Sec.  \ref{sec:5dR4S1} is encoded in the geometry of a holomorphic family of Riemann surfaces equipped with a meromorphic differential \cite{Nekrasov:1996cz, Ganor:1996pc, Gottsche:2006bm}. 
In this subsection, we discuss various aspects of the SW geometry for the 5d theory.

\subsubsection{Seiberg-Witten Curve And Differential}
  
The SW curve $\Sigma_{U,\CR} \subset \mathbb{C}^*\times \mathbb{C}^*$ is defined by the equation
\be\label{SigmaUm}
w+w^{-1}=- \CR^{-2} (\CX + U + \CX^{-1}),
\ee
where $U\in\mathbb{C}$ is identified with the order parameter \eqref{eq:Define-U}. 
The complex structure of $\Sigma_{U,\CR}$ is related to the complexified effective coupling 
\be
\tau=-\frac{1}{2\pi i}\frac{\partial^2 \CF}{\partial a^2}.
\ee

The curve \eqref{SigmaUm} can be transformed into the standard form by introducing $\CY :=-\CR^2 \CX(w-w^{-1})$, 
\be
\label{5dSWcurve}
\CY^2=P(\CX)^2-4\CX^{2}\CR^4,
\ee
with $P(\CX)=\CX^2+U\CX+1$. 
The corresponding SW differential $\lambda$ is 
\be\label{eq:SW-diff1}
R \lambda=\frac{1}{2\pi i }\log(\CX)\,\frac{\mathrm{d}w}{w}.
\ee
This differential is multi-valued due to $\log(\CX)$ and meromorphic due to the singularity in $w$.
Differentiating \eqref{SigmaUm} gives 
\be
\mathrm{d}\log w = (\CX - \CX^{-1}) \frac{\mathrm{d}\CX}{\CY},
\ee
and therefore $\lambda$ is expressed in terms of $\CX$ and $\CY$ as
\be\label{eq:SW-diff2}
R\lambda
=\frac{1}{2\pi i} \log(\mathcal{X})\left(\mathcal{X}-\mathcal{X}^{-1}\right)\frac{\mathrm{d}\mathcal{X}}{\mathcal{Y}}.
\ee
Evidently $\mathcal{X}=0,\infty$ should be excluded so that the SW curve is a punctured elliptic curve. 
Moreover, to ensure single-valuedness of the SW differential, we work on a cyclic cover $\CX=e^{2\pi i s}$ $(s\in \mathbb{C})$. This cyclic cover has an infinite genus but a finite-rank first homology as a module over the ring generated by the action of the deck transformation, where the ring is isomorphic to $\IZ$.

The one-form symmetry $\IZ_2^{(1)}$ of 5d $\mathrm{SU}(2)$ SYM compactified on a circle transforms
\be
\left(\CR^2, U,\CX,w\right) \mapsto \left(\CR^2, -U, - \CX, -w \right).
\ee
This preserves the SW curve but shifts the SW differential,
\be 
\label{eq:shiftlambda}
\lambda \rightarrow \lambda \pm  \frac{1}{2 R}  \frac{\mathrm{d}w}{w}.
\ee
There is also a zero-form $\IZ_4$ symmetry acting by multiplying $\CR$ by fourth roots of unity.
The $\IZ_4$ symmetry is essentially the $\IZ_4$ symmetry discussed in \cite[P. 6]{Seiberg:1994rs}. The formulae for the partition function for manifolds of $b_2^+>1$ are written as sums over the action of this group. 
\footnote{The partition function will transform by an overall phase. It follows from Eqs. \eqref{eq:Zee-QM-1} and \eqref{eq:ModSpaceDim} below that we can cancel that phase by a 4d local counterterm, 
\[
\exp \int_X \left( \overline{w_2(P)}^2 + \frac{3}{4}( e+ s)  \right) \log \CR,
\]
where $e$ and $s$ are the local densities for the Euler character and signature of $X$, respectively, 
and $\overline{w_2(P)}=2\bfmu$ is a choice of integral lift of $w_2(P)$. 
}

The projection $(\CY, \CX) \to \CX$ of \eqref{5dSWcurve} branches at four roots of the equation
\be\label{eq:X-branchlocus}
\left(\CX^2+(U-2\CR^2)\CX+1 \right) \left(\CX^2+(U+2\CR^2)\CX+1 \right)=0.
\ee
In the region where $\text{Re}(U) \gg \vert \CR^2 \vert$, the branch of the square root can be chosen so that the real part of $\sqrt{(U\pm 2\CR^2)^2-4}$ is large and positive, and we can unambiguously label the four branch points as
\be\label{eq:roots}
\begin{split}
\CX^{+\pm}&=\frac{-(U+2\CR^2)\pm\sqrt{(U+2\CR^2)^2-4}}{2}, \\
\CX^{-\pm}&=\frac{-(U-2\CR^2)\pm\sqrt{(U-2\CR^2)^2-4}}{2}.
\end{split}
\ee
The singularities where the branch points collide are labeled as
\be
\label{eq:DiscLocus}
U_1 =2-2\CR^2, \quad 
U_2 =2+2\CR^2, \quad 
U_3 =-2+2\CR^2, \quad
U_4 = -2-2\CR^2. 
\ee
These are permuted by the $\IZ_2^{(1)}$ and $\IZ_4$ symmetries noted above. 
At each singularity, there is a massless particle with charges $\gamma_i$ corresponding to $U_i$, with $\gamma_i$ given (in a specific duality frame) in Eq. \eqref{eq:chargeQuivers} below. 
At $\CR^4=1$, two singularities, either $(U_1,U_3)$ or $(U_2,U_4)$, collide. We describe the physical interpretation in Sec.  \ref{subsubsec:BPS-States-LEET} below. 

In the region where we defined \eqref{eq:roots} we can choose a basis for the first homology by taking the $A$-cycle to encircle only $\CX^{--}$ and $\CX^{+-}$, and the $B$-cycle encircles only $\CX^{-+}$ and $\CX^{--}$, with a suitable orientation. Then the scalar $a$ and its dual $a_D$ are defined by 
\be
\mathfrak{a} = Ra =\oint_A R\lambda,\quad  a_D=\oint_B \lambda.
\ee 
The shift of the SW differential (\ref{eq:shiftlambda}) shifts $a$ as $a\to a \pm  \pi i/R$. This is once again a manifestation of the one-form symmetry. 

Due to the factor $\log \CX$ in the SW differential, $a$ is ambiguous up to the addition of an integral multiple of $2\pi i/R$. As explained in \cite{Banerjee:2018syt}, a single-valued central charge can be formulated by considering the relative homology on the cover of the SW curve. We also include the cycle $C$ corresponding to one of the lifts of a small counterclockwise-oriented circle in the $w$-plane around $w=0$. The local parameter $a^{(I)}$ is then defined by a contour integral around $w=0$. 

We determine $\partial (R \lambda)/\partial U$ keeping $w,\CR$ fixed,
\footnote{It would be desirable to have an \emph{a priori} reason why we should hold $w$ fixed. We are being pragmatic: This is what gives good expressions.}
\be
\left.\frac{\partial (R\lambda) }{\partial U}\right\vert_{w,\CR} =\frac{1}{2\pi i } \frac{\partial \log \CX}{\partial U}\,\frac{\mathrm{d}w}{w},
\ee
Differentiating (\ref{SigmaUm}) with $w$ and $\CR$ fixed gives
\be
\left.\frac{\mathrm{d}(\log \CX)}{\mathrm{d}U} \right\vert_{w,\CR} = - (\CX - \CX^{-1})^{-1},
\ee
and therefore 
\be
\left.\frac{\partial (R\lambda) }{\partial U}\right\vert_{w,\CR} =-\frac{1}{2\pi i }\frac{\mathrm{d}\CX}{\CY}.
\ee
This is thus a multiple of the standard holomorphic differential.

\subsubsection{Modular Parametrization}\label{subsubsec:MoularParametrization}

When writing the Coulomb branch measure below it, will be extremely useful to have a modular parametrization of various quantities that enter the LEEA as functions of the effective coupling $\tau$. In the following, we will demonstrate that the modular parameterization involves a branched cover of $\mathbb{H}/\Gamma^0(4)$. Similar branched covers were encountered in the modular parametrization for $\CN=2$ SQCD \cite{Aspman:2021vhs} and have been applied to BPS state counting \cite{LeFloch:2024cwl}. To derive the modular parametrization, we note that for suitable constants $\alpha,\beta$, a M\"obius transformation of the form \cite[App. A.8]{Gottsche:2006bm}
\be 
x = \alpha \frac{\CX+1}{\CX-1}, \quad y = \beta \frac{\CY}{(\CX-1)^2}
\ee
maps the curve \eqref{SigmaUm} to a standard elliptic curve
\be
\label{hellipticcurve}
y^2=(1-x^2)(1-\CK^2 x^2).
\ee 
Moreover, the N\'{e}ron differentials are related by  
\be
\frac{\mathrm{d}x}{y} = -2 \frac{\alpha }{\beta }\frac{\mathrm{d}\CX}{\CY}.
\ee
We may choose an ordered basis for $H_1$ of this curve such that the $\tilde A$-cycle encircles 1 and $1/\CK$, and the $\tilde B$-cycle encircles $1/\CK$ and $-1/\CK$.
That determines a period matrix $\tilde\tau$, and in terms of that period matrix we have \cite{Whittaker_Watson_1996, Gottsche:2006bm}
\be
\CK=\frac{\vartheta_2(\tilde\tau)^2}{\vartheta_3(\tilde\tau)^2}.
\ee
We will find that the standard weak coupling duality frame is related to $\tilde \tau$ through $\tilde\tau/2 = - 1/\tau$. With this transformation, it follows from the curve (\ref{hellipticcurve}) that the order parameter $U$ has a modular parametrization
\be\label{eq:U-doublecovers-ttu}
U(\tau)^2=-8\CR^2 \ttu(\tau)+4\CR^4+4,
\ee 
with $\ttu$ defined as
\be
\label{ttu}
\mathtt{u}(\tau)=\frac{\vartheta_2(\tau)^4+\vartheta_3(\tau)^4}{2 \vartheta_2(\tau)^2\, \vartheta_3(\tau)^2}.
\ee
The function $\mathtt{u}(\tau)$ is a Hauptmodul for $\Gamma^0(4)$. 

Eq. \eqref{eq:U-doublecovers-ttu} states that the physical Coulomb branch $\mathbb{C}_U$, parametrized by $U$, is a branched double-cover of the $\ttu$-plane,
\be\label{eq: CUcover}
\pi_U: \mathbb{C}_U \rightarrow \mathbb{C}_\ttu,
\ee
where the latter has a modular parametrization by a Hauptmodul for $\Gamma^0(4)$. 
The cover \eqref{eq: CUcover} has a single branch point at  
$\ttu =\frac{1}{2}\left(\CR^2 + \CR^{-2} \right)$ (and a branch point at $\ttu=\infty$ if we try to compactify the $\ttu$ plane). Consequently, we can define the fiber product diagram 
\be\label{eq:DoubleCover-UHP}
\xymatrix{ 
\widetilde{\mathbb{H} }  \ar[r]^{\tilde\pi_{\ttu}} \ar[d]^{\pi}  & \mathbb{C}_U \ar[d]^{\pi_U}  \\ 
\mathbb{H}   \ar[r]^{\pi_\ttu}  &   \mathbb{C}_\ttu  \\ 
}
\ee
where $\pi_{\ttu}(\tau)= \ttu(\tau)$. An important point below is that the one-form symmetry $U\to -U$ manifests itself as the deck transformation of the double-cover on the right-hand side of the diagram.

We now discuss in more detail the location of the branch points. 
The pullback double-cover $\pi: \widetilde{\mathbb{H}} \to \mathbb{H}$  has infinitely many branch points at solutions of the equation
\be 
\ttu(\tau)=\frac{1}{2} \left( \CR^2 + \CR^{-2} \right),
\ee
or equivalently, 
\be\label{eq:tau-CR-bps}
\frac{\vartheta_2(\tau)}{\vartheta_3(\tau)} = \pm \CR, \pm \CR^{-1}.
\ee
A particularly useful set of branch points, which will appear in a standard fundamental domain for $\Gamma^0(8)$, can be written in terms of elliptic integrals. These are solutions of $(\vartheta_2/\vartheta_3)^2=\CR^2$ in the case $|\CR^2|<1$,  
\be
\tau_{\rm bp}=\frac{i}{2}\frac{K_{\rm e}\!\left(\sqrt{1-\CR^4}\right)}{K_{\rm e}\!\left(\CR^2\right)} \mod 4,
\ee 
with $K_{\rm e}$ the complete elliptic integral of the first kind. Assuming $0<\CR^2<1$, the first few terms in the expansion in $\CR$ around $\CR=0$ read
\be\label{eq:SmallRbp}
\tau_{\rm bp}=\frac{4i}{\pi}\log\left(\frac{2}{\CR}\right)-\frac{i}{2\pi} \CR^4 +\CO\left(\CR^8\right) \mod 4.
\ee 
Thus, for these branch points,
\be
\tau_{\rm bp}\to i\infty \quad  \text{ for } \CR\to 0.
\ee
We will return to the Coulomb branch as a branched double-cover of a modular curve in Sec.  \ref{SecFundDom}. 

The modular parametrization will be essential when we write the Coulomb branch measure in Sec.  \ref{subsec:Integrand} below. First of all the discriminant reads
\be
\label{DeltaPhys}
\Delta_{\rm phys}=\prod_{j=1}^4(U-U_j)=64\CR^4 ({\ttu}^2-1).
\ee
We also need to know the modular parametrization of the periods of $\lambda$. 
The integral $\oint_A \partial \lambda/\partial U$ can be explicitly evaluated using elliptic integrals, which gives \cite[App. A]{Gottsche:2006bm} 
\be
\label{dadU}
\frac{\mathrm{d}\mathfrak{a}}{\mathrm{d}U}=\frac{i}{2\CR} \vartheta_2(\tau)\,\vartheta_3(\tau).
\ee
We have, straightforwardly,  
\be
\label{dUdtau}
\frac{\mathrm{d}U}{\mathrm{d}\tau}=-\frac{4\CR^2}{U}\,\frac{\mathrm{d}\ttu}{\mathrm{d}\tau}.
\ee
Combining this with Eq. \eqref{dUdtau} then implies 
\be
\frac{\mathrm{d}\mathfrak{a} }{\mathrm{d}\tau}=-\frac{4\CR^2}{U}\,\frac{\mathrm{d}\ttu}{\mathrm{d}\tau}  \frac{\mathrm{d}\mathfrak{a} }{\mathrm{d}U}.
\ee
Now Eqs. \eqref{eq:da4du} and \eqref{eq:Matone} imply
\be 
\frac{\mathrm{d}\ttu}{\mathrm{d}\tau} = - \frac{i  \pi}{4} \frac{\vartheta_4^8}{(\vartheta_2\vartheta_3)^2}, 
\ee
and then using Eq. \eqref{tsproduct} we arrive at
\be
\label{dadtau}
\frac{\mathrm{d}\mathfrak{a}}{\mathrm{d}\tau}=-\frac{\pi \CR}{4}\frac{1}{U}\,\frac{\vartheta_4(\tau)^9}{\eta(\tau)^3}.
\ee
In the $\tau\to i\infty$ limit, this agrees with (\ref{atau}) derived from the prepotential.
Eq. \eqref{dadtau} is useful in writing the Coulomb branch measure in Sec.  \ref{subsec:Integrand} below.

\subsubsection{An Important Coupling}

An important coupling in the LEET is that between the background gauge field $A^{(I)}$ of the $\mathrm{U}(1)^{(I)}$ symmetry and the dynamical gauge field of the $\mathrm{U}(1)$ low-energy vector multiplet. In terms of SW geometry, this coupling is given by  
\be 
\label{eq:Defv}
v =\frac{R}{4} \frac{\partial a_D}{\partial \log(\CR)},
\ee 
where the derivative is taken with $\mathfrak{a}$ fixed, and hence $U$ is considered as a function of $\mathfrak{a}$ and 
$\CR$.  A useful expression for $v$ as an incomplete elliptic integral was derived in \cite[Sec.  4.4.2 and App. A.5]{Gottsche:2006bm}, 
\footnote{\label{footnote:sign} Although the prepotential $\CF$ is identical to that in \cite{Gottsche:2006bm} in terms of the Cartan variables $a_1$ and $a_2$, we set $a_1=-a_2=a$ while $a_1=-a_2=-a$ is used in \cite{Gottsche:2006bm}. This difference gives rise to a few relative signs between our equations and those in \cite{Gottsche:2006bm}. For example, the right-hand side of Eq. \eqref{vsmallR} differs by a sign from the corresponding equation in \cite{Gottsche:2006bm}.
Due to this sign difference in $v$, the analogues of Eqs. \eqref{tauvid} and \eqref{Ctauv} in \cite{Gottsche:2006bm} (namely Eq. (A.36) and the first equation of Sec.  4.4.3) differ by a sign on the right-hand side. Relatedly, we find a different sign in the relation of $a$ in terms of $q$ based on Eq.  \eqref{eq:tau-functionof-a}, i.e., the right-hand side of \cite[Eq.  (4.8)]{Gottsche:2006bm}. See also Eq.  \eqref{eq:aq}. Another relative sign appears in the expression for $H(a)$ in Eq. \eqref{eq:Ha} and \cite[P. 38]{Gottsche:2006bm}.}
\be
\label{vsmallR}
v=-\frac{1}{\pi} \frac{1}{\vartheta_2(\tau)\,\vartheta_3(\tau)} \int_0^\CR\frac{\mathrm{d}x}{\sqrt{1-2\ttu\,x^2+x^4}}.
\ee
The branch points of the integrand of Eq. \eqref{vsmallR} in the complex $x$-plane are at 
\be 
x^2=   \left(\frac{\vartheta_2(\tau)}{\vartheta_3(\tau)}\right)^2, \quad {\rm and} \quad 
x^2=   \left(\frac{\vartheta_3(\tau)}{\vartheta_2(\tau)}\right)^2, 
\ee
which should be compared to the formula \eqref{eq:tau-CR-bps} for the branch points of $\widetilde{\IH}$ 
over $\IH$, and the integration depends on the choice of contour around these branch points.  
For any value of $\tau$, we can make $\vert \CR\vert$ sufficiently small so that the disk around $x=0$ of circumference $\vert \CR\vert$ does not contain these branch points. Then the integral is unambiguous if we keep the path from $x=0$ to $x=\CR$ within the disk. 
We can describe a region in the $(\tau, \CR)$ space that satisfies this criterion as follows. For $Y$ real, let $(\mathbb{H}/\Gamma^0(4))^{Y}$ be the subset of the fundamental region for $\Gamma^0(4)$ described in Sec.  \ref{SecFundDom} such that ${\rm Im}(\tau) < Y$. Then $\vert {\vartheta_2(\tau)}/{\vartheta_3(\tau)} \vert$ and $\vert {\vartheta_3(\tau)}/{\vartheta_2(\tau)} \vert$ will be bounded away from zero in this region, and we can therefore find a disk $D^Y$ around $\CR=0$ in the $\CR$ plane so that for 
$(\tau, \CR) \in (\mathbb{H}/\Gamma^0(4))^{Y} \times D^Y$ there are no branch points in the $x$-plane within the disk of circumference $\vert \CR\vert$.
Note that if $\vert q \vert$ is small, then  $\vert \CR \vert$ must be sufficiently small so that $(\tau, \CR)$ with 
\be\label{eq;SV-Reg1}
\vert \CR q^{-\frac18} \vert \ll 1 
\ee
will be in the desired region. There is a single-valued definition of $v(\tau,\CR)$ in this region. We can expand the integrand of Eq. \eqref{vsmallR} in powers of $\ttu$ and $x$, exchange sum and integral, evaluate the integral over $x$, and obtain the expansion as a power series in $\CR$ and $\ttu$,
\begin{equation}
\begin{aligned}
\label{vR}
v & =-\frac{1}{\pi} \frac{1}{\vartheta_2(\tau)\vartheta_3(\tau)}
 \sum_{\substack{n\geq 0 \\ n\geq k\geq 0}} \binom{-\frac{1}{2}}{n} \binom{n}{k} \frac{(-2\ttu)^k\,\CR^{4n-2k+1}}{4n-2k+1}  \\ 
 & =  -\frac{1}{\sqrt{\pi}} \frac{1}{\vartheta_2(\tau)\vartheta_3(\tau)}
\sum_{s=0}^\infty \sum_{k=0}^s  \frac{(-2\ttu)^k\,\CR^{2s+1}}{(2s+1)k! \Gamma\left(\frac{s-k}{2} + 1\right) \Gamma\left(\frac{1-s-k}{2}\right)}. 
\end{aligned}
\end{equation}
Note that one of the Gamma functions suppresses terms where $s+k$ is odd. Moreover, in the domain of convergence,
since $\ttu$ is invariant under $\tau \to \tau + 4$, 
\footnote{We will later see that $\tau \to \tau + 4$ is related to the transformation $U\to - U$, but it is important 
that this is not the proper transformation of $v(\tau,\CR)$ under the $\mathbb{Z}_2$ deck transformation of the covering $U \to \ttu$.}
we have 
\be
v(\tau+4, \CR) = - v(\tau, \CR).
\ee

Regarding Eq. \eqref{vR} as a power series in $\CR$, the coefficient of $\CR^{2s+1}$ takes the form
\be 
\left( \vartheta_2(\tau)\vartheta_3(\tau)\right)^{-1} \left( c_{s,s} \ttu^s + c_{s,s-1}\ttu^{s-1} + \cdots + c_{s,0} \right)
\ee 
for some numerical constants $c_{s,k}$. If we attempt to take the limit $q=e^{2\pi i \tau} \to 0$ while keeping $\CR$ fixed, then we must bear in mind that $\ttu(\tau) \sim 2 q^{-1/4}$ diverges, and hence the coefficients in the series expansion in $\CR$ will diverge, causing the series itself to diverge. The function $v(\tau,\CR)$ behaves very differently when $\vert q\vert $ is small.  

In order to describe $v$ when $\vert q \vert$ is very small at fixed $\CR$, we use $2\ttu = K + K^{-1}$ 
with
\be 
K=\frac{\vartheta_2^2(\tau)}{\vartheta_3^2(\tau)}.
\ee 
We use the values for the parameters as discussed in the first paragraph of Sec.  \ref{Monos}, $iq^{1/8}$ and $\CR$ are both real and positive, such that $K$ is real and negative. We then rewrite \eqref{vsmallR} as  
\be\label{eq:v-altint*}
v =-\frac{i}{\pi \vartheta_3(\tau)^2}\int_0^{\CR} \frac{\mathrm{d}x}{\sqrt{\left(1-K x^2\right)\left(x^2-K\right)}}.
\ee 
In the small $q$ limit, we can expand the first factor of the square root as
\be
\frac{1}{\sqrt{1-Kx^2}}=\sum_{\ell=0}^\infty \frac{(2\ell-1)!!}{2^\ell\,\ell!} (Kx^2)^\ell,
\ee 
with radius of convergence $|Kx^2|<1$, and thus for the upper end of the integration domain $|K\CR^2|<1$. Exchanging sum and integral, the terms with $\ell>0$ can be bounded above at fixed $\CR$ by a power $K^{\ell-1/2}$. The integral of the $\ell=0$ term is: 
\be
\int_0^\CR \frac{\mathrm{d}x}{\sqrt{x^2-K}}=\left.\frac{1}{2} \log\left(1+\frac{x}{\sqrt{x^2-K}} \right) -\frac{1}{2}\log\left(1-\frac{x}{\sqrt{x^2-K}}\right)\right\vert_0^{\CR}.
\ee 
We can use the expansion
\be
\frac{\CR}{\sqrt{\CR^2-K}}=\sum_{\ell=0}^\infty \frac{(2\ell-1)!!}{2^\ell \ell!} (K/\CR^2)^\ell,
\ee 
with radius of convergence $|K/\CR^2|<1$. The leading terms in the $\CR$-expansion then give
\be 
v=\frac{i}{2\pi}\left( \log(-K/4)-2\log(\CR)+\cdots\right),
\ee 
where the logarithm is defined in terms of the principal branch. We parametrize $\tau$ at weak coupling as $\tau=-2+8s+iT$, with $s\in \mathbb{Z}$ and $T\gg 0$ a sufficiently large positive number. We then arrive at
\be\label{eq:v-smallq*}
v =-\frac{\tau}{4}-\frac{1}{2} +2s-\frac{i}{\pi}\log(\CR)+\CO\left(q^{\frac14}\right),
\ee 
in agreement with Eq.  (\ref{vinfexp}) below. Considering $v$ as a multi-valued function of $\tau$, we have
\be\label{eq:v-smallq*2}
v =-\frac{\tau}{4}-\frac{1}{2} -\frac{i}{\pi}\log(\CR)+\CO\left(q^{\frac14}\right) \mod 2.
\ee 

\subsubsection{Four-dimensional Limit Of The Seiberg-Witten Geometry}
\label{sec:4dLimitSW}

As discussed in Sec.  \ref{4d5dlimits}, there are two different ways to take the 4d limit of the prepotential. Correspondingly, there are two distinct 4d limits of the SW geometry.

According to Eqs. \eqref{eq:4dLimit-WF},   \eqref{eq:4dLimit-U2}, \eqref{eq:U-doublecovers-ttu}, and \eqref{eq:uttu}, we have the expansion
\be
U= \pm (2 + u R^2)  + \mathcal{O}\left(R^4\right).
\ee
We introduce the following transformation 
\footnote{
The transformation \eqref{eq:trans-X} is different from the one used in \cite[App. A.2]{Gottsche:2006bm}, $\mathcal{X} = \mp(2 - R x)/(2+ R x)$. However, their series expansions in $R$ near $R=0$ coincide up to $\CO\left(R^2\right)$. 
}
\begin{equation}\label{eq:trans-X}
\mathcal{X} = \mp e^{-R x},
\end{equation}
where the sign choice depends on the specific asymptotic behavior of $U$, i.e., as $R\to 0$, $U\to \pm 2$, respectively.
In the limit $R \to 0$, the 5d SW curve \eqref{SigmaUm} reduces to 
\begin{equation}
\label{eq:SWcurve-Reduce4d}
    \pm \Lambda^2 \left(w + w^{-1}\right) =  x^2 - u.
\end{equation}
Setting $y = \Lambda^2 (w - {w}^{-1})$ yields the quartic form
\begin{equation}
  y^2 = (x^2 - u)^2 - 4\Lambda^4,
\end{equation}
which matches the standard 4d SW curve \eqref{4dcurve} reviewed in App. \ref{app4dN=2}.

Meanwhile, in the limit $R \to 0$, the SW differential \eqref{eq:SW-diff1} reduces to
\be
\lambda= \left(-\frac{1}{2\pi i }x + \frac{\log (\mp 1)}{2\pi i R}\right)\frac{\mathrm{d}w}{w}.
\ee
For the $+$ sign, with the branch $\log(1)=0$ this is the standard 4d SW differential associated with \eqref{eq:SWcurve-Reduce4d}. For the $-$ sign, the integration of $\lambda$ around the A-cycle gives 
a shift to $\mathfrak{a}$ by $\pi i$. Comparison of Eqs \eqref{prepotential2} and \eqref{eq:5dPrepotential:U=-2} demonstrates that the shift does not alter the limit for $R\to 0$ from the 5d to the 4d prepotential except for two divergent terms and the sign of the term $2\pi i a^2$. Moreover, both 4d limits are smooth for the physical couplings $\tau$, $v_I$ and $\xi_{II}$.\footnote{See also Sect. \ref{Monos} for more details on the action of the shift on various couplings.}

\subsubsection{BPS States In The Four-dimensional Effective Theory}\label{subsubsec:BPS-States-LEET}

On the Coulomb branch, there are various BPS charged objects whose masses are determined by the central charge $Z$, which is a linear function on the charge lattice $\Gamma$,
\be
Z: \Gamma \rightarrow \mathbb{C},
\ee
where $\Gamma \cong \mathbb{Z}^6$ is a lattice equipped with a Dirac pairing. 
\footnote{It would be desirable to have a geometric interpretation of the full rank-six lattice in terms of the SW curve.}
We introduce a six-dimensional vector $\gamma=[q_m, q_e,q_{DI},q_I,q_{DK}, q_K]$. The monodromy matrices acting on $\gamma$ are elements of $\mathrm{Sp}(3,\mathbb{Z})$. In this paper, we restrict to particles with $q_{DI}=q_{DK}=0$.
The charges $q_m,q_e$ corresponds to the first homology of the unpunctured SW curve.  Let $g$ be the deck transformation, and $C[g]$ the ring of deck transformations. The homology of the (unpunctured) cyclic cover has rank $2$ as a module over $C[g]$. From this perspective, the charge $q_K$ is the sheet number of the cyclic cover, labeling deck transformations. The Dirac-Schwinger-Zwanziger symplectic inner product of charge vectors is given by \eqref{eq:SymplecticProduct} below. 
 
We have seen that the physical quantity $U$ is a function of $\cosh(\fa)$.
The different cycles in the cover correspond to different branches of the logarithm needed to define $\fa$. Adding KK charge $q_K$ corresponds to shifting the cycle on the cover by the deck transformation $g^{q_K}$. With this understood we can write the central charge function $Z$ (for charges with $q_{DI}=q_{DK}=0$) as
\be 
Z(\gamma;a_D,a,a^{(I)},a^{(K)})=q_m\,a_D+q_e\,a+q_I\,a^{(I)}+q_K\,a^{(K)}.
\ee 
Among the occupied charges are four independent charge vectors with simple physical interpretations  \cite{Seiberg:1996bd}:
\begin{itemize}
\item $\gamma=[0,0,0,0,0,q_K]$: KK momentum with $q_K$ units around $S^1$. The contribution of a single unit, $q_K=1$, to the central charge $Z$ is
\be
  \label{mK} 
a^{(K)} =\frac{2\pi i }{R}.
\ee 
In a suitable chamber, there are BPS states with only $q_K$ charge. These are D0-branes from the point of view of type IIA string theory. 
\item $\gamma=[0,2,0,0,0,0]$: W-boson, which contributes $2a$ to the central charge $Z$.
\item $\gamma=[1,0,0,0,0,0]$: Magnetic monopole string wrapping $S^1$, which contributes $a_D$ to $Z$.
\item $\gamma=[0,0,0,1,0,0]$: Instanton particles on Minkowski space $\mathbb{R}^{1,4}$ are believed to exist \cite{Seiberg:1996bd}.
\footnote{The question of existence is quite delicate. The quiver analysis below Eq. \eqref{eq:chargeQuivers} gives a vanishing BPS index for this charge, suggesting that the instanton particle is a marginally unstable bound state. See also the discussion in 
\cite{Closset:2019juk,Longhi:2021qvz}. 
}
They carry $\mathrm{U}(1)^{(I)}$ charge $-\frac{1}{8\pi^2}\int_{\mathbb{R}^4} \text{tr} F\wedge F=1$, where the integral is over any time-slice. Upon compactifying spatial $\mathbb{R}^4$ to $\mathbb{R}^3 \times S^1_R$, 
they become 4d BPS particles with central charge
\be
\label{mCR}
Z_I: =   a^{(I)} = \frac{1}{R} \log(\CR^4).
\ee
Thus, a choice of branch for $a^{(I)}$ is a choice of branch of $\log(\CR^4)$. 
\footnote{The choice of branch is quite delicate. The BPS particles with charges $\gamma_2$ and $\gamma_4$ in Eq. \eqref{eq:chargeQuivers} below should have vanishing mass at the cusps $U_2$ and $U_4$ when $\CR^2=-1$, implying $a^{(I)} = a^{(K)}$ when $\CR^2 = -1$. We will not give a careful general prescription for how to choose this branch.}
\end{itemize}

The BPS particles associated with colliding singularities  (when the singularities in \eqref{eq:DiscLocus} collide) carry mutually local charges. 
The charge of the instanton particle is mutually local with respect to $\gamma_i$. However, there is an interesting order-of-limits issue associated with the order in which we take $U\to U_j$ and $\CR^4 \to 1$ (see Sec.  \ref{Sec:E1Uplane} below). 

The charge vectors associated with massless states depend on the duality frame. We will work in the duality frame where the massless particle at $U_1$ is the 4d monopole with 4d electric-magnetic charge $(1,0)$, and the massless particle at $U_2$ is the 4d dyon with 4d electric-magnetic charge $(-1,2)$. The charges of the massless particles are discussed further in Sec.  \ref{Monos}.

\subsubsection{Couplings And Periods}

We discuss in this subsection various masses and couplings. We express the prepotential $\CF$ as a function of the central charges $a$, $a^{(I)}$ and $a^{(K)}$. From special
geometry, we know that $a_D$ can be obtained from the prepotential $\CF$,
\be
a_D=-\frac{1}{2\pi i} \frac{\partial \CF\left(a,a^{(I)},a^{(K)}\right)}{\partial a}.
\ee
The instanton particle and momentum carry both electric charge under
$\mathrm{U}(1)^{(I)}$ and $\mathrm{U}(1)^{(K)}$, respectively. It is thus natural to
introduce dual observables to $a^{(I)}$ \eqref{mCR} and $a^{(K)}$ \eqref{mK},
\be
a_{D}^{(I)}=-\frac{1}{2\pi i} \frac{\partial \CF}{\partial a^{(I)}},\quad
a_{D}^{(K)}=-\frac{1}{2\pi i} \frac{\partial \CF}{\partial a^{(K)}}.
\ee
Here partial derivatives are taken with respect to one of $a$, $a^{(I)}$ or $a^{(K)}$, while the other two quantities are kept fixed. We combine these observables into the six-dimensional vector $\Pi$,
\be
\label{eq:periodvector}
\Pi=\left(a_D, a,  a_{D}^{(I)}, a^{(I)}, a_{D}^{(K)}, a^{(K)}\right)^\text{T},
\ee
which forms a local system over the $U$-plane, with the symplectic matrix
\be\label{eq:SymplecticProduct}
\Omega=\left(\begin{array}{cccccc} 0 & -1 & 0 & 0 & 0 & 0 \\ 1 & 0 & 0 & 0 & 0 & 0 \\  0 & 0 & 0 & -1 & 0 & 0 \\ 0 & 0 & 1 & 0 & 0 & 0 \\ 0 & 0 & 0 & 0 & 0 & -1 \\ 0 & 0 & 0 & 0 & 1 & 0 \end{array}\right).
\ee 
Furthermore, we define the couplings,
\footnote{The coupling $v$ here corresponds to the quantity 
\[
\frac{\beta h}{2\pi i}=-\frac{\beta}{8\pi i}\frac{\partial^2 \CF_0}{\partial a \partial \log(\Lambda)}
\]
used in \cite[App. A]{Gottsche:2006bm}, and the coupling $\xi_{KK}$ here corresponds to one-half of the coupling in \cite[Eq. (5.31)]{Closset:2022vjj}.}
\be
\begin{aligned}
\label{vandxi}
\tau &:=-\frac{1}{2\pi i} \frac{\partial^2\CF}{\partial a^2},\\
v_I &:=v =-\frac{1}{2\pi i}\frac{\partial^2 \CF}{\partial a \,\partial a^{(I)}}=-\frac{R}{8\pi i}\frac{\partial^2 \CF}{\partial a \,\partial \log\CR},\\
v_K &:=-\frac{1}{2\pi i}\frac{\partial^2 \CF}{\partial a \,\partial a^{(K)}},\\
\xi_{II} &:=-\frac{1}{2\pi i} \frac{\partial^2 \CF}{\partial (a^{(I)})^2}=-\frac{1}{2\pi i}\frac{ R^2}{16} \frac{\partial^2 \CF}{\partial \log(\CR)^2},\\
\xi_{KK} &:=-\frac{1}{2\pi i} \frac{\partial^2 \CF}{\partial (a^{(K)})^2},\\
\xi_{IK} &:=-\frac{1}{2\pi i} \frac{\partial^2 \CF}{\partial a^{(I)} \partial a^{(K)}}.
\end{aligned}
\ee
The coupling $v_I$ will play a very important role in our subsequent analysis and we will usually denote it simply as $v$ to avoid cluttering notation. 
 
To compute monodromies of these couplings around paths in the $U$-plane, we begin with the expansions of $a_D$, $a_{D}^{(I)}$ and $a_{D}^{(K)}$ at fixed $\fa=Ra$ in powers of $\CR$,
\be
\label{aDpert}
\begin{aligned}
Ra_D &=\frac{1}{2\pi i}\left[ -4 \left({\rm
    Li}_2(e^{-2\fa})-\zeta(2)\right)+2\fa\log(\CR^4)-4\pi i \fa-4\fa^2 +\cdots \right],\\
Ra_{D}^{(I)} &=\frac{1}{2\pi i}\left(\fa^2+\cdots\right),\\
Ra_{D}^{(K)} &=\frac{1}{(Ra^{(K)})^2}\left[ 4\left( {\rm
      Li}_3(e^{-2\fa})-\zeta(3)\right)+4\fa {\rm Li_2}(e^{-2\fa})-\fa^2\log(\CR^4)+\frac{2\pi^2\fa}{3}+\frac{4\fa^3}{3}+\cdots\right].
\end{aligned}
\ee
Similarly, the complexified 4d coupling $\tau$ expands around $\CR=0$ as 
\be\label{eq:tau-functionof-a}
\tau=-\frac{1}{2\pi i} \frac{\partial^2\CF}{\partial a^2}
=\frac{1}{2\pi i} \left(8\log(\CR)  -8\log[2 \sinh \fa]  -4\pi i + \CO(\CR^4)  \right),
\ee
where the $\CR^4$-expansion coefficients are rational functions of trigonometric functions of $\mathfrak{a}$. Assuming $\CR$ is real and positive, we deduce the correspondence between two limits for real $a\in \mathbb{R}$ and those for $\tau$, 
\be
\begin{aligned}
a &\to +\infty \quad \longleftrightarrow \quad \tau\to -2+ i\infty, \\
a &\to -\infty \quad \longleftrightarrow \quad \tau \to 2+i\infty.
\end{aligned}
\ee
It follows that 
\be
\label{atau}
\frac{2\pi i \tau}{8} = \begin{cases}
- \fa + \log \CR - \frac{i \pi}{2} +  \CO\left(\CR, e^{-\fa}\right), & {\rm Re}(\fa) \gg 0, \\ 
\fa \pm i \pi + \log \CR - \frac{i \pi}{2} +  \CO\left(\CR,e^{\fa}\right), & {\rm Re}(\fa) \ll 0, 
\end{cases} 
\ee
where the right-hand side is a power series in $\CR$ whose coefficients are rational expressions of trigonometric functions of $e^{\fa}$. Exponentiating Eq. \eqref{atau} and using $q=e^{2\pi i \tau}$ gives 
\be
q^{\frac18}=\begin{cases}  - i \CR e^{-\fa} \left(1 + \CO\left(\CR, e^{-\fa}\right) \right),    &  {\rm Re}(\fa) \gg 0, \\
i \CR e^{\fa} \left( 1 +  \CO\left(\CR, e^{\fa}\right)\right),   &  {\rm Re}(\fa) \ll 0. \\
\end{cases} 
\ee
Thus the left-hand side and the right-hand side are consistent under $\tau\to \tau-4$ and $a\to -a$.

For the Coulomb branch parameter $U$ regarded as a function of $\tau$ and $\CR$, its expansion in $q$ as ${\rm Im}(\tau)\to \infty$ is obtained by taking the square-root in \eqref{eq:U-doublecovers-ttu} and noting that \eqref{ttu} implies that $\ttu(\tau) \sim \frac{1}{8} e^{- 2\pi i \tau/4} + \cdots$,
\be
\label{Uweakcoupling}
U=\pm i \left( \CR\,e^{-2\pi i \tau/8} -\frac{2}{\CR}\,e^{2\pi i \tau/8} +\CO\left(q^{3/8}\right)\right). 
\ee
Defining the square-root by its principal branch and assuming $\CR>0$, we select the $+$ sign above for $0 < {\rm Re}(\tau)<4$ and the $-$ sign for a complementary region in the cylinder $\tau \sim \tau + 8$. In particular, we choose the $-$ sign for $-1/2 < {\rm Re}(\tau)<0$ and $4 < {\rm Re}(\tau)<15/2$ in the region shown in Fig. \ref{FRCG}. 
On the other hand, if we view $U$ as a function of $\fa$ and $\CR$, then we have the expansion \eqref{eq:Exact-formula-U}. 

Furthermore, there are couplings to the flux class $\bfn_I$ of the $\mathrm{U}(1)^{(I)}$ symmetry, which can be interpreted by thinking of $\log(\CR)$ as the scalar of a background superfield \cite[Sec.  3]{Argyres:1996eh}. The leading behavior for large $a$ is
\be
v=R \frac{\partial a_D}{\partial \log(\CR^4)}=-\frac{\tau}{4}+\cdots.
\ee
The expression for $v$ as an incomplete elliptic integral (\ref{vsmallR}) implies that $\tau$ and $v$ satisfy the relation \cite[Eq.  (A.36)]{Gottsche:2006bm}
\footnote{As mentioned in Footnote \ref{footnote:sign}, our formula differs by a sign from \cite[Eq. (A.36)]{Gottsche:2006bm}. An easy check on the sign is obtained by taking the $\CR\to 0$ limit and substituting the leading term from \eqref{vR}.} 
\be
\label{tauvid}
\frac{\vartheta_1(\tau,v)}{\vartheta_4(\tau,v)}=\CR.
\ee
A few key properties of this identity include:
\begin{itemize}
\item The identity (\ref{tauvid}) is left invariant under the monodromy $(\tau,v)\to (\tau-8,v+2)$.
\item Since the right-hand side is independent of $\tau$ and $v$, a solution $v(\tau)$ determining $v$ as a function of $\tau$ cannot take values where $\vartheta_1(\tau,v)$ or $\vartheta_4(\tau,v)$ vanish,
\be 
\label{excludedvs}
v \notin \mathbb{Z}\tau+\mathbb{Z},\quad  v \notin \frac{\tau}{2}+\mathbb{Z}\tau+\mathbb{Z}.
\ee 
\item A series expansion for $v$ at fixed $\CR$ and large ${\rm Im}(\tau)$ can be derived from \eqref{tauvid}. There are multiple branches satisfying $v \sim \pm \tau/4 + \cdots$. In the regime with ${\rm Re}(\fa) > 0$, which is used in the derivation of the monodromy in Sec.  \ref{Monos} below, comparison with Eq. \eqref{atau} shows that we should take the branch  
\begin{align}
\label{vinfexp}
v&=-\frac{\tau}{4}+\frac{1}{2\pi i}\bigg[2\log(\CR)-\pi i+2\left(\CR^2-\CR^{-2}\right)q^{\frac14}+3\left(\CR^4-\CR^{-4}\right)q^{\frac12} \nonumber \\
&\quad + \left(\frac{20}{3}\CR^6-4\CR^2+4\CR^{-2}-\frac{20}{3}\CR^{-6}\right)q^{\frac34}+\cdots \bigg].
\end{align}
The expansion demonstrates that $v+\tau/4$ is antisymmetric under $\CR\leftrightarrow \CR^{-1}$. Moreover, there are no negative powers of $q$.

\item The discussion from Eq. \eqref{eq:v-altint*} to \eqref{eq:v-smallq*2} above shows that the series \eqref{vinfexp} converges in an open region where $q$ is small and real, and $\CR$ is pure imaginary, satisfying 
\be
\left\vert\frac{\vartheta_2^2}{\vartheta_3^2}\right\vert< {\rm min} \left(|\CR|^2,|\CR|^{-2}\right).
\ee
\end{itemize}

We further introduce the couplings $C_{II}$, $C_{KK}$ and $C_{IK}$,
\be 
\label{Cs}
C_{II} :=\exp\left(-\pi i  \xi_{II}\right),\quad 
C_{KK} :=\exp\left(-\pi i  \xi_{KK}\right),\quad 
C_{IK} :=\exp\left(-\pi i \xi_{IK}\right).
\ee 
Hereafter, we simply denote $C_{II}$ by $C$ due to its central role. 
This coupling can be expressed as \cite[P. 45]{Gottsche:2006bm},
\be
\label{Ctauv}
C(\tau,\CR)=\frac{1}{\CR}\frac{\vartheta_1(\tau,v)}{\vartheta_4(\tau)},
\ee
which will be used in Sec.  \ref{sec:evaluate}. Equivalently, using Eq.\eqref{tauvid}, 
\be
\label{Ctauvv2}
C(\tau,\CR)=\frac{\vartheta_4(\tau,v)}{\vartheta_4(\tau)}. 
\ee
An equivalent expression was proposed in \cite[P. 56]{LoNeSha},
which differs from ours by $\tau\to \tau+1$. 

\subsubsection*{Remark} 

As a side remark, we will find below that the order of limits $q\to 0$ and $\CR \to 0$ in various series expansions of physical quantities poses an important physical issue. 

With this in mind, let us consider the expansion of $w=e^{4\pi i v}$. If we first expand around $q\to 0$, we have
\begin{align}
\label{wq}
w&=\CR^4\,q^{-\frac12}+4\CR^2(\CR^4-1)\,q^{-\frac14} \nonumber \\
&\quad +2(1-8\CR^4+7\CR^8) 
+16 \CR^2 (1-4\CR^4+3\CR^8)q^{\frac14} \nonumber \\ 
&\quad + (-\CR^{-4}+92\CR^4-256 \CR^8+165\CR^{12})q^{\frac12}  +\CO\left(q^{\frac34}\right).
\end{align}
Note that there are negative powers of $\CR$ in the coefficients of $q^{1/2}$ and beyond.
On the other hand, empirical evidence suggests a remarkable property: If we consider the expansion of $w$ for small $\CR$ followed by small $q$, the first few terms are 
\be 
\label{wR}
w=1+\left(-2i q^{-\frac18}+\CO\left(q^{\frac38}\right)\right)\CR+\left(-2 q^{-\frac14}+\CO\left(q^{\frac14}\right)\right)\CR^2+\CO\left(\CR^3\right).
\ee
While coefficients in the $\CR$-expansion involve negative powers of $q$, it appears that the coefficients for $\CR^{2n}$ with $n\in\mathbb{Z}_{\geq 2}$ contain only positive powers of $q$. This is related to the following property. 
Let us denote the expansion \eqref{vinfexp} as $v_q$ and the expansion \eqref{vR} as $v_\mathcal{R}$. 
We find that there is a highly nontrivial relation between these two expansions, which we have verified experimentally for $k\in\mathbb{Z}$, $|k|\leq 6$ to orders $\CO\left(q^{(5-k)/4}\right)$ and $\CO\left(\CR^{15}\right)$, 
\begin{equation}
   \text{Ser}_\mathcal{R} \text{Ser}_q \left(e^{2\pi i k v_q} + e^{-2\pi i k v_q}\right) = \text{Ser}_q\text{Ser}_\mathcal{R}
    \left(e^{2\pi i k v_\mathcal{R}} + e^{-2\pi i k v_\mathcal{R}} \right),
\end{equation}
where $\text{Ser}_x$ denotes the series expansion in $x$.

Let us consider another example: the series expansion of $C$. The explicit expression \eqref{Ctauvv2} shows that $C$ is invariant under $\tau\to \tau+4$. Hence, its expansion in $q$ should be a power series in $q^{1/4}$. If we expand $C$ first in $q\to 0$ and then in $\CR \to 0$, we find that
\begin{align} 
\label{Cqexp}
\text{Ser}_\mathcal{R} \text{Ser}_q C(\tau,\CR) &=1+\CR^2 q^{\frac14}+ 2\CR^4 q^{\frac12} 
+\left(-2\CR^2+5\CR^6\right)q^{\frac34} \nonumber \\
&\quad +\left(-10\CR^{4}+14\CR^{8}\right)q+\CO\left(q^{\frac54},\CR^9\right).
\end{align}
Conversely, if we expand first in $\CR \to 0$ and then in $q\to 0$, we obtain from Eq. \eqref{Ctauvv2} that
\begin{align} 
\label{CRexp}
\text{Ser}_q  \text{Ser}_\mathcal{R} C(\tau,\CR)&=1+\left(q^{\frac14}-2q^{3/4}\right) \CR^2+\left(2q^{\frac12}-10q\right)\CR^4 \nonumber \\
&\quad +5q^{\frac34}\CR^6+\CO\left(q^{\frac54},\CR^8\right).
\end{align}
Unlike the expansions for $v(\tau, \CR)$ in Eqs. \eqref{wq} and \eqref{wR}, two expansions \eqref{Cqexp} and \eqref{CRexp} are mutually compatible, and contain only positive powers of $\CR^2$ and $q^{1/4}$. This is checked up to $\CO\left(\CR^{10}\right)$ and $\CO\left(q^2\right)$.

\subsection{Symmetries And Monodromies Of The $U$-plane}
\label{Monos}

In this subsection, we determine the action of the one-form symmetry and monodromies around the singularities. 

To analyze monodromies, we select a base point $U_0=U(\tau_0)$ in the weak-coupling regime of the $U$-plane. Specifically, we take $U_0$ to be a large positive real number. Further imposing the conditions ${\rm Im}(\tau_0) \gg 0$ and $\CR \in (0,1)\subset \mathbb{R}^+$, Eq. \eqref{Uweakcoupling} yields
\be
\tau_0=-2+iT \mod 8,\quad T\gg 0.
\ee
Due to the multi-valuedness of various physical quantities (for example, $a$, $\tau$, $v$) on the $U$-plane, we carefully determine them at $U_0(\tau_0)$. We also choose a base point in the covering of the cylinder $e^{\fa}$ by requiring ${\rm Re}(\fa)\gg 0$. 

The monodromies for the periods $a$, $a_D$ and $a^{(K)}=2\pi i/R$ for this theory was determined by Kanno and Ohta \cite{Kanno:1998vd}  using the Picard-Fuchs equations. In this work, through a combination of the perturbative prepotential, the 4d limit, and general aspects of monodromies, we will derive four independent monodromy transformations ${\bf M}_j$, $j=1,\cdots, 4$, of the period vector $\Pi$ \eqref{eq:periodvector}, which generate the monodromy group. In Sec.  \ref{IntMono}, we will verify that the integrand of the $U$-plane integral is invariant under these monodromies.

\begin{figure}[t]\centering
	\includegraphics[width=0.8\textwidth]{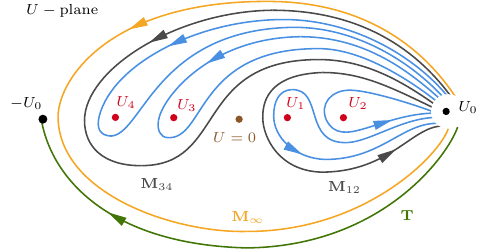} 
\caption{A schematic presentation of the $U$-plane. The red points represent the singularities $U_j$, and the brown dot corresponds to $U=0$. More details about monodromies are explained in Sec.  \ref{Monos}. } \label{UplaneMono}
\end{figure}

\subsubsection*{The Action Of The One-Form Symmetry On Periods}

As mentioned above (see Eq. \eqref{eq:T1form}), the Coulomb branch parameter $U$ transforms under the one-form symmetry $\mathbb{Z}_2^{(1)}$ as $U\to -U$, since the Wilson loop $W_F$ defined in \eqref{DefWF} is in the fundamental representation. In the reduction to the Abelian theory, this one-form symmetry can be viewed as a shift of the Abelian gauge field by a flat connection \cite{Freed:2006yc}. In the Abelian theory the involution changes the local coordinate $\mathfrak{a}$ by a shift by a half-period,
\be
\label{eq:fraOneForm}
\mathfrak{a}\to\mathfrak{a}-i \pi.
\ee
The action of this transformation on the periods $\Pi$ can be computed at weak coupling using the prepotential. In the branch where ${\rm Re}(\mathfrak{a}) > 0$, the polylogarithm is single-valued. Thus, the change in the perturbative prepotential \eqref{prepotential2} arises only from the polynomial terms in $a$, while the instanton part $\CF^{\rm inst}$ \eqref{eq:StructureInstantonExp} remains invariant under the shift \eqref{eq:fraOneForm}. The resulting transformation of the periods is given by 
\be
\begin{aligned} 
a_D &\mapsto a_D+4a-a^{(I)},\\
a &\mapsto a-\frac{1}{2}a_K,\\
a_D^{(I)} &\mapsto a_D^{(I)}-a+\frac{1}{4}a^{(K)},\\
a^{(I)} &\mapsto a^{(I)},\\
a_D^{(K)} &\mapsto \frac{1}{2}a_D+2a-\frac{1}{4}a^{(I)}+a_D^{(K)},\\
a^{(K)} &\mapsto a^{(K)}.
\end{aligned}
\ee
Equivalently, its action on the period vector $\Pi$ can be represented by the symplectic matrix $\bf T$,
\be
\label{Ttrafo}
{\bf T}=\left( \begin{array}{cccccc} 1 & 4 & 0 & -1 & 0 & 0 \\ 0 & 1 & 0 & 0 & 0 & -\frac{1}{2} \\ 0 & -1 & 1 & 0 & 0 & \frac{1}{4} \\ 0 & 0 & 0 & 1 & 0 & 0 \\ \frac{1}{2} & 2 & 0 & -\frac{1}{4} & 1 & 0 \\ 0 & 0 & 0 & 0 & 0 & 1\end{array}\right).
\ee 
The induced transformations of the couplings $\tau$, $v_I$, $v_K$, $\xi_{II}$, $\xi_{IK}$ and $\xi_{KK}$ are
\be
\label{Ttrafotauv}
\begin{aligned}
\tau &\mapsto \tau+4,\\
v_I &\mapsto v_I-1,\\
v_K &\mapsto v_K+\frac{1}{2}\tau+2,\\
\xi_{II} &\mapsto \xi_{II},\\
\xi_{IK} &\mapsto \xi_{IK}+\frac{1}{2}v_I-\frac{1}{4},\\
\xi_{KK} &\mapsto \xi_{KK}+v_K+\frac{1}{4}\tau+1,
\end{aligned}
\ee
which can also be deduced from the expression \eqref{eq:5dPrepotential:U=-2} for the prepotential.

\subsubsection*{Monodromy around $|U|=\infty$}

The monodromy ${\bf M}_\infty$ around $|U|=\infty$ corresponds to the shift 
\be 
\mathfrak{a}  \to \mathfrak{a} +2\pi i. 
\ee
For ${\rm Re}(\fa)\gg 0$, the action of ${\bf M}_\infty$ on the periods can be derived from the prepotential \eqref{prepotential}, following a procedure similar to that used for the action for the one-form symmetry ${\bf T}$. We find
\be
\label{eq:MInftyTrafos}
\begin{aligned}
a_D &\mapsto a_D-8a+2  a^{(I)} -6a^{(K)},\\
a &\mapsto  a+a^{(K)},\\
a_{D}^{(I)} &\mapsto a_{D}^{(I)}+2  a+ a^{(K)},\\
a^{(I)} &\mapsto a^{(I)},\\
a_D^{(K)} &\mapsto -a_D+2a- a^{(I)}+a_D^{(K)}+a^{(K)},\\
a^{(K)} &\mapsto a^{(K)},
\end{aligned}
\ee
which corresponds to the action of the symplectic matrix
\be
\label{eq:Minfty}
{\bf M}_\infty=\begin{pmatrix}
			1& -8 & 0&  2 & 0 & -6 \\
			0&1&0 & 0 & 0 & 1 \\
			0&2  & 1 &0  & 0 & 1 \\
                        0&0&0 & 1 & 0 & 0 \\
                        -1 & 2 & 0 & -1 & 1 & 1\\
			0&0&0 & 0 & 0 & 1 
		\end{pmatrix}.
\ee
The induced transformations of the couplings are
\be\label{eq:Monodromy-At-Infinity}
\begin{aligned}
\tau &\mapsto \tau-8,\\
v_I &\mapsto v_I+2,\\
v_K &\mapsto v_K-\tau+2,\\
\xi_{II} &\mapsto\xi_{II},\\
\xi_{KK} &\mapsto \xi_{KK}-2v_K+\tau-1,\\
\xi_{KI} &\mapsto \xi_{KI}-v_I-1.
\end{aligned}
\ee

The monodromy ${\bf M}_\infty$ can also be interpreted as encircling all strong-coupling singularities. 
As a result, we can decompose ${\bf M}_\infty$ in terms of two monodromies, ${\bf M}_{34}$ followed by ${\bf M}_{12}$, where ${\bf M}_{12}$ and ${\bf M}_{34}$ encircle singularities $\{U_1, U_2\}$ and $\{U_3, U_4\}$, respectively. 
The monodromy ${\bf M}_{12}$ corresponds to the transformation $a\mapsto e^{\pi i} a=-a$. Under this transformation, the instanton part of the prepotential $\CF^{\rm inst}$ \eqref{eq:StructureInstantonExp} remains invariant. For the perturbative parts of the periods given in Eq.  \eqref{aDpert}, we use Eq.  \eqref{Liinv2} for the transformation of the polylogarithms involving Bernoulli polynomials. On the principal branch of the logarithm, the argument of the Bernoulli polynomials becomes
\be 
\frac{1}{2}+\frac{\log(-e^{2\fa})}{2\pi i}=\frac{1}{2}+\frac{\ln(e^{\pi i}\,e^{2\fa})}{2\pi i}=1+\frac{\fa}{\pi i}.
\ee 
Therefore, the periods transform under the action of ${\bf M}_{12}$ as
\be
\begin{aligned}
a_D & \mapsto-a_D+4 a, \\
a &\mapsto e^{\pi i} a=-a, \\
a_D^{(I)} &\mapsto a_D^{(I)}, \\
a^{(I)} &\mapsto a^{(I)}, \\
a_D^{(K)} &\mapsto a_D^{(K)}, \\
a^{(K)} &\mapsto a^{(K)},
\end{aligned}
\ee
which can be represented by a matrix
\be
{\bf M}_{12}=\begin{pmatrix}
			-1& 4 & 0&  0 & 0 & 0 \\
			0& -1&0 & 0 & 0 & 0\\
			0&0 & 1 &0   & 0 & 0 \\
                        0&0&0 & 1 & 0 & 0\\
			0&0&0 & 0 & 1 &0 \\
                        0&0&0 & 0 & 0 &1 \\
		\end{pmatrix}.
\ee
This transformation acts on the effective coupling $\tau$ as 
\be
{\bf M}_{12}:\quad \tau\mapsto \tau-4.
\ee
The monodromy ${\bf M}_{34}$ can be determined from the relation ${\bf M}_\infty={\bf M}_{12}{\bf M}_{34}$, yielding
\be
{\bf M}_{34}=\begin{pmatrix}
			-1& 4 & 0&  -2 & 0 & 2 \\
			0& -1 &0 & 0 & 0 & -1 \\
			0 & 2 & 1 & 0   & 0 & 1 \\
                        0&0&0 & 1 & 0 & 0\\
			-1&2&0 & -1 & 1 & 1 \\ 
                        0&0&0 & 0 & 0 & 1 \\ 
		\end{pmatrix}.
                \ee
This result can also be verified using the transformation $a\mapsto -a-a^{(K)}$ and applying Eqs. \eqref{aDpert} and \eqref{Liinv2}. 
An alternative representation is given as a conjugation,
\be 
{\bf M}_{34}={\bf T}{\bf M}_{12}{\bf T}^{-1}.
\ee 

The results for ${\bf M}_{12}$ and ${\bf M}_{34}$ constrain the individual monodromies ${\bf M}_j$ around a single singularity $U_j$, $j=1,\cdots,4$, through the factorization 
\be 
\label{eq:M12M34}
{\bf M}_{12}={\bf M}_{1}{\bf M}_{2}, \quad {\bf M}_{34}={\bf M}_{3}{\bf M}_{4}.
\ee
To determine the matrices ${\bf M}_j$, we require that their upperleft $2\times 2$ block is the one of pure SW theory, namely the monodromy around the monopole singularity for $j$ odd, and of the dyon singularity for $j$ even. Moreover, we require that the monodromies are in agreement with the Lefshetz formula, i.e. each period shifts by a some multiple of the period of the vanishing cycle. This allows to solve for the matrix form of ${\bf M}_j$. All monodromy matrices determined in the following are elements of $\mathrm{Sp}(3,\mathbb{Z})$.

Ideally, we would like to confirm these monodromies using analytic continuation of Picard-Fuchs solutions, as for example studied in
\cite{Kanno:1998vd, Brini:2021wrm}.

\subsubsection*{Monodromy around $U=U_1$}
The monodromy ${\bf M}_1$ around the strong-coupling singularity $U=U_1$ acts on the periods as
\be
\begin{aligned} 
a_D &\mapsto a_D,\\
a &\mapsto -a_D+a,\\
a_D^{(I)} &\mapsto a_D^{(I)},\\
a^{(I)} &\mapsto a^{(I)},\\
a_D^{(K)} &\mapsto a_D^{(K)},\\
a^{(K)}&\mapsto a^{(K)}.
\end{aligned}
\ee
Expressed as a matrix, we have
\be
{\bf M}_1=\begin{pmatrix}
			1& 0 & 0&  0 & 0 & 0 \\
			-1& 1&0 & 0 & 0 & 0\\
			0&0 & 1 &0   & 0 & 0\\
                        0&0&0 & 1 & 0 & 0\\
			0&0&0 & 0 & 1 & 0 \\
			0&0&0 & 0 & 0 & 1 
                      \end{pmatrix},
                      \ee
which is in agreement with the result in \cite{Kanno:1998vd}. The massless particle at the singularity carries the charge $\gamma_1=[1,0,0,0,0,0]$.

To determine the action of this monodromy on the couplings, we consider the transformed periods $\tilde \Pi=(\tilde a_D,\tilde a,\cdots)$, and compute the $3\times 3$ Jacobian matrix,
\be 
\frac{\partial (a, a^{(I)},a^{(K)})}{\partial (\tilde a, \tilde a^{(I)}, \tilde a^{(K)})}= \left(\frac{\partial (\tilde a, \tilde a^{(I)}, \tilde a^{(K)})}{\partial (a, a^{(I)}, a^{(K)})} \right)^{-1}.
\ee 
For ${\bf M}_1$, this Jacobian evaluates to
\be 
\left( \begin{array}{ccc} \partial a/\partial {\tilde a}\,\, &  \partial a/\partial {\tilde a}^{(I)}\,\, & \partial a/\partial {\tilde a}^{(K)} \\ \partial a^{(I)}/\partial {\tilde a}\,\, &  \partial a^{(I)}/\partial {\tilde a}^{(I)}\,\, & \partial a^{(I)}/\partial {\tilde a}^{(K)} \\ \partial a^{(K)}/\partial {\tilde a}\,\, &  \partial a^{(K)}/\partial {\tilde a}^{(I)}\,\, & \partial a^{(K)}/\partial {\tilde a}^{(K)}\end{array} \right)=\frac{1}{-\tau+1}\left( \begin{array}{ccc} 1\,\, & v_I\,\, & v_K\,\, \\ 0\,\, &  -\tau+1\,\, & 0 \\ 0\,\,& 0 \,\, & -\tau +1\end{array} \right).
\ee 
The transformed coupling $\tilde \tau$ is then obtained as
\be 
\tilde \tau=\frac{\partial \tilde a_D}{\partial a} \frac{\partial a}{\partial \tilde a}+\frac{\partial \tilde a_D}{\partial a^{(I)}} \frac{\partial a^{(I)}}{\partial \tilde a}+\frac{\partial \tilde a_D}{\partial a^{(K)}} \frac{\partial a^{(K)}}{\partial \tilde a}
=\frac{\tau}{-\tau+1},
\ee 
and the transformed coupling $\tilde \xi_{II}$ is
\be 
\tilde \xi_{II}=
    \frac{\partial \tilde{a}^{(I)}_D}{\partial a} \frac{\partial a}{\partial \tilde{a}^{(I)}}+\frac{\partial \tilde{a}_D^{(I)}}{\partial a^{(I)}} \frac{\partial a^{(I)}}{\partial \tilde{a}^{I}}+\frac{\partial \tilde{a}_D^{(I)}}{\partial a^{(K)}} \frac{\partial a^{(K)}}{\partial \tilde{a}^{(I)}}=\frac{v_I^2}{-\tau+1}+\xi_{II}. 
\ee 
All the transformed couplings can be determined in this way,
\be\label{eq:U1-monotmns}
\begin{aligned}
\tau  &\mapsto \frac{\tau}{-\tau+1}, \cr
v_I &\mapsto \frac{v_I}{-\tau+1},\cr
v_K &\mapsto \frac{v_K}{-\tau+1},\cr
\xi_{II} &\mapsto \xi_{II}+\frac{v_I^2}{-\tau+1},\cr
\xi_{KK}  &\mapsto \xi_{KK}+\frac{v_K^2}{-\tau+1},\cr
\xi_{KI} &\mapsto \xi_{KI}+\frac{v_K\,v_I}{-\tau+1}.\cr
\end{aligned}
\ee

\subsubsection*{Monodromy around $U=U_2$}

Proceeding similarly for the strong-coupling singularity $U=U_2$, we find the monodromy ${\bf M}_2$ expressed as matrix,
\be
{\bf M}_2=\begin{pmatrix}
			-1& 4 & 0&  0 & 0 & 0\\
			-1& 3&0 & 0 & 0 & 0 \\
			0&0 & 1 &0   & 0 & 0 \\
                        0&0&0 & 1 & 0 & 0 \\
			0&0&0 & 0 & 1 & 0 \\
                	0&0&0 & 0 & 0 & 1      
		\end{pmatrix}.
\ee
The massless particle at the singularity carries charge $\gamma_2=[-1,2,0,0,0,0]$.

\subsubsection*{Monodromy around $U=U_3$}

The singularity $U_3$ is related to $U_1$ by the $\mathbb{Z}_2^{(1)}$ transformation ${\bf T}$. Therefore, one possibility for a monodromy transformation around $U_3$ is by conjugation of ${\bf M}_1$, 
\be
\widetilde {\bf M}_3={\bf T}{\bf M}_1 {\bf T}^{-1},
\ee
which corresponds to a curve encircling $U_3$ in Fig. \ref{UplaneMono}. However, for our specific purpose of verifying the invariance of the $U$-plane integrand under the monodromy group in Sec.  \ref{Sec:Uplane}, it is more convenient to require that the upperleft $2\times 2$ block of ${\bf M}_3$ matches that of ${\bf M}_1$. Following the approach described below Eq.  (\ref{eq:M12M34}), we obtain
\be
{\bf M}_3=\begin{pmatrix}
			1& 0  & 0&  0 & 0 & 0 \\
			-1& 1&0 & -1 & 0 & 0\\
			1 &0 & 1 &1   & 0 & 0 \\
                        0&0&0 & 1 & 0 & 0 \\
			0&0&0 & 0 & 1 & 0 \\
			0&0&0 & 0 & 0 & 1 \\
                      \end{pmatrix},
\ee
which can be expressed as a conjugation of $\widetilde {\bf M}_3$ by the monodromy ${\bf M}_4$ derived below,
\be
{\bf M}_3={\bf M}_4\widetilde {\bf M}_3 {\bf M}_4^{-1}.
\ee 
The massless particle at this singularity carries charge $\gamma_3=[1,0,0,1,0,0]$. 
The action of this monodromy on the couplings is
\be 
\label{U3trafos}
\begin{aligned} 
\tau &\mapsto \frac{\tau}{-\tau+1},\\
v_I &\mapsto \frac{v_I+\tau}{-\tau+1},\\
v_K &\mapsto \frac{v_K}{-\tau+1},\\
\xi_{II} &\mapsto \xi_{II}+\frac{(v_{I}+1)^2}{-\tau+1},\\
\xi_{KK} &\mapsto \xi_{KK}+\frac{v_{K}^2}{-\tau+1},\\
\xi_{IK} &\mapsto \xi_{IK}+\frac{(v_I+1)v_{K}}{-\tau+1}.
\end{aligned}
\ee

\subsubsection*{Monodromy around $U=U_4$}

Again, following the approach described below Eq.  (\ref{eq:M12M34}), we get
\be  
{\bf M}_4=\begin{pmatrix}
			-1& 4 & 0&  -2 & 0 & 2 \\
			-1& 3&0 & -1 & 0 & 1 \\
			1&-2 & 1 &1   & 0 & -1 \\
                        0&0&0 & 1 & 0 & 0\\
			-1&2&0 & -1 & 1 & 1 \\ 
			0&0&0 & 0 & 0 & 1  
                      \end{pmatrix}.
\ee
This monodromy satisfies 
\be
{\bf M}_1{\bf M}_2{\bf M}_3{\bf M}_4={\bf M}_\infty,
\ee
and 
\be 
{\bf M}_4={\bf T}{\bf M}_1{\bf M}_2 {\bf M}_1^{-1}{\bf T}^{-1}.
\ee
The massless particle at the singularity carries charge $\gamma_4=[-1,2,0,-1,0,1]$. The action on the couplings is
\be 
\begin{aligned} 
\tau &\mapsto \frac{\tau-4}{\tau-3},\\
v_I &\mapsto \frac{v_I+\tau-2}{-\tau+3},\\
v_K &\mapsto \frac{v_K-\tau+2}{-\tau+3},\\
\xi_{II} &\mapsto \xi_{I I}+\frac{\left(v_I+1\right)^2}{3-\tau}, \\
\xi_{IK} &\mapsto \xi_{I K}+\frac{\left(v_K-1\right)\left(v_I+1\right)}{3-\tau}, \\
\xi_{KK} &\mapsto \xi_{K K}+ \frac{\left(v_K-1\right)^2}{3-\tau}.
\end{aligned} 
\ee 

This concludes our derivation of the four generators ${\bf M}_j$ of the monodromy group.

\subsubsection*{Charges and BPS Quivers}

We include here a brief discussion on the BPS quiver and the BPS spectrum, which is relevant for the partial topological twist of this theory described in Sec.  \ref{sec:TopTwist}. 

A BPS quiver can be derived from the massless particles at the singularities \cite{Denef:2002ru, Denef:2007vg, Alim:2011kw, Chuang:2013wt, Closset:2019juk, Beaujard:2020sgs}. The charges of the massless particles at the singularities $U_j$, $j=1,2,3,4$, are given by
\be 
\label{eq:chargeQuivers}
\begin{aligned}
\gamma_1 &  =[1,0,0,0,0,0], \\
\gamma_2 & =[-1,2,0,0,0,0], \\
\gamma_3 & =-\gamma_1{\bf T}^{-1}{\bf M}_4^{-1}=[1,0,0,1,0,0],\\
\gamma_4 & =\gamma_2{\bf M}_1^{-1}{\bf T}^{-1}=[-1,2,0,-1,0,1].
\end{aligned}
\ee
These charges are eigenvectors of their respective monodromy transformations ${\bf M}_j$ with eigenvalue $1$,
\be
\gamma_j {\bf M}_j = \gamma_j.
\ee
The set of charges $\{\gamma_1,\gamma_2,-\gamma_3, -\gamma_4\}$ defines the following quiver:
\begin{center}
\begin{tikzpicture}[inner sep=2mm,scale=1.5]
  \node (a) at ( -1,1) [circle,draw, minimum size=1.3cm] {$\gamma_1$};
  \node (b) at ( 1,1) [circle,draw, minimum size=1.3cm] {$\gamma_2$};
  \node (c)  at ( 1,-1) [circle,draw,  minimum size=1.3cm] {$-\gamma_3$};
  \node (d)  at ( -1,-1) [circle,draw,  minimum size=1.3cm] {$-\gamma_4$};
 \draw [->] (a.10) to node[auto] {$ $} (b.170);
 \draw [->] (a.350) to node[auto] {$ $} (b.190);
 \draw [->] (b.260) to node[auto] {$ $} (c.100);
 \draw [->] (b.280) to node[auto] {$ $} (c.80);
  \draw [->] (c.170) to node[auto] {$ $} (d.10);
\draw [->] (c.190) to node[auto] {$ $} (d.350);
 \draw [->] (d.80) to node[auto] {$ $} (a.280);
 \draw [->] (d.100) to node[auto] {$ $} (a.260);
\end{tikzpicture}
\end{center}
This quiver is isomorphic to the Hirzebruch surface $\mathbb{F}_0$, which occurs in the local Calabi-Yau geometry for the geometric engineering of the theory \cite{Katz:1996fh, Feng:2000mi}. 
The subquiver formed by the nodes $\gamma_1$ and $\gamma_2$ recovers the BPS quiver of 4d $\mathrm{SU}(2)$ SYM.

Alternatively, one may consider BPS quivers with different charges. For example, one can choose the charges $\tilde \gamma_{3}$ and $\tilde \gamma_{4}$ by acting with the inverse operator ${\bf T}^{-1}$ on the charges $\gamma_1$ and $\gamma_2$,
\be
\tilde \gamma_3 = \gamma_1 {\bf T}^{-1}=[1,-4,0,1,0,-2], \quad
\tilde \gamma_4 = \gamma_2 {\bf T}^{-1}=[-1,6,0,-1,0,3].
\ee 
We note that the combination $-\gamma_1-\gamma_2+\tilde \gamma_3+\tilde \gamma_4=[0,0,0,0,0,1]$ corresponds to the charge of the KK momentum. The set $\{\gamma_1,\gamma_2,-\tilde \gamma_3,-\tilde \gamma_4\}$ defines an alternative BPS quiver. 

BPS degeneracies can in principle be determined from the BPS quiver. The combination $\gamma_1-\gamma_3=[0,0,0,-1,0,0]$ carries only a unit of $\mathrm{U}(1)^{(I)}$ charge. Since $\left\langle \gamma_1,\gamma_3\right\rangle=0$, the BPS index of this bound state vanishes, suggesting that it does not correspond to a stable bound state. This might be related to the unresolved issues mentioned in Footnote \ref{foot:quantissues}. We can instead consider the combination $(n+1)\gamma_1+n\gamma_2-\gamma_3=[0,2n,0,-1,0,0]$, which corresponds to the instanton particle with electric charge $2n$. The 3-node quiver with nodes $\gamma_1$, $\gamma_2$ and $\gamma_3$ does not contain oriented loops. The BPS index for small charges can be evaluated using Coulomb \cite{Denef:2002ru, Denef:2007vg, Manschot:2011xc, Kim:2011sc, Manschot:2013sya, Hori:2014tda} or Higgs branch techniques \cite{Reineke_2003, Lee:2013yka, Duan:2020qjy, Longhi:2021qvz, DelMonte:2021ytz}. More generally, one may show that, if nonempty, the complex dimension of the moduli space of semi-stable quiver representations is $2n-1$, and that the cohomology is in degree $(p,p)$, and thus even \cite{Reineke_2003}. As a result, the BPS states are singlets under the $\mathrm{SU}(2)_R$ symmetry \cite{Gaiotto:2010be}. Moreover, half the complex dimension of the moduli space corresponds to the spin of the representation of the $\mathrm{SU}(2)$ little group. 

For generic charges $\gamma=[q_m,q_e,0,q_I,0,q_K]$, we find that the complex dimension of the moduli space of quiver representations is $(q_m)^2-1 \mod 2$. Thus in particular BPS states with $q_m=0$ but generic perturbative charges $q_e,q_I$ and $q_K$, have strictly half integer spins. The explicit Gopakumar-Vafa data in \cite[Table 1]{Closset:2022vjj} confirm this property.

\subsection{Modular Parametrization Of The $U$-plane}
\label{SecFundDom}

In this subsection, we seek to construct a fundamental domain $\CF_{\CR}$ in the upper half-plane $\IH$ so that we can define a single-valued function $U(\tau,\CR)$ on $\CF_{\CR}$ that bijectively maps from $\CF_{\CR}$ to the $U$-plane $\IC_U$. We want to have this construction because it enables us to formulate the $U$-plane integral  as an integral over $\CF_{\CR}$ with the measure expressed in terms of modular functions, significantly simplifying the computations.
Accordingly, we return to the discussion between Eqs. \eqref{hellipticcurve} and \eqref{eq:SmallRbp}.

\begin{figure}[t]
\centering   
	\includegraphics[width=\textwidth]{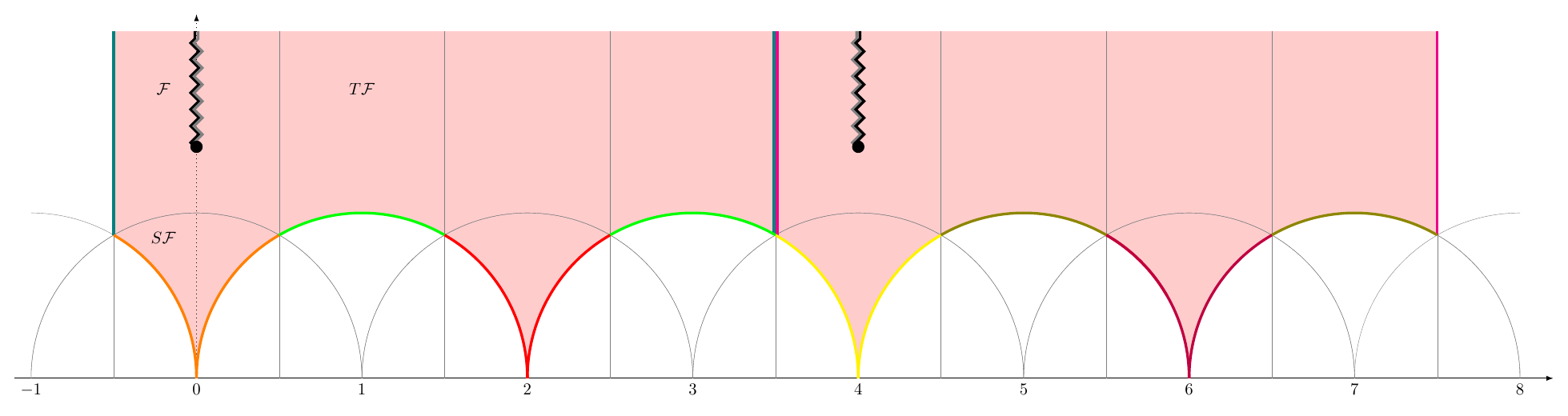} 
	\caption{The integration region for $\tau$, which is equivalent to an integration over the $U$-plane, is a branched double-cover of the modular curve $\mathbb{H}/\Gamma^0(4)$ with a single branch point at the solution for $\tau_0 \mod \Gamma^0(4)$ to the equation $U=0$. We find it convenient to represent this branched double-cover by two adjacent fundamental domains for $\Gamma^0(4)$ in the upper half-plane. Note that at ${\rm Re}(\tau)=7/2$, where the domains are adjacent, the left side of the vertical line is identified with the vertical line at ${\rm Re}(\tau) = -1/2$ while the right side of the vertical line at ${\rm Re}(\tau)=7/2$ is identified with the vertical line at ${\rm Re}(\tau) = 15/2$. A small continuous path in the $\tau$-plane that crosses the line at ${\rm Re}(\tau)=7/2$ does not correspond to a small continuous path in the $U$-plane. In each of the two fundamental domains for $\Gamma^0(4)$ there is a single branch point, and these two points should be identified, together with an identification of each side of the cuts, as indicated by the grey and black colors on each side of the cuts. Note that when written in terms of $\tau$ in the upper half-plane the $U$-plane integrand in Eq. \eqref{UplaneintDef} below jumps discontinuously by a minus sign across the vertical line at ${\rm Re}(\tau) = 7/2$.  }\label{FRCG}
\end{figure}

We construct $\CF_{\CR}$ by first finding a fundamental domain in $\widetilde{\IH}$ for the $\Gamma^0(4)$-covering $\widetilde{\IH} \to \IC_U$, and then mapping that region to $\IH$ via the double cover \eqref{eq:DoubleCover-UHP}. Therefore, the image in $\IH$ consist of two copies of a fundamental domain for $\Gamma^0(4)$ in $\IH$. We can place these two domains next to each other in a $\Gamma^0(4)$-invariant tessellation of $\IH$, as shown in Fig. \ref{FRCG}. 
The branch points of the covering $\pi: \widetilde{\IH} \to \IH$ are located at the solutions to Eq. \eqref{eq:tau-CR-bps}, with one branch point in each fundamental domain for $\Gamma^0(4)$, as indicated by the black dots in Fig. \ref{FRCG}. To guarantee that $U(\tau,\CR)$ is single-valued, we must make a proper choice of branch cuts. Notice that the boundary point $\tau\to i\infty$ is a branch point, since it follows from \eqref{eq:U-doublecovers-ttu} that $U\to -U$ if we let $\tau$ evolve along a straight line segment $\tau\to \tau+4$ for ${\rm Im}(\tau)$ sufficiently large. It is natural to introduce a branch cut from each interior branch point to the boundary branch point $i\infty$, as displayed in Fig. \ref{FRCG}. We identify these two interior branch points, as well as the corresponding sides of the branch cuts, as indicated in Fig. \ref{FRCG}.
In the limit $\CR\to 0$, we see from Eq. \eqref{eq:SmallRbp} that the interior branch points move up towards $i\infty$, and $\CF_{\CR}$ becomes a union of two copies of $\mathbb{H}/\Gamma^0(4)$. This domain has the advantage that one can naturally define local coordinates near the strong-coupling singularities.

To achieve a bijective map to a full copy of the $U$-plane, the boundaries of the two fundamental domains for $\Gamma^0(4)$ should be identified according to the color scheme shown in Fig. \ref{FRCG}. Note particularly that lines along ${\rm Re}(\tau) = 7/2$, at which two domains for $\Gamma^0(4)$ are adjacent, are identified. However, using the function $U(\tau,\CR)$ that we will specify below, a small continuous line segment crossing ${\rm Re}(\tau) = 7/2$ at large ${\rm Im}(\tau)$ in the $\tau$-plane does \emph{not} map to a small continuous path in the $U$-plane. 

Now we can define a single-valued function $U(\tau,\CR)$ on $\CF_{\CR}$, namely the left and right pink regions in Fig. \ref{FRCG}. We define the left region to correspond to the principal branch of the square root and the right region to correspond to the negative principal branch of the square root. Note that 
\be
\lim_{T\to 0} \ttu(2 + i T) = -1.
\ee
With the relation between $U$ and $\ttu$ \eqref{eq:U-doublecovers-ttu}, we obtain 
\begin{equation} 
\label{eq:tauU}
\begin{aligned}  
\tau&\to 0: &  U&\to U_1,\\
\tau&\to 2: &  U&\to U_2,\\
\tau&\to 4: &  U&\to U_3,\\ 
\tau&\to 6: &  U&\to U_4,
\end{aligned}
\end{equation}
with $U_j$ the singularities as in Eq. \eqref{eq:DiscLocus}.

Of course, it is possible to choose a different fundamental domain, which would be a rearranging and mapping of $\CF_\CR$. 
Following the general strategy of \cite{Aspman:2021vhs}, we would get a degree 12 equation for $U$ in terms of the modular invariant $j$-function. In
other words, there is a 12-fold cover of the $U$-plane over the standard keyhole fundamental domain. Since we have already determined Figure \ref{FRCG}, it is most natural to rearrange this domain.
 
\begin{figure}[t]
\centering
\includegraphics[width=1.0\textwidth]{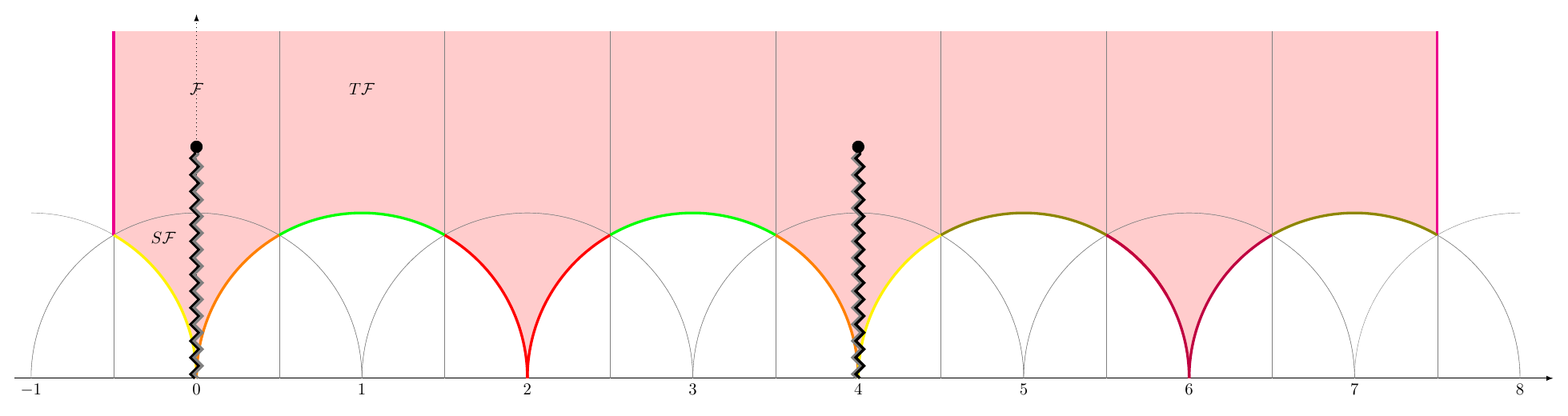} 
\caption{Fundamental domain $\CF_\CR$. Boundary arcs with the same color are identified. The black dots represent branch points and the zig-zag lines branch cuts. The left/right (black/grey) of the left branch cut is identified with the right/left (black/grey) side of the right branch cut. In the limit $\CR\to 0$ the branch points move up.}\label{FR}
\end{figure}

Alternatively, we can naturally obtain a different fundamental domain by changing the choice of branch cuts. In particular, instead of branch cuts extending to $i\infty$ we can cut the domain in Fig. \ref{FRCG} vertically at $\tau=0$, $\tau=7/2$ and $\tau=4$, and then exchange the regions between $[-1/2,0]$ and $[7/2,4]$. The result is the fundamental domain in Fig. \ref{FR}. Now the left and right regions of the branch cuts are not separated by a line. In terms of the square root function, the branch cuts are now along the negative real axis of the argument of the square root. The images of the branch cuts in the $U$-plane in Fig. \ref{UplaneMono} run from $U=0$ to the singularities $U_1$ and $U_3$.

At $\CR^4= 1$, the branch points collide with the real axis, and $\CF_{\CR}$ becomes a fundamental domain for $\Gamma^0(8)\subset \mathrm{SL}_2(\mathbb{Z})$, or a subgroup of $\mathrm{SL}_2(\mathbb{Z})$ conjugate to $\Gamma^0(8)$, where $\Gamma^0(8)$ is the congruence subgroup defined as
\be 
\Gamma^0(8)=\left\{ A\in \mathrm{SL}_2(\mathbb{Z}) \left| A=\left(\begin{array}{cc} a & b \\ c & d\end{array}\right),\quad b=0\mod 8 \right. \right\}.
\ee 
In particular for $\CR^2=-1$, we can identify the $U$-plane with the modular curve $\mathbb{H}/\Gamma^0(8)$. Since $\Gamma^0(4) = \Gamma^0(8) \amalg \Gamma^0(8)\cdot T^4$, 
\be
\mathbb{H}/\Gamma^0(8) \rightarrow \mathbb{H}/\Gamma^0(4)
\ee
is an unbranched double-cover away from the branch cut. The domain $\mathbb{H}/\Gamma^0(8)$ is made manifest in the expression for $U(\tau)$ for  $\CR^2=-1$. The expression \eqref{eq:U-doublecovers-ttu} simplifies in this case such that $U$ can be written as a Hauptmodul for $\Gamma^0(8)$ \cite{Closset:2021lhd},
\be
\label{UR2=-1}
  U(\tau)=\mp 2\frac{\vartheta_2(\tau)^2+\vartheta_3(\tau)^2}{\vartheta_2(\tau)\, \vartheta_3(\tau)},
\ee
where the overall sign reflects the double-cover structure. 

On the other hand, for the choice in Sec.  \ref{Monos} that $\CR$ is real, $\CR^2=1$, and the formula for $U$ simplifies to
\be
\label{UR2=1}
U(\tau)=\mp 2i\frac{\vartheta_2(\tau)^2-\vartheta_3(\tau)^2}{\vartheta_2(\tau)\, \vartheta_3(\tau)}.
\ee
This is not a Hauptmodul of $\Gamma^0(8)$, but rather of a subgroup conjugate to $\Gamma^0(8)$, which we will denote by  
\be 
\tilde \Gamma^0(8)= T^{-2} \Gamma^0(8) T^2 :=\left\{ A\in \mathrm{SL}_2(\mathbb{Z})| A=T^{-2}BT^{2},  B\in \Gamma^0(8) \right\}.
\ee 
Note that we could also write $\tilde \Gamma^0(8) = T^2 \Gamma^0(8) T^{-2}$, because $T^4$ normalizes $\Gamma^0(8)$ in $\Gamma^0(4)$.

At $\CR^2=1$, singularities $U_1$ and $U_3$ merge, the cusps at $\tau=0$ and $\tau=4$ join together, and the periodicity in the local modular parameter $\tau_1=-1/\tau$ becomes 2. That is to say $U$ \eqref{UR2=1} is periodic under $\tau_1\to \tau_1+2$. This can also be seen from the identifications in Figs. \ref{FRCG} and \ref{FR} for $\CR^2=1$. The cusps at $\tau=2$ and $\tau=6$ have width 1 as for $\CR^2<1$. That is to say, if we define $\tau_2$ through $\tau=2-1/\tau_2$, then $U$ \eqref{UR2=1} is periodic under $\tau_2\to \tau_2+1$, and similarly for $\tau_6\to \tau_6+1$.

\section{Partial Topological Twisting On $X\times S^1$}
\label{sec:TopTwist}

We consider the theory on $X\times S^1$, where $X$ is, as before, a smooth, compact, oriented four-manifold without boundary. 
Let $L$ denote the lattice in $H^2(X,\mathbb{R})$ obtained by embedding $H^2(X,\mathbb{Z}) \rightarrow H^2(X,\mathbb{R})$. The kernel of this embedding is the torsion subgroup of
$H^2(X,\mathbb{Z})$. The intersection form on $X$ is denoted by
\footnote{We will use the same notation $B$ for the case when the coefficients are $\IZ_2$ or $\IR$.}
\be 
B: H^2(X,\mathbb{Z})\times H^2(X,\mathbb{Z}) \rightarrow \mathbb{Z},
\ee
with signature $(b_2^+,b_2^-)$. 
For simplicity (and only for simplicity) we assume $X$ is simply connected. As explained below, anomaly cancellation will require that $X$ admits an almost complex structure. 
There is a mild topological condition for this: If $X$ is simply connected, or if $H^2(X)$ has indefinite signature, then a sufficient topological condition for the existence of an almost complex structure is
$b_2^+ + b_1 = 1 \mod 2$ \cite{Donaldson90, Scorpan}. 

The ``zero-form''  global symmetry group of the 5d SYM theory on $\mathbb{R}^5$
is expected to be
\be 
\frac{\mathrm{Spin}(5) \times \mathrm{SU}(2)_R }{\IZ_2}  \times  \mathrm{U}(1)^{(I)},
\ee
where the $\IZ_2 =\langle (-1, -1) \rangle$ is the diagonal subgroup of the product of the centers $Z(\mathrm{Spin}(5)) \times Z(\mathrm{SU}(2)_R)$. 
Note that the supersymmetries of the theory 
are invariant under $(\mathrm{Spin}(5)\times \mathrm{SU}(2)_R)/\langle (-1,-1)\rangle$. 
The $\IZ_2$ subgroup we quotient by has trivial intersection with the $\mathrm{U}(1)^{(I)}$ factor. 
This might seem a bit surprising. We make this choice because of the representation content of the BPS states found after  compactification along a circle. 
As shown in the quiver analysis at the end of Sec.  \ref{Monos}, the BPS states transform in the representation $\rho_{hh} \otimes \mathfrak{h}$, where $\rho_{hh}$ is the ``half-hypermultiplet'' representation of $({\bf 2};{\bf 1}) \oplus ({\bf 1};{\bf 2})$ of $\mathrm{Spin}(3) \times \mathrm{SU}(2)_R$, 
and is a singlet under $\mathrm{U}(1)^{(I)}$, and $\mathfrak{h}$ is of the form $({\bf v};{\bf 1})^{(q_I)}$ with ${\bf v}$ an even-dimensional (a.k.a. half-integer) representation of $\mathrm{Spin}(3)$ for all instanton charges $q_I$.  

Now we turn to the dynamical 5d fields on general five-manifolds. The fields in the vector multiplet of 5d SYM on Euclidean spin manifolds are sections of bundles with structure group $( \mathrm{Spin}(5) \times \mathrm{SU}(2)_R)/\mathbb{Z}_2 \times G$.
\footnote{All fields are in the adjoint representation of $G$. When $G$ has a nontrivial center there are distinct theories obtained by choosing different groups $G$ with the same adjoint group.  
}
In this paper we restrict to gauge group $G=\mathrm{SU}(2)$ or $\mathrm{SO}(3)$.

We would like to evaluate correlation functions of the theory on five-manifolds of the form $X \times S^1$, where $S^1$ carries a spin structure but $X$ does not need to, and indeed we wish to include manifolds $X$ which do not even admit any spin structure. 
This is achieved through a suitable notion of topological twisting using a carefully chosen background gauge field for the $\mathrm{SU}(2)_R$ symmetry.  
The definition of the twisting is slightly subtle since the twisted theory should be independent of any
spin structure (or even spinnability) of $X$ but should depend on the spin structure of $S^1$. 

According to our viewpoint, topological twisting is based on a choice of background fields in the ``physical theory'', implemented via transfer of the structure group along a twisting homomorphism. For more discussion of this perspective, see  \cite{Moore:2024vsd}.
In our case the relevant homomorphism is 
\be 
\varphi: \mathrm{Spin}(1) \times \mathrm{SO}(4) \rightarrow  (\mathrm{Spin}(5) \times \mathrm{SU}(2)_R)/\mathbb{Z}_2.
\ee
Here $\mathrm{Spin}(1)$ is identified with $\mu_2$, the multiplicative group of order $2$. Let $\zeta: \mathrm{Spin}(4) \hookrightarrow \mathrm{Spin}(5)$ be an embedding so that the vector representation $\mathbf{5}$ of $\mathrm{Spin}(5)$ pulls back to a vector representation $\mathbf{4}$ plus a trivial representation $\mathbf{1}$ of $\mathrm{Spin}(4)$.  
\footnote{The commutant of the image of $\zeta$ in $\mathrm{Spin}(5)$ is $\{ \pm 1, \pm \gamma_5\}$, which is a subgroup of the image of $\zeta$ itself.
This implies that there exists no embedding $\zeta': \mathrm{Spin}(1) \times \mathrm{Spin}(4) \hookrightarrow \mathrm{Spin}(5)$ so that $\zeta$ is given by the composition 
\be 
\mathrm{Spin}(4) \hookrightarrow \mathrm{Spin}(1)\times \mathrm{Spin}(4) \hookrightarrow \mathrm{Spin}(5). 
\ee
Put differently, on a general product manifold $X\times S^1$ we cannot decompose every 5d spinor as a tensor product of 1d and 4d spinors. 
To see this obstruction, note that if $\{\gamma^i\}$ gives a four-dimensional representation of $Cl_5$, then 
$\zeta ( \pm \gamma\cdot n_1 \cdots \gamma\cdot n_{2k})
=\pm \gamma\cdot n_1 \cdots \gamma\cdot n_{2k}$, where $n_i$ are 4d unit vectors. 
If $\pm \gamma\cdot v_1 \cdots \gamma\cdot v_{2k}$  were in the commutant (with $v_i$ 5d unit vectors), 
then $v_i$ must either be 4d unit vectors or $\pm e_5$, which is in contradiction with the fact that $\gamma_5 \in \mathrm{Spin}(4)$.
Thus, we need to introduce the extra $\mathrm{Spin}(1)$ factor in the domain of $\varphi$ by hand.}
If $(u_1, u_2) \in \mathrm{SU}(2)\times \mathrm{SU}(2) \cong \mathrm{Spin}(4)$ is a lifting of $R\in \mathrm{SO}(4)$, then the expression $[(\zeta(u_1,u_2),u_2)]\in (\mathrm{Spin}(5) \times \mathrm{SU}(2)_R)/\mathbb{Z}_2 $ is well defined. For $\xi\in \mu_2$ and $R\in \mathrm{SO}(4)$, we define $\varphi$ as 
\be\label{eq:PartialTopTwistMap}
\varphi(\xi, R) = [(\zeta(\xi u_1,\xi u_2),u_2)] =  [(\zeta(u_1,u_2),\xi u_2)]
\in (\mathrm{Spin}(5) \times \mathrm{SU}(2)_R)/\mathbb{Z}_2.
\ee
Due to the quotient, this definition does not depend on the choice of a lifting of $R$.
The background connection for the principal $(\mathrm{Spin}(5) \times \mathrm{SU}(2)_R)/\mathbb{Z}_2$-bundle is then chosen as the image of the Levi-Civita connection for $\mathrm{SO}(4)$ with the natural mapping of connections arising from the reduction of the structure group.  

The supercharges in 5d $\CN=1$ theory are in the representation $(\bf 4;\bf 2)_{\IR}$ of 
$(\mathrm{Spin}(5) \times \mathrm{SU}(2)_R)/\mathbb{Z}_2$. 
The real structure exists because the representation $(\bf 4;\bf 2)$ is a tensor product of pseudoreal representations of $\mathrm{Spin}(5)$ and $\mathrm{SU}(2)_R$. As a vector space, the real dimension of this representation is $8$. 
Its pullback under $\varphi$ is a real representation of $\mathrm{SO}(4)$, which is decomposed into a direct sum of irreducible representations 
\be
(\mathbf{1},\mathbf{1})\oplus (\mathbf{3}, \mathbf{1}) \oplus  (\mathbf{2},\mathbf{2}).
\ee 
Each component transforms in the nontrivial representation of $\mu_2$. 
The $(\mathbf{1},\mathbf{1})$ component of the supercharges is denoted by $\bar\CQ$. 

Similarly, the fermions decompose under $\mathrm{SO}(4)$ as 
\be
(\mathbf{1},\mathbf{1})\oplus (\mathbf{3}, \mathbf{1}) \oplus  (\mathbf{2},\mathbf{2}),
\ee 
which correspond to a scalar $\eta$, a self-dual two-form $\chi$, and a one-form $\psi$ on $X$, respectively. 
Again, all components transform nontrivially under  $\mu_2$. 
Thus, omitting the pullbacks from the projections to the two factors, the fermions are valued in $\Gamma[L\otimes \Omega^0(X)]$, $\Gamma[L\otimes \Omega^{2,+}(X)]$, and $\Gamma[L \otimes \Omega^1(X)]$, respectively, where $L\to S^1$ is the real line bundle associated with the spin double-cover.

The pullback by $\varphi$ of the bosonic fields  decompose into irreducible representations of $\mathrm{SO}(4)$ as
\be
(\mathbf{2},\mathbf{2}) \oplus (\mathbf{1},\mathbf{1}) \oplus (\mathbf{1},\mathbf{1})\oplus (\mathbf{3},\mathbf{1}).
\ee
The first two components correspond to the gauge field
\be
A_m \mathrm{d}x^m = A_\mu \mathrm{d}x^\mu + A_5 \mathrm{d}x^5,
\ee 
where $m=1,\cdots, 5$ and $\mu=1,\cdots, 4$. We will denote the one-form gauge field on $X$ by 
$A_X = A_\mu \mathrm{d}x^\mu$. 
The third component corresponds to a real adjoint scalar $\sigma$, and the last component corresponds to an auxiliary field $D$, which is an adjoint-valued self-dual two-form on $X$. 

This twisting preserves the scalar supercharge $\bar \CQ$, and extends the action \eqref{mixed CS action} to  five-manifolds $X\times S^1$ for any oriented four-manifold $X$, regardless of whether $X$ is spin. 
The twisted fields transform under $\bar \CQ$ as 
\footnote{\label{footnote:SD} The self-dual part $F_+$ and the anti-self-dual part $F_-$ of the field strength two-form $F_X$ on $X$ are defined using the Hodge dual $*$ as $F^{\pm}=F_{\pm} = \frac{1}{2} \left(F_X \pm * F_X\right)$.}
\begin{equation}\label{eq:Qbar-nonabelian}
    \begin{aligned}
    \left[ \mathcal{\bar Q},  A_X\right] &= -  \psi,  \\
    \left[ \mathcal{\bar Q},A_5\right] &= \eta,  \\
    \left[\mathcal{\bar Q},   \sigma\right] &=-i\eta, \\
    \left\{\mathcal{\bar{Q}}, \psi \right\} &= 2 (D_{A_X} \sigma - i \iota_{\partial_5}F ), \\
    \{\mathcal{\bar Q} ,\eta \} &= -2 D_5 \sigma, \\
    \left\{ \mathcal{\bar Q}, \chi \right\}  &= -2i F_+  -i D,   \\
    \left[\mathcal{\bar Q}, D\right] &= 2(D_{A_X}\psi)_+ 
    -2 [D_5 -\sigma,\chi],
\end{aligned}
\end{equation}
satisfying 
\begin{equation}
    \mathcal{\bar Q}^2
    = 
    2 \delta_{\sigma+ i A_5} + 2i \partial_5.
\end{equation}
Here $\delta_{\sigma+ i A_5}$ is the gauge variation defined by
\begin{equation}
    \delta_{\sigma+i A_5} A_m = 
    -D_m(\sigma+i A_5), 
    \quad 
    \delta_{\sigma+iA_5} \varphi = -i[\sigma+i A_5, \varphi],
\end{equation}
where $\varphi$ represents all fields except $A_m$.

A canonical way to construct observables involves the $(\mathbf{2}, \mathbf{2})$ component of the supercharges $\CK = \CK_\mu \mathrm{d}x^\mu$, which obeys 
\begin{equation}
    \{\mathcal{\bar Q}, \mathcal{K}\} = D_{A_X}.
\end{equation}
The action of $\CK$ is explicitly given by
\begin{equation}
    \begin{aligned}
        \left[\mathcal{K}, A_X \right] &= \frac{i}{2} \chi, \\
        \left[\mathcal{K}, A_5\right] &= -\frac{i}{4}\psi, \\
        \left[\mathcal{K}, \sigma\right] &= \frac{1}{4}\psi, \\
        \left\{\mathcal{K}, \psi\right\} &= D -2 F_{-}, \\
        \left\{
        \mathcal{K}, \eta
        \right\} & = 
        -\frac{i}{2}
        (D_\mu \sigma + i F_{\mu 5})
        \mathrm{d}x^\mu, \\
        \left\{\mathcal{K},\chi\right\} 
        & = \frac{i}{8}\epsilon_{\nu\mu\rho\sigma} (D^\nu \sigma + i F^{\nu 5}) \mathrm{d}x^\mu \wedge \mathrm{d}x^\rho \wedge \mathrm{d}x^\sigma, \\
        \left[\mathcal{K}, D\right] &=
         -\frac{i}{8} \epsilon_{\gamma\tau\mu\nu}\left(D^\gamma \eta + 2 D^\rho \chi_\rho{}^\gamma + [D_5 + \sigma, \psi^\gamma]\right)\mathrm{d}x^\tau \wedge \mathrm{d}x^\mu \wedge \mathrm{d}x^\nu.
    \end{aligned}
\end{equation}

\section{Partition Function Of The Partially Twisted Theory And Its Relation To K-theoretic Donaldson Invariants}
\label{sec:KtheoreticPF}

In this section, we consider the 5d $\CN=1$ SYM on $X \times S^1$ with a partial topological twist realized by turning on background $\mathrm{SU}(2)_R$-symmetry gauge fields, as defined in Sec.  \ref{sec:TopTwist}. We will argue that the partition function of this theory is equivalent to that of a SQM with one real supercharge (denoted $\bar \CQ$) and target space given by the moduli space of instantons on $X$. 
Therefore, when the $S^1$ direction is endowed with a nonbounding spin structure (a.k.a. Ramond or periodic boundary condition), the partition function computes the Witten index of this SQM. 
This index corresponds to the most basic example of the ``K-theoretic Donaldson invariants'' of $X$. 
While this correspondence goes back to work of Nekrasov (see \cite[Appendix]{Nekrasov:1996cz}), we will address several new aspects such as global anomalies and their cancellation.

\subsection{First Reduction To Effective Supersymmetric Quantum Mechanics}
\label{subsec:redu2SQM}

The background fields $V^{(I)}$ associated with the $\mathrm{U}(1)^{(I)}$ symmetry will play a very important role in our discussion. We begin by putting all the fermions in the multiplet to zero. We will also assume that the $\mathrm{U}(1)^{(I)}$-bundle is pulled back from $X$ and that the gauge field  $A^{(I)}$ is flat but we will 
allow the holonomy around $S^1$ in Eq. \eqref{eq:thetadef} to be non-vanishing. 
(We will later generalize to allow nontrivial periods of $F^{(I)}$.) 
Possible holonomy around cycles in $X$ will play no role. 
In particular, we have 
\be\label{eq:BackgndFirstRed}
F^{(I)} = 0, \quad \partial_\mu \oint_{S^1} A^{(I)} =0, 
\quad \partial_5 A^{(I)}_\mu = 0, 
\ee
 
Moreover, we take the same background specified near \eqref{eq:sigmaI-5dcoupling}: 
\begin{equation}\label{eq:BackgndFirstRed-2}
\sigma^{(I)} = -\frac{8\pi^2}{ g^2_{5d}}, 
\quad (D^{AB})^{(I)} = 0, \quad \lambda^{(I)}_A=0. 
\end{equation}
These conditions allow us to simplify \eqref{mixed CS action}, yielding
\begin{equation}
S_\text{twisted} = \frac{i }{8\pi^2}\int_{X\times S^1} A^{(I)} \wedge \text{tr}(F \wedge F) 
-\frac{ \sigma^{(I)}}{8\pi^2}\int_{X\times S^1}  \mathrm{d}^5 x\, \sqrt{g} \mathcal{L}^\text{twisted}_\text{SYM}, 
\end{equation}
where $\mathcal{L}^\text{twisted}_\text{SYM}$ is given by
\begin{equation}
\label{eq:LSYMtwisted}
\begin{aligned}
    \mathcal{L}^\text{twisted}_\text{SYM} 
     &= \tr\bigg(\frac{1}{2} F^{\mu\nu} F_{\mu\nu} + F_{\mu 5} F^{\mu 5} - \frac{1}{4} D_{\mu\nu} D^{\mu\nu}
 + D_\mu \sigma D^\mu \sigma 
    + D_5\sigma D^5\sigma\\
    &\quad + 2i \chi^{\mu\nu} D_{\mu}\psi_\nu 
    -\frac{i}{2} \chi^{\mu\nu}\left[D_5 -\sigma, \chi_{\mu\nu}\right]
    -\frac{i}{2} \eta \left[D_5-\sigma,\eta\right] 
    + i \psi^\mu D_\mu \eta  +  \frac{i}{2} \psi^\mu \left[D_5+\sigma,\psi_\mu\right] \bigg).
\end{aligned}
\end{equation}
Introducing 
\begin{equation}
\mathcal{V}=
\frac{1}{g^2_{5d}}\int \mathrm{d}^5 x\, \sqrt{g}
\text{tr}\bigg(  \frac{i}{4}   \chi^{\mu\nu}   (2 F_{\mu\nu}-  D_{\mu\nu}) -  \frac{i}{2} \psi^\mu F_{\mu 5}   +\frac{1}{2}   \psi^\mu D_\mu \sigma   -\frac{1}{2}    \eta D_5 \sigma \bigg),
\end{equation}
we obtain 
\be 
-\frac{ \sigma^{(I)}}{8\pi^2}\int \mathrm{d}^5 x\, \sqrt{g} \mathcal{L}^\text{twisted}_\text{SYM} - 
 \bar{\mathcal{Q}} \mathcal{V}  = - \frac{1}{4g_{5d}^2} \int_{X\times S^1} \mathrm{d}^5 x\,  \sqrt{g} \text{tr} F_{\mu\nu} F_{\rho\sigma} \epsilon^{\mu\nu\rho\sigma}.
\ee
Although $\text{tr} F_{\mu\nu} F_{\rho\sigma} \epsilon^{\mu\nu\rho\sigma}$ generally depends on $x^5$, its integral over $X$ gives the $x^5$-independent instanton charge. 
Therefore, for $A^{(I)}$ with only a component along $S^1$, the action becomes
\be
\label{eq:ParialTwistedAction}
S_\text{twisted}=
\frac{i\theta}{8\pi^2} \int_X \text{tr}(F\wedge F) 
- \frac{R}{g^2_{5d}} \int_X 
\text{tr}(F\wedge F)+ \bar{\mathcal{Q}} \mathcal{V}.
\ee
Recalling \eqref{CR4}, we can write  
\begin{equation}\label{eq:Stwisted-nI=0}
S_\text{twisted} = \bar{\mathcal{Q}} \mathcal{V} + 
(\log \mathcal{R}^4 ) \frac{1}{8\pi^2} \int_X F \wedge F.
\end{equation}
It follows from \eqref{eq:Stwisted-nI=0} that the components $T_{\mu\nu}$ of the energy-momentum tensor along $X$ are $\bar\CQ$-exact. 
Hence, the theory has topological invariance along $X$ directions. We are allowed to modify the $\bar{\mathcal{Q}}$-exact term in \eqref{eq:Stwisted-nI=0} while keeping the partition function unchanged. On the other hand, the theory is not topological along $S^1$. 

To make localization manifest, we rescale the 4d metric $g_{\mu\nu} \to \lambda g_{\mu\nu}$ and the whole $\bar\CQ$-exact term $\bar\CQ V \to \lambda \bar\CQ V$, obtaining
\begin{equation}
\begin{aligned}
\label{twisted Q exact}
\bar\CQ \mathcal{V} \rightarrow 
\frac{1}{g^2_{5d}}\int \mathrm{d}^{5} x  \sqrt{g} \text{tr}\Bigg[ & \lambda^{-1}\left(\left(F_{+}^{\mu \nu}\right)^{2} 
+ 2i \chi^{\mu\nu} D_{\mu}\psi_\nu 
- \frac{i}{2} \chi_{\mu \nu}\left[D_{5}-\sigma, \chi^{\mu \nu}\right]  
-\frac{1}{4} D_{\mu\nu} D^{\mu\nu}
\right)\\
+&\left(D_{\mu} \sigma D^{\mu} \sigma+F_{\mu 5} F^{\mu 5}+ i \psi_{\mu} D^{\mu} \eta + \frac{i}{2} \psi_{\mu}\left[D_{5}+\sigma, \psi^{\mu} \right] \right) \\
+&\lambda\left(D_{5} \sigma D^{5} \sigma
-\frac{i}{2}\eta\left[D_{5}-\sigma, \eta\right] \right)\Bigg].
\end{aligned}
\end{equation}
We shrink the size of $X$ by taking $\lambda\to 0$. Then the terms in the third line vanish as $\lambda\to 0$. The 4d component of the 5d gauge field (hitherto denoted $A_X$ is forced to be selfdual. 
Henceforward in this subsection, to keep the notation from getting too heavy, we denote $A_X$ simply by $A$. On the
BPS locus, $A$ therefore satisfies the 4d equations:  
\begin{equation}\label{localization locus}
    F_+ = 0,\quad \left(D_{A}\psi\right)_{+}=0, \quad D = 0,
\end{equation}
where the middle equation comes from demanding that the locus of localization preserves
the $\bar\CQ$-symmetry. 

The space of isomorphism classes of gauge fields satisfying the first equation in 
Eq. \eqref{localization locus} is the famous moduli space of anti-self-dual (ASD) connections, 
a.k.a. the moduli space of instantons on $X$. We denote this moduli space by $M_{\bfmu}$. It has 
infinitely many connected components labeled by a nonnegative instanton number: 
\be
M_{\bfmu} = \bigsqcup_{k=0}^\infty M_{\bfmu,k},
\ee
where $M_{\bfmu,k}$ is the component with fixed instanton charge
\be
k = -\frac{1}{8\pi^2}\int_X \text{tr}(F\wedge F).
\ee
We also fix the 't Hooft flux $\bfmu \in H^2(X,\mathbb{Q})$, where $2\bfmu$ is an integral lift of the second Stiefel-Whitney class $w_2(P)$ for a principal $\mathrm{SO}(3)$-bundle $P\rightarrow X$.

The terms in the second line of \eqref{twisted Q exact} generate the action of a supersymmetric sigma model on $S^1$ with target space the moduli space of instantons on $X$.
To see this, let us focus on a component $M_{\boldsymbol{\mu},k}$ of the moduli space and parameterize it by local coordinates $\{Z^I\}$. 
Consider infinitesimal deformations of an ASD connection $A(x)$,
\be
A(x) \rightarrow A(x) + \delta_I A.
\ee
The deformed connection is still ASD to the first order in the deformation if $\delta_I A$ satisfies
\be
D_A^+  \delta_I A = 0.
\ee
In addition we do not want to include trivial solutions arising from infinitesimal gauge transformations. This can be achieved by requiring that the deformations are orthogonal to any infinitesimal gauge transformations, which leads to the gauge condition
\be\label{gauge fixing}
D_A * \delta_I A=0. 
\ee
A family of solutions $A(x,Z)$ to the ASD equations
satisfying this gauge condition can be constructed by setting  
\be\label{univ connection}
\delta_I A = \frac{\partial}{\partial Z_I} A(x,Z) - D_A\alpha_I,
\ee
where $\alpha_I$ can be understood as the compensating gauge symmetry, which is needed to ensure \eqref{gauge fixing}. 
We define locally in $M_{\bfmu,k}$ a universal connection on the universal bundle $\mathbb{E}$ over $M_{\bfmu,k}\times X$,
\be\label{universal connection}
{\mathbb A} = A(x,Z) + \alpha_I \mathrm{d}Z^I.
\ee

We now perform the standard collective coordinate quantization with $A_\mu(x,t) = A_\mu(x, Z(t))$, where $t=x^5$ is the time coordinate of the SQM. Using \eqref{univ connection}, one can write
\be\label{Fmu5}
F_{5\mu} =(\delta_I A_\mu)\dot Z^I + D_\mu (\alpha_I \dot Z^I - A_5).
\ee
The bosonic part of the second line of \eqref{twisted Q exact}  becomes
\begin{equation}
   \int_{X \times S^1}  \mathrm{d}^5 x\, \sqrt{g}  \tr \left( F_{\mu 5} F^{\mu 5} + D_\mu \sigma D^\mu \sigma \right)  =
    \int_{S^1} \mathrm{d}t\, g_{IJ} \dot Z^I \dot Z^J + \int_{X\times S^1}  \mathrm{d}^5 x\,  \sqrt{g}  \tr \left(D_\mu \Phi D^\mu \bar\Phi \right),
\end{equation}
where $g_{IJ}$ is the metric defined on $M_{\boldsymbol{\mu},k}$
\begin{equation}
    g_{IJ} = 
    \int_{X}
    \tr(\delta_I A 
    \wedge * \delta_J A).
\end{equation}
Here, $\Phi$ and $\bar \Phi$ are independent scalars defined as
\begin{equation}\label{eq:cc-Phi}
    \begin{aligned}
    \Phi &:=   
    \sigma+ i A_5 - i 
    \alpha_I \dot Z^I, 
    \\
    \bar\Phi & :=  
    \sigma - i A_5 
    + i 
    \alpha_I \dot Z^I.
    \end{aligned}
\end{equation}
Integrating over $\eta$ imposes the condition
\be
D^\mu\psi_\mu =0,
\ee
which, together with \eqref{localization locus}, 
ensures that $\psi$ transforms in the tangent bundle of $M_{\boldsymbol{\mu},k}$. 
Introducing fermionic collective coordinates $\zeta^I$ via
\be\label{psi expansion}
\psi(x,t) = \zeta^I(t) \delta_I A, 
\ee
the fermionic part of the action reads
\begin{equation}
    \begin{aligned}
     &\int_{X \times S^1}
    \mathrm{d}^5 x \, \sqrt{g} 
    \tr\left(\frac{i}{2}
    \psi^\mu[D_5 + \sigma, \psi_\mu]
    \right) \\
    =& 
    \int_{S^1} \mathrm{d}t\,
    g_{IJ}
    \bigg(
    \frac{i}{2} 
    \zeta^I \nabla_t \zeta^J
    \bigg)
    + \int_{X_4\times S^1} \mathrm{d}^5 x \,
    \sqrt{g} 
    \tr\left(\frac{i}{2}\psi^\mu [\bar\Phi,\psi_\mu]\right),
    \end{aligned}
\end{equation}
where 
$\nabla_t\zeta^J=\partial_t \zeta^J+\Gamma_{I K}^J \dot{Z}^K \zeta^I$, and the Christoffel symbols on $T_AM_{\boldsymbol{\mu},k}$ are
\begin{equation}
    \Gamma_{IK,J}=
    \int_{X_4}
    \mathrm{d}^4 x\, \sqrt{g_4}\tr( \delta_I A_\mu
      \nabla_K(\delta_J A^\mu)),
\end{equation}
with the covariant derivative $\nabla_I$ on 
$T_A M_{\boldsymbol{\mu},k}$ defined by
\begin{equation} 
  \nabla_I = \partial_I -i \alpha_I.
\end{equation}
The twisted action now reduces to
\begin{equation}
\int \mathrm{d}t\, g_{IJ} \left(\dot Z^I \dot Z^J + 
\frac{i}{2}
\zeta^I \nabla_t \zeta^J\right) +
\int \mathrm{d}t\int_{X} \mathrm{d}^4x \, \sqrt{g} \text{tr}\left(
D^\mu \Phi D_\mu \bar \Phi 
+ \frac{i}{2}\psi^\mu[\bar\Phi, \psi_\mu]\right).
\end{equation}
The equation of motion of $\bar\Phi$,
\begin{equation}
    D_\mu  D^\mu \Phi 
    = \frac{i}{2}[\psi_\mu,\psi^\mu],
\end{equation}
can be solved locally \cite{Nekrasov:1996cz} 
\begin{equation}\label{eq:Phi-To-Zeta}
    \Phi = \frac{i}{2}[\nabla_I, \nabla_J]
    \zeta^I \zeta^J.
\end{equation}
Integrating out $\bar\Phi$ gives the $\mathcal{N}=1$ 
supersymmetric sigma model,
\begin{equation}
    \int_{S^1}\mathrm{d}t 
     \,g_{IJ}(\dot Z^I \dot Z^J + \frac{i}{2}
    \zeta^I \nabla_t\zeta^J).
\end{equation}

In the discussion above we have not gone into details of the one-loop determinants. By topological symmetry they should cancel: The effective action in one dimension should be local and it is uniquely determined by supersymmetry. Of course, a more careful check that the determinants really do cancel would be highly desirable.

\subsection{There Can Be Global Anomalies}\label{subsec:CanBeAnomalies}

When the target space of a SQM is not oriented and spin, anomalies can arise because the fermion determinant cannot be globally defined as a function on the loop space of the target. See \cite{Witten:1985mj,AST_1985_131_43_0, Distler:2010an,Freed:2014iua} for discussions of this phenomenon. 

The moduli space of instantons in general has nontrivial first and second homotopy groups and nontrivial first and second cohomology groups. 
\footnote{The generalized Atiyah-Jones conjecture \cite{Atiyah:1978, Taubes:1989, ballico202225} says, roughly speaking, 
that in the limit of large $k$, $M_{k}$  approximates $\CA/\CG$ topologically: There is a homotopy equivalence up to the $n(k)$-skeleton where $n(k)$ grows linearly with $k$. 
On the other hand  
\be 
\pi_{n}( \CA_k/\CG_k ) \cong \pi_{n-1}(\CG_k) 
\ee
where $\CA_k$ is the space of connections with instanton charge $k$ and $\CG_k$ is the group of automorphisms of the gauge bundle. In general   $\pi_0(\CG_k)$ and $\pi_1(\CG_k)$ are nonzero.
For example, for $X=S^4$ we have $\pi_0(\CG_k) \cong \pi_4(G)$ and $\pi_1(\CG_k) \cong \pi_5(G)$. For $\mathrm{SU}(2)$-bundles on $S^4$ the homotopy groups were  determined precisely in the proof of Hurtubise et. al. of the Atiyah-Jones conjecture for $X=S^4$ \cite{Boyer1992TheAC, tian1996atiyah}.
}
Famously, although the moduli spaces of instantons are not always simply connected, Donaldson has shown that they are orientable. However, it was shown in \cite{KAMIYAMA201035}
that even for $\mathrm{SU}(2)$ instantons on $S^4$, the moduli space $M_k$ is spin if and only if the instanton charge $k$ is odd \cite{KAMIYAMA201035}. 
The question of when the moduli space of instantons is spin has recently been re-examined in \cite{FHMW}.
(It is important in \cite{FHMW} that one actually works with the moduli stack of instantons. We ignore subtleties of stacks in this paper.) 
One outcome is that when there is a nontrivial two-sphere in the moduli space, then the moduli space of $\mathrm{SO}(3)$-instantons on an almost complex manifold $X$ is not spin if 
\be\label{eq:ModSpaceSpinCond}
B\left(w_2(X), w_2(P)\right) \neq 0,
\ee
where $P$ is the $\mathrm{SO}(3)$-bundle over $X$. 
This would appear to pose a fundamental problem since we certainly want to consider manifolds and bundles for which \eqref{eq:ModSpaceSpinCond} holds. We will resolve the problem in   Sec.  \ref{subsec:AnomalyCancellation}. We will show that the issue boils down to defining a suitable 
 Spin$^c$ structure on the moduli space of instantons. 
We describe the Spin$^c$ structure as a product of a spin bundle (which fails to exist because of a $\IZ_2$ gerbe) tensored with a ``square-root'' of a line bundle over $M_{k}$ with odd first Chern-class. As a preliminary step, we give a mathematical argument that the smooth locus of the instanton moduli space $M_{\bfmu,k}$ is Spin$^c$ when $X$ admits an almost complex structure. To begin with, we note that if $X$ admits a complex structure, then there is a canonically induced complex structure on $M_{\bfmu,k}$.
\footnote{See \cite{Bursztyn:2012rj,Witten:2024yod} for nice explanations and generalizations of this well-known fact.} 
The complex manifold $M_{\bfmu,k}$ is automatically Spin$^c$, because $w_2$ admits an integral lift, namely the first Chern class of the canonical class. 

The analysis when $X$ is only an almost complex manifold is more subtle, since it is no longer clear that there is an induced almost complex structure on $M_{\bfmu,k}$. Nevertheless, we claim that a \emph{continuous deformation} of the tangent bundle $TM_{\bfmu,k}$ admits an almost complex structure. 
Note that the tangent space to $M_{\bfmu,k}$ at smooth points is the kernel (modulo gauge transformations) of the map $\sigma_A: \Omega^1(X; \mathrm{ad} P) \to \Omega^{2,+}(X; \mathrm{ad} P)$ defined by $\sigma_A(\alpha)=P^+(D_A \alpha)$, where $P^+$ is the projector onto self-dual two-forms. 
When $X$ is generalized from having a complex structure to having an almost complex structure, only the $(2,0)$ and $(0,2)$ components of the equations $P^+(D_A \alpha)=0$ are deformed,
\begin{align} 
(2,0)\text{-component:} \quad & \partial \alpha^{(1,0)} + [A^{(1,0)}, \alpha^{(1,0)}] +  N^{\bar i}_{jk} \alpha_{\bar i} e^j \wedge e^k = 0, \\
(0,2)\text{-component:} \quad & \bar\partial \alpha^{(0,1)} + [A^{(0,1)}, \alpha^{(0,1)}] + N^{i}_{\bar j , \bar k } \alpha_{i}  e^{\bar j} \wedge e^{\bar k} = 0,
\end{align}
where $\{e^i\}$ is an orthonormal basis for $T^{1,0}X$, and $N$ is the Nijenhuis tensor.
The new terms involving $N$ lead to a compact deformation of $\sigma_A$, which induces a smooth deformation of the tangent bundle $TM_{\bfmu,k}$.  
Then, since the third integral Stiefel-Whitney class $W_3$ is a $\mathbb{Z}_2$ class, this suffices to ensure that $M_{\bfmu,k}$ admits a Spin$^c$ structure.

\subsection{``Line Bundle'' Induced By Background Flux}
\label{subsec:ChernSimonsLineBundle}

We are certainly interested in cases where the moduli space of instantons is not spinnable, as characterized by the condition \eqref{eq:ModSpaceSpinCond}. 
To explore such scenarios, we relax the condition \eqref{eq:BackgndFirstRed}, allowing $F^{(I)}$ to be pulled back from $X$ with non-vanishing periods. 
When $F^{(I)}$ carries topologically nontrivial flux, we turn on the 3d Chern-Simons observables \cite{LoNeSha}, 
\be\label{eq:MixedObs}
S = \frac{i}{8\pi^2} \int F^{(I)}\wedge CS_3(A) + \text{(supersymmetric completion)},
\ee
where $CS_3(A) = \text{tr} \left(AdA + \frac23 A^3\right)$, and $A$ is the 5d gauge 
connection. 
Indeed, the first term in \eqref{eq:MixedObs} corresponds to the electric coupling with the $\mathrm{U}(1)^{(I)}$ current in \eqref{mixed CS action} after integration by parts. As discussed above, when both $A^{(I)}$ and $A$ are topologically nontrivial, \eqref{eq:MixedObs} is defined as an element of $\IR/\IZ$ via differential cohomology. 
Crucially, this coupling naturally defines a line bundle $\CL^{(I)}$ with connection on the moduli space of instantons. 
\footnote{More precisely, if the ’t Hooft flux is non-vanishing, then $(\CL^{(I)})^2$ is well defined, while $\CL^{(I)}$ may only make sense as a twisted K-theory class, as we demonstrate below. 
 }
The background vector multiplet $V^{(I)}$ defines a line bundle with connection over 
the four-manifold $X$ and hence defines a differential cohomology class in $\check H^2(X)$. The universal connection induces a differential cohomology class in $\check H^4(X \times \CA/\CG)$. Taking the cup product and integrating over $X$ yields an element in $\check H^2(\CA/\CG)$, namely, a line bundle with connection over $\CA/\CG$. After restricting to the moduli space $M_{\bfmu,k}$, this gives the desired line bundle $\CL^{(I)}$.

To give an explicit description of the connection on $\CL^{(I)}$, we revisit the collective coordinate derivation of the SQM action, now in the presence of non-vanishing $F^{(I)}$.  
The other fields in the background multiplet $V^{(I)}$ are taken to be  
\begin{equation}\label{eq:BackgndFirstRed-3}
\sigma^{(I)} = -\frac{8\pi^2}{ g^2_{5d}}, 
  \quad \lambda^{(I)}_A=0,\quad  D^{(I)} = F^{(I)}_+,
\end{equation}
where the value of $D^{(I)}$ is fixed due to supersymmetry.
Then the bosonic part of the action \eqref{mixed CS action} becomes
\be
S\vert_{\rm bosonic} = S_{\text{YM}} + S_\Delta,
\ee
where 
\be \label{bosonic S delta}
S_\Delta = \frac{i}{8\pi^2}\int F^{(I)} \wedge CS_3 (A) - \frac{1}{8\pi^2} \int \mathrm{d}^5 x\, \sqrt{g} F_{mn}^{(I)} \text{tr} (\sigma F^{mn}). 
\ee 
Notice that the term $\sigma D^{(I)AB} D_{AB}$ in \eqref{mixed CS action} is dropped, since $D$ is set to zero by localization to moduli space of instantons. 
Using the anti-self-dual condition \eqref{localization locus} and the following identity for any closed two-form $F^{(I)}$,  
\be
\int_{X\times S^1} F^{(I)} \wedge CS_3\left(A-i\sigma \mathrm{d}x^5\right) = \int_{X\times S^1} F^{(I)} \wedge CS_3(A) -2i \int_{X\times S^1} F^{(I)} \wedge \text{tr}\left(F\sigma \mathrm{d}x^5\right),
\ee
we can rewrite \eqref{bosonic S delta} as
\be
S_\Delta|_{\text{bosonic}} = \frac{i}{8\pi^2} \int_{X\times S^1} F^{(I)} \wedge  CS_3\left(A-i\sigma \mathrm{d}x^5\right). 
\ee 
Decomposing $A = A_X + A_5 \mathrm{d}x^5$, and applying the collective coordinate ansatz (see the discussion around \eqref{universal connection}), we obtain
\be
A- i\sigma \mathrm{d}x^5 = A_X+(A_5-i\sigma)\mathrm{d}x^5  =  A_X + \alpha_I \dot Z^I \mathrm{d}x^5.
\ee
Therefore, the effect of $S_\Delta$ in the SQM is to introduce a phase for the particle moving on the moduli space of instantons along a path $\gamma$,
\be\label{eq:SDeltaphase}
e^{-S_\Delta} \rightarrow   e^{-  i\int_\gamma \CA^{(I)}  },
\ee 
where $\CA^{(I)}$ is expressed in terms of the universal connection ${\mathbb A}$  \eqref{universal connection} as
\be
\CA^{(I)} = \frac{1}{8\pi^2}\int_X F^{(I)} \wedge CS_3(\mathbb{A}).
\ee 
Such a connection leads to a Chern-Weil representative of the first Chern class of the form
\be\label{connection on M}
\frac{1}{2\pi} \mathrm{d}_{M} \CA^{(I)} = \int_X \frac{F^{(I)}}{2\pi}  \wedge \frac{\text{tr}({\mathbb F} \wedge {\mathbb F})}{8\pi^2}.
\ee 
This corresponds to the Donaldson $\mu$-map evaluated on the homology class 
\be
S = \text{PD}(c_1(L^{(I)})) \in H_2(X,\mathbb{Z}),
\ee
where $L^{(I)}$ is defined in \eqref{eq:U(1)-I-Supermultiplet} and $\overline{c_1(L^{(I)})}= \bfn_I$,
\be\label{eq:c1LI-muS}
\overline{c_1(\CL^{(I)})} = - \mu_D(S).
\ee 
Here the overline indicates the exclusion of the torsion part, and the sign is consistent with \eqref{eq:p1Universal}. See App. \ref{app:DonaldsonMuMap} for the definition of the Donaldson $\mu$-map $\mu_D$.

It is important to note that the background flux $\bfn_I$ must be properly chosen to induce a well-defined line bundle $\CL^{(I)}$ over the moduli space. To see this, let $P$ be the principal $\mathrm{SO}(3)$-bundle over $X$. We can write
\be\label{eq:LiftY}
\frac{i}{8\pi^2}\int_{X\times S^1} F^{(I)} \wedge \text{tr}\left( A\wedge dA + \frac23 A^3\right) = \frac{i}{8\pi^2}\int_{X_6} F^{(I)} \wedge \text{tr}(F\wedge F),
\ee
where $X_6$ is a six-manifold with $\partial X_6 = X \times S^1$. 
\footnote{Note that $\Omega^{\mathrm{SO}}_5(B\mathrm{SO}(3)) = \mathbb{Z}_2\oplus \mathbb{Z}_2$ is nontrivial.
We assume, for convenience, that the relevant bordism obstructions vanish. With more work one could instead use differential cohomology directly in 5d.}
In order for the theory to be independent of a choice of such $X_6$, we need
\be\label{quantization M6}
\frac{1}{8\pi^2} \int_{Y} F^{(I)} \wedge \text{tr}(F\wedge F)\in 2\pi \mathbb{Z},
\ee
for any closed six-manifold $Y$. 
For an $\mathrm{SO}(3)$-bundle $P$ extended to $\widetilde P$ over $Y$, 
\be\label{SO3 w2}
\frac{1}{8\pi^2} \text{tr} (F\wedge F) = \frac14 c_1(\widetilde P)^2 \mod 1,
\ee
where $c_1(\widetilde P)$ is an integral lift of the 't Hooft flux.  
This implies that the proper quantization of \eqref{eq:LiftY} puts a nontrivial constraint on $\bfn_I$, 
\be\label{eq:simpler-constraint}
\int_{Y}  \bfn_I \cup c_1(\widetilde P)^2 \in 4 \mathbb{Z}.
\ee
We specialize to $Y = X \times S^2$ and  
\be
c_1(\widetilde{P}) =  \Omega + c_1(P),
\ee
where $\Omega$ is a generator of $H^2(S^2,\mathbb{Z})$.
Evaluating \eqref{eq:simpler-constraint} for this configuration leads to the constraint
\be
\int_{X}  \bfn_I \cup c_1(P) \in 2 \mathbb{Z},
\ee
or, expressed in terms of $\bfmu$,
\be\label{nI condition}
 B\left( \bfn_I, 2\bfmu  \right) \in 2 \mathbb{Z}.
\ee
When this condition is not satisfied, our ``line bundle $\CL^{(I)}$'' does not actually exist, although its square still exists (as we have mentioned in a footnote above). 

\subsection{But The Anomalies Can Be Cancelled}\label{subsec:AnomalyCancellation}

In Sec.  \ref{subsec:CanBeAnomalies}, we established that the SQM on the instanton moduli space $M$ exhibits a global anomaly when the condition \eqref{eq:ModSpaceSpinCond} holds. 
The anomaly manifests in the path integral through the problematic term
\be\label{eq:Pfaff-instSQM}
{\rm Pfaff}\left(\slashed{D}_{\phi^*(TM)} \right).
\ee
In the Hamiltonian formulation, the anomaly stems from a nontrivial gerbe obstruction that prevents the global definition of the chiral spin bundle $S$ over $M$. 
Consequently, the Hilbert space $L^2(S)$ does not exist.

In Sec.  \ref{subsec:ChernSimonsLineBundle}, we demonstrated that a background vector multiplet induces a line bundle $\CL^{(I)}$ with connection $\CA^{(I)}$ over $M$. However, $c_1(\CL^{(I)})$ might not be properly quantized. The problematic term in the path integral measure for the SQM is 
\be\label{eq:ParTrans-instSQM}
\exp \left( -i  \oint_{S^1} \CA^{(I)} \right) ,
\ee
and in the Hamiltonian interpretation correspond to a change of the Hilbert space to 
$L^2(S \otimes  \CL^{(I)})$. 
The relevant gerbes cancel leading to a well-defined product when 
\be\label{eq:AnomalyCancelCriterion}
B\left(w_2(P), w_2(X)  + \bfn_I\right) = 0  \mod 2. 
\ee
In this case the ambiguities in \eqref{eq:Pfaff-instSQM} and 
\eqref{eq:ParTrans-instSQM} cancel, and the product 
\be\label{eq:Product-instSQM}
{\rm Pfaff}(\slashed{D}_{\phi^*(TM)} ) \exp \left( -i  \oint_{S^1} \CA^{(I)} \right) 
\ee
is a well-defined function on the loop space of the moduli space of instantons. 

The global anomaly cancellation condition \eqref{eq:AnomalyCancelCriterion} manifests concretely in many explicit results to be presented below. 

\subsubsection*{Remarks}
\begin{enumerate}
\item We conjecture that the ``line bundle'' $\CL^{(I)}$ over moduli space 
(where only $(\CL^{(I)})^{\otimes 2}$ is well defined) is such that $S(TM) \otimes \CL^{(I)}$ is 
well defined. In other words, it is defining a Spin$^c$ structure on the moduli space of instantons. 
We can view the map $H^2(X,\IZ) \rightarrow H^2(M,\IZ)$ given by $\bfn_I \to \overline{2c_1(\CL^{(I)}) } = -2\mu_D(S)$
as a map of $H^2(X,\IZ)$ to the set of characteristic classes of Spin$^c$ structures on the moduli space of instantons.  
\item The criterion \eqref{eq:AnomalyCancelCriterion} is in harmony with the observation that the $U$-plane integral vanishes from an anomaly under the involution $U \to - U$ coming from  $\mathbb{Z}_2^{(1)}$. See Eq. \eqref{OmegT} below. This is an example of the original remark of Witten's \cite{witten19822} that 
in the case of a global anomaly the sum over gauge copies leads to a vanishing partition function.
The novelty in this case is that the vanishing is not from gauge copies but from a $\mathbb{Z}_2^{(1)}$ one-form symmetry.
\item In the case where $X$ is complex the moduli space $M$ has an induced complex structure. 
Then $K_M \mod 2 = w_2(M)$.
\footnote{In order to keep the equations from being too cumbersome we will often identify a line bundle with its first Chern class and we will also identify an element of $H^2(X,\mathbb{Z})$ with its Poincar\'e dual in $H_2(X,\mathbb{Z})$. Note that we are ignoring torsion.}
On the other hand, the Donaldson $\mu$-map satisfies  $\mu_D(K_X)=K_M/2$ \cite[Sec.   1.3]{Gottsche:2006bm}\cite[Proposition 8.3.1]{HuyLen10}. 
Hence a probe of whether $M$ is spin or not is  the integrality of the Donaldson surfaces observables Poincar\'e dual to $K_X$.  The integrality of the surface observables for $X=\mathbb{CP}^2$ and $\mu=0$ \cite[Theorem 4.2]{ellingsrud1995wall} is compatible with these moduli spaces being spin, while the negative powers of $2$ for the surface observables with $\mu=1/2$ \cite[Theorem  4.4]{ellingsrud1995wall} shows that these moduli spaces are definitely non-spin. See also \cite[Table 3]{Korpas:2019cwg}.
\end{enumerate}

\subsection{The Partition Function}
\label{sec:ThePartitionFunction}

The partition function is an expansion in nonnegative powers of $\CR$, 
\be\label{eq:Zee-QM-1}
Z_{\bfmu,\bfn_I}^J(\CR)=\sum_{l}  d_{\bfmu,\bfn_I}(l)\,\CR^l.
\ee 
The coefficients are defined as generalized Witten indices
\be\label{eq:WittenIndex}
d_{\bfmu,\bfn_I}(l)={\rm Tr}_{\CH}(-1)^F e^{-RH},
\ee
where the trace is over the Hilbert space $\CH$, i.e., the space of all $L^2$-integrable sections of $S\otimes\CL^{(I)}$ over $M_{\bfmu,k}$, and the Hamiltonian $H= \slashed{D}^2$ is the square of the supersymmetry operator. 
Hence $d_{\bfmu,\bfn_I}(l)$ is formally the Dirac index of $M_{\bfmu,k}$.  
The factor $\CR^{l}$ comes from the topological term in \eqref{eq:Stwisted-nI=0}.
Multiplying the partition function by an overall factor $\CR^{- 3 \left(  \frac{ \chi + \sigma}{4} \right)}$ arising from the usual counterterms involving the Euler character and signature densities, we have \cite{Atiyah:1978wi}
\begin{align}\label{eq:ModSpaceDim}
l &=  4k - 3 \left(  \frac{ \chi + \sigma}{4} \right) = \frac{1}{2}{\rm vdim}_{\mathbb{R}}(M_{\bfmu,k}) \nonumber\\
& = -4\bfmu^2- 3\chi_{\rm h} \mod 4, 
\end{align}
where $\chi_{\rm h}:= (\chi + \sigma)/4$. 
The value of $k$ ensures that the power of $\CR$ is nonnegative.

Since the moduli space of instantons is noncompact and singular, one should be cautious about the identification of the Witten index with the Dirac index. 
The Witten index should be identified with the $L^2$-index of the Dirac operator, because wavefunctions in the Hilbert space of the SQM are $L^2$-normalizable.  
Since the index of the Dirac operator is often discussed in the mathematical framework of K-theory, 
the generalized Donaldson invariants under investigation are called - following Nekrasov - the ``K-theoretic Donaldson invariants''.
  
The Atiyah-Singer index theorem associates a cohomology class to the Dirac operator.
After introducing observables (as in Sec.  \ref{sec:OtherObs}), the relevant class integrated over the instanton moduli space $M_{\bfmu,k}$ takes the form
\be\label{eq:KDW-indexdensity}
\int_{M_{\bfmu,k} }   e^{\mu_D(x)} \hat{A}(M_{\bfmu,k}), \quad x\in H_2(X),
\ee
which contrasts with the standard Donaldson invariants, which lack the $\hat{A}$-genus factor,
\be\label{eq:KDW-indexdensity2}
\int_{M_{\bfmu,k} }  e^{\mu_D(x)}.
\ee
However, interpreting \eqref{eq:KDW-indexdensity} as the $L^2$-index of the Dirac operator is subtle. Since $M_{\bfmu,k}$ is both noncompact and singular, it is challenging to demonstrate that the usual evaluation of the index of the Dirac operator in terms of a cohomological integral is correct. In the case of the standard Donaldson invariants, prior work demonstrates that intersection numbers of $\mu_D(x)$ can be supported away from singularities through careful analysis \cite[Chap.  9]{Donaldson90}\cite{DONALDSON1990257}. However, there is no analogous result for the K-theoretic Donaldson invariants.
 
It is interesting to relate the partition function $Z_{\bfmu,\bfn_I}^J(\CR)$ to the generating series for holomorphic Euler characteristics of moduli stacks of sheaves on algebraic surfaces. 
The latter were studied by G\"ottsche, Nakajima, and Yoshioka (GNY). For projective surfaces $X$, GNY define \cite[Eq.  (1.5)]{Gottsche:2006bm}
\be  
\chi_{c_1}^H\left(L_{{\rm GNY}},\Lambda_{{\rm GNY}}\right) =   \sum_{d\geq 0} \chi(M^X_H(c_1,d),{\cal{O}}(\mu(L_{{\rm GNY}})) \Lambda_{{\rm GNY}}^d,
\ee
where $M^X_H(c_1,d)$ is a suitably compactified moduli stack of rank-two semistable sheaves on $X$, with Chern classes determined by $(c_1,d)$ and stability condition determined by $H$. 

Recall that for any K\"ahler manifold $N$, there exists a canonical Spin$^c$ structure, which provides an isomorphism of vector bundles \cite{friedrich2000dirac},
\be 
\alpha: S\otimes K_N^{-\frac12} \, \xlongrightarrow{\sim} \, \bigoplus_{k} \Lambda^{0,k} T^*N,
\ee
where $S$ is the Dirac spin bundle, and $K_N$ is the canonical bundle.
If $N$ is not spin, then $S$ does not exist, but the tensor product $S\otimes K_N^{-1/2}$ is globally well defined, since the gerbe obstructing the existence of $S$ is canceled by the choice of a square root of the canonical bundle $K_N$.
Under this isomorphism, the conjugation by $\alpha$ of the chiral Dirac operator acting on sections of $S\otimes K_N^{-1/2}$ becomes the Dolbeault operator $\bar\partial$ acting on $\Omega^{0,*}(X)$, and hence 
\be\label{eq:EqualIndex}
{\rm Ind}(\slashed{D}_{L} ) = {\rm Ind}(\bar\partial_{\tilde L}),
\ee
where 
\be\label{eq:IdentifyLineBundles}
\tilde L \cong L \otimes K_N^{\frac12}. 
\ee
As a manifestation of this relationship, the $\hat{A}$-genus and the Todd genus on a complex
manifold are related by 
\be
\label{eq:ShiftIndexDensities}
\hat A(TN)  = e^{-\frac{1}{2} c_1(T^{1,0}N)} \mathrm{Td}(T^{1,0}N). 
\ee

We now apply the above results to determine the relation between $Z_{\bfmu,\bfn_I}^J(\CR)$ and $\chi_{c_1}^H(L_{{\rm GNY}},\Lambda_{{\rm GNY}})$ when $X$ is an algebraic surface. The Donaldson-Uhlenbeck-Yau theorem identifies a compactification of $M_{\bfmu,k}$ with the moduli space of semistable sheaves $M^X_H(c_1,d)$, where the period point $J$ determines a stability condition $H$, the integral lift 
\footnote{Note that our integral lift of the 't Hooft flux $2\bfmu$ can be shifted by twice a lattice vector. Correspondingly, the moduli spaces $M^X_H(c_1,d)$ can be identified if we shift $c_1$ by twice a lattice vector via tensor product of the rank-two sheaf with a line bundle. }
of $2\bfmu$ is identified with $c_1$, and instanton charge $k$ determines $d$. 
Applying \eqref{eq:EqualIndex} with $N= M_{\bfmu,k}$, we identify the generating functions 
\be\label{eq:GNY-series-Our-series}
Z^J_{\bfmu,\bfn_I}(\CR) = \chi_{c_1}^H(L_{{\rm GNY}},\Lambda_{{\rm GNY}}),
\ee
where 
\footnote{The power series in \cite{Gottsche:2006bm} is written in both 
$\Lambda_{{\rm GNY}}$ and $(\beta\Lambda)_{{\rm GNY}}$, and we should 
identify $\beta_{{\rm GNY}} =R_{{\rm ours}}$ and $\Lambda_{{\rm GNY}} = 
\Lambda_{{\rm ours}}$ so $(\beta \Lambda)_{{\rm GNY}} = \CR$. Note that in some parts of \cite{Gottsche:2006bm}, $\beta_{{\rm GNY}}$ is set to one. }
\be
\Lambda_{{\rm GNY}} = {\cal{R}},
\ee
and
\footnote{We note another unfortunate clash of notation. On the left-hand side $\mu$ is a map, 
defined in \cite{Gottsche:2006bm}, from a line bundle on $X$ to a line bundle on $M_{\bfmu,k}$. The subscript on $M$ on the right-hand side is the 't Hooft flux. }
\be\label{eq:L-gny-CLI}
\mu(L_{{\rm GNY}}) \cong \CL^{(I)} \otimes K_{M_{\bfmu,k}}^{\frac12}.
\ee
 
Note that we have made a leap here: We are conjecturing that the $L^2$-index on $M$ is an index of a Spin$^c$ Dirac operator on $M^X_H(c_1,d)$.
We take the first Chern class of \eqref{eq:L-gny-CLI}. Using  \eqref{eq:c1LI-muS}, $\mu_D(K_X) = K_M/2$, and the relation \cite[sentences under Eq. (1.5)]{Gottsche:2006bm}
\be 
\mu_D(c_1(L_{{\rm GNY}})) = c_1(\mu(L_{{\rm GNY}}) ),
\ee
we conclude that 
\be\label{eq:c1-Lgny-bfnI}
c_1(L_{{\rm GNY}}) = -\bfn_I + c_1(K_X). 
\ee

This dictionary can be confirmed by comparing K-theoretic Donaldson invariants derived via the $U$-plane integral with results in \cite{Gottsche:2006bm}. 
Specifically, our explicit expressions \eqref{eq:Phi-half-zero-CP2n} and \eqref{eq:Phi-zero-CP2n} match the GNY expression \eqref{GNYseries}, modulo a minor discrepancy noted under Eq. \eqref{PhiGNYP20}. 
The proof of the identification is technically nontrivial, and is detailed in App. \ref{ResultsGNY}.

\subsection{Some Other Observables}
\label{sec:OtherObs}

The partially topologically twisted SYM admits various topological observables supported on cycles in $X \times S^1$. 
While a systematic analysis of these observables and their correlation functions lies beyond our scope, we briefly outline some key properties. 

A basic observable is the supersymmetric circular Wilson loop operator $W_{\mathfrak{R}}$,
\be
\label{DefWilsonLoopR}
W_{\mathfrak{R}}(p) =  \text{tr}_{\mathfrak{R}}\,\text{P}\!\exp \left[\oint_{p\times S^1} (\sigma + i A_5 )\mathrm{d}x^5\right],
\ee
where $\mathfrak{R}$ is a representation of the gauge group $\mathrm{SU}(2)$, and $p\in X$ is a point.
This is $\bar \CQ$-closed by \eqref{eq:Qbar-nonabelian}, and shifting the point $p\in X$ deforms the observable by a $\bar \CQ$-exact expression.  
If ${\mathfrak{R}}$ is the fundamental representation, $W_{\mathfrak{R}}(p)$ reduces to the observable defined earlier in \eqref{DefWF}.
Using the supercharge $\cal{K}$, one constructs descendant observables using the standard descent procedure, 
\be
W_{\mathfrak{R}}^{(j)} = \mathcal{K}^j W_{\mathfrak{R}}, \quad j=1,\cdots, 4,
\ee
and for a $j$-cycle $\Sigma^{(j)} \in X$, the integral 
\be
W_{\mathfrak{R}}^{(j)} \left(\Sigma^{(j)}\right) = \int_{\Sigma^{(j)}} W_{\mathfrak{R}}^{(j)}
\ee
is $\bar \CQ$-closed and depends only on the homology class of $\Sigma^{(j)}$.  

Upon reduction to SQM with target space $M_{\bfmu,k}$, $W_{\mathfrak{R}}(p)$ corresponds to an insertion in the SQM path integral as
\be
 \text{tr}_{\mathfrak{R}}\,\text{P}\! \exp \left(  \int_{\gamma} i \CA_p  + \Phi \mathrm{d}t\right),
\ee
where $\gamma$ is a path in $M_{\bfmu,k}$, and $\CA_p\in \Omega^1(M_{\bfmu,k})$ is the evaluation at $p$ of the universal connection $\mathbb{A} = A_X + \alpha$,
\be
\CA_p=\mathbb{A}/p.
\ee 
When we integrate out $\Phi$ it is identified with a quadratic expression in the 
fermions as in \eqref{eq:Phi-To-Zeta}.  
Accordingly, the expression \eqref{eq:WittenIndex} is modified as follows. Associated to the universal principal $G$-bundle (here $G=\mathrm{SU}(2)$ or $\mathrm{SO}(3)$ )  is an associated bundle $V_{\mathfrak{R}}$ over $X \times \CA/\CG$. 
Pulling back to $\{ p \} \times \CA/\CG$, we obtain a vector bundle $V_{{\mathfrak{R}}, p} $ over $\CA/\CG$ with connection inherited from the universal connection.  With this connection we have 
\be\label{eq:ChernCharVRp}
\text{ch}(V_{{\mathfrak{R}},p}) = \left.\left(\text{tr}_{\mathfrak{R}} e^{ i\mathbb{F}}\right)\right/p.
\ee 
In Eq. \eqref{eq:WittenIndex}  we  replace the spinor bundle 
$S^+ \otimes \CL^{(I)}$ by   $S^+\otimes\CL^{(I)}\otimes  V_{{\mathfrak{R}},p}$. 
With multiple insertions 
\be\label{eq:Many-Ws}
\langle  W_{\mathfrak{R}}(p_1) \cdots    W_{\mathfrak{R}}(p_k) \rangle, 
\ee
we replace the spinor bundle $S^+ \otimes \CL^{(I)}$ by its tensor product with $\otimes_i W_{\mathfrak{R}}(p_i)$. 
 
We expect that these observables can be related to expressions such as  \eqref{eq:KDW-indexdensity} 
above but we will not discuss the precise relation in detail here. 

In the LEET, $W_{{\mathfrak{R}}}(p)$ becomes $U$ in the case where ${\mathfrak{R}}$ is the fundamental representation. For other irreducible representations ${\mathfrak{R}}$ of $\mathrm{SU}(2)$,  $W_{{\mathfrak{R}}}(p) $  will be a polynomial in $U$ which is a quantum deformation of a Tchebyshev polynomial. We can expand $U$ as a power series in $\CR$ with coefficients which are polynomials in $u$. Insertions of $u$ correspond to insertions of $\mu_D(p)$ in the moduli space intersection numbers. This is compatible with the expansion of the Chern character \eqref{eq:ChernCharVRp}. 

The surface operator corresponding to $U$ is very similar to the 4d case,
\be 
\CO(S) = \int_S  \frac{1}{\sqrt{8}} \frac{\mathrm{d}U}{\mathrm{d}a} (F_- + D) +\frac{1}{32} \frac{\mathrm{d}^2U}{\mathrm{d}a^2} \psi^2.
\ee
Note that $\mathrm{d}U/\mathrm{d}a$ is almost the same as that in the 4d case when expressed in terms of $\tau$. 
The expression including these observables in the $U$-plane is very similar to the familiar 4d case. As in that case, one will need to introduce contact terms when formulating the generating function for this observable. 

Finally, we can couple the 5d theory with 3d $\CN=4$ theories with $G$ symmetry on $S \times S^1$, extending the topological surface defects supported on $S$ in the 4d case. Again, the investigation of these interesting generalizations is beyond the scope of this paper.

\section{$U$-plane  Integral}
\label{Sec:Uplane}

In this section, we consider the LEET on $X$ of 5d $\mathrm{SU}(2)$ SYM theory on $X\times S^1$, with a partial topological twist along $X$.
For $b_2^+=1$, the path integral of the resulting theory can be evaluated by generalizing \cite{Witten:1995gf, Moore:1997pc, LoNeSha}.
Analogous to Donaldson-Witten theory in 4d, the path integral reduces to an integral over zero modes. 
After integrating over the fermionic and auxiliary field zero modes, the residual path integral simplifies to an integral over the $U$-plane. 
We will analyze this integral in detail, including its dependence on the period point $J$.

\subsection{Effective Action For The Twisted Theory}

To incorporate couplings to background $\bfn_{I}$ and $\bfn_{K}$ fluxes, we consider the $U$-plane theory as a specialization of a $\mathrm{U}(1)^N$ $\CN=2$ supersymmetric theory with $N-1$ nondynamical vector multiplets. This is analogous to the setup for background flavor fluxes \cite{Manschot:2021qqe, Aspman:2022sfj}. The low-energy effective Lagrangian of the $\mathrm{U}(1)^N$ theory restricted to zero modes simplifies to \cite{Marino:1998bm}
\footnote{The coefficient of the $\eta^j\chi^k$ term differs from \cite[Eq.  (3.4)]{Marino:1998bm}, since the convention for the local coordinate $a$ here differs by $\sqrt{2}$. See also \cite[Footnote 2]{Manschot:2019pog}.}
\be
\label{LEETL}
\CL_0=\frac{i}{16\pi} \left( \bar \tau_{jk} F_+^j\wedge F^k_+ +
\tau_{jk}F_-^j\wedge F^k_- \right)-\frac{1}{8\pi}y_{jk} D^j\wedge D^k
- \frac{i}{4\pi} \bar \CF_{jkl}\eta^j\chi^k\wedge (D+F_+)^l,
\ee
where the indices $j,k,l=1,\cdots,N$, and
\be
\tau_{jk}=-\frac{1}{2\pi i}\frac{\partial^2 \CF}{\partial {a^j}\partial {a^k}}, \quad 
y_{jk}={\rm Im}(\tau_{jk}).
\ee
The $\bar {\mathcal Q}$ and $\mathcal{K}$ transformations in 4d conventions are 
\be
\begin{aligned}
    [\mathcal{\bar Q}, a] & = 0, \quad
    [\mathcal{\bar Q}, \bar a]  =     i \sqrt{2} \eta, \quad 
    [\mathcal{\bar Q}, A]  =     \psi, \quad 
     [\mathcal{\bar Q}, D] =  (\mathrm{d} \psi)^+,  \\
    \{ \mathcal{\bar Q}, \psi \} & =  4 \sqrt{2} \mathrm{d} a, \quad
    \{ \mathcal{\bar Q}, \chi \} = i(F_+ -D), \quad 
    \{ \mathcal{\bar Q}, \eta \} =0,
\end{aligned}
\ee
and
\be
\begin{aligned}
        [\mathcal{K}, \bar a ] &= 0, \quad 
        [\mathcal{K}, a]  = \frac{1}{4\sqrt{2}} \psi, \quad
        [\mathcal{K},A]  = -2i \chi, \quad 
         [\mathcal{K},D ]  =  \frac{3i}{4} * \mathrm{d} \eta  + \frac{3i}{2} \mathrm{d}\chi,  \\
        \{\mathcal{K},\psi\}  &= 2 (D+ F_-),\quad 
        \{ \mathcal{K}, \eta \} = 
        -\frac{i\sqrt{2}}{2} \mathrm{d} \bar a, \quad
        \{ \mathcal{K}, \chi\} =  \frac{3 \sqrt{2}i}{4} * \mathrm{d} \bar a.
    \end{aligned}
\ee
The LEET of interest involves a dynamical $\mathrm{U}(1)$ with field strength $F$, and two background $\mathrm{U}(1)$ global symmetry groups, $\mathrm{U}(1)^{(I)}$ and $\mathrm{U}(1)^{(K)}$, with multiplets $(\eta^{(I)},\chi^{(I)},a^{(I)},D^{(I)},F^{(I)})$ and  $(\eta^{(K)},\chi^{(K)},a^{(K)},D^{(K)},F^{(K)})$, respectively. 
We denote the cohomology classes of the fluxes by
\footnote{Our $F$ is related to the $\mathrm{U}(1)$ flux $F_{\text{CM}}$ in \cite{Closset:2021lhd} by $F=2F_{\text{CM}}$, while $F^{(I)}=F^I_{\text{CM}}$. }
\be
\bfk=\left[ \frac{F}{4\pi} \right]\in \bfmu+L,\quad \bfn_I=\left[ \frac{F^I}{2\pi}
\right],\quad \bfn_K=\left[ \frac{F^K}{2\pi} \right].
\ee
Note the different normalizations of the dynamical and global $\mathrm{U}(1)$'s. In the weak-coupling regime when the $\mathrm{SU}(2)$ gauge group ($\bfmu=0$) is broken to its diagonal Cartan subgroup, the field strength $F$ of the low-energy dynamical $\mathrm{U}(1)$ is related to that of the $\mathrm{SU}(2)$ gauge group by
\be 
F_{\mathrm{SU}(2)}=\frac{1}{2}\left(\begin{array}{cc} F & 0 \\ 0 & -F \end{array}\right).
\ee 
The bosonic low-energy effective Lagrangian for the zero modes reads
\bea\label{LEEA bos zero}
\CL_0|_{\text{bos}} = & \frac{i}{16\pi}\left(\bar\tau F_+\wedge F_+ + \tau F_- \wedge F_-\right) - \frac{y}{8\pi} D\wedge D \\
&+ \frac{i}{4\pi}\left(\bar v_I F_+^{(I)} \wedge F_+ + v_I F_-^{(I)} \wedge F_-\right) - \frac{ \text{Im}(v_I)}{2\pi} D^{(I)}\wedge D \\
&+ \frac{i}{4\pi} \left(\bar\xi_{II} F_+^{(I)}\wedge F_+^{(I)} + \xi_{II} F_-^{(I)}\wedge F_-^{(I)}\right) - \frac{\text{Im}(\xi_{II})}{2\pi} D^{(I)}\wedge D^{(I)}\\
&+ \frac{i}{4\pi}\left(\bar v_{K} F_+^{(K)} \wedge F_+ + v_K F_-^{(K)}\wedge F_-\right) - \frac{ \text{Im}(v_K)}{2\pi} D^{(K)}\wedge D\\
&+ \frac{i}{4\pi} \left(\bar\xi_{KK} F_+^{(K)}\wedge F_+^{(K)} + \xi_{KK} F_-^{(K)}\wedge F_-^{(K)}\right) - \frac{\text{Im}(\xi_{KK})}{2\pi} D^{(K)}\wedge D^{(K)}\\
&+ \frac{i}{2\pi} \left(\bar\xi_{KI} F_+^{(K)}\wedge F_+^{(I)} + \xi_{KI} F_-^{(K)}\wedge F_-^{(I)}\right) - \frac{ \text{Im}(\xi_{KI})}{\pi} D^{(K)}\wedge D^{(I)}.
\eea
We can consistently set the zero modes in the nondynamical $\mathrm{U}(1)^{(I)}$ and $\mathrm{U}(1)^{(K)}$ multiplets as $\eta^{(I)}=\chi^{(I)}=\eta^{(K)}=\chi^{(K)}=0$ and
\be
\label{eq:DI-LEET}
D^{(I)}=F^{(I)}_{+},\quad D^{(K)}=F^{(K)}_{+},
\ee
ensuring that the scalar supercharge is independent of background fluxes. 

For algebraic surfaces $X$, we can use the effective Lagrangian above to confirm the identification of parameters in this present paper with those in \cite{Gottsche:2006bm} given in \eqref{eq:c1-Lgny-bfnI}. Note that the exponential of the coupling $-\pi i \xi_{II}=\frac{R^2}{32}
\partial^2\CF/\partial \log(\CR)^2$ appears in the wall-crossing expression. By comparing with the wall-crossing expression in \cite[Corollary 4.2]{Gottsche:2006bm}, we determine
$\bfn_I$ to be
\be
\label{nIGNY}
\pm  \bfn_I=K_X+c_1(v)+\frac{{\rm rk}(v)}{2}(c_1-K_X),
\ee
where $c_1$ is a lift of $w_2(P)$ to $H^2(X,\mathbb{Z})$, $v$ is a certain auxiliary K-theory class over $X$, and our comparison only gives the identification up to sign on the left-hand side.    
For $v$ determined by a line bundle $L_{\rm GNY}$ over $X$ as in \cite[Eq.  (1.4)]{Gottsche:2006bm}, we have ${\rm rk}(v(L_{\rm GNY}))=0$, and \eqref{nIGNY} becomes
\be 
\pm \bfn_I = K_X + c_1\left(v\left(L_{\rm GNY}\right)\right) = K_X + c_1\left(L_{\rm GNY}^{-1}\right) = K_X - c_1\left(L_{\rm GNY}\right),
\ee
which reproduces \eqref{eq:c1-Lgny-bfnI} if we choose the plus sign on the left-hand side. 

While it is straightforward to incorporate $\bfn_K\neq 0$, we will restrict to $\bfn_K=0$ for simplicity. Then the bosonic effective action on the $U$-plane reduces to \footnote{Here and in the following, we use $\bfx^2=B(\bfx,\bfx)$ for $\bfx\in H^2(X,\mathbb{R})$.}
\begin{align}
\label{effactionevaluated}
 e^{-\int_X \CL_0|_{\text{bos}} } &\sim \exp\left(-\pi i \bar \tau \bfk_+^2-\pi i \tau
   \bfk_-^2 - 2 \pi i \bar v_I B(\bfk_+,\bfn_I) -2\pi i  v_I B(\bfk_-,\bfn_I)\right) \nonumber \\
 &\quad \times \exp\left(-\pi i \xi_{II}\, \bfn_I^2-2\pi \frac{{\rm
     Im}(v_I)^2}{y} \bfn_{I,+}^2 \right).
\end{align}

\subsection{Integrand}\label{subsec:Integrand}

As mentioned above, the local form of the measure for the $U$-plane integral comes from reducing the IR theory to its zero modes \cite{Witten:1995gf, Moore:1997pc}. 
Beyond the exponentiated action (\ref{effactionevaluated}), there are a few additional contributions: First, the path integral involves a phase $e^{\pi i B(\bfk,K)}$, where $K$ is a characteristic vector of the lattice $L$. This phase is particularly important for duality transformations of the integrand of the path integral. 
It is attributed to integrating out the massive fermionic modes in \cite{Witten:1995gf}, and is derived in 4d SYM theory as the infinite-mass limit of the theory with an adjoint hypermultiplet in \cite{Manschot:2021qqe}. 
Second, there are topological couplings to the Euler characteristic $\chi$ and signature $\sigma$ of $X$, expressed as $A^{\chi}\,B^{\sigma}$ \cite{Witten:1995gf, Moore:1997pc, Manschot:2019pog, Nakajima:2005fg}.
Therefore, the local measure takes the form
\footnote{There is an unfortunate clash of notation in this equation: We use $\chi$ both for the Euler characteristic of $X$ and a fermionic zero mode, trusting that no confusion will ensue. }
\be 
\label{eq:localmeasure}
K_U\,  \mathrm{d}a \wedge \mathrm{d}\bar a\, A^{\chi} B^{\sigma} \int \mathrm{d}\eta\, \mathrm{d} \chi\, \mathrm{d}D\, e^{\pi i B(\bfk,K)} \,e^{-\int_X \CL_0},
\ee
where $K_U$ is a normalization constant, $\eta, \chi, D$ are the zero modes,  $\CL_0$ \eqref{LEETL} is the effective Lagrangian of the low-energy $\mathrm{U}(1)$ gauge theory, and $A,B$ are couplings to the background graviton in the LEET. 

For 5d $\mathrm{SU}(2)$ SYM theory, the topological couplings $A,B$ are unchanged as functions of $\tau$ compared to 4d $\mathrm{SU}(2)$ SYM \cite{Nakajima:2005fg, Gottsche:2006bm, Closset:2021lhd}, 
\be 
A=\alpha\, R^{-1}\,\left( \frac{\mathrm{d}U}{\mathrm{d}a}\right)^{\frac{1}{2}} ,\quad
B=\beta\, \Lambda^{\frac12}\,\Delta_{\rm phys}^\frac{1}{8},
\ee
where $\alpha$ and $\beta$ are numerical constants. The factors $R^{-1}$ and $\Lambda^{1/2}$ are included so that the 4d limit \eqref{4dlimit} of $A$ and $B$ agree with those of 4d $\mathrm{SU}(2)$ SYM, up to numerical factors which can be absorbed into $\alpha$ and $\beta$. We fix them
such that Eq. \eqref{nuR} holds. We define the measure $\nu_R$ as 
\be 
\label{nuRdef}
\nu_R(\tau)=\Lambda^{-3} 2\sqrt{2}\pi i\,A^{\chi}\, B^{\sigma}\, \frac{\mathrm{d}a}{\mathrm{d}\tau}.
\ee 
Using Eq. \eqref{dUdtau}, we find that the measure $\nu_R$ is simply related to the measure $\nu$ of 4d SYM in \cite[Eq.  (4.4)]{Korpas:2019cwg},
\be 
\label{nuR}
\nu_R(\tau)=\frac{2\nu(\tau)}{U}=-\frac{i}{4U}\frac{\vartheta_4(\tau)^{13-b_2}}{\eta(\tau)^9}.
\ee

Due to the branch cut for ${\rm Im}(\tau)>{\rm Im}(\tau_{\rm bp})$, the weak-coupling limit, $\tau\to i\infty$, of the measure requires careful treatment. We have for the limit $\tau\to i\infty$ of $U$ infinitesimally right to the branch cut, 
\be 
\lim_{\eps\to 0}\,\lim_{\tau\to \epsilon+i\infty} U(\tau)= i\CR\,q^{-\frac18}+\CO\left(q^{\frac18}\right).
\ee 
 As a result, $\nu_R$ diverges in this limit as
\be
\label{nuRq}
\lim_{\tau\to i\infty} \nu_R(\tau)=
-\frac{1}{4\CR}\,q^{-\frac14}-\frac{1+\CR^4}{2\CR^3}+\CO\left(q^{\frac14}\right).
\ee

On the other hand for $\CR\to 0$, the argument of the square root in Eq.  (\ref{nuR}) is dominated by $4$, yielding the expansion in this regime 
\be
\label{nuRR}
\lim_{\CR\to 0} \nu_R(\tau)=\left(-\frac{i}{8}\, q^{-\frac38} + \CO\left(q^{\frac18}\right)\right)+\left(-\frac{i}{64}q^{-\frac58}+\CO\left(q^{-\frac18}\right)\right)\CR^2+\CO\left(\CR^4\right).
\ee
The leading term reproduces the measure for the 4d SYM.

For simplicity, we first consider $\bfn_I=0$.
From the exponentiated action (\ref{effactionevaluated}) and integrating over the fermionic modes $\eta$ and $\chi$, we find that the sum over $\mathrm{U}(1)$ fluxes then takes the form of a Siegel-Narain theta series,
\be
\Psi^J_\bfmu(\tau,\bar \tau)=\frac{i}{2\sqrt{2y}} \sum_{\bfk\in
  L+\bfmu} B(\bfk,J) \, e^{\pi i B(\bfk,K)}\,q^{-\bfk_-^2/2}\,\bar q^{\bfk_+^2/2}.
\ee
The subscript $\bfmu$ of $\Psi^J_\bfmu$ is half an integral lift of $w_2(P)$ modulo $2$. 
This function vanishes if $e^{2\pi i B(\bfmu,K)}=1$. Moreover, for a different choice of characteristic vector, $K\to K+2\zeta$, $\zeta\in L$, the phase $e^{\pi i B(\bfk,K)}$ changes by $e^{2\pi i B(\bfmu,\zeta)}$. $\Psi^J_\bfmu$ is thus a function of the characteristic vector $K$ of the lattice $L$ modulo $4$, i.e., $K$ is an integral lift of $w_2(TX)$ modulo $4$ and is not necessarily the canonical class $K_X$.  A topological term $\exp[\frac{\pi i}{2}((2\bfmu)^2-B(2\bfmu,K))]$ can be included such that correlation functions only depend on $K$ modulo $2$ \cite{Moore:1997pc}, but at the expense that the path integral depends on $\overline{w_2(P)}$ modulo $4$. This can be interpreted in terms of mixed anomalies and anomaly inflow, analogous to the 4d theory \cite{Cordova:2018acb, Wang:2018qoy}.

We formulate the $U$-plane integral with $\bfn_I=0$ as
\be
\Phi^J_{\bfmu}(\CR)=K_U\,\int_{\CF_\CR} \mathrm{d}\tau\wedge\mathrm{d}\bar \tau\,\nu_R(\tau)\,\Psi^J_\bfmu(\tau,\bar \tau).
\ee
Since $\chi+\sigma=4$, scaling $K_U$, $\alpha$ and $\beta$ according to 
\be\label{eq:scalingKU}
(K_U,\alpha,\beta)\sim (\zeta^{-4}\,K_U,\zeta\,\alpha,\zeta\,\beta)
\ee
gives an equivalent integral \cite{Moore:2017cmm, Manschot:2019pog}.
For 4d SYM theory, \cite{Korpas:2019cwg} used the scaling \eqref{eq:scalingKU} to normalize the $u$-plane integral to unity. Since the $U$-plane integral provides a double cover of the $u$-plane integral in the 4d limit, we set
\be
\label{KU}
K_U=\frac{1}{2}.
\ee
Correlation functions should then agree in the limit $\CR\to 0$.

When the $\mathrm{U}(1)^{(I)}$ background flux $\bfn_I$ is non-vanishing, we need to further include its coupling to $v_I$ \eqref{vandxi}. To this end, we introduce the Siegel-Narain theta series $\Psi^J_\bfmu$ with elliptic argument $\bfz\in L\otimes \mathbb{C}$,
\begin{align}
\label{DefPsi}
\Psi^J_\bfmu(\tau,\bar \tau,\bfz ,\bar \bfz ) 
& =\exp(-2\pi y \bfb_+^2) \sum_{\bfk\in
  L+\bfmu} \partial_{\bar \tau} \left(\sqrt{2y}B(\bfk+\bfb,J)
\right)\, e^{\pi i B(\bfk,K)}\,q^{-\bfk_-^2/2}\,\bar q^{\bfk_+^2/2} \nonumber \\
&\quad  \times \exp\!\left( -2\pi i B(\bfk_-,\bfz) -2\pi i B(\bfk_+,\bar \bfz)\right),
\end{align}
where $\bfb={\rm Im}(\bfz)/y$. Comparing with Eq. \eqref{effactionevaluated}, we deduce that the elliptic argument $\bfz$ is identified with $v_I \bfn_I$.

We summarize a few properties of $\Psi_\bfmu^J$ for generic $\tau$ and $\bfz$, which will be important for the analysis of the $U$-plane integral in following subsections. Under the generators of $\Gamma^0(4)$, $\Psi_\bfmu^J$ transforms as
\be
\label{Psitrafos}
\begin{aligned}
S^{-1}TS: & \quad \Psi^J_\bfmu\left(\frac{\tau}{-\tau+1},\frac{\bar
    \tau}{-\bar \tau+1},\frac{\bfz}{-\tau+1},\frac{\bar \bfz}{-\bar
    \tau+1}\right)  \\
   & =(-\tau+1)^{b_2/2}(-\bar
\tau+1)^2\exp\left(\frac{\pi i \bfz^2}{-\tau+1} -\frac{\pi
    i}{4}\sigma(X)\right) \Psi^J_\bfmu(\tau,\bar \tau,\bfz,\bar \bfz),  \\
T^4:& \quad  \Psi^J_\bfmu(\tau+4,\bar \tau+4,\bfz,\bar \bfz)  =e^{2\pi i B(\bfmu,K)}\,\Psi^J_\bfmu(\tau,\bar \tau,\bfz,\bar \bfz).  
\end{aligned}
\ee
Additionally, we list its transformations under reflections and shifts of $\bfz$:
\begin{itemize}
\item Reflection $\bfz\to -\bfz$,
\begin{equation}\label{ellparamminus}
\Psi^J_\bfmu(\tau,\bar \tau,-\bfz,-\bar \bfz)=-e^{2\pi i B(\bfmu,K)}\,\Psi^J_\bfmu(\tau,\bar \tau,\bfz,\bar \bfz).
\end{equation} 
\item Shift of the elliptic variable, $\bfz\to \bfz+\bfnu$ with $\bfnu\in L$, 
\be\label{zshift}
\Psi^J_\bfmu(\tau,\bar \tau,\bfz+\bfnu,\bar \bfz+\bfnu)=e^{-2\pi i B(\bfnu,\bfmu)}\, \Psi^J_{\bfmu}(\tau,\bar \tau,\bfz,\bar \bfz).
\ee
\item Shift $\bfz\to \bfz+\bfnu\tau$ with $\bfnu\in L \otimes \mathbb{R}$,
\be\label{zshifttau}
\Psi^J_\bfmu(\tau,\bar \tau, \bfz+\bfnu\tau, \bar \bfz+\bfnu \bar \tau  )=e^{2\pi i B(\bfz,\bfnu)} q^{\frac{1}{2}\bfnu^2} e^{-\pi i B(\bfnu,K)}\, \Psi^J_{\bfmu+\bfnu}(\tau,\bar \tau,\bfz,\bar \bfz).
\ee
We can restrict to $\bfnu\in L/2$ if ${\bfmu+\bfnu} \in L/2$ is required. For $\bfnu\in L$,
$\bfmu+\bfnu\simeq \bfmu \in (L/2)/L$.
\end{itemize}
Note that $\Psi^J_\bfmu$ is bounded for generic $J^2>0$ and $\tau\to i\infty$,
\be
\label{Psibounded}
\begin{split}
|\Psi^J_\bfmu(\tau,\bar \tau,\bfz,\bar \bfz)| &\leq \exp(-2\pi y \bfb_+^2) \sum_{\bfk\in
  L+\bfmu} \partial_{y} \left(\sqrt{2y}B(\bfk+\bfb,J)
\right) \\
& \times \exp\left[\pi y (\bfk+\bfb)_-^2-\pi y (\bfk+\bfb)_+^2-\pi y(\bfb_-^2-\bfb_+^2)\right]\\
&< \infty.
\end{split}
\ee

The $U$-plane integral with the coupling to $\bfn_I$ is then formally defined as the integral 
\be
\label{UplaneintDef}
\Phi^J_{\bfmu,\bfn_I}(\CR,\bar \CR)=K_U\,\int_{\CF_\CR} \mathrm{d}\tau\wedge \mathrm{d}\bar
\tau\,\nu_R(\tau)\,C^{\bfn_I^2}\,\Psi^J_\bfmu(\tau,\bar \tau,v_I\bfn_I,\bar v_I\bfn_I),
\ee
where the $\bar \CR$ dependence on the left-hand side enters through $\bar v_I$, and $C^{\bfn_I^2}$ is a contact term with $C$ given in Eq. \eqref{Ctauv}. We deduce from Eq. \eqref{ellparamminus} that
\footnote{It would be nice to deduce this from discrete symmetries of the IR theory. }
\be 
\label{Phin-n}
\Phi_{\bfmu,-\bfn_I}^J(\CR,\bar \CR)=- e^{2\pi i B(\bfmu,K)}\,\Phi_{\bfmu,\bfn_I}^J(\CR,\bar \CR).
\ee 

In addition to the background flux for $\mathrm{U}(1)^{(I)}$, we can also introduce a background flux for $\mathrm{U}(1)^{(K)}$. Besides the contact term $C_{KK}$, this also gives rise to the mixed contact term $C_{IK}$ (\ref{Cs}). Then the $U$-plane integral  becomes
\be 
\label{DefUplaneIntnIK}
\begin{split}
\Phi^J_{\bfmu,\bfn_I,\bfn_K}(\CR,\bar \CR)&=K_U\,\int_{\CF_\CR} \mathrm{d}\tau\wedge \mathrm{d}\bar
\tau\,\nu_R(\tau)\,C_{II}^{\bfn_I^2}\,C_{KK}^{\bfn_K^2}\,C_{IK}^{2 B(\bfn_I,\bfn_K)}\\
&\quad \times\Psi^J_\bfmu(\tau,\bar \tau,v_I\bfn_I+v_K\bfn_K,\bar v_I\bfn_I+\bar v_K\bfn_K).
\end{split}
\ee

Eqs. \eqref{UplaneintDef} and \eqref{DefUplaneIntnIK} raise a number of issues: 

\begin{enumerate} 

\item The integrand is not obviously single-valued on the $U$-plane. Physically the integrand is an observable that must be single-valued. 

\item The integrals, taken literally, are highly divergent. We must ask if there is a sensible definition of the integral. 

\item The integral is, \emph{a priori} a function of both $\CR$ and $\bar{\CR}$, while formal arguments from the localization of path integrals predict it is holomorphic in $\CR$. We must therefore investigate whether there is a holomorphic anomaly in the $\CR$-dependence. 

\end{enumerate}

\subsection{The Integrand And Monodromies}
\label{IntMono}

As mentioned above, an important consistency requirement is that the integrands defined in Eqs. \eqref{UplaneintDef} and \eqref{DefUplaneIntnIK} must be single-valued on the $U$-plane. As explained in Sec.  \ref{SecFundDom}, we have modeled the $U$-plane as the domain $\CF_R$ with appropriate identifications. In this subsection, we verify that the integrand is indeed invariant under the monodromies ${\bf M}_j$ discussed in Sec.  \ref{Monos}, which generate the monodromy group. 
The $U$-plane integrand comprises several terms: $C_{II}$ \eqref{Ctauv}, $C_{IK}$, $C_{KK}$ \eqref{Cs}, $\nu_R$ \eqref{nuR}, and $\Psi_\bfmu^J$ \eqref{DefPsi}. Using Eq.  \eqref{Psitrafos} and the transformation properties of modular forms reviewed in App. \ref{App:MForms}, we demonstrate that the integrand is indeed single-valued under the monodromy group for $\bfmu\in L/2$ and $\bfn_I,\bfn_K\in L$. For simplicity, we restrict our analysis to special cases:
\begin{itemize}
\item Case I: $\bfn_I \in L$, $\bfn_K=0$, which is our primary case of interest;
\item Case II: $\bfn_I=0$, $\bfn_K\in L$;
\item Case III: $\bfn_I,\bfn_K\in L$, which is the general case.
\end{itemize}

\subsubsection*{Case I: $\bfn_I \in L$, $\bfn_K=0$}
\paragraph{Monodromy around $U_\infty$:}
The action of the monodromy ${\bf M}_\infty$ on the couplings is listed in Eq.  \eqref{eq:Monodromy-At-Infinity}, in particular $\tau\mapsto \tau-8$,
$v_I\mapsto v_I+2$ with ${\rm Im}(\tau)>{\rm Im}(\tau_{\rm bp})$. The terms in the integrand transform as
\begin{itemize}
\item $\nu_R(\tau)\to e^{4\pi i}\,\nu_R(\tau)=\nu_R(\tau)$,
\item $C_{II}\to \,C_{II}$,
\item $\Psi^J_\bfmu(\tau,\bar \tau,v_I\bfn_I,\bar v_I\bfn_I)\to e^{-4 \pi i B(\bfmu,\bfn_I)}\Psi^J_\bfmu(\tau,\bar \tau,v_I\bfn_I,\bar v_I\bfn_I)$,
\end{itemize}
where we assumed for the last transformation that $2\bfn_I\in L$, which is weaker than the actual physical domain $\bfn_I\in L$. Thus, the conditions for the integrand to be single-valued under the monodromy ${\bf M}_\infty$ are 
\be
\label{Uinfconst}
e^{-4\pi i  B(\bfmu,\bfn_I)}=1,\quad {\rm and} \quad 2\bfn_I \in L,
\ee 
which are always satisfied for all $\bfmu\in L/2$ and $\bfn_I\in L$.

\paragraph{Monodromy around $U_1$:}
Using the transformations of the couplings \eqref{eq:U1-monotmns}, we deduce that the monodromy ${\bf M}_1$ changes the terms of the $U$-plane integrand as:  
\begin{itemize}
\item $\nu_R(\tau)\to e^{\frac{1}{4}\pi i \sigma}(-\tau+1)^{2-b_2/2}\,\nu_R(\tau)$,
\item $C_{II}\to \,\exp\!\left(-\pi i\frac{ v_I^2}{-\tau+1}\right)\,C_{II}$,
\item $\Psi_\bfmu^J\to (-\tau+1)^{\frac{1}{2}b_2}(-\bar
  \tau+1)^2\,\exp\!\left(\pi i\frac{ v_I^2\bfn_I^2}{-\tau+1}-\frac{\pi i \sigma}{4} \right)\,\Psi_\bfmu^J$. 
\end{itemize}
Since $\mathrm{d}\tau\wedge \mathrm{d}\bar \tau$ has weight $(-2,-2)$, we confirm that the integrand is single-valued for this monodromy.

\paragraph{Monodromy around $\{U_1,U_2\}$:}
The action of the monodromy ${\bf M}_{12}={\bf M}_1 {\bf M_2}$ on the couplings gives $\tau\to \tau-4$, $v_I\to - v_I$ and $\xi_{II}\to \xi_{II}$ while ${\rm Im}(\tau)<{\rm Im}(\tau_{\rm  bp})$. In this domain of $\tau$, $U$ does not change sign for $\tau\to \tau-4$. The various terms of the integrand then transform as
\begin{itemize}
\item $\nu_R(\tau)\to e^{3\pi i}\,\nu_R(\tau)=-\nu_R(\tau)$,
\item $C_{II}\to \,C_{II}$,
\item $\Psi^J_\bfmu\to -\Psi^J_\bfmu$,
\end{itemize}
which demonstrates that the integrand is single-valued for this monodromy. Since invariance under monodromy ${\bf M}_1$ has been shown above, this also establishes invariance under monodromy ${\bf M}_2$.

\paragraph{Monodromy around $U_3$:}
The action of this monodromy on the couplings is given in Eq. \eqref{U3trafos}. The terms in the integrand transform as
\begin{itemize}
\item $\nu_R(\tau)\to e^{\frac{1}{4}\pi i \sigma}\,(-\tau+1)^{2-b_2/2}\,\nu_R(\tau)$,
\item $C_{II}\to \,\exp\!\left(-\pi i \frac{(v_I+1)^2}{-\tau+1} \right)\,C_{II}$,
\item For $\bfn_I\in L$, $\Psi^J_\bfmu\to (-\tau+1)^{\frac{1}{2}b_2}(-\bar \tau+1)^2\,
  \exp\!\left(\pi i \frac{(v_I+1)^2\bfn_I^2}{-\tau+1}-\frac{\pi i \sigma}{4}
  \right)\, \Psi^J_\bfmu$.
\end{itemize}
The condition $\bfn_I\in L$ is stronger than Eq. (\ref{Uinfconst}) above. For all $\bfn_I\in L$, the integrand is thus single-valued for this monodromy.

Since the monodromy ${\bf M}_4$ around $U_4$ can be composed from
monodromies around the other singularities, the above demonstrates that the integrand is single-valued under the monodromy group generated by ${\bf M}_\infty$, ${\bf M}_1$, ${\bf M}_2$ and ${\bf M}_3$. 

\subsubsection*{Case II: $\bfn_I=0$, $\bfn_K\in L$}
\paragraph{Monodromy around $U_\infty$:}
With the action on the couplings of ${\bf M}_\infty$ given in \eqref{eq:Monodromy-At-Infinity}, we find that the terms in the integrand transform as
\begin{itemize}
\item $\nu_R(\tau)\to e^{4\pi i}\,\nu_R(\tau)=\nu_R(\tau)$,
\item $C_{KK}\to \,q^{-\frac{1}{2}} e^{2\pi i v_K+\pi i}C_{KK}$,
\item $\Psi^J_\bfmu(\tau,\bar \tau,v_K\bfn_K,\bar v_K\bfn_K)\to
  e^{-2\pi i \bfn_K^2v_K}\,q^{\frac{1}{2}\bfn_K^2}(-1)^{\bfn_K^2}\,\Psi^J_\bfmu(\tau,\bar \tau,v_K\bfn_K,\bar v_K\bfn_K).$
\end{itemize}
With $\bfn_K\in L$, the integrand is single-valued under this monodromy.

\paragraph{Monodromy around $U_1$:}
The action of ${\bf M}_1$ on the couplings (\ref{eq:U1-monotmns}) shows that the action on $v_K$ and $\xi_{KK}$ is identical to the action on $v_I$ and $\xi_{II}$. Therefore, it follows directly from Case I discussed above that the integrand is single-valued for $\bfn_K\in L$ and $\bfn_I=0$.

\paragraph{Monodromy around $\{U_1,U_2\}$:}
For this monodromy, single-valuednes of the integrand also follows directly from Case I discussed above.

\paragraph{Monodromy around $U_3$:}
From the action of ${\bf M}_3$ on the couplings (\ref{U3trafos}), it follows that in this case
\begin{itemize}
\item $\nu_R(\tau)\to e^{\pi i \sigma/4}\,(-\tau+1)^{2-b_2/2}\,\nu_R(\tau)$,
\item $C_{KK}\to \,\exp\!\left(-\pi i \frac{v_K^2}{-\tau+1} \right)\,C_{KK}$,
\item $\Psi^J_\bfmu\to (-\tau+1)^{\frac{1}{2}b_2}(-\bar \tau+1)^2\,
  \exp\!\left(\pi i \frac{v_K^2\bfn_K^2}{-\tau+1}-\frac{\pi i \sigma}{4}  \right)\, \Psi^J_\bfmu$. 
\end{itemize}
The integrand is thus also single-valued with respect to this monodromy.

In conclusion, the integrand is single-valued for $\bfn_I=0$, $\bfn_K\in L$.

\subsubsection*{Case III: $\bfn_I,\bfn_K\in L$}
\paragraph{Monodromy around $U_\infty$:} Employing the same approach as before, we find that the transformations of various terms in the integrand under the action of ${\bf M}_\infty$ are given by
\begin{itemize}
\item $\nu_R(\tau)\to e^{4\pi i}\,\nu_R(\tau)=\nu_R(\tau)$,
\item $C_{II}\to \,C_{II}$,
\item $C_{KK}\to \,q^{-\frac12}\,e^{2\pi i v_K+\pi i}\,C_{KK}$,
\item $C_{IK} \to e^{\pi i v_I}\, C_{IK}$,
\item for $\bfn_I\in L/2$ and $\bfn_K\in L$,
\be 
\begin{split} 
&\Psi^J_\bfmu(\tau,\bar \tau,v_I\bfn_I+v_K\bfn_K,\bar v_I\bfn_I+\bar v_K\bfn_K)\\
\to & (-1)^{B(\bfn_K,K)}e^{-4\pi i B(\bfmu,\bfn_I)-\pi i B( v_I\bfn_I+2v_K\bfn_K,\bfn_K)} q^{\frac{1}{2}\bfn_K^2}\\
& \times
\Psi^J_\bfmu(\tau,\bar \tau,v_I\bfn_I+v_I\bfn_I,\bar v_I\bfn_I+\bar v_K\bfn_K).
\end{split}
\ee 
\end{itemize}
For $\bfn_I,\bfn_K\in L$, the integrand is single-valued under the monodromy ${M}_\infty$.

\paragraph{Other monodromies:}
For monodromies around $U_1$ and $\{U_1,U_2\}$, it follows directly from the analysis for Case I that the integrand is single-valued. For monodromy around $U_3$, we use the action on the couplings \eqref{U3trafos} to verify that the integrand is single-valued for $\bfn_I,\bfn_K\in L$.

We therefore conclude that the integrand is single-valued under the monodromy group for general $\bfn_I,\bfn_K\in L$.

\subsection{Vanishing Of The $U$-plane Integral Due To One-form Symmetry}
\label{sec:UInt1Form}

In Sec.  \ref{UPlaneIntDefs} we will give a careful definition of the rather singular $U$-plane integral \eqref{UplaneintDef}. Prior to this, however, we note an important formal property (which will remain valid under the careful definition given below). 

We first focus on the case $\bfn_K=0$. We regard the measure on the $U$-plane as a $(1,1)$-form. When pulled back via $\tilde\pi_{\ttu}$ to $\widetilde{\mathbb{H}}$ as in \eqref{eq:DoubleCover-UHP} it is of the form $\pi^*(\Omega_{\bfmu})/U$, where $\Omega_{\bfmu}$ is a $(1,1)$ form on the upper half-plane that is invariant under $\Gamma^0(8)$. Explicitly, using Eqs. \eqref{nuR} and \eqref{UplaneintDef} we find
\be
\label{eq:OmegaDef}
\Omega_{\bfmu} = -\frac{i}{4}\, \mathrm{d}\tau \wedge\mathrm{d}\bar \tau\,  \frac{\vartheta_4(\tau)^{13-b_2}}{\eta(\tau)^9}\, C^{\bfn_I^2}\, \Psi_\bfmu^J(\tau,\bar\tau, v_I \bfn_I, \bar v_I \bfn_I).
\ee
The measure $\pi^*(\Omega_{\bfmu})/U$ is almost invariant under  $\Gamma^0(4)$ but 
can transform by a sign under the deck transformation of $\widetilde{\mathbb{H}}$.
The deck transformation of the cover $\IC_{U} \to \IC_{\ttu}$ pulls back to a transformation on $\widetilde{\mathbb{H}}$ which covers $\tau \to \tau+4$. Using Eqs. \eqref{eq:DoubleCover-UHP} and \eqref{Ttrafotauv} we find that the deck transformation leads to the transformation 
$\tau \to \tau +4$ and $v_I \to v_I-1$.
Applying Eqs. \eqref{Psitrafos} and \eqref{zshift} for $\Psi_\bfmu^J$ and Eq. \eqref{Ttrafotauv} for $C=\exp(-\pi i \xi_{II})$, it is straightforward to show that $\Omega_{\bfmu}$ transforms under 
$\tau \to \tau + 4 $ and $v_I \to v_I-1$ as 
\be 
\label{OmegT}
\Omega_{\bfmu} \to - \exp\left(i \pi   B(2\bfmu,\overline{w_2(X)} -\bfn_I) \right) \Omega_{\bfmu}. 
\ee
Therefore, since $U\to - U$ under the deck transformation, we observe that when $B(2\bfmu, \overline{w_2(X)} - \bfn_I)$ is odd, the integral over the $U$-plane equals zero. Thus, precisely when the condition \eqref{eq:AnomalyCancelCriterion} for global anomaly cancellation fails, the $U$-plane integral  vanishes due to cancellation between copies related by the one-form symmetry discussed in Sec.  \ref{Monos} and Sec. \ref{subsec:AnomalyCancellation}.

We continue by exploring the action of ${\bf T}$ with both $\bfn_I,\bfn_K\in L$. Let $\pi^*(\Omega_\bfmu)/U$ be the integrand of \eqref{DefUplaneIntnIK}, where $\Omega_\bfmu$ now includes the couplings $C_{IK}$ and $C_{KK}$. Then, acting with ${\bf T}$ results in 
\be 
\Omega_{\bfmu}\to -\exp\left(\pi i B(2\bfmu+\bfn_K,K+\bfn_I)-\frac{1}{2}\pi i B(\bfn_I,\bfn_K)\right) \Omega_{\bfmu+\frac{1}{2}\bfn_K}.
\ee 
Thus, the action of ${\bf T}$ is not a symmetry if $\bfn_K\notin 2L$, which is expected since the circle bundle is not a direct product for $\bfn_K\neq 0$. Interestingly, for $\bfn_K\in 2L$, the action preserves the integrand up to a sign. The background fluxes $\bfn_I$ and $\bfn_K$ should be chosen such that the sign equals $+1$ for a non-vanishing path integral. It would be valuable to understand this condition from first principles in terms of anomalies.

Finally, we remark that one can include observables as in Sec.  \ref{sec:OtherObs}. 
The observables $U$ and $\CO(S)$ are odd under the one-form symmetry. Therefore, for a non-anomalous theory, the partition function is an even function on the homology of $X$.

\subsection{Two Definitions Of The $U$-plane Integral}
\label{UPlaneIntDefs}

Eq. \eqref{UplaneintDef} makes it clear that the $U$-plane integral is a linear combination of integrals of the form
\be 
\label{genIntf}
\CI_f=\int_{\CD} \mathrm{d}\tau\wedge \mathrm{d}\bar \tau\,y^{-s}\,f(\tau,\bar \tau),
\ee 
where $s\in \mathbb{Z}/2$, $f$ is a non-holomorphic modular form of weight $(2-s,2-s)$, and $\CD\in \mathbb{H}$ denotes the integration domain. Without a proper definition, such integrals are known to be divergent in many cases of interest \cite{Dixon:1990pc, Harvey:1995fq, Borcherds:1996uda}. 

We now discuss two distinct definitions of the $U$-plane integral, which roughly correspond to: 
\begin{enumerate}
\item expanding the integrand in a power series in small $\CR$ and $\bar \CR$ before integrating over $\tau$,
\item integrating over the domain $\CF_{\CR}$ at a fixed, finite $\CR$ and $\bar \CR$.
\end{enumerate}
Before stating these definitions carefully, we will briefly review a few aspects of integrals over modular fundamental domains. More details can be found in \cite{Bringmann:2016, Korpas:2019ava}.

\begin{figure}[t]\centering
\includegraphics[width=0.7\textwidth]{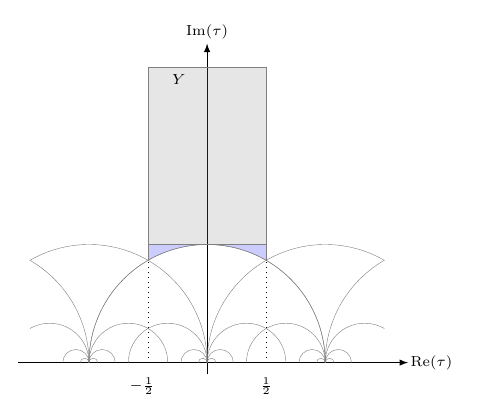} 
\caption{Plot of the domain $\CF^Y$ as the union of two purple regions and grey rectangle. The union of the two purple regions is $\CF^1$.} \label{Fig:FY}
\end{figure}

We start by defining the integral
\be 
\label{defL}
L_{m,n,s}(\CD)=\int_\CD \mathrm{d}\tau \wedge \mathrm{d}\bar \tau\,y^{-s}\,q^m\bar q^n,
\ee 
where $s\in \mathbb{Z}/2$, $m,n\in \mathbb{R}$, $m-n\in \mathbb{Z}$, and $\CD\in \mathbb{H}$ is a domain in the upper half-plane. An example of a compact domain is $\CF^Y$ with $Y\in [1,\infty)$, defined as the domain bounded by the following four arcs
\be
\begin{aligned}
&(1) \quad \tau=\frac{1}{2} +i y, & y&\in \left[\frac{1}{2}\sqrt{3},Y\right], \\
&(2) \quad \tau=x+iY, & x&\in \left[-\frac{1}{2},\frac{1}{2}\right], \\
&(3) \quad \tau=-\frac{1}{2} +i y, & y&\in \left[\frac{1}{2}\sqrt{3},Y\right], \\
&(4) \quad \tau=i\,e^{i\varphi}, & \varphi&\in \left[-\frac{1}{6}\pi,\frac{1}{6}\pi\right].
\end{aligned}
\ee
The domain is displayed in Fig. \ref{Fig:FY}.
We furthermore define the noncompact domain
\be 
\CF^{\infty} = \lim_{Y\to \infty} \CF^Y,
\ee
which is the standard keyhole fundamental domain for $\mathbb{H}/\mathrm{SL}(2,\mathbb{Z})$. A general $\mathrm{SL}(2,\mathbb{Z})$ image of $\CF^{\infty}$ is denoted by $\CF$. If the domain $\CD$ in (\ref{defL}) is noncompact, the integral $L_{m,n,s}(\CD)$ may not be well defined. This is for example the case for $m,n<0$ and $\CD=\CF^{\infty}$. 

We next consider the integral \eqref{defL} with $\CD=\CF^\infty$ as the limit $Y\to \infty$ of integrals over the domain $\CF^Y$,
\be
\label{Lmnslimit}
L_{m,n,s}(\CF^{\infty})=\lim_{Y\to \infty} L_{m,n,s}(\CF^Y). 
\ee 
This limit converges for $m+n>0$ with arbitrary $s\in \mathbb{R}$, and for $m+n=0$ with $s>1$. In these cases, the limit evaluates to
\be 
L_{m,n,s}(\CF^{\infty})=L_{m,n,s}(\CF^1)-2i\,\delta_{m,n} E_{s}(4\pi m),
\ee 
where $E_l(z)$ is the generalized exponential integral defined by
\be
\label{defEellz}
E_l(z)=\begin{dcases} z^{l-1} \int_z^\infty e^{-t} t^{-l}\,\mathrm{d}t, & \quad \text{ for } z\in \mathbb{C}^*, l\in \mathbb{Z}/2,\\
\frac{1}{l-1}, & \quad \text{ for } z=0, l\neq 1, \\
0, & \quad \text{ for } z=0, l=1.
\end{dcases}
\ee 
The contour for $E_l$ is chosen so that for $z\in \mathbb{R}^+$, 
\be 
{\rm Im}(E_l(-z))=\lim_{\delta\to 0} {\rm Im}(E_l(-z-i\delta))=\frac{\pi\,z^{l-1}}{\Gamma(l)}, \quad l\geq 1.
\ee 

To properly regularize $L_{m,n,s}(\CF^{\infty})$ for all $m,n,s$, including divergent cases, we define $L^{\rm r}_{m,n,s}$ by \cite{Bringmann:2016, Korpas:2019ava}
\be 
\label{LrDef}
L^{\rm r}_{m,n,s}(\CF^{\infty})=L_{m,n,s}(\CF^1)-2i\,\delta_{m,n} E_{s}(4\pi m).
\ee 
This definition is motivated by regularizing the divergence for $Y\to \infty$, or as an analytic continuation in the parameters $m,n,s$. A distinguishing feature of this regularization is that if the integrand in \eqref{genIntf} can be expressed as a total $\bar \tau$-derivative, 
\be 
\partial_{\bar \tau} \widehat h(\tau,\bar \tau)=y^{-s}\,f(\tau,\bar \tau),
\ee 
with $\widehat h$ a non-holomorphic modular form of weight $(2,0)$ and non-singular on the interior of $\CF^{\infty}$. Then if the holomorphic part $h(\tau)$ of $\widehat h(\tau,\bar \tau)$ has the Fourier expansion, 
\be
h(\tau)=\sum_{n}d(n)q^n,
\ee
it can be demonstrated that the regularization \eqref{LrDef} applied to the integral \eqref{genIntf} evaluates to  the constant term of $h$ \cite{Bringmann:2016},
\be 
\label{CId0}
\CI^{\rm r}_f=d(0).
\ee 
This generalizes earlier definitions for restricted domains of the parameters $m,n$ and $s$ \cite{Dixon:1990pc, Harvey:1995fq, Borcherds:1996uda}. We will motivate and demonstrate later that K-theoretic Donaldson invariants are reproduced using this prescription.  

\subsubsection{Definition 1: Integration Over Formal $\CR$-Expansion}

In the first definition, we treat $\Phi_\bfmu^J$ as a formal power series in $\CR$ and $\bar \CR$
\be 
\label{PhiRbR}
\Phi_{\bfmu,\bfn}^J(\CR,\bar \CR)=\sum_{l_1,l_2} D_{l_1,l_2} \,\CR^{l_1} \bar \CR^{l_2},
\ee
where $D_{l_1,l_2}$ are determined by expanding the integrand of $\Phi_{\bfmu,\bfn}^J$ in $\CR$ and $\bar \CR$, and integrating the coefficients over the appropriate integration domain in the small-$\CR$ limit. 

Let us consider the $\CR$-expansion of each ingredient of the $U$-plane integrand in Eq.  \eqref{UplaneintDef}:
\begin{itemize}
\item  For $\nu_R$, the $\CR$-expansion gives a series whose coefficients contain powers of the Hauptmodul $\ttu$ \eqref{ttu}. Consequently, the exponents of $q$ are {\it not} bounded below in the small $\CR$ expansion, though they are bounded for fixed power of $\CR$.
\item For the coupling $C$, we see experimentally that its $\CR$-expansion \eqref{CRexp} has only positive powers of $\CR$ and $q$, at least up to $\CO\left(\CR^{10}\right)$.  
\item For the sum over fluxes $\Psi_\bfmu^J$, if $\tau$ and $v$ are treated as independent, then it is clear that the exponents of $q$ and $\bar q$ are positive. However, the $\CR$-expansion of the coupling $v$ \eqref{vsmallR} implies that $\Psi_\bfmu^J$ generally involves both negative and positive powers of $q$ and $\bar q$. These powers are bounded below for fixed powers of $\CR$ and $\bar \CR$.
\end{itemize}
From the expansions near the strong-coupling cusps $\tau_j$, we see that no negative powers of $q_j$ or $\bar q_j$ arise near these cusps. 

Regarding the integration domain, the discussion in Sec.  \ref{SecFundDom} demonstrates that the integration domain for infinitesimal $\CR$ becomes the fundamental domain $\mathbb{H}/\Gamma^0(4)$. More precisely, the domain consists of two copies of $\mathbb{H}/\Gamma^0(4)$, whose contributions add or subtract depending on the sign of the integrand under the Deck transformation \eqref{OmegT}. Then, it is straightforward to reduce the integral to an integral over a single copy of $\mathbb{H}/\Gamma^0(4)$.

As a result, the integrand can be further expanded in ${\rm Im}(\tau)=y$, $q$, and $\bar q$, such that the coefficients $D_{l_1,l_2}$ take the form
\be 
D_{l_1,l_2}=\sum_{m,n,s} \tilde D_{l_1,l_2,m,n,s}\,L_{m,n,s}(\mathbb{H}/\Gamma^0(4)),
\ee 
with $L_{m,n,s}$ as in Eq. \eqref{defL}. For fixed $l_1, l_2$, $m$ and $n$ are bounded below. Furthermore, $s>0$ for $\Phi_\bfmu^J(\CR,\bar \CR)$. Since $\CR$ and $\bar \CR$ are modular-invariant, the integrand
\be
\sum_{l_1,l_2,m,n,s} \tilde D_{l_1,l_2,m,n,s}\,y^{-s}q^m\bar q^n,
\ee
transforms properly on the domain $\mathbb{H}/\Gamma^0(4)$.

Using modular transformations, each copy of the fundamental domain $\CF \subset\mathbb{H}/\Gamma^0(4)$ can be mapped back to the standard keyhole fundamental domain $\CF^{\infty}$. Consequently, $D_{l_1,l_2}$ takes the form
\be 
D_{l_1,l_2}=\sum_{m,n,s} D_{l_1,l_2,m,n,s}\,L_{m,n,s}(\CF^{\infty}), 
\ee 
for some coefficients $D_{l_1,l_2,m,n,s}$. As $L_{m,n,s}(\CF^{\infty})$ is not always well defined for all values of $m,n, s$, we then \emph{define} the $U$-plane integral in terms of $L^{\rm r}_{m,n,s}$ \eqref{LrDef} as 
\be 
\label{PhiIntDef}
\Phi_{\bfmu,\bfn}^J(\CR,\bar \CR)=\sum_{l_1,l_2 \atop m,n,s} D_{l_1,l_2,m,n,s} \,\CR^{l_1} \bar \CR^{l_2}\,L^{\rm r}_{m,n,s}(\CF^{\infty}).
\ee 
We refer to this definition as {\it Definition 1}. We will use Stokes' theorem to evaluate the $\bar{\tau}$ integral, and then the remaining integral is computed in terms of the small-$q$ expansion, $\tau\to i\infty$. As a result, the evaluation of the $U$-plane integral in Sec.  \ref{Sec:FormEvInt} and Sec. \ref{Sec:DEvalU} using this definition will take the form ${\rm Coeff}_{q^0}{\rm Ser}_\CR[\cdots]$, with the dots an appropriate anti-derivative of the integrand.

\subsubsection{Definition 2: Integration With Fixed $\CR$}

As mentioned above, one may define the $U$-plane integral differently from {\it Definition 1}. The approach for {\it Definition 2} of the $U$-plane integral (\ref{UplaneintDef}) is to view $\CR$ and $\bar \CR$ as fixed in the integrand. {\it Definition 2} for $\Phi_{\bfmu,\bfn}^J$ then reads,
\be\label{eq:UplaneInt-FixedCR}
    \Phi_{\bfmu,\bfn}^J(\CR,\bar \CR)= \sum_{m,n,s} C_{m,n,s}(\CR, \bar \CR)\, L_{m,n,s}(\CF_\CR),
\ee 
with coefficients $C_{m,n,s}(\CR, \bar \CR)$. Unlike {\it Definition 1}, evaluation of $\Phi_{\bfmu,\bfn}^J$ using {\it Definition 2} takes the form ${\rm Ser}_\CR{\rm Coeff}_{q^0}[\cdots]$. 
  
Let us consider the behavior of the different ingredients of the integrand for this definition:
\begin{itemize} 
\item  For $\nu_R(\tau)$, we deduce from the $q$-expansion (\ref{nuRq}) that the exponents of $q$ are bounded below by $-1/4$.
\item For the coupling $C$, it is clear that the $q$-expansion (\ref{Cqexp}) gives rise to a series whose exponents are bounded below. 
\item For $\Psi^J_\bfmu$, since it is bounded (\ref{Psibounded}), we deduce from the small $q$-expansion for $v$ (\ref{vinfexp}) that this expansion for $\Psi^J_\bfmu$ only involves positive powers of $q$ and $\bar q$. 
\end{itemize}

\subsubsection{Which Definition Is Correct?}\label{subsubsec:WhichCorrect}

We have presented two ways to give meaning to the highly divergent $U$-plane integral. 
Roughly speaking, in Eq. \eqref{PhiIntDef} we first expand in $\CR$ and then evaluate the integral over $\tau$, whereas in Eq. \eqref{eq:UplaneInt-FixedCR} we hold $\CR$ fixed and evaluate the integral over $\tau$. 

\begin{figure}[t]\centering
\includegraphics[width=0.8\textwidth]{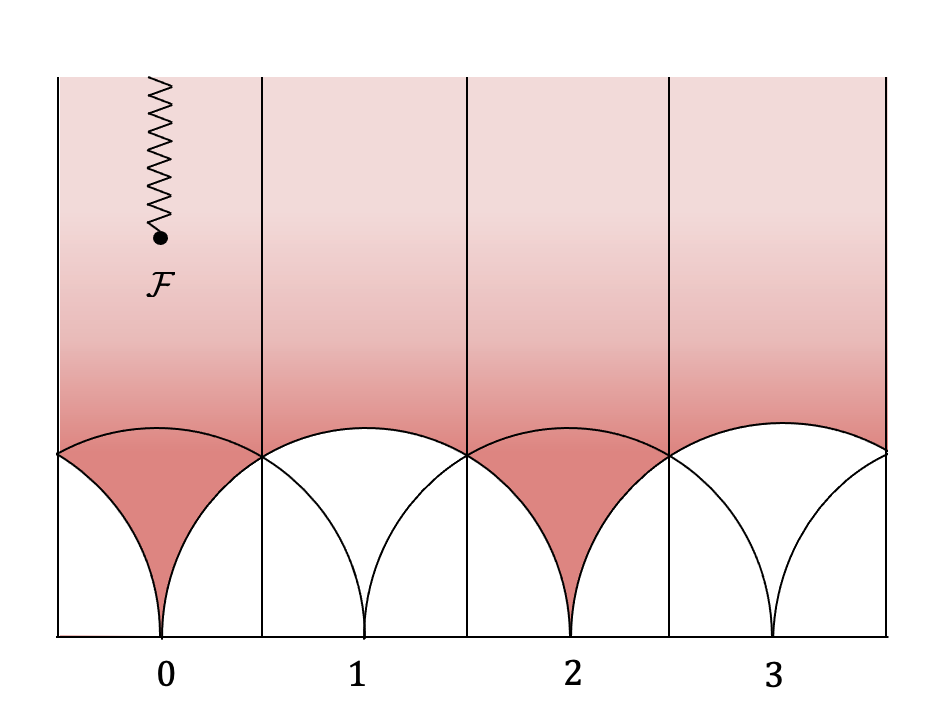} 
\caption{The red dark region represents the region where the physics is essentially 4d, and the light red region corresponds to the region where the physics is essentially 5d. } \label{Fig:5d-4d}
\end{figure}

The physically correct definition turns out to be {\it Definition 1} \eqref{PhiIntDef}. This follows from careful analysis of the 4d limit defined in Sec.  \ref{sec:5dR4S1}. In this limit the singularities $U_1, U_2$ collide near $U=2$ and the singularities $U_3, U_4$ collide near $U=-2$. A scaling region around $U=2$ gives a single copy of the 4d $u$-plane, and that around $U=-2$ gives another. Focusing on $U\approx 2$, we have 
\be 
\frac{U-2}{R^2} \rightarrow u(\tau)  
\ee
as $R\to 0$ and $U\to 2$ at fixed $\tau$. The physics near $U=0$ and in the region $\vert U^2 - 4\vert \gg 1 $ is effectively 5d. But $U=0$ corresponds to the branch points, which from Eq. \eqref{eq:SmallRbp}
behaves as
\be 
\tau_{\rm bp} \sim  \frac{ 4i}{\pi} \log \CR^{-1} 
\ee
in the $\CR \to 0$ limit. Thus, to recover the physics of the 4d effective theory, we should evaluate the $U$-plane focusing on the scaling regions around $U_1, U_2$ and $U_3, U_4$, and in particular should integrate in the region ${\rm Im}(\tau) \ll {\rm Im}(\tau_{\rm bp})$. 
This is naturally achieved by taking $\CR \to 0$ before integrating over $\tau$, which is the definition \eqref{PhiIntDef}. See Fig. \ref{Fig:5d-4d}.
This aligns qualitatively with the reduction of 4d to 3d supersymmetric field theories \cite{Aharony:2013dha}. 
Moreover, we find below that explicit evaluation using {\it Definition 1} agrees with the work of \cite{Gottsche:2006bm}, while explicit evaluation using {\it Definition 2} produces a 
Laurent series in $\CR$ with negative powers of $\CR$, which does not match the general form \eqref{eq:Zee-QM-1}. Compare, for example, the expressions for wall-crossing with the small $\CR$ expansion \eqref{DeltaPhiP1P1}, and with fixed $\CR$ \eqref{eq:Rfixed}.

There are other considerations, for example those in  \cite{Closset:2018bjz,BenettiGenolini:2020doj,Bertolini:2025wyj}, which suggest that it should be possible to give yet another definition of the partition function that has $\CR \to 1/\CR$ symmetry. It would be interesting to try to extract such a result from a definition of the $U$-plane integral. 
(Note that some of our expressions, such as \eqref{vinfexp}, do indeed exhibit such a symmetry.)

\subsection{Independence Of $\bar \CR$}
With the definition \eqref{PhiIntDef}, we can study the $\bar \CR$-dependence of the integral. Using \cite[Eq.  (5.43)]{Manschot:2021qqe}, we derive for the $\bar \CR$ dependence of $\Psi_\bfmu^J$,
\be 
\left.\frac{\mathrm{d}\Psi_\bfmu^J(\tau,\bar \tau,\bfz,\bar \bfz)}{\mathrm{d}\bar \CR}\right\vert_{\tau,\bar \tau \text{ fixed}}
=\partial_{\bar \tau} \left( \frac{i}{\sqrt{y}}
B(\partial_{\bar \CR}\bar \bfz,J)\,\Psi^J_\bfmu[1](\tau,\bar \tau,\bfz,\bar \bfz) \right),
\ee 
where we used that $\bfz$ is independent of $\bar \CR$. On the right-hand side, the function $\Psi_\bfmu^J[1]$ is defined in \cite[Eq. (B.3)]{Korpas:2019ava}; it is a Siegel-Narain theta series as $\Psi_\bfmu^J$ \eqref{DefPsi} but with kernel equal to 1. 

Similarly, we can show that $\Phi_\bfmu^J$ is independent of $\bar \CR$. We find
\be 
\frac{\partial \Phi_{\bfmu,\bfn}^J(\CR,\bar \CR)}{\partial \bar \CR}= K_U B(\bfn,J)\int_{\CF_R} \mathrm{d}\tau\wedge \mathrm{d}\bar \tau\,\nu_R(\tau)\,C^{\bfn^2}\, \frac{\mathrm{d}}{\mathrm{d} \bar \tau}  \left( \frac{i}{2\sqrt{y}}
\frac{\partial\bar v}{\partial \bar \CR} \, \Psi^J_\bfmu[1](\tau,\bar \tau,\bfn v,\bar \bfn \bar v)\right).
\ee 
We apply the definition \eqref{PhiIntDef} to this integral. We expand the integrand as a series in $\CR$. From the explicit total $\bar \tau$ derivative we deduce that this does not have a constant term $d(0)\, q^0\bar q^0$, since all terms are multiplied by $y^{-s}$ with $s\geq 1/2$. As a result,  we arrive at
\be 
\frac{\partial \Phi_{\bfmu,\bfn}^J(\CR,\bar \CR)}{\partial \bar \CR}=0.
\ee

\subsection{Formal Evaluation Of The Integral}
\label{Sec:FormEvInt}
With the result that $\Phi_\bfmu^J$ is independent of $\bar \CR$, we can formally evaluate the integral. The series (\ref{PhiIntDef}) truncates to $l_2=0$, and the exponent of $\bar q$, $n$, is bounded below by 0. As a result, the classic regularization of \cite{Dixon:1990pc, Harvey:1995fq, Borcherds:1996uda} is applicable.

The integral can be formally evaluated via an anti-derivative using mock modular forms \cite{MR2605321, ZwegersThesis, Bringmann:2309148} as discussed in
\cite{Moore:1997pc,Malmendier:2008db, Malmendier:2010ss, Korpas:2017qdo, Moore:2017cmm, Korpas:2019cwg, Manschot:2021qqe}. In Sec.  \ref{Sec:DEvalU}, we will demonstrate the existence of an anti-derivative $\widehat G_{\bfmu,\bfn}$ satisfying
\be
\label{eq:antiderPsi}
\frac{\mathrm{d}\widehat G_{\bfmu, \bfn}(\tau,\bar \tau, z,  \bar z)}{\mathrm{d}\bar \tau}=\Psi_\bfmu(\tau,\bar \tau, \bfn z, \bfn \bar z),
\ee 
where $z\in \mathbb{C}$ is an elliptic argument which may depend on $\tau$. Then, $\widehat G_{\bfmu,\bfn}(\tau,\bar \tau, z, \bar z)$ decomposes as
\be
\widehat G_{\bfmu,\bfn}(\tau,\bar \tau, z, \bar z)=G_{\bfmu,\bfn}(\tau, z)+\Delta G_{\bfmu,\bfn}(\tau,\bar \tau, z, \bar z),
\ee
where $G_{\bfmu,\bfn}(\tau, z)$ is holomorphic in $\tau$ and $z$ away from singularities, and $\Delta G_{\bfmu,\bfn}$ is
\be 
\begin{split}
\Delta G_{\bfmu,\bfn}(\tau,\bar \tau, z, \bar z)& = -\frac{1}{2}\sum_{\bfk \in L+\bfmu} \sqrt{2y}B(\bfk+\bfn b,J)\,E_{\frac{1}{2}}\left(2\pi y \left(\bfk+\bfn b\right)_+^2\right)\\
&\quad \times \,e^{\pi i B(\bfk,K)}\,q^{-\frac{1}{2}\bfk^2}\,e^{-2\pi iB(\bfk,\bfn z)},
\end{split}
\ee
where $b={\rm Im}(z)$ and $E_{1/2}$ is defined in Eq. \eqref{defEellz}. For suitable $G_{\bfmu,\bfn}(\tau,z)$, the function $\widehat G_{\bfmu,\bfn}(\tau,\bar \tau, z, \bar z)$ exhibits good modular transformation properties and gives rise to a well-defined anti-derivative of the $U$-plane integrand. Suitable choices for $\mathbb{CP}^2$ are given in Eqs. \eqref{G12n} and \eqref{G0n} below. Note that, while the right-hand side of Eq. (\ref{eq:antiderPsi}) is a function of $\bfn z$, we have split the dependence on the lhs in $\bfn$ and $z$ separately. Indeed, Eq. \eqref{eq:UGNYcomp} gives an example where the anti-derivative is not a function of $nv$, but of $n$ and $v$ separately. See \cite[Sec.  6]{Manschot:2021qqe} for a similar discussion on the integration of the $u$-plane integral for $\CN=2^*$ supersymmetric gauge theory in the presence of a background flux.

Once $\widehat G_{\bfmu,\bfn}$ is determined, we consider the integration over the domain displayed in Fig. \ref{FRCG}. As discussed in Sec.  \ref{UPlaneIntDefs}, we consider the compact domain obtained by introducing a cutoff $Y$ in the local coordinate near each cusp. Moreover, in making the $\CR$-expansion, the branch points move beyond the cutoff $Y$, so the integration domain becomes two adjacent copies of the cutoff domain $(\mathbb{H}/\Gamma^0(4))^Y$. The integral thus becomes 
\be 
\begin{split}
\Phi_{\bfmu,\bfn}^{J}(\CR,\bar \CR) & =K_U\,
 \lim_{Y\to \infty} \int_{(\mathbb{H}/\Gamma^0(4))^Y_{\mathrm{L}}} \mathrm{d}\tau\wedge \mathrm{d}\bar \tau\,{\rm Ser}_\CR\!\left[\nu_R(\tau) C^{\bfn^2} \frac{\mathrm{d}}{\mathrm{d}\bar \tau} \widehat G_{\bfmu,\bfn}(\tau,\bar \tau,v, \bar v)\right]\\
& \quad + K_U \lim_{Y\to \infty} \int_{(\mathbb{H}/\Gamma^0(4))^Y_{\mathrm{R}}} \mathrm{d}\tau\wedge \mathrm{d}\bar \tau\,{\rm Ser}_\CR\!\left[\nu_R(\tau) C^{\bfn^2} \frac{\mathrm{d}}{\mathrm{d}\bar \tau} \widehat G_{\bfmu,\bfn}(\tau,\bar \tau, v, \bar v)\right],
\end{split}
\ee 
where $(\mathbb{H}/\Gamma^0(4))^Y_{\mathrm{L}}$ denotes the standard fundamental domain for $\Gamma^0(4)$ with weak-coupling region extending from ${\rm Re}(\tau)=-1/2$ to $7/2$, and $Y$ is a cutoff for the asymptotic region near each cusp; $(\mathbb{H}/\Gamma^0(4))^Y_{\mathrm{R}}$ denotes the fundamental domain for $\Gamma^0(4)$ with weak-coupling region extending from ${\rm Re}(\tau)=7/2$ to $15/2$.
The discussion around Eq. \eqref{OmegT} then demonstrates that this reduces to
\be 
\begin{split}
\Phi_{\bfmu,\bfn}^{J}(\CR,\bar \CR)&=K_U\, \left(1+(-1)^{B(2\bfmu,\overline{w_2(X)}+\bfn)}\right)\\
&\quad \times \lim_{Y\to \infty} \int_{(\mathbb{H}/\Gamma^0(4))^Y} \mathrm{d}\tau\wedge \mathrm{d}\bar \tau\,{\rm Ser}_\CR\!\left[\nu_R(\tau) C^{\bfn^2} \frac{\mathrm{d}}{\mathrm{d}\bar \tau} \widehat G_{\bfmu,\bfn}(\tau,\bar \tau, v, \bar v)\right].
\end{split}
\ee 
As established in the previous subsection, $\Phi_{\bfmu,\bfn}^{J}(\CR,\bar \CR)=\Phi_{\bfmu,\bfn}^{J}(\CR)$ is independent of $\bar \CR$. Hence, we can set $\bar \CR=0$ in the integrand, producing
\be 
\begin{split}
\Phi_{\bfmu,\bfn}^{J}(\CR)&=K_{U,\bfmu,\bfn} \lim_{Y\to \infty} \int_{(\mathbb{H}/\Gamma^0(4))^Y} \mathrm{d}\tau\wedge \mathrm{d}\bar \tau\,\, {\rm Ser}_\CR\!\left[\nu_R(\tau)\, C^{\bfn^2} \frac{\mathrm{d}}{\mathrm{d}\bar \tau} \widehat G_{\bfmu,\bfn}(\tau,\bar \tau, v, 0)\right],
\end{split}
\ee 
where we have defined
\be
\label{KUmun}
K_{U,\bfmu,\bfn}=K_U \, \left(1+(-1)^{B(2\bfmu,\overline{w_2(X)} +\bfn)}\right)\in \{0,1\}.
\ee

In the limit $Y\to \infty$, $\Delta G_{\bfmu,\bfn}(\tau,\bar \tau, v, \bar v)$ vanishes. As a result, this prescription leads to making an $\CR$-expansion of the holomorphic part and subsequently extracting the $q^0$ term, denoted by ${\rm Coeff}_{q^0}{\rm Ser}_\CR[\cdots]$. The integral then evaluates to contributions from the three cusps,
\be
\begin{split} 
\Phi_{\bfmu,\bfn}^J(\CR)&=\left\{\begin{array}{cr}
0, & B(2\bfmu,\overline{w_2(X)}+\bfn)={\rm odd},\\[\medskipamount]
4\,{\rm Coeff}_{q^0}\,{\rm Ser}_\CR\left[\nu_R(\tau)\, C(\tau)^{\bfn^2} G_{\bfmu,\bfn}(\tau, v)\right] &\\[\medskipamount]
+\,{\rm Coeff}_{q_1^0}\,{\rm Ser}_\CR\left[\tau_1^{-2}\,\nu_R(\tau)\, C(\tau)^{\bfn^2} G_{\bfmu,\bfn}(\tau, v)\right]& B(2\bfmu,\overline{w_2(X)}+\bfn)={\rm even,}\\[\medskipamount]
+\,{\rm Coeff}_{q_2^0}\,{\rm Ser}_\CR\left[\tau_2^{-2}\,\nu_R(\tau)\, C(\tau)^{\bfn^2} G_{\bfmu,\bfn}(\tau, v)\right], & \end{array}\right.
\end{split}
\ee 
where $q_j=e^{2\pi i \tau_j}$ for $j=1,2$, with $\tau_1=-1/\tau$ and $\tau_2=-1/(\tau-2)$. The coefficient $4$ corresponds to the periodicity of the weak coupling cusp for $\tau\to i\infty$ of $\mathbb{H}/\Gamma^0(4)$. We will discuss in detail how to evaluate ${\rm Coeff}_{q^0_j}$ in Sec.  \ref{b2Min>0}.

\subsection{Wall-Crossing}\label{sec:wall-crossing}

Similar to Coulomb branch integrals for other 4d topologically twisted quantum field theories \cite{Moore:1997pc, Marino:1998bm, Labastida:1998sk, Manschot:2021qqe, Aspman:2023ate}, the $U$-plane integral exhibits the phenomenon of wall-crossing: $\Phi_{\bfmu,\bfn}^J$ is locally constant as function of the period point $J$, but changes discontinuously across walls of marginal stability.

To determine the change across a wall of marginal stability, we express the difference $\Psi^J_\bfmu-\Psi^{J'}_\bfmu$ as a total derivative of an indefinite theta function $
\widehat \Theta^{JJ'}_\bfmu$ \cite{Korpas:2017qdo, Korpas:2019cwg},
\be\label{eq:Psi-Difference-TD}
\begin{split}
\Psi^J_\bfmu(\tau,\bar \tau,\bfz,\bar \bfz)-\Psi^{J'}_\bfmu(\tau,\bar
\tau,\bfz,\bar \bfz)=\partial_{\bar \tau} \widehat
\Theta_\bfmu^{J,J'}(\tau,\bar \tau,\bfz,\bar \bfz),
\end{split}
\ee
where $\widehat \Theta^{JJ'}_\bfmu$ is defined by
\be
\label{eq:DefIndefTheta}
\begin{split}
\widehat \Theta^{JJ'}_\bfmu(\tau,\bar \tau,\bfz,\bar \bfz)&=\sum_{\bfk\in
  L+\bfmu}
\frac{1}{2}\left[E(\sqrt{2y}B(\bfk+\bfb,J))-E(\sqrt{2y}B(\bfk+\bfb,J'))\right] \\
&\quad \times \,e^{\pi i B(\bfk,K)}\,q^{-\frac{1}{2}\bfk^2}e^{-2\pi iB(\bfk,\bfz)},
\end{split}
\ee
with $\bfb={\rm Im}(\bfz)/y$ and $E$ the error function defined in Eq. \eqref{Eerror}. 
$\widehat \Theta^{JJ'}_\bfmu$ transforms under the generators $T$ and $S$ of the modular group as
\be
\label{eq:IndefThetaTrafo}
\begin{aligned}
\widehat \Theta^{JJ'}_\bfmu(\tau+1,\bar \tau+1,\bfz,\bar \bfz) &=\exp\left[\pi i   (\bfmu^2-B(K,\bfmu))\right]\widehat \Theta^{JJ'}_\bfmu\left(\tau,\bar \tau,\bfz+\bfmu-\frac{K}{2},\bar \bfz+\bfmu-\frac{K}{2}\right),\\
\widehat \Theta^{JJ'}_\bfmu \left(-\frac{1}{\tau},-\frac{1}{\bar \tau},\frac{\bfz}{\tau},\frac{\bar\bfz}{\bar \tau}\right) &=i(-i\tau)^{\frac{b_2}{2}} \exp\left(\pi i B(\bfmu,K)-\frac{\pi i \bfz^2}{\tau}\right)\\
&\quad\times \widehat \Theta_{\frac{K}{2}}^{JJ'}\left(\tau,\bar \tau,\bfz-\bfmu+\frac{K}{2},\bar \bfz-\bfmu+\frac{K}{2}\right), 
\end{aligned}
\ee
where $b_2$ is the dimension of $L$.

We also define the holomorphic part
\be
\label{TJJ'}
\begin{split}
\Theta^{JJ'}_\bfmu(\tau,\bfz)&=\sum_{\bfk\in
  L+\bfmu}
\frac{1}{2}\left[\sgn(B(\bfk+\bfb,J))-\sgn(B(\bfk+\bfb,J'))\right] \\
&\quad \times\,e^{\pi i B(\bfk,K)} \,q^{-\frac{1}{2}\bfk^2}e^{-2\pi iB(\bfk,\bfz)}.
\end{split}
\ee

From Eq. \eqref{eq:Psi-Difference-TD} we see that the \emph{difference} between $U$-plane integrals for two values of $J$ can be computed by integration by parts. We now investigate the contributions of the different cusps in Sec.  \ref{subsubsec:WeakCpl-WC} and \ref{subsubsec:StrongCpl-WC}.

\subsubsection{Wall-Crossing From The Weak-Coupling Cusp}\label{subsubsec:WeakCpl-WC}

The wall-crossing contributed from the weak-coupling cusp (in the region $\tau\to i\infty$) is
\be
\begin{aligned}
\Delta \Phi^{JJ'}_{\bfmu,\bfn,\infty}(\CR) & =\Phi^J_{\bfmu,\bfn,\infty}(\CR)-\Phi^{J'}_{\bfmu,\bfn,\infty}(\CR) \\
& =K_U\lim_{Y\to \infty}
\int_{iY-\frac12}^{iY+\frac72} \mathrm{d}\tau\, {\rm Ser}_{\CR}\left[\nu_R(\tau)\,C^{\bfn^2}\,\widehat \Theta^{JJ'}_{\bfmu}(\tau,\bar \tau,\bfn
  v,\bfn \bar v)\right]\\
& + K_U\lim_{Y\to \infty}
\int_{iY+\frac72}^{iY+\frac{15}{2}} \mathrm{d}\tau\, {\rm Ser}_{\CR}\left[\nu_R(\tau)\,C^{\bfn^2}\,\widehat \Theta^{JJ'}_{\bfmu}(\tau,\bar \tau, \bfn
  v,\bfn \bar v)\right],
\end{aligned}
\ee 
where the two integrals correspond to the weak-coupling boundaries of the two $\mathbb{H}/\Gamma^0(4)$ domains. This simplifies to
\be\label{eq:WeakCoupl-WC-Integral}
\begin{split}
\Delta \Phi^{JJ'}_{\bfmu,\bfn,\infty}(\CR)
& = K_{U,\bfmu,\bfn} \lim_{Y\to \infty}
\int_{iY-\frac12}^{iY+\frac72} \mathrm{d}\tau\, {\rm Ser}_\CR\left[\nu_R(\tau)\,C^{\bfn^2}\,\widehat \Theta^{JJ'}_{\bfmu}(\tau,\bar \tau,\bfn
  v,\bfn \bar v)\right],\\
\end{split}
\ee
with $K_{U,\bfmu,\bfn}$ defined as in \eqref{KUmun}.
The non-holomorphic part vanishes in the limit $Y\to \infty$, allowing us to write the result as
\be\label{eq:Physical-WC}
\Delta \Phi^{JJ'}_{\bfmu,\bfn,\infty}(\CR)=4 K_{U,\bfmu,\bfn}\,{\rm Coeff}_{q^0}{\rm Ser}_\CR\left[\nu_R(\tau)\,C^{\bfn^2}\,\Theta^{JJ'}_{\bfmu}(\tau,\bfn v) \right].
\ee

\subsubsection*{Remark} 

Eq. \eqref{eq:Physical-WC} gives a formula for the weak-coupling wall-crossing of the 
partition function --- but where are the walls of discontinuity as a function of $J$? 
The location of the walls in $H^2(X,\IR)$ provide a compelling illustration of the subtleties described in Sec.  \ref{UPlaneIntDefs}. Recall that $\bfb={\rm Im}(\bfz)/y={\rm Im}(v)\,\bfn/y$ in Eq.  (\ref{TJJ'}). Thus a vector $\bfk$ contributes to (\ref{TJJ'}) only if
\be\label{eq:tau-Walls}
B\left(\bfk+\frac{{\rm Im}(v)}{y}\,\bfn, J\right)>0 \quad \text{and} \quad B\left(\bfk+\frac{{\rm Im}(v)}{y}\,\bfn, J'\right)<0, \quad \text{or vice versa},
\ee
except for boundary cases where an inequality is saturated. In App. \ref{App:ShiftedWalls}, we provide an elegant physical interpretation of such walls using the effective action coupled to the background $\mathrm{U}(1)^{(I)}$ vector multiplet.  

For the wall-crossing formula we must take limits in \eqref{eq:WeakCoupl-WC-Integral}. 
In the bracketed expression in Eq.  \eqref{eq:Physical-WC}, we expand in positive powers of $\CR$ around $\CR=0$ at fixed $\tau$, and then extract the $q^0$ term at each order in the $\CR$-expansion. The resulting expressions also exhibit wall-crossing behavior, but now the positions of the walls are given by taking $\CR \to 0$ and keeping $\tau$ fixed in the expression \eqref{eq:tau-Walls}. Looking 
back to Eq. \eqref{vR} we see that the condition becomes 
\footnote{If one tries to expand the sign function in powers of $\CR$ using Eq. \eqref{vR} one will get a series involving $\delta$-functions and their derivatives which does not make sense. However, with the 
definition we have given we should expand the integrand of \eqref{eq:WeakCoupl-WC-Integral} in powers of $\CR$ first. Therefore we consider the expansion of the error functions appearing in the definition of  $\widehat{\Theta}$. So we write 
$E(\sqrt{2y}B(\bfk + \bfb, J)) = E(\sqrt{2y}B(\bfk,J)) + \sqrt{2y} B(\bfb,J) E'(\sqrt{2y}B(k,J)) + \cdots $ because from \eqref{vR} we see that $\bfb$ is a series beginning with a positive power of $\CR$. 
Next, the derivatives of the error function are continuous as a function of $J$, so with the order of limits in question, they will not lead to discontinuity. Therefore we need only keep the first 
term $E(\sqrt{2y}B(\bfk,J))$ in computing the discontinuity.}
\be
\label{muWalls}
B(\bfk, J)>0 \quad \text{and} \quad B(\bfk, J')<0, \quad \text{ or vice versa}.
\ee
So the walls are located at solutions $J$ of $B(\bfk, J)=0$ where $\bfk \in L + \bfmu$. 
This is essentially the $\mu$-stability condition. Both the positions of the walls and the resulting wall-crossing formula agree with those in \cite{Gottsche:2006bm}. 

On the other hand, if we had taken the definition \eqref{eq:UplaneInt-FixedCR} of the integral, where the large ${\rm Im}(\tau)$ limit is taken first, then from the asymptotic behavior \eqref{vinfexp} of $v$ in the limit of large ${\rm Im}(\tau)$ keeping $\CR$ fixed, we would deduce that ${\rm Im}(v)/y=-1/4+\cdots$. Therefore, the vectors $\bfk$ contributing to the wall-crossing satisfy 
\be
B\left(\bfk-\frac{\bfn}{4}, J\right)>0 \quad \text{and} \quad B\left(\bfk-\frac{\bfn}{4}, J'\right)<0, \quad \text{ or vice versa},
\ee
That is, the walls would be located at those $J$ such that $B(\bfk-\bfn/4, J)=0$ for $\bfk \in L + \bfmu$. This  clearly differs from the walls in Eq.  \eqref{muWalls}. This difference highlights the order-of-limits sensitivity in our analysis.

\subsubsection{Wall-Crossing From Strong-Coupling Cusps}\label{subsubsec:StrongCpl-WC}

In addition to the change of the $U$-plane integral at the weak-coupling cusp, the contributions to the $U$-plane integral from the strong-coupling singularities $U_j$ $(j=1,\cdots, 4)$ also depend discontinuously on the period point $J$. However, these wall-crossing behaviors are absent in the full partition function $Z^J_{\bfmu,\bfn}$, since physical arguments show that they are canceled by opposite, but otherwise identical, contributions involving SW invariants to the path integral \cite{Moore:1997pc}. 
Consequently, the wall-crossing behavior of $\Phi_{\bfmu,\bfn}^J$ at $U_j$ will play a crucial role in the determination of the partition function for four-manifolds of SW simple type. Conjecturally, all simply-connected four-manifolds with $b_2^+>1$ are of SW simple type.  

To derive the strong-coupling contributions, we introduce local couplings $\tau_j$, $v_j$, and $C_j$ near each singularity $U_j$. Their definitions and $q_j$-expansions are given in App. \ref{app:ExpCusps}.

\paragraph{Contribution from $U_1$:} 
The contribution from the cusp $U_1$ reads,
\be
\begin{split}
  \label{WCU1}
\Delta \Phi^{JJ'}_{\bfmu,\bfn,1}(\CR) &:=   
\left[\Phi^{J}_{\bfmu,\bfn}(\CR)-\Phi^{J'}_{\bfmu,\bfn} (\CR) \right]_1 \\
&=-\frac{1}{4}\lim_{Y\to \infty}\int_{-\frac12+iY}^{\frac12+iY}\mathrm{d}\tau_1\,\frac{\vartheta_2(\tau_1)^{11+\sigma}}{\eta(\tau_1)^9}\,\frac{1}{U}\,C_1(\tau_1,v_1)^{\bfn^2}\\
&\quad \times (-1)^{B(\bfmu,K)} \sum_{\bfk\in L+\frac{1}{2}K} \frac{1}{2}\left[
  \sgn(B(\bfk,J))-\sgn(B(\bfk,J')) \right] \\
&\quad \times\,q_1^{-\frac{1}{2}\bfk^2}\,e^{-2\pi i B(\bfk,\bfn) v_1}\,e^{-2\pi i B(\bfk,\bfmu)}.
\end{split}
\ee 
We can write this as a residue in the local coordinate $a_1$ using Eq.  (\ref{da1dtau1}),
\be
\begin{split}
\label{WCU1res}
\Delta \Phi^{JJ'}_{\bfmu,\bfn,1}(\CR) &=\frac{2}{\Lambda} \, \underset{a_1=0}{\rm Res}\left[  \frac{\vartheta_2(\tau_1)^{2+\sigma}}{\eta(\tau_1)^6}\,C_1(\tau_1,v_1)^{\bfn^2}\right.\\
&\quad \times (-1)^{B(\bfmu,K)} \sum_{\bfk\in L+\frac{1}{2}K} \frac{1}{2}\left[
  \sgn(B(\bfk,J))-\sgn(B(\bfk,J')) \right] \\
&\left.\quad \times\,q_1^{-\frac{1}{2}\bfk^2}\,e^{-2\pi i B(\bfk,\bfn) v_1}\,e^{-2\pi i B(\bfk,\bfmu)}\right].
\end{split}  
\ee

\paragraph{Contribution from $U_2$:}
The contribution from the cusp $U_2$ reads,
\be
\begin{split}
  \label{WCU2}
\Delta \Phi^{JJ'}_{\bfmu,\bfn,2}(\CR) &:= \left[\Phi^{J}_{\bfmu,\bfn}(\CR)-\Phi^{J'}_{\bfmu,\bfn}(\CR)
\right]_2 \\
&=-\frac{1}{4} e^{-\frac{3\pi i}{2}-2\pi i \bfmu^2}\lim_{Y\to \infty}\int_{-\frac12+iY}^{\frac12+iY}\mathrm{d}\tau_2\,\frac{\vartheta_2(\tau_2)^{11+\sigma}}{\eta(\tau_2)^9}\,\frac{1}{U}\\
&\quad \times (-1)^{B(\bfmu,K)} \,C_2(\tau_2,v_2)^{\bfn^2} \sum_{\bfk\in L+\frac{1}{2}K} \frac{1}{2}\left[
  \sgn(B(\bfk,J)-\sgn(B(\bfk,J')) \right] \\
&\quad \times\,q_2^{-\frac{1}{2}\bfk^2}\,e^{-2\pi i B(\bfk,\bfn) v_2}\,e^{-2\pi i B(\bfk,\bfmu)}.
\end{split}
\ee
Note that the additional phase factor $e^{-3\pi i/2-2\pi i \bfmu^2}$ equals
$e^{\pi i \,l/2}$, where $l$ (\ref{eq:ModSpaceDim}) is the dimension of instanton moduli space for $\chi+\sigma=4$.
We can again write this as a residue in the local coordinate $a_2$,
\be
\begin{split}
\label{WCU2res}
\Delta \Phi^{JJ'}_{\bfmu,\bfn,2}(\CR)&=-\frac{2}{\Lambda}e^{-2\pi i \bfmu^2}  \underset{a_2=0}{\rm Res}\left[  \frac{\vartheta_2(\tau_2)^{2+\sigma}}{\eta(\tau_2)^6}\,C_2(\tau_2,v_2)^{\bfn^2}\right.\\
&\quad \times (-1)^{B(\bfmu,K)} \sum_{\bfk\in L+\frac{1}{2}K} \frac{1}{2}\left[
  \sgn(B(\bfk,J)-\sgn(B(\bfk,J')) \right] \\
&\left.\quad \times\,q_2^{-\frac{1}{2}\bfk^2}\,e^{-2\pi i B(\bfk,\bfn) v_2}\,e^{-2\pi i B(\bfk,\bfmu)}\right].
\end{split}  
\ee

\paragraph{Contribution from $U_3$ and $U_4$:}
The neighborhoods of $U_3$ and $U_4$ are related to those of $U_1$ and $U_2$ via the transformations \eqref{Ttrafo} and \eqref{Ttrafotauv}. As discussed in Sec.  \ref{IntMono}, the contributions differ by the sign,
\be
(-1)^{B(2\bfmu,K- \bfn)}.
\ee 
We thus have the contributions from $U_j$ for $j=3,4$,
\be
\label{WCU34U12}
\Delta \Phi^{JJ'}_{\bfmu,\bfn,j}(\CR) =(-1)^{B(2\bfmu,K- \bfn)}\Delta \Phi^{JJ'}_{\bfmu,\bfn,j-2}(\CR).
\ee

\subsubsection{Special Examples: $X=\mathbb{CP}^1\times \mathbb{CP}^1$} 

To illustrate the weak-coupling wall-crossing behavior with explicit examples, we examine the simple case $X=\mathbb{CP}^1\times \mathbb{CP}^1$. Notably, the SW invariants and strong-coupling contributions vanish identically, so the partition function is entirely given by the $U$-plane integral, and its wall-crossing behavior is completely governed by Eq.  \eqref{eq:Physical-WC}. 

The $\mathrm{U}(1)^{(I)}$ background flux for $\mathbb{CP}^1\times \mathbb{CP}^1$ takes the form
\be
\bfn=(n_1,n_2).
\ee
Using the properties of $\mathbb{CP}^1\times \mathbb{CP}^1$ and Eq. \eqref{Phin-n}, we find the relations
\begin{equation}
    \Delta\Phi^{JJ'}_{\bfmu,(n_1,n_2)}(\CR) =
    -\Delta\Phi^{JJ'}_{\bfmu,(-n_1,-n_2)}(\CR), \quad 
    \Delta\Phi^{JJ'}_{\bfmu,(n_1,n_2)}(\CR) =
    \Delta\Phi^{JJ'}_{\bfmu,(n_2,n_1)}(\CR).
\end{equation}
We further choose
\be
\bfmu=(\frac12,\frac12),\quad J=(1 + \delta,1),\quad J'=(\epsilon,1),
\ee
with $0 < \delta \ll 1$ and $0<\epsilon<1$. For a given $\bfn$ and a given power of $\CR$, the parameter $\epsilon$ must lie in an appropriate window specified in Eq. \eqref{DeltaPhiP1P1} below.  
The partition function vanishes when $n_1+n_2$ is odd, since $K_{U,\bfmu,\bfn}$ given in \eqref{KUmun} vanishes in this case. Employing the prescription of first taking the small $\CR$ expansion, we can list a number of examples for even $n_1+n_2$:
\begin{equation}
\label{DeltaPhiP1P1}
\Delta\Phi^{JJ'}_{\bfmu,\bfn}(\CR)=\left\{\begin{array}{lll}
\CO(\CR^{9}) ,& \bfn = (0,0), 
& 0 < \epsilon < 1, 
\\
-4\CR^{3}+\CO(\CR^{9}),
& \bfn = (0,2), 
& 0 < \epsilon < 1,
\\ 
-35\CR^{3}+28\CR^{7}+\CO(\CR^{9}), & \bfn = (0,4),& 
0< \epsilon < \frac{1}{3},\\
84 \cR^3 +64 \cR^7+\CO(\CR^{9}) , & \bfn = (1,-5), & 
0 < \epsilon < \frac{1}{5},\\
20 \CR^3 +\CO(\CR^{9}), & \bfn = (1,-3),& 
0 < \epsilon < \frac{1}{5},\\
\CR^3+ \CO(\cR^9), & \bfn = (1,-1), & 
0 < \epsilon < \frac{1}{3},\\
\CO(\CR^{9}), & \bfn = (1,1),& 
0< \epsilon < 1, \\           
56\CR^{3}+120\CR^{7}+\CO(\CR^{9}), & \bfn = (2,-4),& 0 < \epsilon < \frac{1}{5},\\
10 \CR^3 +\CO(\cR^9), & \bfn = (2,-2),& 0 < \epsilon < \frac{1}{5},\\
-\CR^{3}-\CR^{7}+\CO(\CR^{9}), & \bfn = (2,2),& 0 < \epsilon < \frac{1}{5},\\ 
-20\CR^{3}+\CO(\CR^{9}), & \bfn = (2,4),& 0 < \epsilon < 1,\\
 120\CR^{3}+1100\CR^{7}+\CO(\CR^{9}), & \bfn = (3,-5),& 0 < \epsilon < \frac{1}{5},\\ 
 35 \CR^3 +56 \CR^7 +\CO(\mathcal{R}^9), &\bfn = (3,-3),& 0 < \epsilon < \frac{1}{5},\\
 -4\CR^{3}-10\CR^{7}+\CO(\CR^{9}), & \bfn = (3,3), 
 & 0 < \epsilon < 
 \frac{1}{5},\\
 84\CR^{3}+603\CR^{7}+
       \CO(\CR^9), & \bfn = (4,-4),
       & 0 < \epsilon < \frac{1}{5},\\
-10\CR^{3}-54\CR^{7}+\CO(\CR^{9}), & \bfn = (4,4),& 0< \epsilon < \frac{1}{5}.\\ \end{array}\right.
\end{equation}

\subsubsection*{Remarks}
\begin{enumerate}
\item We can consider \eqref{eq:Physical-WC} with $K_{U,\bfmu,\bfn}$ replaced by $1$, even in the case that the anomaly cancellation condition is violated and $K_{U,\bfmu,\bfn}$ vanishes. Take $0 < \epsilon < \frac{1}{5}$ and $\bfn=(0,-1)$ as an example, we get
\be
\label{DeltatPhiP1P1}
\frac{5}{16}\CR^3+\frac{91}{2048}  \CR^7+\CO(\CR^{9}).
\ee
Note that the coefficients in the $\CR$-expansion are not integers. If we regard the result as a generating function of an integral of an index density over a moduli space, for which some evidence is provided by \cite{Fantechi_2010}, then it is possible that the non-integrality of the expansion coefficients is related to the fact that the moduli space of instantons is not spin.  Similar nonintegrality is familiar from integrating the $\hat A$ genus over a compact non-spin manifold.
\item In Sec.  \ref{UPlaneIntDefs}, we discuss two distinct definitions of the $U$-plane integral and argue that only one of them is physically viable.
To illustrate the problematic outcomes that arise from evaluating the integral over $\tau$ while holding $\CR$ fixed, consider $\Delta\Phi^{JJ'}_{\bfmu,\bfn}(\CR)$ in the chamber $0< \epsilon < \frac{1}{5}$ with $\bfn=(4,-2)$. This definition would give
\be
\label{eq:Rfixed}
4  \left(\CR^{-5}-4\CR^{-1}+4 \CR^{15}-\CR^{19}\right).
\ee
Such an expression is in contradiction with the general expectation of a power series in nonnegative powers of $\CR$, as required by \eqref{eq:Zee-QM-1}. 
\end{enumerate}

\section{Direct Evaluation Of The $U$-plane Integral In Some Important Cases}
\label{Sec:DEvalU}

In this section, we evaluate the $U$-plane path integral for four-manifolds with $b_2^+=1$. We start with the case with $(b_2^+,b_2^-)=(1,0)$, which includes $X=\mathbb{CP}^2$. 
\footnote{Since $H^2(\mathbb{CP}^2,\mathbb{Z})\simeq \mathbb{Z}$, each element in $H^2(\mathbb{CP}^2,\mathbb{Z})$ can be represented by an integer multiplying the hyperplane class $H$, which is the standard positive generator of $H^2(\mathbb{CP}^2,\mathbb{Z})$. Thus, we will use an integer $K$ to represent $K\,H\in H^2(\mathbb{CP}^2,\mathbb{Z})$, and similarly, we use an integer $n$ to represent the $\mathrm{U}(1)^{(I)}$ flux $\bfn=nH$.} 
In Sec.  \ref{b2Min>0}, we extend the computation to more general cases with $b_2^->0$.

\subsection{Evaluation For $X=\mathbb{CP}^2$}
\label{sec:evaluate}

The $U$-plane integral for $\mathbb{CP}^2$ can be evaluated using methods similar to those for 4d $u$-plane integrals \cite{Korpas:2017qdo, Korpas:2019cwg, Manschot:2021qqe}. To this end, we will express $\Psi^J_\bfmu$ as a total derivative, relying on various properties of modular and mock modular forms, which we review in App. \ref{App:MForms}. Since $b_2=b_2^+=1$, the period point $J$ is unique for $\mathbb{CP}^2$, and there are no wall-crossing effects. We therefore omit $J$ as superscript from $\Psi_\bfmu^J$.

We start by factoring out a phase from $\Psi_\mu$ and introducing the function $f_\mu$, 
\be
\Psi_\mu(\tau,\bar \tau,v n,\bar v n)=-i (-1)^{\mu(K-1)}
f_{\mu,n}(\tau,\bar \tau,v,\bar v),
\ee
where $K$ is a characteristic vector (an odd number for $\mathbb{CP}^2$)
appearing in the definition of $\Psi_\mu$ \eqref{DefPsi}.
We then aim to determine a suitable function $\widehat G_{\mu,n}$ such that  
\be
f_{\mu,n}(\tau,\bar \tau,v,\bar v )= \frac{\mathrm{d}\widehat G_{\mu,n}(\tau,\bar
\tau, v, \bar v)}{\mathrm{d}\bar \tau}.
\ee
 This function $\widehat G_{\mu,n}$ should have suitable transformation properties and be
non-singular on the fundamental domain for $\tau$. Moreover, the ambiguity to shift $\widehat G$ by a holomorphic modular form does not change the result, owing to the fact the $\CF^{\infty}$ is simply-connected.

We first consider the case $n=0$. In this case, $f_{\mu,0}$ vanishes or can be expressed in terms of the Dedekind eta function $\eta$ \eqref{DefEta},
\be 
f_{\mu,0}(\tau,\bar \tau)=\left\{ \begin{array}{ll} 0, &\quad \mu=0\mod \mathbb{Z}, \\ -\frac{i}{2\sqrt{2y}}\,\overline{\eta(\tau)^3}, & \quad  \mu=\frac{1}{2}\mod \mathbb{Z}. \end{array} \right.
\ee 
For $\mu=0$, a suitable anti-derivative is thus $\widehat G_{0,0}=0$, while a suitable anti-derivative for $\mu=1/2$ is given by a familiar function from the theory of mock modular forms,  
\be 
\widehat G_{\frac{1}{2},0}(\tau,\bar \tau)=G_{\frac{1}{2},0}(\tau)-\frac{i}{2}\int_{-\bar \tau}^{i\infty} \frac{\eta(w)^3}{\sqrt{-i(w+\tau)}}dw,
\ee 
where $G_{1/2,0}(\tau)$ is the McKay-Thompson series $H^{(2)}_{1A,2}$ \cite{Cheng:2012tq} given by
\be 
\begin{split}
   G_{\frac{1}{2},0}(\tau)&= -\frac{1}{\vartheta_4(\tau)}\sum_{n\in \mathbb{Z}} \frac{(-1)^n q^{\frac{n^2}{2}-\frac{1}{8}}}{1-q^{n-\frac{1}{2}}}\\ 
&=2q^{\frac{3}{8}}\left(1+3\,q^{\frac{1}{2}}+7\,q+14\,q^{\frac{3}{2}}+\cdots\right).
\end{split}
\ee 
While $G_{1/2,0}$ is non-vanishing, the anomaly condition below (\ref{OmegT}) implies that $\Phi_{1/2,0}=0$ due to the cancellation between the two copies.

For general $n$, we note that $f_{\mu,n}$ satisfies
\be
f_{\mu,n}(\tau,\bar \tau,z,\bar z)=\frac{1}{2}  e^{\pi i \nu} \,q^{-\frac{\nu^2}{2}}w^{-\frac{1}{2}n\nu} 
\frac{\mathrm{d}R(\tau,\bar\tau, nz+\nu\tau, n\bar z +\nu\bar \tau)}{\mathrm{d}\bar \tau},\quad \nu=\mu-\frac12,
\ee
with the non-holomorphic function $R$ defined in Eq.  \eqref{muR}. To determine a suitable holomorphic part, we set \cite[Eq. (5.50)]{Korpas:2019cwg}
\be 
\label{eq:hatGhatM}
\widehat G_{\frac{1}{2},n}(\tau,\bar \tau,v,\bar v)=-i\,\widehat M\left(\tau,\bar \tau,nv+\frac{\tau}{2},n\bar v+\frac{\bar \tau}{2},\frac{\tau}{2},\frac{\bar \tau}{2}\right),
\ee 
where $\widehat M$ is given in \eqref{mucomplete}. The holomorphic part of $\widehat G_{1/2,n}$ 
is obtained by replacing the error function $E$ in the non-holomorphic term of $\widehat M$ with its value in the limit $y\to \infty$, i.e., the sign of its argument. The expansion \eqref{vR} shows that $v$ is $\CO(\CR)$ in the small $\CR$ expansion. Thus, the holomorphic part is given by the specialization of Eq. \eqref{Mtuv},
\be 
\label{G12nM}
G_{\frac{1}{2},n}(\tau,v)=-i M\left(\tau,nv+\frac{\tau}{2},\frac{\tau}{2}\right),
\ee 
which evaluates to the series,
\be 
\label{G12n}
G_{\frac{1}{2},n}(\tau,v)=-\frac{e^{\pi i nv}}{\vartheta_4(\tau)}
\sum_{l\in \mathbb{Z}} \frac{(-1)^l\,q^{\frac{l^2}{2}-\frac{1}{8}}}{1-e^{2\pi i n v}q^{l-\frac{1}{2}}}.
\ee 
Using the anomaly condition below (\ref{OmegT}) and substituting $K_U=1/2$, $K=3$ and $\nu_R(\tau)$ from (\ref{nuR}), we obtain for $\mu=1/2$
\be\label{eq:Phi-half-zero-CP2n}
\Phi_{\frac{1}{2},n}(\CR)=\left\{\begin{array}{ll} \frac{1}{2}{\rm Coeff}_{q^0}{\rm Ser}_{\CR}\left[ \frac{\vartheta_4(\tau)^{12}}{\eta(\tau)^9}  \left({1-2\CR^2\ttu(\tau)+\CR^4}\right)^{-\frac{1}{2}} C^{n^2}  G_{\frac{1}{2},n}(\tau,v) \right], & \quad  n\text{ is odd},\\
0, &\quad n\text{ is even}.\end{array}  \right. 
\ee 
For small odd values of $n$, we obtain 
\be
\label{PhiU12}
\Phi_{\frac12,n}(\CR)=\left\{\begin{array}{lr} 
1+\CO\left(\CR^{13}\right),& n=\pm 1, \\
1+\CR^4+\CR^8+\CR^{12}+\CO\left(\CR^{13}\right), & n=\pm 3,\\ 
1+6\CR^{4}+21\CR^{8}+56\CR^{12}+\CO\left(\CR^{13}\right), & n=\pm 5, \\
1+21\CR^{4}+210\CR^{8}+1401\CR^{12}+\CO\left(\CR^{13}\right), & n=\pm 7,\\ 
1+55\CR^{4}+1310\CR^{8}+19432\CR^{12}+\CO\left(\CR^{13}\right), & n=\pm 9. \\  \end{array}\right.
\ee 
The series reproduce the corresponding Donaldson invariant as $\CR\to 0$, while the symmetry $n\leftrightarrow -n$ is consistent with Eq. \eqref{Phin-n}. Moreover, the integrality of the $\CR$-expansion coefficients is highly nontrivial from the $q$-series expression \eqref{eq:Phi-half-zero-CP2n}.
Closed expressions for these coefficients as rational functions in $\CR$ are derived by G{\"o}ttsche in \cite[Theorem  1.3]{Gottsche:2016}. We list a few of them in Eqs. \eqref{PhiG12rational} and \eqref{PhiG0rational}.

Returning to $\mu=0$, we observe that the anomaly condition does not impose the restriction $n\in \mathbb{Z}$. We find 
\be\label{eq:Phi-zero-CP2n}
\Phi_{0,n}(\CR)=-\frac{1}{2}{\rm Coeff}_{q^0}{\rm Ser}_{\CR}\left[ \frac{\vartheta_4(\tau)^{12}}{\eta(\tau)^9} \left({1-2\CR^2\ttu(\tau)+\CR^4}\right)^{-\frac{1}{2}} C^{n^2} G_{0,n}(\tau,v) \right],
\ee 
with \cite[Eq.  (5.62)]{Korpas:2019cwg}
\be 
\label{G0n}
\begin{split} 
G_{0,n}(\tau,v)&=\frac{1}{\eta(\tau)^3}\sum_{\substack{k_1 \in \mathbb{Z} \\ k_2\in \mathbb{Z}+\frac{1}{2}}}
\frac{1}{2}\left[{\rm sgn}(k_1+k_2)-{\rm sgn}(k_1)\right] k_2\,e^{\pi i (k_1+k_2)}e^{2\pi i n v k_1}q^{\frac{1}{2}(k_2^2-k_1^2)}\\
&=\frac{i}{2}-\frac{i}{\vartheta_4(\tau)}\sum_{l\in \mathbb{Z}}\frac{(-1)^l q^{\frac{l^2}{2}}}{1-e^{2\pi i vn} q^l} -\frac{i}{\vartheta_4(\tau,nv)}\left.\partial_\rho {\rm ln} \left(\frac{\vartheta_1(\tau,\rho)}{\vartheta_4(\tau,\rho)} \right)\right\vert_{\rho=nv}.
\end{split}
\ee 
The form of this function is somewhat different from that of Eq. \eqref{G12n}. An alternative expression \cite[Eq.  (5.53)]{Korpas:2019cwg}, which is very similar to \eqref{G12n},  can be used to evaluate $\Phi_{0,n}(\CR)$. However, this alternative is less favorable, since it has a pole at vanishing elliptic variable. Note that $G_{0,n}$ is odd under $n\to - n$, as easily seen from the first line of \eqref{G0n} by changing the signs of $k_1, k_2$ and using the fact that $k_1+k_2$ is half-integral. 

For small $n$, explicit expressions for $\Phi_{0,n}(\CR)$ as power series expansions in $\CR$ are given by
\be
\label{PhiP20}
\Phi_{0,n}(\CR)=\left\{\begin{array}{lr}
\frac{15}{2}\CR-21\CR^5-56\CR^9-126\CR^{13}+\CO\left(\CR^{17}\right), & \quad n=5,\\ 
6\CR-6\CR^5-10\CR^9-15\CR^{13}+\CO\left(\CR^{17}\right), & \quad n=4,\\ 
\frac{9}{2}\CR-\CR^5-\CR^9-\CR^{13}+\CO\left(\CR^{17}\right), & \quad n=3,\\ 
3\CR+\CO\left(\CR^{17}\right), & \quad n=2,\\ 
\frac{3}{2}\CR+\CO\left(\CR^{17}\right) , & \quad n=1,\\ 
0 , & n=0,\\ 
-\frac{3}{2}\CR+\CO\left(\CR^{17}\right), & \quad n=-1,\\ 
-3\CR+\CO\left(\CR^{17}\right), & \quad n=-2,\\ 
-\frac{9}{2}\CR+\CR^5+\CR^9+\CR^{13}+\CO\left(\CR^{17}\right), & \quad n=-3,\\ 
-6\CR+6\CR^{5}+10\CR^{9}+15\CR^{13}+\CO\left(\CR^{17}\right), & \quad n=-4, \\ 
-\frac{15}{2}\CR+21\CR^{5}+56\CR^{9}+126\CR^{13}+\CO\left(\CR^{17}\right), & \quad n=-5,\\ 
-9\CR+56\CR^{5}+230\CR^{9}+770\CR^{13}+\CO\left(\CR^{17}\right), &\quad n=-6.
\end{array}\right.
\ee

\subsubsection*{Remarks} 
\begin{enumerate}
\item Although unphysical, one may substitute an even integer for $n$ into the expression for odd $n$ in Eq. \eqref{eq:Phi-half-zero-CP2n}. For example, with $n=0$, one finds
\be
\label{PhiU12odd}
1-\frac{7}{128}\CR^4-\frac{49}{4096}\CR^8+\CO\left(\CR^{9}\right).
\ee  
This expansion differs significantly from those in Eq. \eqref{PhiU12} by its rational rather than integer coefficients, a behavior also observed in the wall-crossing term for an anomalous choice of the flux $\bfn$ (\ref{DeltatPhiP1P1}).
\item As discussed in Sec.  \ref{UPlaneIntDefs} there are two competing definitions of the $U$-plane integral. 
The above results were obtained using the definition \eqref{PhiIntDef}, which involves expanding first in $\CR$ around zero and then taking the coefficient of $q^0$. 
If we adopt the definition  \eqref{eq:UplaneInt-FixedCR} (``holding  $\CR$ fixed and integrating over $\tau$''), then the results become inconsistent with the UV interpretation of the path integral as a generating function of the twisted Dirac indices. 
For example, for $(\mu,n)=(1/2,3)$, this definition yields $-\CR^{-4}$, which is physically unacceptable.
\item There is an odd feature of the result \eqref{PhiP20}: The first term in the $\CR$-expansion is half-integral when $n$ is odd. 
We believe that this is an indication that the integration is performed over a moduli space with singularities, or rather a moduli \emph{stack}, with an extra automorphism group present at the lowest instanton charge for $\mu=0$.  Note also that the sign of the first term is opposite to that of all the higher terms. It would be valuable to have a deeper understanding of this phenomenon. 
\end{enumerate}

\subsubsection*{Comparison with results by G\"ottsche-Nakajima-Yoshioka}
We compare here the results obtained from the $U$-plane evaluation, given in Eqs. \eqref{eq:Phi-half-zero-CP2n} and \eqref{eq:Phi-zero-CP2n}, with those of \cite{Gottsche:2006bm, Gottsche2015, Gottsche:2016} for $\mathbb{CP}^2$. 
For ease of comparison, we rewrite the results of \cite{Gottsche:2006bm} as Eqs. \eqref{PhiGNYM0} and \eqref{PhiGNYM12}, respectively, in App. \ref{ResultsGNY}. Comparing $\Phi_{\mu,n}(\CR)$ and $\Phi^{\rm GNY}_{\mu,n-3}(\CR)$, we observe that the expressions are related by replacing $G_{\mu,n}(\tau,v)$ in Eqs. \eqref{eq:Phi-half-zero-CP2n} and \eqref{eq:Phi-zero-CP2n} by
\be
\label{eq:UGNYcomp}
G_{\mu,n}(\tau,v)+\Delta_{\mu,n}(\tau,v),
\ee
where
\be
\begin{split}
  \Delta_{0,n}(\tau,v)&= -\frac{1}{\CR}\frac{\eta(\tau)^3\,\vartheta_1(\tau,(n-1)v)}{\vartheta_4(\tau)\,\vartheta_1(\tau,nv)\,\vartheta_4(\tau,(n-1)v)}\\
&\quad \left.-\frac{i}{\vartheta_4(\tau,nv)} \partial_\rho \ln\!\left( \frac{\vartheta_1(\tau,\rho)}{\vartheta_4(\tau,\rho)}\right)\right|_{\rho=nv}-\frac{\vartheta^+(\tau,v)}{\vartheta_1(\tau,v)}, \\
  \Delta_{\frac{1}{2},n}(\tau,v)&=-\CR \frac{\eta(\tau)^3\,\vartheta_1(\tau,(n-1)v)}{\vartheta_4(\tau,nv)\,\vartheta_4(\tau,(n-1)v)\,\vartheta_4(\tau)}.
  \end{split}
\ee
The terms in $\Delta_{\mu,n}$ which contain only only modular functions $\eta$ and $\vartheta_j$ do not contribute to the $q^0$ term. This is the case because, together with the other terms of the integrand, they lead to a modular form of weight 2 for $\Gamma^0(4)$. This can be converted into a modular form under $\mathrm{SL}(2,\mathbb{Z})$ through modular transformations. The latter does not have a constant term, because there is no one-dimensional cohomology on $\mathbb{H}/\mathrm{SL}(2,\mathbb{Z})$. 
Consequently, our result for $\Phi_{1/2,n}$ is in agreement with \cite{Gottsche:2006bm}.

On the other hand, comparing $\Phi_{0,n}(\CR)$ and $\Phi^{\rm GNY}_{0,n-3}(\CR)$, we observe a discrepancy involving $\vartheta^+(\tau,v)$, which is neither a modular form nor a mock modular form. It is an example of a ``partial'' theta function. This term may contribute to the final result. In fact, while the terms of order $\CO(\CR^5)$ to $\CO(\CR^{13})$ agree with the results of \cite{Gottsche:2006bm} listed in \eqref{PhiGNYP20}, the linear term in $\CR$ does not, due to $\vartheta^+(\tau,v)$. 
Although we have not excluded the possibility that the term with $\vartheta^+(\tau,v)$ contributes to a high power of $\CR$, numerical evidence strongly suggests that the term with $\vartheta^+(\tau,v)$ only affects the linear term. Moreover, the linear and higher-order terms of Eq. \eqref{PhiP20} agree with the non-equivariant limit of the equivariant invariants listed in Table \ref{P2eqK0} in Sec.  \ref{Sec:SusyLoc}, which are evaluated using a completely different method.

We expect that the geometric explanation for the discrepancy with \cite{Gottsche:2006bm} is due to a different treatment of the singularities mentioned in Remark 3 above about the half-integrality of the linear term. Leaving aside this minor discrepancy, the remaining issue is to match the overall signs of Eqs. \eqref{eq:Phi-half-zero-CP2n} and \eqref{eq:Phi-zero-CP2n} with those of Eqs. \eqref{PhiGNYM0} and \eqref{PhiGNYM12} using the relation \eqref{Phin-n}. We therefore confirm that $\Phi_{\mu,n}$ agrees with $\Phi^{\rm GNY}_{\mu,-n-3}$, precisely as in Eq.  \eqref{eq:c1-Lgny-bfnI}.

\subsection{Manifolds With $b_2^->0$}
\label{b2Min>0}

We now extend our analysis to four-manifolds with $b_2^->0$. The key new feature is that for sufficiently large $b_2^-$, the $U$-plane integral receives contributions from the strong-coupling singularities. Their jumps under wall-crossing are cancelled by the SW contributions discussed in the next section. 

To set up the integral for $b_2^->0$, recall that the intersection form of a lattice with signature $(1,b_2^-)$ can be brought into two canonical forms, depending on whether the lattice is odd or even. For an odd lattice, the associate quadratic form $Q$ takes the form
\be
\label{Qodd}
\left< 1\right>\oplus b_2^-\left<-1\right>,
\ee
while for a even lattice, $Q$ is of the form
\be
\label{Qeven}
\left(\begin{array}{cc} 0 & 1 \\ 1 & 0 \end{array}\right)\oplus n Q_{E_8},
\ee
where $Q_{E_8}$ is the Cartan matrix for the Lie group $E_8$.

Choosing a period point $J$ that is consistent with the decompositions \eqref{Qodd} and (\ref{Qeven}), it is straightforward to evaluate the integral as discussed in \cite{Korpas:2019cwg} for 4d $\CN=2$ $\mathrm{SU}(2)$ SYM. We will discuss here the case of odd lattices. Let $J=(1,0,\cdots,0)\in L$ with respect to the basis \eqref{Qodd}, and similarly $\bfmu=(\mu_1,\bfmu_-)$, $\bfn=(n_1,\bfn_-)$ and $K=(K_1,K_-)$.
The sum over fluxes then reads
\be 
\Psi_\bfmu^{J}(\tau,\bar \tau,v,\bar v)=-i(-1)^{\mu_1(K_1-1)}f_{\mu_1,n_1}(\tau,\bar \tau,v_1,\bar v_1)\,\Theta_\bfmu(\tau,v \bfn_-),
\ee 
where
\be 
\Theta_{\bfmu_-}(\tau,\bfz_-)=\sum_{\bfk_-\in L_-+\bfmu_-} (-1)^{B(\bfk_-,K_-)}q^{-\frac{1}{2}\bfk_-^2}e^{-2\pi i B(\bfz_-,\bfk_-)}.
\ee 
This theta series can be expressed as a product of $\vartheta_1$ and $\vartheta_4$, depending on the choice of $\bfmu_-$. For the evaluation, we also need the dual $\Theta_{D,\bfmu_-}$ \cite[Eq.  (5.48)]{Korpas:2019cwg}, defined as 
\be 
\Theta_{D,\bfmu_-}(\tau,\bfz)=\sum_{\bfk_-\in L_-+K_-/2}(-1)^{B(K_-,\bfmu_-)} q^{-\frac{1}{2}\bfk_-^2}e^{-2\pi i B(\bfz,\bfk_-)}.
\ee 
We further define functions $G_{D,\mu,n}$ dual to $G_{\mu,n}$,
\be 
\begin{split}
\widehat G_{D,\mu,n}(\tau,\bar \tau,z,\bar z)&=-(-i\tau)^{-\frac{1}{2}}\exp\left(\frac{\pi i n^2 z^2}{\tau}\right) \widehat G_{\mu,n}\left(-\frac{1}{\tau},-\frac{1}{\bar \tau},\frac{z}{\tau},\frac{\bar z}{\bar \tau}\right)\\
&=-i\,\widehat M\left(\tau,\bar \tau,nz+\frac{1}{2},n\bar z+\frac{1}{2},\frac{1}{2},\frac{1}{2}\right).
\end{split}
\ee 
Taking the holomorphic part for small ${\rm Im}(z)$, we obtain
\be 
\label{GDDef}
\begin{split}
G_{D,\mu,n}(\tau,z)&=-\frac{e^{\pi i n z}}{\vartheta_2(\tau)}\sum_{l\in \mathbb{Z}}\frac{q^{\frac{1}{2}l(l+1)}}{1-(-1)^{2\mu}e^{2\pi i nz} q^{l}}\\
&\quad -\frac{\delta_{\mu,0}}{\vartheta_2(\tau,nz)}\partial_\rho \left.{\rm ln}\left(\frac{\vartheta_1(\tau,\rho)}{\vartheta_2(\tau,\rho)}\right) \right|_{\rho=nz}.
\end{split}
\ee 
 
The $U$-plane integral evaluates to a sum over contributions from the three cusps of $\mathbb{H}/\Gamma^0(4)$,
\be
\Phi^J_{\bfmu,\bfn}(\CR)=-i(-1)^{\mu_1(K_1-1)}\sum_{j\in \{\infty,1,2\}}\Phi^J_{\bfmu,\bfn,j}(\CR),
\ee
where each contribution combines the cusp in $(\mathbb{H}/\Gamma^0(4))_{L}^{Y}$ and  $(\mathbb{H}/\Gamma^0(4))_{R}^{Y}$, as discussed in Sec.  \ref{Sec:FormEvInt}. 
With $K_{U,\bfmu,\bfn}$ given in Eq. \eqref{KUmun}, the contributions then evaluate to
\begin{align}
\Phi^J_{\bfmu,\bfn,\infty}(\CR)&=4K_{U,\bfmu,\bfn}\,{\rm Coeff}_{q^0}{\rm Ser}_{\CR}\left[ \nu_R(\tau)\,C^{\bfn^2}\,  G_{\mu_1,n_1}(\tau,v)\Theta_\bfmu(\tau,v \bfn_-)\right],\\
\Phi^J_{\bfmu,\bfn,1}(\CR)&=K_{U,\bfmu,\bfn}\,{\rm Coeff}_{q_1^0}{\rm Ser}_{\CR}\left[ \nu_{R,1}(\tau_1)\,C_1(\tau_1,v_1)^{\bfn^2}  G_{D\mu_1,n_1}(\tau_1,v_1)\Theta_{D\bfmu}(\tau_1,v_1 \bfn_-)\right],\\
\Phi^J_{\bfmu,\bfn,2}(\CR)&=K_{U,\bfmu,\bfn}\left(ie^{-2\pi i \bfmu^2}\right)\, {\rm Coeff}_{q_2^0}{\rm Ser}_{\CR}\left[ \nu_{R,2}(\tau_2)\,C_2(\tau_2,v_2)^{\bfn^2}  G_{D\mu_1,n_1}(\tau_2,v_2)\Theta_{D\bfmu}(\tau_2,v_2 \bfn_-)\right].
\end{align}
Here $\nu_{R,1}$ is the dual measure near the cusp 1 at $\tau=0$, 
\be 
\label{eq:nuDR}
\begin{split} 
\nu_{R,1}(\tau_1)&=(-i\tau_1)^{b_2/2-2}\nu_{R}\left(-\frac{1}{\tau_1}\right)\\
&=-\frac{i}{4} \frac{\vartheta_2(\tau_1)^{13-b_2}}{\eta(\tau_1)^9} \left({-8\CR^2 \ttu_1(\tau_1)+4\CR^4+4}\right)^{-\frac{1}{2}},
\end{split}
\ee 
where $\ttu_1$ is defined by $\ttu_1(\tau_1)=\ttu(-1/\tau_1)$, and is expressed in terms of theta functions as
\be 
\label{ttuD}
\ttu_1(\tau_1)=\frac{\vartheta_3(\tau_1)^4+\vartheta_4(\tau_1)^4}{2\vartheta_3(\tau_1)^2\vartheta_4(\tau_1)^2}.
\ee 
Similarly, $\nu_{R,2}$ is the dual measure near the cusp 2 at $\tau=2$, 
\be 
\label{eq:nuDR2}
\begin{split} 
\nu_{R,2}(\tau_2)&=-\frac{i}{4} \frac{\vartheta_2(\tau_2)^{13-b_2}}{\eta(\tau_2)^9} \left({-8\CR^2
    \ttu_2(\tau_2)+4\CR^4+4}\right)^{-\frac{1}{2}},
\end{split}
\ee 
with $\ttu_2(\tau_2)=\ttu(2-1/\tau_2)=-\ttu_1(\tau_2)$. 
For the dual couplings $v_1$, $C_1$, $v_2$, $C_2$, we refer to Eqs. \eqref{eq:v1app}, \eqref{eq:C1app}, \eqref{eq:v2app}, and \eqref{eq:C2app} in App. \ref{app:ExpCusps}.

\section{Seiberg-Witten Contributions For $b_2^+>1$}
\label{sec:SWConts}

In this section, we derive partition functions for four-manifolds with $b_2^+>1$ making use of the cancellation of the wall-crossing of the $U$-plane integral at strong-coupling cusps and of the SW invariants. This approach was first carried out in \cite{Moore:1997pc} for Donaldson invariants, and has been successfully extended to various other invariants \cite{Marino:1998uy, Marino:1998bm, Manschot:2021qqe, Aspman:2023ate}. Here, we demonstrate that this approach reproduces known results for K-theoretic Donaldson invariants of algebraic surfaces with $b_2^+>1$, as derived by G\"ottsche, Kool, and Williams in
\cite{Gottsche:2019vbi}.

\subsection{Wall-crossing Of $U$-plane And Seiberg-Witten Contributions}
\label{SWCont}
To derive the SW contributions from the $U$-plane contribution $\Phi_{\bfmu,\bfn}^J$, recall that the full partition function $Z_\bfmu^J$ reads
\be
Z_{\bfmu,\bfn}^J=\Phi_{\bfmu,\bfn}^J+\sum_{j=1}^4 Z^J_{\mathrm{SW},\bfmu,\bfn,j}.
\ee 
The SW moduli space is a union of projective spaces. The
contribution from each strong-coupling singularity $U_j$ takes the form
\be
\label{ZJj}
Z^J_{\mathrm{SW},\bfmu,\bfn,j}=\sum_{c} {\rm SW}(c;J)\underset{a_j=0}{\rm Res}\left[a_j^{-1-n(c)}\,e^{-S_{\mathrm{SW},j}} \right],
\ee
where the sum is over the characteristic classes $c$ of Spin$^c$ structures on $X$, ${\rm SW}(c;J)$ is the SW invariant, $a_j$ is the local coordinate near the singularity, and $n(c)=(c^2-(2\chi+3\sigma))/8$ is half the real dimension of the SW moduli space. A useful relation for ${\rm SW}(c;J)$ is
\be
\label{SW-c}
{\rm SW}(c;J)=(-1)^{\chi_{\rm h}}\,{\rm SW}(-c;J).
\ee
The term $e^{-S_{\mathrm{SW},j}}$ is the exponentiated action near the singularity, which is a product of the couplings $\CA_j, \cdots,\CF_{\bfmu,j}$ and a normalization factor $\kappa_j$,
\be
\label{eq:SWconts}
e^{-S_{\mathrm{SW},j}} =\kappa_j\,\CA_j^\chi\,\CB_j^\sigma\,\CC_j^{\bfn^2}\,\CD_j^{B(\bfn,c)}\,\CE^{c^2}_j\,\CF_{\bfmu,j}.
\ee
Here the $\CF_{\bfmu,j}$ is a shorthand for three couplings,
\be
\CF_{\bfmu,j}=f_{1,j}^{\bfmu^2}\,f_{2,j}^{B(\bfn,\bfmu)}\,f_{3,j}^{B(c,\bfmu)},
\ee
with the $f_{i,j}$ being roots of unity.

The cancellation of wall-crossing between $\Phi_\bfmu^J$ and $Z^J_{\mathrm{SW},j,\bfmu}$ requires
\be
\left[ \Phi_{\bfmu,\bfn}^{J}-\Phi_{\bfmu,\bfn}^{J'}\right]_j=Z^{J'}_{\mathrm{SW},\bfmu,\bfn,j}-Z^{J}_{\mathrm{SW},\bfmu,\bfn,j}.
\ee
The couplings in Eq. \eqref{eq:SWconts} can be determined by comparing with the wall-crossing formula for the $U$-plane integral. The SW contributions depend on $J$ due to the wall-crossing of ${\rm SW}(c;J)$,
\be
{\rm SW}(c;J^+)-{\rm SW}(c;J^-)=-(-1)^{n(c)},
\ee
where $J^\pm$ are two period points separated by a wall of marginal stability for the class $c$, i.e., $B(c,J) =0$. This condition can be written using the sign function as
\be
\begin{split}
\left[ \Phi_{\bfmu,\bfn}^{J}-\Phi_{\bfmu,\bfn}^{J'}\right]_j&=\kappa_j \sum_c
(-1)^{n(c)}\frac{1}{2} \left[\sgn(B(c,J)-\sgn(B(c,J'))\right]\\
&\quad \times \underset{a_j=0}{\rm Res}\left[ a_j^{-1-n(c)} \CA_j^\chi\,\CB_j^\sigma\,\CC_j^{\bfn^2}\,\CD_j^{B(\bfn,c)}\,\CE^{c^2}_j\,\CF_{\bfmu,j} \right].
\end{split}
\ee
Substituting $\chi+\sigma=4$, we obtain
\be 
\label{eq:USWwallcross}
\begin{split}
\left[ \Phi_{\bfmu,\bfn}^{J}-\Phi_{\bfmu,\bfn}^{J'}\right]_j&=-\kappa_j \sum_c
(-1)^{\frac{1}{8}(c^2-\sigma)}\,\frac{1}{2} \left[\sgn(B(c,J)-\sgn(B(c,J'))\right]\\
&\quad \times \underset{a_j=0}{\rm Res}\left[ a_j^{-\frac{1}{8}(c^2-\sigma)} \CA_j^4\,(\CB_j/\CA_j)^\sigma\,\CC_j^{\bfn^2}\,\CD_j^{B(\bfn,c)}\,\CE^{c^2}_j\,\CF_{\bfmu,j} \right].
\end{split}
\ee 
From this equation, we will deduce in the next subsection the explicit form of the couplings $\CA_j,\cdots, \CF_{\bfmu,j}$ in terms of the couplings near each cusp $U_j$ determined in App. \ref{app:ExpCusps}. 

\subsection{Derivation Of The Partition Function}

The normalization factor $\kappa_j$ in Eq.  \eqref{eq:SWconts} can be determined by comparison with an independent result for a specific four-manifold. Here we take it to be the K3 surface, for which the only non-vanishing SW invariant is ${\rm SW}(0)=1$. The cohomology of the moduli spaces of instantons is isomorphic to that of Hilbert schemes of points, whose generating function of $\chi_y$-genera is known \cite{Gottsche1990}. 
In the limit $y\to 0$, this gives the generating function of holomorphic Euler characteristics,
\be 
\label{resultK3}
Z_{\bfmu,0}(\CR)=\frac{1}{2}\left[\left(1-\CR^2\right)^{-2}+(-1)^{2\bfmu^2}\left(1+\CR^2\right)^{-2}\right].
\ee 

\subsubsection{Universal Form Of Contributions From Strong-coupling Cusps}

We now proceed by determining the contribution from each strong-coupling cusp $U_j$ for $j=1,\cdots,4$.

\subsubsection*{Contribution from $U_1$}
We determine the universal couplings $\CA_1, \CB_1,\cdots$ in terms of couplings near $U_1$ by comparing Eq.  (\ref{eq:USWwallcross}) with the wall-crossing of the $U$-plane integral at $U_1$, Eq.  (\ref{WCU1res}). With the identification $c=2\bfk$, we find 
\begin{align}
\CA_1 &=2^{\frac{1}{4}} \kappa_1^{-\frac{1}{4}}\Lambda^{-\frac{1}{4}} e^{\frac{\pi i}{4}} \left( \frac{\vartheta_2(\tau_1)^2}{\eta(\tau_1)^6}\right)^{\frac{1}{4}},\\
\CB_1 &=e^{\frac{\pi i}{8}} \vartheta_2(\tau_1)\CA_1 a_1^{-\frac{1}{8}},\\
\CC_1 &=C_1(\tau_1,v_1),\\
\CD_1 &= e^{-\pi i v_1},\\
\CE_1 &=e^{-\frac{\pi i}{8}} a_1^{\frac{1}{8}} q_1^{-\frac{1}{8}},\\
\CF_1 &=e^{\pi i B(K-c,\bfmu)},
\end{align}
with $a_1$, $v_1$, and $C_1$ given in App. \ref{app:ExpCusps}. The leading terms in their $q_1$-expansions are
\begin{align}
\CA_1&=2^{\frac{3}{4}} \kappa_1^{-\frac{1}{4}}\Lambda^{-\frac{1}{4}} e^{\frac{\pi i}{4}}+\cdots,\\
\CB_1&= 2^{\frac{7}{4}} \kappa_1^{-\frac{1}{4}}\Lambda^{-\frac{3}{8}} e^{\frac{\pi i}{4}}\left(\frac{32}{1-\CR^2} \right)^{-\frac{1}{8}}+\cdots,\\
\CC_1&=\left(1-\CR^2\right)^{-\frac{1}{2}}+\cdots,\\
\CD_1&=\sqrt{\frac{1-\CR}{1+\CR}}+\cdots,\\
\CE_1&=\Lambda^{\frac{1}{8}} \left(\frac{32}{1-\CR^2}\right)^{\frac{1}{8}}.
\end{align}

We specialize to manifolds of SW simple type, for which by definition only classes with $n(c)=0$ contribute. The contribution (\ref{ZJj}) from the $U_1$ singularity is thus
\be
\begin{split}
Z_{\bfmu,\bfn,1}(\CR)&=K_U\,\kappa_1^{1-\chi_{\rm h}}(-1)^{\chi_{\rm h}}2^{2\chi+3\sigma}\left(1-\CR^2\right)^{-\frac{1}{2}\bfn^2-\chi_{\rm h}}\\
&\quad \times \sum_{c} {\rm SW}(c)
\left( \frac{1-\CR}{1+\CR}\right)^{\frac{1}{2}B(\bfn,c)} e^{\pi i
  B(K-c,\bfmu)},
\end{split}
\ee
where we have reinstated the overall normalization $K_U$
(\ref{KU}). This can be simplified slightly using Eq.  (\ref{SW-c}),
\be
\begin{split}
Z_{\bfmu,\bfn,1}(\CR)&=K_U\,\kappa_1^{1-\chi_{\rm h}} 2^{2\chi+3\sigma} \left(1-\CR^2\right)^{-\frac{1}{2}\bfn^2-\chi_{\rm h}}\\
&\times \sum_{c} {\rm SW}(c)
\left( \frac{1+\CR}{1-\CR}\right)^{\frac{1}{2}B(\bfn,c)} e^{\pi i   B(K+c,\bfmu)}.
\end{split}
\ee
We fix $\kappa_1$ by comparing this result for $\bfn=0$ with that of the K3 surface (\ref{resultK3}), where $K=0$, $\chi=24$, and $\sigma=-16$. 
The contributions from $U_1$ and $U_3$ are identical for $\bfn=0$ due to Eq.  \eqref{WCU34U12}. Therefore, matching the coefficient of $(1-\CR^2)^{-2}$ in Eq.  (\ref{resultK3}) gives $2K_U\kappa_1^{1-(\chi+\sigma)/4}=1/2$. With $K_U=1/2$ (\ref{KU}), this gives $\kappa_1=2$. Thus, the final expression for the contribution from $U_1$ is
\be
\label{ZSWjmun}
\begin{split}
Z_{\bfmu,\bfn,1}(\CR) &=2^{2\chi+3\sigma-\chi_{\rm h}} \left(1-\CR^2\right)^{-\frac{1}{2}\bfn^2-\chi_{\rm h}} \\
&\quad \times \sum_{c} {\rm SW}(c)
\left( \frac{1+\CR}{1-\CR}\right)^{\frac{1}{2}B(\bfn,c)} e^{\pi i B(K+c,\bfmu)}.
\end{split}
\ee

\subsubsection*{Contribution from $U_2$}
We determine the universal couplings $\CA_2,\CB_2,\cdots$ by comparing Eq.  (\ref{eq:USWwallcross}) with the wall-crossing of the $U$-plane integral, Eq.  (\ref{WCU2res}). We obtain
\begin{align}
\CA_2&=2^{\frac{1}{4}} \kappa_2^{-\frac{1}{4}}\Lambda^{-\frac{1}{4}}\left( \frac{\vartheta_2(\tau_2)^2}{\eta(\tau_2)^6}\right)^{\frac{1}{4}},\\
\CB_2&=e^{\frac{\pi i}{8}}\vartheta_2(\tau_2)\CA_2 a_2^{-\frac{1}{8}}, \\
\CC_2&=C_2(\tau_2,v_2),\\
\CD_2&= e^{-\pi i v_2},\\
\CE_2&=e^{-\frac{\pi i}{8}}\,a_2^{\frac{1}{8}} q_2^{-\frac{1}{8}},\\
\CF_2&=e^{\pi i B(K-c,\bfmu)}e^{-2\pi i \bfmu^2},
\end{align}
with $a_2$, $v_2$, and $C_2$ given in App. \ref{app:ExpCusps}. These couplings have the following leading terms in their $q_2$-expansions,
\begin{align}
\CA_2 &= 2^{\frac{1}{4}} \kappa_2^{-\frac{1}{4}} +\cdots,\\
\CB_2 &= 2^{\frac{7}{4}}\,\kappa_2^{-\frac{1}{4}}\Lambda^{-\frac{3}{8}}\,e^{\frac{\pi i}{16}}\left(\frac{32}{1+\CR^2} \right)^{-\frac{1}{8}}+\cdots,\\
\CC_2 &=\left(1+\CR^2\right)^{-\frac{1}{2}}+\cdots,\\
\CD_2 &=\sqrt{\frac{1+i\CR}{1-i\CR}}+\cdots,\\
\CE_2 &=\Lambda^{\frac{1}{8}}e^{-\frac{\pi i}{16}} \left(\frac{32}{1+\CR^2}\right)^{\frac{1}{8}}.
\end{align}
Similarly to the discussion for $U_1$, we find by comparison with the result for K3 surface \eqref{resultK3} that $\kappa_2=2$. Combining all the terms and using Eq.  (\ref{SW-c}), we obtain the contribution from $U_2$,
\be
\begin{split}\label{ZSWjmun2}
  Z_{\bfmu,\bfn,2}(\CR)&=2^{2\chi+3\sigma-\chi_{\rm h}}\left(1+\CR^2\right)^{-\frac{1}{2}\bfn^2-\chi_{\rm
        h}} e^{-3\pi i \chi_{\rm h}/2-2\pi i \bfmu^2}\\
 &\quad \times  \sum_{c} {\rm SW}(c)
\left( \frac{1-i\CR}{1+i\CR}\right)^{\frac{1}{2}B(\bfn,c)} e^{\pi i
  B(K+c,\bfmu)}.
\end{split}
\ee

\subsubsection*{Contribution from $U_3$ and $U_4$}
From Eq. \eqref{WCU34U12}, it is straightforward to find that the contributions from the other two cusps, $U_3$ and $U_4$, are given by
\be\label{ZSWjmun34}
Z_{\bfmu,\bfn,j}(\CR)=(-1)^{B(2\bfmu,K- \bfn)}Z_{\bfmu,\bfn,j-2}(\CR),\quad j=3,4.
\ee 

Thus, the normalization factors $\kappa_j$ are equal to $2$ for all $j$.

\subsubsection{Full Partition Function For $b_2^+>1$}
For $b_2^+>1$, the contribution from the $U$-plane integral vanishes. The full partition function is given by
\be
\label{fullZSW}
  Z_{\bfmu,\bfn}(\CR)=\sum_{j=1}^4   Z_{\bfmu,\bfn,j}(\CR)= \left\{ \begin{array}{ll}  2\sum_{j=1}^2   Z_{\bfmu,\bfn,j}(\CR),  & \quad (-1)^{B(2\bfmu,K- \bfn)}=1, \\ 0,  & \quad (-1)^{B(2\bfmu,K- \bfn)}=-1.  \end{array}\right.
\ee
For $(-1)^{B(2\bfmu,K- \bfn)}=1$, the four contributions can be written as
\be
Z_{\bfmu,\bfn,j}(\CR)= 
\left( \omega^{3\chi_h + 4\bfmu^2}\right)^{j-1} \,Z_{\bfmu,\bfn,1}(\omega^{j-1}\CR),
\ee
where $\omega = e^{- \frac{i \pi}{2}}$ is a primitive fourth root of unity. The series $Z_{\bfmu,\bfn,1}(\CR)$ is a power series in $\CR$ with all nonnegative integer powers. The sum over the four cusps projects onto the series that contains only terms whose power of $\CR$ is $-\left( 3\chi_h + 4\bfmu^2 \right) \mod 4$, consistent with the definition of the partition function as a $\CR$-series given in Eqs. \eqref{eq:Zee-QM-1} and \eqref{eq:ModSpaceDim}.

\subsection{Comparison With Results By G\"ottsche-Kool-Williams}
We compare here our results with the calculation of K-theoretic Donaldson invariants for algebraic surfaces with $b_2^+>1$ in \cite{Gottsche:2019vbi}. Their Conjecture 1.1 states that the virtual holomorphic Euler characteristic twisted by $\mu_D(L)$, denoted $\chi(M(2\bfmu,k),\CO(\mu_D(L))$, is given by the coefficient of $\CR^{l(k)}$, with $l(k)$ the dimension of $M_{\bfmu,k}$ \eqref{eq:ModSpaceDim}, of the generating function
\footnote{Note that in our convention, a SW basic class satisfies $c^2=2\chi+3\sigma$, which is different from that in \cite{Gottsche:2019vbi}.}
\be
\label{algbp}
\begin{aligned}
G_{\bfmu,L}(\CR, X)&=2^{2-\chi_{\rm h}+K_X^2} \left(1-\CR^2\right)^{-\frac{1}{2}(L-K_X)^2-\chi_{\rm h}} \\
&\quad \times \sum_{c} {\rm SW}(c) (-1)^{B(\bfmu,c+K_X)}\,\left(\frac{1+\CR}{1-\CR}\right)^{\frac{1}{2}B(c,K_X-L)},
\end{aligned}
\ee
where $L$ abbreviates $c_1(L)$. By extracting the appropriate powers of $\CR$ modulo 4, we get
\be
\label{GKWsum}
\sum_k \chi(M(2\bfmu,k),\CO(\mu_D(L)))\CR^{l(k)}
=\frac{1}{4}\sum_{l=0}^3 e^{-\frac{1}{2}\pi i
  l (4\bfmu^2+3\chi_{\rm h})} G_{\bfmu,L}(e^{-\frac{1}{2}\pi i l}\,\CR,X).
\ee
Using Eq.  (\ref{SW-c}), one can deduce that
\be
\label{Geo}
G_{\bfmu,L}(-\CR, X)=(-1)^{\chi_h+B(2\bfmu, K_X)} G_{\bfmu,L}(\CR, X),
\ee 
which implies that $G _{\bfmu,L}(\CR, X)$ contains only even or odd powers of $\CR$. As a result, Eq.  (\ref{GKWsum}) reduces to a sum of two terms,
\be
\label{GKWpartition}
\frac{1}{2}G_{\bfmu,L}(\CR,X)+\frac{1}{2}e^{\frac{\pi i}{2} (4\bfmu^2+3\chi_{\rm h})}G_{\bfmu,L}(i\CR,X).
\ee 
This reproduces exactly the first line of Eq.  (\ref{fullZSW}) under the identifications $K=K_X$ and $\bfn=-L+K_X$ \eqref{eq:c1-Lgny-bfnI}. Ref. \cite{Gottsche:2019vbi} does not distinguish between the two cases in Eq.  (\ref{fullZSW}), which we have commented on before.
The derivation from the $U$-plane integral predicts that the result (interpreted in terms of the $L^2$ index of the Dirac operator) holds more broadly for almost-complex four-manifolds.

\section{The Path Integral Of The $E_1$ Superconformal Field Theory On $X\times S^1$}
\label{Sec:E1Uplane}

The 5d $\mathrm{SU}(2)$ SYM studied in previous sections emerges as a LEET of the strongly coupled $E_1$ superconformal field theory perturbed by a relevant operator \cite{Seiberg:1996bd}. The $E_1$ theory has an $\mathrm{SU}(2)=E_1$ global symmetry group, which is broken to $\mathrm{U}(1)^{(I)}$ in the non-conformal theory. If the parameter $\CR^4$ (\ref{CR4}) of the theory on $X\times S^1$ is tuned to $1$, some massive states for $\CR^4\neq 1$ become massless. In addition to the Coulomb branch, the $E_1$ theory also has a Higgs branch. In this section, we discuss the path integral of the $E_1$ theory, uncovering a few novel strong-coupling phenomena that remain to be better understood.

Due to the appearance of extra massless particles, various couplings determined in App.  \ref{app:ExpCusps} become singular in the limit $\CR^4 \rightarrow 1$. 
We therefore recompute these couplings starting from Eq.  (\ref{tauvid}), obtaining finite expressions. This is, again, a manifestation of the order-of-limits issue persistent  throughout this paper.  For the four solutions to $\CR^4=1$, $v$ satisfies
\be
\begin{aligned}
\CR&=i: &    v&=-\frac{\tau}{4}  & \text{or}& \quad  \frac{\tau}{4}+1 &\mod  2\IZ+\tau\IZ, \\
\CR&=-1: &   v&=-\frac{\tau}{4}+\frac{1}{2}  & \text{or}& \quad  \frac{\tau}{4} +\frac{1}{2} &\mod 2\IZ+\tau\IZ, \\
\CR&=-i: &   v&=-\frac{\tau}{4}+1  &\text{or}& \quad  \frac{\tau}{4} &\mod 2\IZ+\tau\IZ, \\
\CR&=1: &    v&=-\frac{\tau}{4}-\frac{1}{2}  &\text{or}& \quad  \frac{\tau}{4} -\frac{1}{2} &\mod 2\IZ+\tau\IZ.
\end{aligned}
\ee
Many solutions of $v$ give identical results, since the $U$-plane integrand is invariant under shifts $v\to v+ 2m_1 + \tau m_2$ $(m_1,m_2\in \mathbb{Z})$. We resolve this ambiguity by fixing $v$ via the expansion of the prepotential given in Eq. \eqref{vinfexp}.
The resulting BPS equations are the perturbed BPS equations detailed in App. \ref{App:ShiftedWalls}.

In this section, we focus on the case $\CR=i$, leaving the analysis of other cases for future work.

\subsection{The $U$-plane Integral}

For $\CR=i$, the order parameter $U$ given in Eq. \eqref{eq:U-doublecovers-ttu} reduces to Eq. \eqref{UR2=-1}, which is a Hauptmodul for the congruence subgroup $\Gamma^0(8)$.
The fundamental domain has three strong-coupling cusps: 
\be
\begin{aligned}
&U_1\text{ at } \tau=0:  &\text{width}&=1,  & U(0)&=4, \\
&U_2\text{ at } \tau=2:  &\text{width}&=2,  & U(2)&=0, \\
&U_3\text{ at } \tau=4:  &\text{width}&=1,  & U(4)&=-4.
\end{aligned}
\ee
The singularities $U_2$ and $U_4$ merge, and their corresponding massless particles, with charges spanned by $\gamma_2$ and $\gamma_4$ \eqref{eq:chargeQuivers}, are mutually local.
Near the singularities, the local parameters are given by
\be 
\label{eq:vjR41}
\begin{aligned}
&U_1:& \tau&=-\frac{1}{\tau_1}, & v&=\frac{v_1}{\tau_1},  &v_1&=\frac{1}{4},\\
&U_2:& \tau&=2-\frac{1}{\tau_2}, & v&=\frac{v_2}{\tau_2}, &v_2&=\frac{1-2\tau_2}{4}, \\
&U_3:& \tau&=4-\frac{1}{\tau_3}, & v&=\frac{v_3}{\tau_3}, &v_3&=\frac{1-4\tau_3
}{4}.
\end{aligned}
\ee 
With $v=-\tau/4$, the coupling $C$ (\ref{Ctauv}) becomes
\be 
\label{CR-i}
C(\tau)=\frac{\vartheta_4\left(\tau,-\frac{1}{4}\tau\right)}{\vartheta_4(\tau)}.
\ee 
Its behavior near the cusps is
\be
\begin{split} 
C_1(\tau_1)&=e^{-\frac{\pi iv_1^2}{\tau_1}}C\left(-\frac{1}{\tau_1}\right)
=\frac{\vartheta_2\left(\tau_1,\frac{1}{4}\right)}{\vartheta_2(\tau_1)}
=\frac{1}{\sqrt{2}}-\sqrt{2}q_1+\CO\left(q_1^2\right),\\
C_2(\tau_2)&=e^{-\frac{\pi iv_2^2}{\tau_2}}C\left(2-\frac{1}{\tau_2}\right)
=e^{\frac{\pi i}{4}} q_2^{-\frac{1}{8}} \frac{\vartheta_3\left(\tau_2,\frac{1}{4}\right)}{\vartheta_2(\tau_2)}
=\frac{1}{2}e^{\frac{\pi i}{4}} q_2^{-\frac{1}{4}}+\CO\left(q_2^{\frac{3}{4}}\right),\\
C_3(\tau_3)&=e^{-\frac{\pi i v_3^2}{\tau_3}}C\left(4-\frac{1}{\tau_3}\right)
=\frac{1}{\sqrt{2}}e^{\frac{\pi i}{2}} q_3^{-\frac{1}{2}}+\CO\left(q_3^{\frac{1}{2}}\right).
\end{split} 
\ee 
The negative power of $q_2$ in $C_2$ is particularly significant, as it gives rise to nontrivial contributions from this strong-coupling cusp.

For $\CR=i$, Eq.  (\ref{dadtau}) becomes
\be 
\frac{\mathrm{d}a}{\mathrm{d}\tau}=\frac{\pi}{4R} \frac{\vartheta_4(\tau)^8}{\vartheta_2(\tau)^2+\vartheta_3(\tau)^2}.
\ee 
Near the singularities $U_j$, this gives
\be 
\begin{split}
\left(\frac{\mathrm{d}a}{\mathrm{d}\tau} \right)_1&=\frac{\pi i}{4R}\frac{\vartheta_2(\tau_1)^8}{\vartheta_3(\tau_1)^2+\vartheta_4(\tau_1)^2}= \frac{32\pi i}{R}  q_1+\cdots \\
\left(\frac{\mathrm{d}a}{\mathrm{d}\tau} \right)_2&= \frac{\pi i}{4R}\frac{\vartheta_2(\tau_2)^8}{\vartheta_3(\tau_2)^2-\vartheta_4(\tau_2)^2}=\frac{8\pi i}{R}  q_2^{\frac{1}{2}} + \cdots \\
\left(\frac{\mathrm{d}a}{\mathrm{d}\tau} \right)_3&= \frac{\pi i}{4R}\frac{\vartheta_2(\tau_3)^8}{\vartheta_3(\tau_3)^2+\vartheta_4(\tau_3)^2}=\frac{32\pi i}{R}  q_3 +\cdots.
\end{split} 
\ee 
After integrating, we obtain the leading terms for $a_j$,
\be
\begin{split} 
a_1&=\frac{16}{R}q_1+\cdots,\\
a_2&=\frac{8}{R}q_2^{\frac{1}{2}}+\cdots,\\
a_3&=\frac{16}{R}q_3+\cdots.
\end{split}
\ee 

Crucially, the limit of the $U$-plane as $\CR$ approaches a fourth root of unity is 
\emph{not} the same as the integral of the limit of the integrand as $\CR$ approaches a fourth root of unity. The former typically has singularities, 
as seen explicitly in Sec.  \ref{sec:SWConts}, while the latter yields  
\be 
\Phi^J_{\bfmu,\bfn}(i)=-\frac{i}{8} \int_{\mathbb{H}/\Gamma^0(8)} \mathrm{d}\tau\wedge \mathrm{d}\bar \tau\frac{\vartheta_4(\tau)^{12-b_2}}{\vartheta_2(\tau)^2+\vartheta_3(\tau)^2} \frac{1}{\eta(\tau)^6}\,C(\tau)^{\bfn^2} \Psi_\bfmu^J\left(\tau,\bar \tau,-\frac{\bfn\tau}{4},-\frac{\bfn \bar \tau}{4}\right),
\ee 
where we substituted $v=-\tau/4$. Using the properties of $\Psi^J_\bfmu$ \eqref{Psitrafos} and \eqref{zshift}, we verify that the integrand changes under the transformation $\tau\to \tau+4$ by the sign factor,
\be
e^{\pi i B\left(2\bfmu,K_X+\bfn\right)}.
\ee 
This is identical to the sign arising from the $\mathbb{Z}_2^{(1)}$ symmetry discussed in Sec.  \ref{sec:UInt1Form}. Thus, the $U$-plane integrals vanish when $B\left(2\bfmu,K_X+\bfn\right)$ is odd.

\subsection{Evaluation For $X=\mathbb{CP}^2$}
We apply the techniques developed in Sec.  \ref{Sec:Uplane} to evaluate the $U$-plane integral for four-manifolds with $b_2^+=b_2=1$, specifically $X=\mathbb{CP}^2$ and $\mu=\tfrac{1}{2}$. Using the anti-derivative from Eqs. \eqref{G12n} for $\Psi_\frac{1}{2}$, we obtain 
\be 
\Phi_{\frac{1}{2},n}=\frac{1}{8} \sum_{j\in \rm{cusps}} n_j {\rm Coeff}_{q^0}\left[\frac{\vartheta_4(\tau)^{11}}{\vartheta_2(\tau)^2+\vartheta_3(\tau)^2} \frac{1}{\eta(\tau)^6}\,C(\tau)^{n^2}\,G_{\frac{1}{2},n}\left(\tau,-\frac{\tau}{4}\right) \right],
\ee 
with $n_j$ the width of the cusp, $j=1,2,3,\infty$ and $q_\infty=q$. The function $G_{\frac{1}{2},n}$ is again the holomorphic part of $\widehat G_{\frac{1}{2},n}$ \eqref{eq:hatGhatM}, obtained by replacing the error function $E$ in the non-holomorphic completion with the sign of its argument. For $v=-\tau/4$ and odd $n$, this gives
\be 
\label{eq:G12nRi}
\begin{split}
G_{\frac{1}{2},n}\left(\tau,-\frac{\tau}{4}\right)&=-\frac{q^{-\frac{n}{8}}}{\vartheta_4(\tau)}\sum_{l\in \mathbb{Z}} \frac{(-1)^l q^{\frac{1}{2}l^2-\frac{1}{8}}}{1-q^{l-\frac{n}{4}-\frac{1}{2}}}\\
&\quad +\sum_{l\in \mathbb{Z}+\frac{1}{2}} \frac{1}{2}\left[\sgn(l) -\sgn\left(l-\frac{n}{4}\right)\right](-1)^{l-\frac{1}{2}}  q^{-\frac{1}{2}l^2+\frac{1}{4}nl}\\
&=(-1)^{\frac{n^2-1}{8}}q^{\frac{n^2+7}{32}}\left(1+2q^{\frac{1}{4}}+3q^{\frac{1}{2}}+5q^{\frac{3}{4}}+\CO\left(q\right)\right).
\end{split}
\ee 
The coefficients of this $q$-series match those of a second order mock modular form, denoted by $B(q)$ in \cite{McIntosh_2007}.
The contribution from the weak-coupling cusp for $\mu=\frac{1}{2}$ is 
\be 
\Phi_{\frac{1}{2},n,\infty}={\rm Coeff}_{q^0}\left[ \frac{\vartheta_4(\tau)^{11}}{\vartheta_2(\tau)^2+\vartheta_3(\tau)^2} \frac{1}{\eta(\tau)^6}\,C(\tau)^{n^2}\,G_{\frac{1}{2},n}\left(\tau,-\frac{\tau}{4}\right)\right].
\ee 
Values of $\Phi_{\frac{1}{2},n,\infty}$ with insertions of $U^l$ for $n=1,3,5, 7$ are tabulated in Table \ref{PtObsE1}.

\begin{table}[t]
\centering
\renewcommand{\arraystretch}{2.2}
\begin{tabular}{|c|c|c|c|c|c|c|c|c|}
\hline 
$l$ & $\Phi_{\frac{1}{2},1,\infty}$ & $\Phi_{\frac{1}{2},1,2}$ & $\Phi_{\frac{1}{2},3,\infty}$ & $\Phi_{\frac{1}{2},3,2}$ & $\Phi_{\frac{1}{2},5,\infty}$ & $\Phi_{\frac{1}{2},5,2}$ & $\Phi_{\frac{1}{2},7,\infty}$ & $\Phi_{\frac{1}{2},7,2}$  \\
\hline 
0 & 1 & 0 & 0 & $\frac{1}{2^3}$ & 0 & $\frac{1}{2^{15}}$ & 0 & $-\frac{1509}{2^{39}}$  \\  
 2 & 5 & 0 & $-1$ & 0 & 0 & $\frac{1}{2^{10}}$ & 0 & $-\frac{121}{2^{31}}$  \\ 
4 & 45 & 0 & $-5$ & 0 & 0 & $\frac{1}{2^4}$ & 0 &  $-\frac{83}{2^{27}}$ \\ \hline 
\end{tabular}
\caption{Values of $\Phi_{\frac{1}{2},n,\infty}[U^l]$ and $\Phi_{\frac{1}{2},n,2}[U^l]$ for $X=\mathbb{CP}^2$ and $l=0,2,4$. $\Phi_{\frac{1}{2},n,*}[U^l]$ vanishes for $l$ odd.}
\label{PtObsE1}
\end{table} 

For the contribution from strong-coupling cusps,
we determine the holomorphic parts of
\be 
\tau_j^{-\frac12}\exp\left(\frac{\pi i n^2 v_j^2}{\tau_j}\right)\widehat G_{\frac{1}{2},n}\!\left(2j-2-\frac{1}{\tau_j},\frac{v_j}{\tau_j}\right),
\ee 
for the three cusps for $j=1,2,3$. It turns out that these can all be expressed in terms of $G_{D,\frac{1}{2},n}$ \eqref{GDDef}. For cusp 1, we thus find
\be
\begin{split}
\label{GDtau1}
G_{D,\frac{1}{2},n}(\tau_1, v_1)=-\frac{q_1^{-\frac{1}{8}}}{2\sqrt{2}}\left(1+q_1+q_1^2-2q_1^3-q_1^4+\CO\left(q_1^5\right)\right),
\end{split}
\ee 
This is independent of the value of odd $n$. It then follows that this cusp does not contribute for any odd $n$, regardless of $U^l$ insertions, since the leading term of the full integrand is $\CO\left(q_1\right)$.

For cusp 2, the holomorphic part of $\widehat G_{D,1/2,n}(\tau_2,v_2)$ for odd $n$ is,
\be 
\begin{split}
G_{D,\frac{1}{2},n}(\tau_2,v_2)&=-\frac{e^{\frac{1}{4}\pi i n}q_2^{-\frac{n}{4}}}{\vartheta_2(\tau_2)}\sum_{l\in \mathbb{Z}} \frac{q_2^{\frac{1}{2}l(l+1)}}{1+e^{\frac{1}{2}\pi i n}q_2^{l-\frac{n}{2}}} \\
&\quad + \sum_{l\in \mathbb{Z}+\frac{1}{2}} \frac{1}{2}\left[\sgn(l)-\sgn\left(l-\frac{n}{2}\right)\right](-1)^{l-\frac{1}{2}}e^{-\frac{1}{2}\pi i nl}q_2^{\frac{1}{2}nl-\frac{1}{2}l^2}\\
&=e^{-\frac{3\pi i}{4}}q_2^{\frac{n^2+4}{8}}\left(1-2q_2+4q_2^2-6q_2^3+\CO\left(q_2^4\right)\right).
\end{split}
\ee 
The coefficients match those of a second order mock modular form, denoted by $A(q)$ in \cite{McIntosh_2007}. Notably the power of the overall coefficient increases with $n$ but this is cancelled by the negative power of $C_2$ giving rise to a non-vanishing contribution from this cusp,
\be 
\Phi_{\frac{1}{2},n,2}={\rm Coeff}_{q_2^0}\left[ \frac{i}{4}\frac{\vartheta_2(\tau_2)^{11}}{\vartheta_3(\tau_2)^2-\vartheta_4(\tau_2)^2} \frac{1}{\eta(\tau_2)^6}\,C_2(\tau_2)^{n^2}\,G_{D,\frac{1}{2},n}\left(\tau_2,\frac{1-2\tau_2}{4}\right)\right].
\ee 
We give the values of $\Phi_{1/2,n,2}$ for various $n$ and insertions of the order parameter $U$ in Table \ref{PtObsE1}.

For cusp 3, we obtain
\be 
G_{D,\frac{1}{2},n}(\tau_3,v_3)=\frac{i(-1)^{\frac{n+1}{2}}}{2\sqrt{2}}q_3^{\frac{1}{2}n^2-\frac{1}{8}}\left(1+q_3+q_3^2-2q_3^3-q_3^4+\CO\left(q_3^5\right)\right).
\ee
where the coefficients of the $q_3$-series agree with those of the $q_1$-series \eqref{GDtau1} up to the prefactor. Hence, there is no contribution from cusp 3 since the leading term of the full integrand is $\CO\left(q_3\right)$.

The non-vanishing contribution from the strong-coupling cusp 2 is a novel feature of this theory. In other theories, strong-coupling contributions vanish due to the absence of SW invariants of $\mathbb{CP}^2$.\footnote{However, see \cite{Okonek:1996hd} for discussions of non-vanishing SW invariants of $\mathbb{CP}^2$, which may be relevant here.} We give below an interpretation in terms of shifted multi-monopole equations. Alternatively, this might be attributable to Higgs branch contributions in the $E_1$ theory --- a possibility that we leave for future investigations.

\subsection{Wall-crossing And Seiberg-Witten Contributions}

We now analyze the wall-crossing behavior for a general four-manifold with $b_2^+=1$. Using $\widehat \Theta_\bfmu^{JJ'}$ \eqref{eq:DefIndefTheta}, we find the difference between the $U$-plane integrals with period points $J$ and $J'$ is given by
\be
\begin{aligned}
\Delta \Phi^{JJ'}_{\bfmu,\bfn} &= \sum_{j\in \rm cusps} \left( \Phi^{J}_{\bfmu,\bfn,j} - \Phi^{J'}_{\bfmu,\bfn,j} \right) \\
&= \frac{1}{8} \sum_{j\in \rm cusps} n_j 
{\rm Coeff}_{q_j^0}\left[ \frac{\vartheta_4(\tau)^{12-b_2}}{\vartheta_2(\tau)^2+\vartheta_3(\tau)^2} \frac{1}{\eta(\tau)^6} C(\tau)^{\bfn^2}\, \widehat \Theta^{JJ'}_{\bfmu}\left(\tau,\bar\tau,-\frac{\bfn\tau}{4},-\frac{\bfn\bar \tau}{4}\right) \right].
\end{aligned}
\ee 

To determine the contribution from each cusp, we make the appropriate modular transformations. 
\begin{itemize}
\item The indefinite theta function at cusp 1 is 
\be
\tau_1^{-\frac{1}{2}b_2} \exp\left(\frac{\pi i \bfn^2 v_1^2}{\tau_1}\right) \widehat \Theta_\bfmu^{JJ'}\left(-\frac{1}{\tau_1},-\frac{1}{\bar \tau_1},\frac{\bfn v_1}{\tau_1}, \frac{\bfn v_1}{\bar \tau_1}\right).
\ee
Using Eq.  \eqref{eq:IndefThetaTrafo}, this becomes
\be 
\exp\left(\frac{\pi i\sigma}{4}+\pi i B(\bfmu,K)\right)\widehat \Theta^{JJ'}_{\frac{K}{2}}\left(\tau_1,\bar \tau_1,\bfn v_1-\bfmu+\frac{K}{2},\bfn v_1-\bfmu+\frac{K}{2}\right).
\ee 
\item The indefinite theta function at cusp 2 is
\be 
\tau_2^{-\frac{1}{2}b_2} \exp\left(\frac{\pi i \bfn^2 v_2^2}{\tau_2}\right) \widehat \Theta_\bfmu^{JJ'}\left(2-\frac{1}{\tau_2},2-\frac{1}{\bar \tau_2},\frac{\bfn v_2}{\tau_2}, \frac{\bfn \bar v_2}{\bar \tau_2}\right),
\ee 
which evaluates to
\be 
\begin{split}
\exp\left(\frac{\pi i\sigma}{4} + 2\pi i \bfmu^2-\pi i B(\bfmu,K)\right)\widehat \Theta^{JJ'}_{\frac{K}{2}}\left(\tau_2,\bar \tau_2,\bfn v_2-\bfmu+\frac{K}{2},\bfn \bar v_2-\bfmu+\frac{K}{2}\right).\\
\end{split}
\ee 
\item The indefinite theta function at cusp 3 is
\be
\tau_3^{-\frac{1}{2}b_2} \exp\left(\frac{\pi i \bfn^2 v_3^2}{\tau_3}\right) \widehat \Theta_\bfmu^{JJ'}\left(4-\frac{1}{\tau_3},4-\frac{1}{\bar \tau_3},\frac{\bfn v_3}{\tau_3}, \frac{\bfn \bar v_3}{\tau_3}\right),
\ee
which evaluates to 
\be
\exp\left(\frac{\pi i \sigma}{4} -\pi i B(\bfmu,K)\right) \widehat \Theta_{\frac{K}{2}}^{JJ'}\left(\tau_3,\bar \tau_3,\bfn v_3-\bfmu+\frac{K}{2},\bfn \bar v_3 -\bfmu+\frac{K}{2}\right).
\ee 
\end{itemize}

The wall-crossing at cusps 1 and 3 can be ``absorbed'' by the standard  wall-crossing of SW invariants, ${\rm SW}(c)$. In fact, the couplings at these cusps given in App.  \ref{app:ExpCusps} can be specialized to $\CR=i$ without encountering a divergence. On the other hand, the dependence of $v_2$ on $\tau_2$ leads to the condition of walls,
\be\label{eq:NewWalls}
B\left(\bfk+\frac{{\rm Im}(v_2)}{y_2}\bfn, J\right)=B\left(\bfk-\frac{\bfn}{2}, J\right) = 0,   
\ee
for $\bfk \in \half K + H^2(X,\mathbb{Z})$ and $v_2$ as in Eq.  (\ref{eq:vjR41}). Note that for $\bfn\not=0 \mod H^2(X,2\mathbb{Z})$, these do \emph{not} correspond to the usual walls associated with Spin$^c$ structures familiar from SW theory. Consequently, this wall-crossing of the $U$-plane integral cannot be canceled by standard SW invariants. 

Physically, this arises because two monopoles, with charges $\gamma_2$ and $\gamma_4$ \eqref{eq:chargeQuivers}, become massless near $U_2$. Therefore, we should consider multi-monopole equations similar to those discussed in \cite{bryan1996, Kanno:1998qj, Dedushenko:2017tdw, Doan:2019, Aspman:2022sfj, Aspman:2023ate}. Our equations will be a variant of the equations studied in previous references. 

To derive our multi-monopole equations, note that the massless monopole with charge $\gamma_2$ is coupled to the flux $\bfk$, or equivalently Spin$^c$ structure $c=2\bfk$, while the massless monopole $\gamma_4$ carries $\mathrm{U}(1)^{(I)}$ charge and is coupled to the flux $\bfk-\bfn$, or equivalently the Spin$^c$ structure $2(\bfk-\bfn)=c-2\bfn$. 
Following a derivation analogous to App. \ref{App:ShiftedWalls}, the equations for the Spin$^c$ connection and the monopole fields $M_2$ and $M_4$ are specialization of Eq. \eqref{eq:multiMonoEqs},
\be 
\label{eq:multimoeqs}
\begin{split} 
& F_+(A) + F_+\left(A-2A^{(I)}\right) + M_2\bar M_2+M_4\bar M_4=0, \\
&\slashed D_A M_2=0, \\
&\slashed D_{A-2A^{(I)}} M_4=0,
\end{split} 
\ee 
where we substituted ${\rm Im}(v_2)/y_2=-1/2$. The equations are  invariant under two $\mathrm{U}(1)$ symmetries, 
\be
\begin{aligned}
&\mathrm{U}(1)_+: &\left(M_2,M_4\right)&\to \left(e^{i\theta}M_2,e^{i\theta}M_4\right), \\
&\mathrm{U}(1)_-: &\left(M_2,M_4\right)&\to \left(e^{i\zeta}M_2,e^{-i\zeta}M_4\right),
\end{aligned}
\ee
where $\mathrm{U}(1)_+$ is gauged, and $\mathrm{U}(1)_-$ is a global symmetry. For $\bfn=0$, the equations exhibit an $\mathrm{SU}(2)$ symmetry acting on $(M_2,M_4)$, with $\mathrm{U}(1)_-$ as a subgroup.

The (virtual) complex dimension $m(c_1,c_2)\in \mathbb{Z}$ of the two-monopole moduli space is given as a special case of Eq. \eqref{eq:Nmondim}. Hence, the moduli space of solutions for \eqref{eq:multimoeqs} modulo gauge transformations has dimension
\be
m(c,c-2\bfn)=\frac{1}{4}\left(2\bfn^2-2B(\bfn,c)+c^2-\chi-2\sigma\right).
\ee 
It is natural to expect that the wall-crossing occurs when the \emph{sum} of the characteristic classes of the two Spin$^c$ structures is orthogonal to $J$, i.e., $B(c,J) + B(c-2\bfn,J) = 0$, which simplifies to
\be\label{eq:NewWalls-3}
B(c-\bfn, J)=0. 
\ee
We note that this is precisely the location of the walls we get from the $U$-plane integral in Eq. \eqref{eq:NewWalls}. We stress again that if $\bfn$ is not even, these are \emph{not} the walls associated with Spin$^c$ structures. 
\footnote{In mathematical treatments, it is common to perturb the first equation describing the SW invariants by a closed self-dual 2-form $\eta$, which shifts the walls by the cohomology class of $\eta$. }

The question now arises regarding the contribution from the LEET of the two massless monopole hypermultiples $M_2$ and $M_4$ coupled to the $\mathrm{U}(1)_+$ vector multiplet. 
The path integral localizes to an integral over the moduli space of solutions to \eqref{eq:multimoeqs}. However, much like the multi-monopole moduli spaces explored in \cite{bryan1996, LoNeSha, Dedushenko:2017tdw, Doan:2019, Aspman:2022sfj, Aspman:2023ate}, this moduli space is noncompact. Consequently, the standard definition of SW invariants becomes ill-defined. To extract meaningful invariants, we move slightly away from $\CR=i$. For small $(\CR - i)$, the monopoles are light, and the integral may be replaced by an equivariant integral with respect to the $\mathrm{U}(1)_-$ global symmetry, with $(\CR-i)$ serving as the equivariant parameter. However, the localization formulae of \cite{LoNeSha, Dedushenko:2017tdw} do not apply directly, as their walls of marginal stability differ from Eq. \eqref{eq:NewWalls-3}.
\footnote{For $\CR^4\neq 1$, Eqs. \eqref{ZSWjmun}, \eqref{ZSWjmun2}, \eqref{ZSWjmun34} and subsequent results in this work can probably be understood in the framework of equivariant localization developed in \cite{LoNeSha, Dedushenko:2017tdw}. Specifically, the monopole masses may be identified with $(\CR-\omega)$, where $\omega$ denotes a fourth-root of unity.}

We conjecture that there is a generalization of the standard SW invariants, 
\be
{\rm SW}_N(c_1, \cdots, c_N;J),
\ee
associated with the multi-monopole equations \eqref{eq:multiMonoEqs2} in Sec.  \ref{App:ShiftedWalls}, exhibiting simple wall-crossing behavior at the walls \eqref{eq:NewWalls-N}. 
We will leave further investigation of this conjecture to future work. 
\footnote{A similar conjecture was made in \cite{Aspman:2023ate}. }
If such invariants exist, then it is natural to conjecture that the 
Higgs branch contribution can be expressed in terms of  ${\rm SW}_2(c_1,c_2;J)$ in a way analogous to the previous cases,
\be 
\sum_{c\in K_X+H^2(X,2\mathbb{Z})} {\rm SW}_2(c,c-2\bfn;J)\,{\rm Res}_{a_2=0}\left[ a_2^{-1-m(c,c-2\bfn)} e^{-S_{E,2}}\right],
\ee 
with $e^{-S_{E,2}}$ given by universal couplings, 
\be 
\label{eq:E2ufunctions}
e^{-S_{E,2}}=\kappa_2 \CA_2^\chi \CB_2^\sigma \CC_2^{\bfn^2} \CD_2^{B(c,\bfn)} \CE_2^{c^2}\CF_{\bfmu,2}.
\ee 
For $\bfn\neq 0$, this extends discussions of the multi-monopole contributions in 4d $\mathrm{SU}(2)$, $N_f=2$ theory in \cite{Kanno:1998qj, Dedushenko:2017tdw, Aspman:2022sfj}. It will be interesting to explore whether the two-monopole invariants ${\rm SW}_2(c_1,c_2;J)$ exist and whether they capture more refined information of four-manifolds than the well-known SW invariants ${\rm SW}(c;J)$. It is intriguing that for $c_1\neq c_2$, these invariants are potentially non-vanishing for manifolds whose standard SW invariants are known to vanish, such as $\mathbb{CP}^2$, and that the walls of marginally stability are not necessarily associated with Spin$^c$ structures.

\section{Supersymmetric Localization}
\label{Sec:SusyLoc}

When $X$ is a smooth toric surface that admits a $T=\IC^* \times \IC^*$ toric action, we equip it with a K\"ahler metric such that the compact subgroup $\mathrm{U}(1) \times \mathrm{U}(1)$ acts isometrically.
\footnote{The existence of such metrics follows from constructing $X$ via a symplectic reduction of $\IC^r$ with its standard K\"ahler metric under a unitary $\mathrm{U}(1)^{r-2}$ action. However, it should be noted that such metrics are not unique, since sufficiently small invariant symmetric tensor deformations produce additional valid metrics. }
This symmetry enables an alternative derivation of the wall-crossing formula through supersymmetric localization of the 5d theory. We deform the 5d theory \eqref{mixed CS action} by putting it in the $\Omega$-background associated with the $\mathrm{U}(1)\times \mathrm{U}(1)$ symmetry \cite{Ne}. The $\Omega$-background is characterized by two equivariant parameters, $\e_1$ and $\e_2$, which were introduced in Refs \cite{Moore:1997dj,Moore:1998et, Lossev:1997bz},  and used in Ref. \cite{Moore:1998et} to regularize integrals over moduli spaces of instantons on $\mathbb{R}^4$.
The action of the theory in the $\Omega$-background can be derived either by extending the formal substitution procedure explained in \cite{Lossev:1997bz, Nekrasov:2010ka}, or via the off-shell supergravity approach as in \cite{Hosseini:2018uzp}.
See Eqs. \eqref{eq:TwistedOmegaAction}, \eqref{eq:TwistedOmegaAction2} in App. \ref{app:SQMinOmega} for explicit formulae for the action. Throughout this section, we restrict ourselves to $\bfn_K=0$. 
\footnote{
The generalization to non-vanishing $\bfn_K$ should be interesting. Some aspects of the problem have been addressed in \cite{Closset:2021lhd}, and the result should be related to \cite{Lockhart:2012vp}.
}

The partition function in the $\Omega$-background of $X\times S^1$, a.k.a. the K-theoretic $T$-equivariant partition function on $X$, is denoted by 
\be\label{eq:Z-p-eps}
Z_{\bfmu, \mathfrak{p}^{(I)}}^{\e_1,\e_2}(\CR),
\ee
which depends on the $\Omega$-deformation parameters $\epsilon_1,\epsilon_2$ associated with the toric action $T$, and a $\chi$-tuple of integers $\mathfrak{p}^{(I)}$ determining the background flux $\bfn_I$ via \eqref{nfrakpIl}. 
The partition function also depends on the metric on $X$ due to an anomaly in the odd symmetries $\bar \CQ_{(i)}$. 
In the $\epsilon_1, \epsilon_2 \to 0$ limit, \eqref{eq:Z-p-eps} reduces to the partition function $Z^J_{\bfmu,\bfn_I}(\CR)$, which depends on the metric only through the period point $J$. 
When $\epsilon_1, \epsilon_2\neq 0$, although our computations involve some important ingredients (such as in \eqref{eq:g-with-nonzero-h}) which depend on more data in the metric on $X$, explicit evaluation reveals that the dependence of the partition function \eqref{eq:Z-p-eps} on the metric on $X$ is still only through the period point $J$. Furthermore, we derive a wall-crossing formula for the $J$-dependence in Eq. \eqref{equivariant wall crossing}. 
On the other hand, the dependence of \eqref{eq:Z-p-eps} on $\epsilon_1, \epsilon_2$ reflects sensitivity to the 5d metric, which we define in Eq. \eqref{eq:metric-with-v} using the metric on $X$ together with the vector field generating the $\mathrm{U}(1) \times \mathrm{U}(1)$ action on $X$. This is not surprising since the theory we are dealing with is not a topological field theory in five dimensions. 

As first proposed by Nekrasov in \cite{nekrasov2006localizing} and later developed in \cite{Gottsche:2006tn, Gottsche:2006bm, Gasparim:2009sns, Bershtein:2015xfa, Bershtein:2016mxz, Bonelli:2020xps}, for $X$ a compact toric surface, the partition function \eqref{eq:Z-p-eps} can be written schematically as
\be
Z_{\bfmu, \mathfrak{p}^{(I)}}^{\e_1,\e_2} (\CR) = \sum_{\text{fluxes}} \int \mathrm{d}a \prod_{\ell} \CZ \left(a^{(\ell)},\epsilon_1^{(\ell)},\epsilon_2^{(\ell)},\Lambda^{(\ell)},R\right),
\ee
where the product is over all fixed loci of the toric action, and $\CZ$ is the (K-theoretic) Nekrasov partition function on $\mathbb{C}^2$. 
The K-theoretic version was used in \cite{Gottsche:2006bm} to derive the wall-crossing formula of K-theoretic Donaldson invariants. See also \cite{Closset:2022vjj} for related developments. 

The novelty of this section is to provide a careful path integral derivation of the wall-crossing formula via toric localization in 5d gauge theory. Our results exactly reproduce the wall-crossing formula obtained via the $U$-plane integral approach in Sec.  \ref{sec:wall-crossing}. In Sec.  \ref{sec:equivariantKtheoreticP2}, we will compute the equivariant K-theoretic partition function for $X=\mathbb{CP}^2$.

Analogous to the non-equivariant case discussed in Sec.  \ref{sec:ThePartitionFunction}, the partition function \eqref{eq:Z-p-eps} admits an interpretation using collective coordinate decomposition detailed in App. \ref{app:SQMinOmega}. The partition function is computed from SQM on the moduli space of instantons, yielding a sum over all instanton sectors similar to \eqref{eq:Zee-QM-1},
\be\label{eq:Zee-QM-Equiv}
Z_{\bfmu, \mathfrak{p}^{(I)}}^{\e_1,\e_2}(\CR) =\sum_{l}  d_{\bfmu,\mathfrak{p}^{(I)} }^{\e_1,\e_2}(l)\,\CR^l,
\ee 
where $l$ is the same as in Eq. \eqref{eq:ModSpaceDim}. In particular, the expansion is a series in nonnegative powers of $\CR$.  

In order to interpret the expansion coefficients
$d_{\bfmu,\mathfrak{p}^{(I)} }^{\e_1,\e_2}(l)$, we observe that the $\mathrm{U}(1)\times \mathrm{U}(1)$ isometry of $X$ induces a $\mathrm{U}(1)\times \mathrm{U}(1)$ action on the moduli space of instantons, because the pullback under the $\mathrm{U}(1)\times \mathrm{U}(1)$ takes an instanton connection to an instanton connection. This induces a $\mathrm{U}(1)\times \mathrm{U}(1)$ isometry of the metric on the moduli space. In the present case the moduli space is K\"ahler, so the isometry is generated by a Hamiltonian vector field with moment map $\mu_{\epsilon_1, \epsilon_2}$ which is linear in $\epsilon_1, \epsilon_2$. We write $\mu_{\e_1, \e_2} = \e_1 \mu_1  + \e_2 \mu_2$ where $\mu_i$ are the moment maps for the two generators. 
The $\Omega$-background is an $X$-bundle over $S^1$ that twists with an action of $\mathrm{U}(1)\times \mathrm{U}(1)$ around the circle. 
This induces a twist of the SQM partition function on the $S^1$ by the $\mathrm{U}(1)\times \mathrm{U}(1)$ action on the instanton  moduli space. From the results in App. \ref{app:SQMinOmega} 
we learn
that the expansion coefficients in $\CR$ around $\CR=0$ are given by the 
$L^2$-character-valued Witten index: 
\be
\label{eq:dChar}
d_{\bfmu,\mathfrak{p}^{(I)} }^{\e_1,\e_2}(l) = {\rm 
Tr}_{{\cal H}_{k,\bfmu}} (-1)^F e^{-R ( H + \epsilon_1 J_1 + \epsilon_2 J_2 )},
\ee
where $l$ and $k$ are again related as in \eqref{eq:ModSpaceDim},
$H= \slashed{D}^2$, and $J_i$ is a linear combination of the charge $\mu_{\epsilon_i}$ with 
the Cartan generator of $R$-symmetry. The dependence on $\mathfrak{p}^{(I)}$ enters in $J_i$ since one must lift the $\mathrm{U}(1)\times \mathrm{U}(1)$ action on the moduli space to an action on 
sections of $S^+ \otimes \CL^{(I)}$.

\subsection{Supersymmetry On Toric K\"ahler Surfaces}

\subsubsection{Topological Twist And Supersymmetry Transformations On Toric K\"ahler surfaces}

In Sec.  \ref{sec:TopTwist}, we describe how to implement the partial topological twist for 5d $\CN=1$ SYM on $X\times S^1$. For $X$ a toric K\"ahler surface, we can further incorporate the $\Omega$-deformation, which replace the original 5d product metric by
\be\label{eq:metric-with-v}
ds^2 = e^1 e^{\bar 1} + e^2 e^{\bar 2} + (\mathrm{d}x^5 + v_\mu dx^\mu)^2 =  e^1 e^{\bar 1} + e^2 e^{\bar 2} + (\hat e^5)^2.
\ee
Here $\{e^a, e^{\bar a},a=1,2\}$ denotes the complex frame on $X$. This metric describes an $X$-bundle over $S^1$ with monodromy given by a $\mathrm{U}(1)\times \mathrm{U}(1)$ action on $X$, diagonalized in the basis $\{e^1, e^2\}$. The associated Killing vector field is
\be 
v = v^m \frac{\partial}{\partial x^m} =v^\mu \frac{\partial}{\partial x^\mu} + \frac{\partial}{\partial x^5},
\ee
where $m$ runs from $1$ to $5$, and $v^\mu$ depends linearly on $\epsilon_{1},\epsilon_{2}$.
Detailed geometric data of $X$ and the precise definition of $\Omega$-deformation with respect to the $\mathrm{U}(1)\times \mathrm{U}(1)$ isometry on $X$ are provided in App. \ref{sec:Toric Geometry}.

We decompose the positive spinor bundle as
\be
S_+ \cong K_X^{\frac12} \oplus  K_X^{-\frac12}.
\ee 
The partial topological twist (roughly) identifies $E_R \cong S_+$, so on a K\"ahler surface the supersymmetry transformation parameters $\bar\xi_A^{\dot\alpha}$ become a section of 
\be
\left(K_X^{\frac12} \oplus  K_X^{-\frac12}\right)^{\otimes 2} \cong K_X \oplus K_X^{-1} \oplus \mathcal{O} \oplus \mathcal{O}.
\ee
We will denote the two scalar supercharges preserved on $X$ by $\bar \CQ_{(1)}$ and $\bar \CQ_{(2)}$. They can be written explicitly as
\be
\bar \CQ_{(1)} = \bar\zeta^{\dot\alpha}_{(1)A} {\bar Q}_{\dot\alpha}^A, \quad \bar \CQ_{(2)} = \bar\zeta^{\dot\alpha}_{(2)A} {\bar Q}_{\dot\alpha}^A,
\ee
where 
\be
\bar\zeta^{\dot\alpha}_{(1)A} = \left(\begin{array}{cc}1 & 0 \\ 0 & 0\end{array}\right),\quad 
\bar\zeta^{\dot\alpha}_{(2)A} = \left(\begin{array}{cc}0 & 0 \\ 0 & 1\end{array}\right).
\ee
The complex structure over $X$ can be constructed via
\be\label{def: complex-structure-I}
I_{\mu\nu} := 2i \bar\zeta^{A}_{(1)} \bar\sigma_{\mu\nu} \bar\zeta_{(2)A}.
\ee
It is convenient to introduce the twisted variables for the $\CN=1$ vector multiplet $V = (A_m,\sigma,\lambda_A, D_{AB})$, where we adopt the conventions in \cite{Closset:2022vjj}
\begin{equation}
    \begin{aligned}
     \bar\lambda^{(0,0)}_{(1)} 
    & :=  
    \bar\zeta_{(1)}^A 
    \bar\lambda_A
    , \\
    \bar\lambda^{(0,0)}_{(2)} 
    & :=  
    \bar\zeta_{(2)}^A 
    \bar\lambda_A,\\
    \bar\lambda^{(0,2)}_{\mu\nu} &:= 
    2 
    \bar\zeta^A_{(2)} 
    \bar\sigma_{\mu\nu} \bar\lambda_A 
    + i 
    \bar\zeta^A_{(2)} 
    \bar\lambda_A
    I_{\mu\nu},
    \\
    \bar\lambda^{(2,0)}_{\mu\nu} &:= 
    2 
    \bar\zeta^A_{(1)} 
    \bar\sigma_{\mu\nu} \bar\lambda_A 
    - i 
    \bar\zeta^A_{(1)} 
    \bar\lambda_A
    I_{\mu\nu}, 
    \\
    \lambda_\mu^{(0,1)}
    &:= 
    \bar\zeta_{(2)A} 
    \bar\sigma_\mu \lambda^A,
    \\
    \lambda_\mu^{(1,0)} 
    &
     :=  
    \bar\zeta_{(1)A} 
    \bar\sigma_\mu \lambda^A
    ,\\
    D^{(0,2)}_{\mu\nu} & :=   
    -2 i 
    \bar\zeta_{(2)}^A 
    \bar\sigma_{\mu\nu} 
    \bar\zeta_{(2)}^B
    D_{AB},\\
    D^{(2,0)}_{\mu\nu} & :=   
    -2 i 
    \bar\zeta_{(1)}^A 
    \bar\sigma_{\mu\nu} 
    \bar\zeta_{(1)}^B
    D_{AB}, \\
    D^{(0,0)} & := 
    \bar\zeta_{(2)}^A 
    D_{AB} 
    \bar\zeta^B_{(1)}
    =
    -\bar\zeta_{(1)}^A 
    D_{AB} 
    \bar\zeta^B_{(2)}
    =D_{12},
    \\
    F^{(0,0)} & := 
    -i
    \bar\zeta_{(2)}^{A} 
    \bar\sigma^{\mu\nu}
    \bar\zeta_{(1)A} 
    F_{\mu\nu}
    = \frac{1}{2}
    I^{\mu\nu} F_{\mu\nu}.
    \end{aligned}
\end{equation}
We find it convenient to decompose the two-form $F$ as
\be
F= F^{(2,0)}+F^{(0,2)}+F^{(1,1)}+F^{(1,0)}\wedge \hat e^5 +F^{(0,1)}\wedge \hat e^5.
\ee
Note particularly that $F^{(0,0)}$ is the contraction of the $(1,1)$-part of the field strength $F$ with the K\"ahler form. 
We also define
\be\label{1d auxiliary}
    \mathcal{D}^{(0,0)} =  D^{(0,0)}+F^{(0,0)},
\ee 
which can be thought of as the auxiliary field for a 1d $\CN=2$ vector multiplet.

Using the above notation, the supersymmetry transformations for $\bar \CQ_{(1)}$ and $\bar \CQ_{(2)}$ are
\begin{equation}
    \begin{aligned}
    \,[\mathcal{\bar Q}_{(1)} , \sigma ] & = -i \bar\lambda_{(1)}^{(0,0)},\\
    [\mathcal{\bar Q}_{(1)} , A ] &= -
    \lambda^{(1,0)}+\bar\lambda_{(1)}^{(0,0)}  \hat e^5 ,\\
    \{\mathcal{\bar Q}_{(1)} , \lambda^{(0,1)}\}
    & = 2 \bar\partial_A \sigma + 2i F^{(0,1)},\\
    \{\mathcal{\bar Q}_{(1)} , \lambda^{(1,0)}\}
    & = 0,\\
    \{ \mathcal{\bar Q}_{(1)}, \bar\lambda^{(0,2)}
    \} 
    & =  -4i F^{(0,2)},\\
    \{ \mathcal{\bar Q}_{(1)}, \bar\lambda^{(2,0)}
    \} 
    & =  -i D^{(2,0)},
    \\
    \{ \mathcal{\bar Q}_{(1)}, \bar\lambda^{(0,0)}_{(1)}
    \} 
    & = 0, 
     \\
    \{ \mathcal{\bar Q}_{(1)}, \bar\lambda^{(0,0)}_{(2)}
    \} 
    & = 
    - v^m D_m \sigma
    -\mathcal{D}^{(0,0)},\\
    [\mathcal{\bar Q}_{(1)}
    , D^{(0,2)}
    ]
    &= 4 \bar\partial_A 
    \lambda^{(0,1)}
    + 
    2\left[\sigma,\bar\lambda^{(0,2)}\right] 
    -2 v^m D_m \bar\lambda^{(0,2)},
    \\
    [\mathcal{\bar Q}_{(1)}
    , D^{(2,0)}
    ]
    &= 0,\\
    [\mathcal{\bar Q}_{(1)}
    , F^{(0,0)}]
    &= -I^{\mu\nu} 
    D_\mu \lambda_\nu^{(1,0)},
    \\
    [\mathcal{\bar Q}_{(1)}
    , D^{(0,0)}]
    & = I^{\mu\nu} 
    D_\mu \lambda_\nu^{(1,0)}
    - i \left[\sigma,\bar\lambda_{(1)}^{(0,0)}\right]
    +i v^m D_m  \bar\lambda_{(1)}^{(0,0)},
    \\
    [\mathcal{\bar Q}_{(1)}
    , \mathcal{D}^{(0,0)}]
    & = 
    -i \left[\sigma,\bar\lambda_{(1)}^{(0,0)}\right]
    +i v^m D_m  \bar\lambda_{(1)}^{(0,0)},
    \end{aligned}
\end{equation}
and
\begin{equation}
    \begin{aligned}
    \,[\mathcal{\bar Q}_{(2)} , \sigma ] & = -i \bar\lambda^{(0,0)}_{(2)},\\
    [\mathcal{\bar Q}_{(2)} , A ] &= -
    \lambda^{(0,1)} + \bar\lambda_{(2)}^{(0,0)}\hat e^5,
    \\
    \{\mathcal{\bar Q}_{(2)} , \lambda^{(0,1)} \} &= 0, \\
    \{\mathcal{\bar Q}_{(2)} , \lambda^{(1,0)} \} &= 
    2\partial_A \sigma + 2 i F^{(1,0)}
    ,\\
    \{\mathcal{\bar Q}_{(2)} , \bar\lambda^{(0,2)} \} &= - i  
    D^{(0,2)},\\
    \{\mathcal{\bar Q}_{(2)} , \bar\lambda^{(2,0)} \} &= -4 i  
    F^{(2,0)},\\
    \{\mathcal{\bar Q}_{(2)} , \bar\lambda^{(0,0)}_{(1)} \} &= -v^m D_m  \sigma 
    +\mathcal{D}^{(0,0)},\\
    \{\mathcal{\bar Q}_{(2)} , \bar\lambda^{(0,0)}_{(2)} \} &= 0,
    \\
    [\mathcal{\bar Q}_{(2)} , D^{(0,2)} ] &  = 0, 
    \\
    [\mathcal{\bar Q}_{(2)} , D^{(2,0)} ] &  = 
    4 \partial_A 
    \lambda^{(1,0)}+ 
    2\left[\sigma,\bar\lambda^{(2,0)}\right] -2 v^m D_m  \bar\lambda^{(2,0)}, 
    \\
    [\mathcal{\bar Q}_{(2)}
    , F^{(0,0)}]
    &= -I^{\mu\nu} 
    D_\mu \lambda_\nu^{(0,1)},
    \\
    [\mathcal{\bar Q}_{(2)}
    , D^{(0,0)}]
    & = I^{\mu\nu} 
    D_\mu \lambda_\nu^{(0,1)}
    +i\left[\sigma,\bar\lambda^{(0,0)}_{(2)}\right]
    -i v^m D_m  \bar\lambda^{(0,0)}_{(2)}, \\
    [\mathcal{\bar Q}_{(2)}
    , \mathcal{D}^{(0,0)}]
    & = 
    i\left[\sigma,\bar\lambda^{(0,0)}_{(2)}\right]
    -i v^m D_m  \bar\lambda^{(0,0)}_{(2)}.
\end{aligned}
\end{equation}

\subsubsection{BPS Configurations}\label{subsubsec:BPS configurations}

The twisted action on a K\"ahler surface can be written as a $\bar \CQ$-exact term plus a topological term, as in \eqref{eq:ParialTwistedAction}. In the notation above,
\begin{equation}\label{eq:5d-Action-Toric-With-Epsilon}
\begin{aligned}
S_\text{5d}=&\frac{1}{g^2_{5d}} \int \mathrm{d}^5 x \, \sqrt{g_{5}} \bar{\mathcal{Q}}_{(2)}\bar{\mathcal{Q}}_{(1)}\text{tr}\left[\frac{1} {32}\epsilon^{\mu\nu\rho\sigma}\bar\lambda_{\mu\nu}^{(2,0)}\bar\lambda_{\rho\sigma}^{(0,2)} -\frac{i}{4}\left(\sigma-iv^\mu A_\mu\right)\left(\mathcal{D}^{(0,0)}-2\mathcal{F}^{(0,0)}\right) \right]\\
&-\frac{1}{2g^2_{5d}}\int \text{tr}( F_4\wedge F_4)\wedge \hat e^5,
\end{aligned}
\end{equation}
whose bosonic part in the $\bar\CQ$-exact terms reads
\begin{equation}
\begin{aligned}
S_\text{5d}|_\text{bos}= \frac{1}{g^2_{5d}} \int \sqrt{g_{5}}\mathrm{d}^5x\, \text{tr}\bigg( &\frac{1}{2}\epsilon^{\mu\nu\rho\sigma} F^{(2,0)}_{\mu\nu} F^{(0,2)}_{\rho\sigma} +\frac{1}{2} \left(F^{(0,0)}\right)^2 -\frac{1}{32}\epsilon^{\mu\nu\rho\sigma}D^{(2,0)}_{\mu\nu}D^{(0,2)}_{\rho\sigma}\\
&-\frac{1}{2} \left(D^{(0,0)}\right)^2 + \frac{1}{2}D_m\sigma D^m\sigma +\frac{1}{2}v^m v_n F_{l m}F^{l n} \bigg).
\end{aligned}
\end{equation}
Here
\begin{align}
\frac{1}{2}\epsilon^{\mu\nu\rho\sigma}F^{(2,0)}_{\mu\nu}F^{(0,2)}_{\rho\sigma} 
& = (F_{13}-F_{24})^2 + (F_{14}+F_{23})^2, \\ 
-\frac{1}{32}\epsilon^{\mu\nu\rho\sigma}D^{(2,0)}_{\mu\nu}D^{(0,2)}_{\rho\sigma} & = \frac{1}{2}D_{11}D_{22},
\end{align}
and $v$ is defined in Eq. \eqref{eq:metric-with-v} and is linear in $\epsilon_1,\epsilon_2$. \footnote{The precise choice of the action \eqref{eq:5d-Action-Toric-With-Epsilon} differs from that in some references, e.g., \cite{Bonelli:2020xps}, which may be related to the puzzle that we discuss in Sec.  \ref{sec: puzzle}.} 

Imposing reality conditions on the physical fields (along with pure imaginary auxiliary fields) and scaling the 
exact terms, the 5d path integral localizes to the BPS configurations:
\bea\label{eq:BPS-eq-v1}
D^{(0,2)} = D^{(2,0)} = F^{(2,0)} = F^{(0,2)} = 0,
\eea
and
\be\label{eq:BPS-eq-v2}
D_m\sigma =0,\quad F^{(0,0)}=D^{(0,0)}=0,\quad \iota_v F=0. 
\ee

We now analyze some consequences of these equations. 
The equations $F^{(2,0)} = F^{(0,2)} = F^{(0,0)} = 0$ imply that 
\be 
F_4 := F^{(2,0)} + F^{(0,2)} + F^{(1,1)} = \frac12 F_{\mu\nu} dx^\mu dx^\nu
\ee
is ASD. The equation $\iota_v F=0$, when expressed in components, yields the following two conditions
\be
v^\mu F_{\mu 5} = 0, \quad  F_{5\mu } + v^\nu F_{\nu \mu} = 0. 
\ee
For simplicity we now focus on $G=\mathrm{SU}(2)$. If $\sigma \neq 0$, 
the first equation of \eqref{eq:BPS-eq-v2} implies that the gauge bundle splits,
\be
\label{eq:Fdiagnal}
F_{\mathrm{SU}(2)} = \left(\begin{array}{cc} \frac{1}{2}F_{\mathrm{U}(1)} & 0 \\ 0 & - \frac{1}{2} F_{\mathrm{U}(1)} \end{array}\right),\quad (A_5)_{\mathrm{SU}(2)}=
\left(\begin{array}{cc} 
(A_5)_{\mathrm{U}(1)} & 0 \\ 0 & -  (A_5)_{\mathrm{U}(1)} \end{array}\right),
\ee
and $\sigma$ is constant and diagonal,
\be
\label{eq:asigmadiagnal}
\sigma_{\mathrm{SU}(2)} = \left(\begin{array}{cc} \sigma_{\mathrm{U}(1)} & 0 \\ 0 & - \sigma_{\mathrm{U}(1)} \end{array}\right),
\ee
where $\sigma_{\mathrm{U}(1)}$ is a constant. Note that equations $F_{\mathrm{U}(1)}^{(2,0)} = F^{(0,2)}_{\mathrm{U}(1)} = F^{(0,0)}_{\mathrm{U}(1)} = 0$ do not have solutions for generic choice of metric. Relaxing the condition that the auxiliary fields $D^{(0,0)}$ is pure imaginary, the condition $F^{(0,0)}_{\mathrm{U}(1)}=D^{(0,0)}_{\mathrm{U}(1)}=0$ is replaced by
\be\label{eq:DOO-auxfield}
\CD^{(0,0)}_{\mathrm{U}(1)} =   D^{(0,0)}_{\mathrm{U}(1)}+F^{(0,0)}_{\mathrm{U}(1)} = 0,
\ee
which allows non-vanishing $F^{(0,0)}$.
In the following, all the bosonic fields are understood to be the $\mathrm{U}(1)$ fields and we will omit the subscript $\mathrm{U}(1)$. 

In order to proceed we will make the assumption that the solutions to the BPS equations involve only connections which are pulled back from $X$. 
In this case, the BPS locus has infinitely many connected components labeled by a $\chi$-tuple of integers $\mathfrak{p} \in \IZ^\chi$. 

Let us first consider the equation $\iota_v F =0$. If $A_\mu dx^\mu$ is pulled back from a connection on $X$, then we can choose a gauge such that $\partial_5 A_\mu=0$, and hence 
\be\label{eq:iota-v-F4}
\iota_v F_4 = 2\mathrm{d}A_5, 
\ee
where the factor of $2$ comes from  different conventions in the diagonalizations given in Eqs. \eqref{eq:Fdiagnal} and \eqref{eq:asigmadiagnal}. 
The remaining component of the equation $\iota_v F=0$, namely, $v^\mu \partial_\mu A_5 =0$, follows directly from \eqref{eq:iota-v-F4}.
We define an equivariant extension of $F_4$,
\be\label{eq:Feqdyn}
F_{\text{eq}}:=  F_4+2A_5.
\ee
Then $\left[F_{\text{eq}}/(2\pi)\right]$ is an equivariant cohomology class in the  equivariant cohomology group  $H^2_{\mathrm{U}(1)\times \mathrm{U}(1)}(X)$, with $A_5$ playing the role of the zero-form component of $F_\text{eq}$.
Let $\{D_s^{\text{eq}}, s = 1, \cdots, \chi\}$ be a basis for this equivariant cohomology group. Then we can expand 
\be\label{eq:Def-mathfrakpell}
 \left[\frac{F_\text{eq}}{2\pi}\right]  = \sum_{s=1}^{\chi} {\frak p}_s D_s^{\text{eq}}, \quad \mathfrak{p}_s\in \mathbb{Z}
\ee
which defines a $\chi$-tuple of ``equivariant $\mathrm{U}(1)$ fluxes'' 
\be
\mathfrak{p} = (\mathfrak{p}_1,\cdots, \mathfrak{p}_{\chi}).
\ee
There is a canonical map $H^2_{\mathrm{U}(1)\times\mathrm{U}(1)}(X) \to H^2(X)$ that sends $D^{\rm eq}_s \to D_{s}$. Hence, we have 
\be\label{eq:Delleq-to-Dell}
D_s^\text{eq} = D_s + \nu_s,
\ee
where $\nu_s$ is a linear combination of $\epsilon_1$ and $\epsilon_2$. 
Taking the non-equivariant limit of \eqref{eq:Def-mathfrakpell} yields  
\be
\label{kfrakpl}
\bfk  = \left[\frac{F_4}{4\pi}\right] = \frac{1}{2} \sum_{s=1}^{\chi} {\frak p}_s D_s  \in L+\bfmu.
\ee
It is important to note that, for a given $\bfk$, $\mathfrak{p}$ satisfying \eqref{kfrakpl} is not uniquely determined due to the relations among the divisors given in Eq. \eqref{relation-divisor}. Two $\chi$-tuples of equivariant fluxes $\mathfrak{p}$ and $\mathfrak{p}'$ are equivalent if the following condition is satisfied for all $s = 1, \cdots, \chi$, 
\begin{equation}
    \mathfrak{p}_s - \mathfrak{p}'_s = \sum_{i=1,2} m_s n^i_s, \quad m_1, m_2 \in \mathbb{Z},
\end{equation}
where $n_s^i$ are combinatorial coefficients defining the toric geometry. For further details, see Eq. \eqref{equivalence relation for p}  and the surrounding discussion in App. \ref{sec:Toric Geometry}.

In the localization computation, we need the values of $\nu_s$ at the $\mathrm{U}(1)\times \mathrm{U}(1)$ fixed points.
As explained in App. \ref{sec:Toric Geometry}, at the $\ell$-th fixed point, exactly two divisors $D_{\ell}$ and $D_{\ell+1}$ intersect, and the explicit values of $\nu_s$ is given by
\be\label{eq:nu-at-fixedpoints}
\nu_s^{(\ell)}=-\frac{i}{2\pi}\left(\epsilon_1^{(\ell)} \delta_{s, \ell}+\epsilon_2^{(\ell)} \delta_{s-1, \ell}\right),
\ee 
where $\epsilon^{(\ell)}_{1}, \epsilon^{(\ell)}_{2}$ are defined in Eq. \eqref{def: epsilonl}.
Evaluating both sides of \eqref{eq:Def-mathfrakpell} at the $\ell$-th fixed point, we obtain the value of $A_5$ at the $\ell$-th fixed point,
\be\label{eq:A5-fp-value}
A_5^{(\ell)}=-\frac{i}{2}\left(\epsilon_1^{(\ell)} \mathfrak{p}_{\ell}+\epsilon_2^{(\ell)} \mathfrak{p}_{\ell+1}\right) + a_5,
\ee
where $a_5$ is a globally defined real constant which is not fixed by Eq. \eqref{eq:iota-v-F4}.
The classical Coulomb parameter evaluated at the $\ell$-th fixed locus reads
\be \label{eq: a-shift}
a^{(\ell)}=a+\frac12 \left(\epsilon_1^{(\ell)} \mathfrak{p}_{\ell}+\epsilon_2^{(\ell)} \mathfrak{p}_{\ell+1}\right).
\ee
Here $a=\sigma + ia_5$ is a globally defined constant with periodicity $a\sim a + 2\pi i/R$, which serves as the equivariant parameter for the $\mathrm{U}(1)$ gauge action on the moduli space. An important property is that the difference $a^{(\ell)} - a^{(\ell')}$ for any pair $\ell,\ell'$ depends only on the equivalence class $[\mathfrak p]$, as defined in \eqref{equivalence relation for p}. A proof of this statement is given in App. \ref{sec:Toric Geometry}.

We also require the background vector multiplet $V^{(I)}$ to satisfy BPS equations in order to preserve supersymmetry. Then we can likewise introduce a $\chi$-tuples of background equivariant $\mathrm{U}(1)^{(I)}$-fluxes 
\be
\mathfrak{p}^{(I)} = \left(\mathfrak{p}^{(I)}_1,\cdots, \mathfrak{p}^{(I)}_{\chi}\right),
\ee
where $\mathfrak{p}^{(I)}_s$ are defined by
\be\label{eq:nfrakp}
\left[ \frac{F_\text{eq}^{(I)}}{2\pi} \right]
= \left[\frac{F_4^{(I)}+A_5^{(I)}}{2\pi}\right] =  \sum_{s=1}^{\chi} {\mathfrak p}^{(I)}_s D_s^\text{eq}.
\ee
Note the difference in the relative factors between definitions \eqref{eq:Feqdyn} and \eqref{eq:nfrakp}. 
The non-equivariant limit of this cohomology class is 
\be
\label{nfrakpIl}
\bfn_I= \left[\frac{F_4^{(I)}}{2\pi}\right] = \sum_{s=1}^{\chi} {\mathfrak p}^{(I)}_s D_s.
\ee
The value of $A_5^{(I)}$ at the $\ell$-th fixed point is
\begin{equation}
   (A_5^{(I)})^{(\ell)} =  -i\left(\epsilon_1^{(\ell)} \mathfrak{p}^{(I)}_{\ell} +
    \epsilon_2^{(\ell)} \mathfrak{p}^{(I)}_{\ell+1}\right) + a_5^{(I)},
\end{equation}
where ${\mathfrak p}_{\ell}^{(I)}$ are the values of the background fluxes localized at the $\ell$-th fixed locus, as in the Eq. \eqref{nfrakpIl} for $\bfn_I$ above. 
Thus, the background parameter shifts as
\begin{equation}
    (a^{(I)})^{(\ell)} = a^{(I)} +  \epsilon_1^{(\ell)} \mathfrak{p}^{(I)}_{\ell} +  \epsilon_2^{(\ell)} \mathfrak{p}^{(I)}_{\ell+1}.
\end{equation}
From \eqref{mCR}, this is equivalent to a shift in $\Lambda$ (also see \eqref{eq:ZnI})
\begin{equation}
    \Lambda^{(\ell)} = e^{-\frac{1}{4}n_I^{(\ell)}}\Lambda,
\end{equation}
where 
\be
\label{eq:nIl}
n_I^{(\ell)} = - R\left(\epsilon_1^{(\ell)} {\mathfrak p}_{\ell}^{(I)} + \epsilon_2^{(\ell)} {\mathfrak p}_{\ell+1}^{(I)}\right).
\ee

For a fixed configuration of fluxes, the path integral reduces via localization to a finite-dimensional integral over the zero modes $(a,\bar a, \lambda_{(1)}^{(0,0)}, \lambda_{(2)}^{(0,0)})$. To study the wall-crossing formula, it is useful to turn on a non-BPS zero mode for the auxiliary field, 
\be \label{def: non-bps-h}
\CD^{(0,0)} = 2ih,
\ee
where $h$ is a constant. The integration contour for $h$ will be specified below. After introducing $h$, the Coulomb branch zero modes, $V_0 =(a,\bar a, \lambda_0,\bar\lambda_0, h)$, retain the residual supersymmetry,
\be
\label{residual susy}
\begin{aligned}
[\bar\CQ_{(1)},a] &= 0, & [\bar\CQ_{(1)},\bar a] &= 2\lambda_0, & 
\{\bar\CQ_{(1)},\lambda_0\} &=0, &  
\{\bar\CQ_{(1)},\bar\lambda_0\} &=-4h,  &
[\bar \CQ_{(1)}, h]&=0,\\
[\bar\CQ_{(2)},a] &= 0, & [\bar\CQ_{(2)},\bar a] &= 2\bar\lambda_0, & 
\{\bar\CQ_{(2)},\lambda_0\} &=4h, & 
\{\bar\CQ_{(2)},\bar\lambda_0\} &=0, &
[\bar \CQ_{(2)}, h]&=0.
\end{aligned}
\ee
These supersymmetry transformation laws can be derived from the following off-shell variations
\bea
\delta A_5 &= i\bar\epsilon_{(1)} \bar\lambda_{(1)}^{(0,0)} + i\bar\epsilon_{(2)} \bar\lambda_{(2)}^{(0,0)},\\
\delta\sigma & = \bar\epsilon_{(1)} \bar\lambda_{(1)}^{(0,0)} + \bar\epsilon_{(2)} \bar\lambda_{(2)}^{(0,0)},\\
\delta \bar\lambda_{(1)}^{(0,0)} &= -i\bar\epsilon_{(2)} \left(D_5\sigma - \CD^{(0,0)}\right),\\ 
\delta \bar\lambda_{(2)}^{(0,0)} &= -i\bar\epsilon_{(1)} \left(D_5\sigma + \CD^{(0,0)}\right),\\ 
\delta \CD^{(0,0)} &= -\bar\epsilon_{(1)} \left[D_5-\sigma,\bar\lambda_{(1)}^{(0,0)}\right] + \bar\epsilon_{(2)} \left[D_5-\sigma,\bar\lambda_{(2)}^{(0,0)}\right], 
\eea
where $\delta = i \bar\epsilon_{(1)} \bar \CQ_{(1)} + i\bar \epsilon_{(2)} \bar \CQ_{(2)}$. Note that these are the transformations of a 1d $\CN=2$ vector multiplet, but we are \emph{not} passing to the SQM collective coordinates at this stage.

\subsection{Localization} \label{sec: localization}

\subsubsection{Reduction Of The Partition Function To A Finite-Dimensional Integral}

The localization of the path integral to the pseudo-BPS locus (referred as ``pseudo'' when $h\not=0$), as described in the previous subsection, allows us to write the partition function as a sum of finite-dimensional integrals over the zero modes. The resulting expression --- analogues of which have proven to be extremely useful in related contexts --- is of the form: 
\be\label{partition ftn zero}
Z_{\bfmu, \mathfrak{p}^{(I)}}^{\e_1,\e_2}(\CR) = K_{\Omega}
\sum_{\boldsymbol{k}\in L + \boldsymbol{\mu}} \int_{\Gamma} \mathrm{d}h \int_{\frak M} \mathrm{d}a \mathrm{d}\bar a \int \mathrm{d}\lambda_0 \mathrm{d}\bar\lambda_0 \, g_{\mathfrak{p},\mathfrak{p}^{(I)}}^{\e_1,\e_2} (a,\bar a, \lambda_0, \bar\lambda_0,h),
\ee 
We begin by defining the various elements in this expression. The overall constant $K_{\Omega}$ will be determined by matching with the wall-crossing formula. The result turns out to be
\be 
K_{\Omega} = -\frac{R}{4\pi^2}.
\ee
The contour $\Gamma$ for the $h$-integral is an important input into our computation and will be discussed in detail in Sec.  \ref{subsubsec:g-function-singularities-contours}. The domain $\frak M$ for the integral over the Coulomb branch with measure $\mathrm{d}a \mathrm{d}\bar a$ is   $\mathfrak{t}\otimes\IC/\Lambda$, where $\Lambda$ is the co-character lattice, i.e. the lattice of elements $x\in \mathfrak{t}$ such that $\exp(2\pi x) = 1$. 
In the rank-one case studied in the present paper, this corresponds to integrating over a cylinder. 

In order to define the integrand $g_{\mathfrak{p},\mathfrak{p}^{(I)}}^{\e_1,\e_2} (a,\bar a, \lambda_0, \bar\lambda_0,h)$,  
let us consider the decomposition of the fields $V=V_0 + g_{5d}V'$, with $V'$ the quantum fluctuation orthogonal to the zero mode $V_0$. For the gauge fields on $X$, we have
\be
A_\mu = A_\mu^{\mathfrak t} +g_{5d}A_\mu'.
\ee
Here $A^{\mathfrak{t}} = A_{\mathrm{U}(1)}\,\sigma_3/2$, where $A_{\mathrm{U}(1)}$ is a connection with the curvature in the cohomology class \eqref{kfrakpl}, and $A'$ denotes the fluctuation. Similarly, we decompose
\be
\sigma = \sigma^{\mathfrak t} + g_{5d}\sigma',\quad A_5 = A_5^{\mathfrak t} + g_{5d}A_5'
\ee
where $\sigma^{\mathfrak t} = \sigma_{\mathrm{U}(1)}\,\sigma_3$ is a constant, and $A_5^{\mathfrak t} = (A_5)_{\mathrm{U}(1)}\,\sigma_3$ has a nontrivial profile over $X$ whose value at the fixed points are given by \eqref{eq:A5-fp-value}. For the auxiliary field $\CD^{(0,0)}$, we set
\be
\CD^{(0,0)} = 2ih + g_{5d}\CD^{'(0,0)}
\ee
Then the integrand is the path integral over all the nonzero modes $V'$ at fixed equivariant flux:
\be\label{eq:g-full-defined}
g_{\mathfrak{p},\mathfrak{p}^{(I)}}^{\e_1,\e_2} (a,\bar a, \lambda_0, \bar\lambda_0,h) := \int [dV'] e^{-S_{5d}[V_0+ g_{5d} V']}.
\ee
In principle, a proper definition of $g_{\mathfrak{p},\mathfrak{p}^{(I)}}^{\e_1,\e_2}$ should be given using a UV-complete theory, while 5d SYM is certainly not. Here, we are working with the LEET valid beyond the length scale set by the coupling $g^2_{5d}$. We expect this to suffice for the computation of topological quantities in the partially twisted theory. 

While the function $g_{\mathfrak{p},\mathfrak{p}^{(I)}}^{\e_1,\e_2} (a,\bar a, \lambda_0, \bar\lambda_0,h)$ depends on the entire $\chi$-tuple of integers $\mathfrak{p}$, we will argue below that, after performing the zero-mode integrals, each summand depends only on the equivalence $[\mathfrak{p}]$, also denoted $\bfk$. Therefore, we claim that the summation over the fluxes must be over the $(\chi-2)$-tuple of integers $\boldsymbol{k}$, which are in one-to-one correspondence with these equivalence classes.

Precisely at the BPS locus, $h=0,\lambda_0=\bar\lambda_0=0$, the integrand is expected to reduce to a meromorphic function in $a$,
\be\label{eq:g-fun-specialized-h=0}
g_{\mathfrak{p},\mathfrak{p}^{(I)}}^{\e_1,\e_2} (a,\bar a, 0, 0, 0) = g_{\mathfrak{p},\mathfrak{p}^{(I)}}^{\e_1,\e_2} (a).
\ee
Moreover, the function on the right-hand side of \eqref{eq:g-fun-specialized-h=0} is independent of the metric on $X$ used to construct \eqref{eq:metric-with-v}.  
It can be computed via equivariant localization with respect to the equivariant group $\BT=\mathrm{GL}(1,\mathbb{C})\times T$ which is the product of the maximal torus of the complexified gauge group and the toric action $T$. The path integral then localizes to the fixed points $M^{\BT}_{\bfmu,k} \subset M_{\bfmu,k}$, labeled by $2\chi$-tuples of Young diagrams,
\be
M^{\BT}_{\bfmu,k} \quad \longleftrightarrow \{Y_{a}^{(\ell)},\ell=1,\cdots, \chi,a=1,2\}, \quad  \sum_{\ell,a}\left|Y_{a}^{(\ell)}\right| = d,
\ee
where $d$ is related to the instanton charge $k$ by \footnote{See App. \ref{app:Loc Hilbert Schemes} for this relation.}
\be
k=c_2 - \frac14 c_1^2 = d-\bfk^2.
\ee
As derived in App. \ref{app:Loc Hilbert Schemes}, the meromorphic function $g_{\mathfrak{p},\mathfrak{p}^{(I)}}^{\e_1,\e_2}(a)$ can be written as a product over the fixed points,
\be\label{ga}
g_{\mathfrak{p},\mathfrak{p}^{(I)}}^{\e_1,\e_2} (a) = \CR^{-1} \exp\left(- R\sum_{\ell=1}^{\chi}  \frac{\left(\epsilon_1^{(\ell)}+\epsilon_2^{(\ell)}\right)^3}{48\epsilon_1^{(\ell)}\epsilon_2^{(\ell)}} \right)\prod_{\ell=1}^{\chi} 
Z(a^{(\ell)}, \epsilon_1^{(\ell)},\epsilon_2^{(\ell)}, n_I^{(\ell)},\Lambda,R). 
\ee
Here $Z$ represents the K-theoretic Nekrasov partition function \cite{Ne, Nekrasov:2003rj}, incorporating the additional coupling to the background $\mathrm{U}(1)^{(I)}$ flux \cite{Gottsche:2006bm}. As demonstrated below, in the non-equivariant limit (\ref{nekrasov conjecture holo}), the integrand becomes a function only of the linear combinations $\bfk$ and $\bfn$ derived from $\mathfrak{p}$ and $\mathfrak{p}^{(I)}$. 
Expressions closely related to \eqref{ga} have previously appeared in \cite{Nakajima:2003pg, Nakajima:2003uh, nekrasov2006localizing, Gottsche:2006bm, Bershtein:2015xfa, Bershtein:2016mxz, Hosseini:2018uzp}.

\subsubsection{Singularity Structure Of The Integrand And Regularization Of The Integral}\label{subsubsec:g-function-singularities-contours}

The definition of the expression \eqref{partition ftn zero} involves several subtleties.  First, the precise form of the integrand $g_{\mathfrak{p},\mathfrak{p}^{(I)}}^{\e_1,\e_2} (a,\bar a, \lambda_0, \bar\lambda_0,h)$ as a function of the zero-mode multiplet $V_0=(a,\bar a, \lambda_0,\bar\lambda_0,h)$ is challenging to compute, as it receives contributions from the non-BPS modes. Even when the integrand can be determined explicitly, as in certain lower-dimensional analogs, the definition of the integral still requires  careful treatment. In this section, we conjecture the existence of a well-defined zero-mode integral \eqref{partition ftn zero} and comment on the choice of contours, motivated by computations of indices in lower-dimensional gauge theories \cite{Hori:2014tda,Benini:2013nda,Benini:2015noa,Closset:2015rna}.   

Before addressing the bosonic integral, we integrate out the fermionic zero modes using supersymmetric Ward identities for $g_{\mathfrak{p},\mathfrak{p}^{(I)}}^{\e_1,\e_2} (a,\bar a, \lambda_0, \bar\lambda_0,h)$ as follows. 
The residual supersymmetry algebra \eqref{residual susy} implies that
\be
\begin{aligned}
0 &= \bar\CQ_{(1)}g_{\mathfrak{p},\mathfrak{p}^{(I)}}^{\e_1,\e_2} = \left( 2\lambda_0 \frac{\partial }{\partial\bar a} -4h \frac{\partial }{\partial \bar\lambda_0} \right)g_{\mathfrak{p},\mathfrak{p}^{(I)}}^{\e_1,\e_2},\\
0 &= \bar\CQ_{(2)}g_{\mathfrak{p},\mathfrak{p}^{(I)}}^{\e_1,\e_2} = \left(2\bar\lambda_0 \frac{\partial }{\partial \bar a} +4h \frac{\partial }{\partial \lambda_0} \right)g_{\mathfrak{p},\mathfrak{p}^{(I)}}^{\e_1,\e_2}.
\end{aligned}
\ee
It follows that
\be
2h \frac{\partial}{\partial\lambda_0} \frac{\partial}{\partial\bar\lambda_0} g_{\mathfrak{p},\mathfrak{p}^{(I)}}^{\e_1,\e_2} = \frac{\partial}{\partial \bar a} g_{\mathfrak{p},\mathfrak{p}^{(I)}}^{\e_1,\e_2}.
\ee
Thus, we can integrate out the fermionic zero modes and rewrite \eqref{partition ftn zero} (still only formally defined) as  
\be\label{total D loc}
Z_{\bfmu, \mathfrak{p}^{(I)}}^{\e_1,\e_2}(\CR) = \lim_{ g_{5d}\rightarrow 0}
\frac{K_{\Omega}}{2}
\sum_{[\mathfrak{p}]}    \int_\Gamma  \frac{\mathrm{d}h}{h}\int_{\frak M } \mathrm{d}a \mathrm{d}\bar a \, \partial_{\bar a} g_{\mathfrak{p}, \mathfrak{p}^{(I)}}^{\e_1,\e_2} 
(a,\bar a, 0, 0,h).
\ee 
 
Now we wish to integrate by parts with respect to $\bar a$. As we will see in Sec.  \ref{sec: instanton contribution}, the function $g_{\mathfrak{p},\mathfrak{p}^{(I)}}^{\e_1,\e_2} (a)$ has potential singularities at
\be\label{a cylinder sing}
\alpha a^{(\ell)} +  \e_1^{(\ell)} m + \e_2^{(\ell)} n=0 \mod \frac{2\pi i}{R}\mathbb{Z},
\ee
for $\alpha=\pm 2$, $m,n\in\mathbb{Z}$, and toric patch labels $\ell=1,\cdots, \chi$.  
These loci correspond to points where the weight of the $\BT$-action vanishes, leading to new zero modes in the path integral. The specific set of integer pairs $(m,n)$ for which singularities occur depends on the toric geometry of the manifold $X$. See App. \ref{poles-and-zeros} for further discussion. 

We expect that the path integral in a given chamber will eventually localize to the genuine BPS locus $h=0$, with potential singularity at \eqref{a cylinder sing}. Let $\Delta_\delta$ be the union of small disks of radius $\delta$ around these singularities. (Note that we do not cut out disks around the singularities of the non-holomorphic integrand \eqref{eq:g-full-defined}). We decompose the integration domain as
\be
 \frak M = \frak M\backslash \Delta_\delta \, \cup \, \Delta_\delta,
\ee
where $\Delta_\delta$ can be further decomposed into its connected components, 
\be
\Delta_\delta = \bigcup_{(n,m,\ell,\alpha)} \Delta_\delta^{(m,n,\ell,\alpha)},
\ee
with $\alpha=\pm 2$ as in \eqref{a cylinder sing}. 
Suppose the existence of a limit $g^2_{5d}\rightarrow 0$ together with $\delta\rightarrow 0$ such that the contribution from $\Delta_\delta$ can be ignored, as justified in lower-dimensional examples \cite{Benini:2013nda,Hori:2014tda,Benini:2015noa}.
Then, integration by parts gives
\be\label{zero mode contour}
Z_{\bfmu, \mathfrak{p}^{(I)}}^{\e_1,\e_2}(\CR) =\lim_{\substack{\delta\rightarrow 0\\ g_{5d}\rightarrow 0}}\frac{K_{\Omega}}{2}\sum_{[\mathfrak{p}]}    \int_\Gamma  \frac{\mathrm{d}h}{h}\int_{\partial\left(\frak M\backslash \Delta_\delta\right)} \mathrm{d}a\, g_{\mathfrak{p}, \mathfrak{p}^{(I)}}^{\e_1,\e_2} 
(a,\bar a, 0, 0,h).
\ee 
The new integration domain is a union of contours around the bulk and the asymptotic regions of the cylinder,
\be
\partial\left(\frak M\backslash \Delta_\delta\right) = \partial\frak M - \partial \Delta_\delta.
\ee
The integral over the boundary of the cylinder, $\partial\frak M$, will reproduce the 
weak-coupling contribution to wall-crossing from the $U$-plane integral, as shown in Sec. 
\ref{sec:loc wall-crossing} and Sec.  \ref{sec: loc to U} below. Here we comment on the contributions from the integrals around $\partial \Delta_\delta$. 

Defining the $h$-contour requires understanding the analytic structure of the non-holomorphic integrand $g_{\mathfrak{p}, \mathfrak{p}^{(I)}}^{\e_1,\e_2} (a,\bar a, 0, 0,h)$ for nonzero $h$. Since this structure is not protected by supersymmetry, explicit analysis is difficult. However, we conjecture the existence of a well-defined contour $\Gamma$ with the following properties:
\begin{enumerate}
 \item To ensure convergence, $\Gamma$ asymptotes to the real line.
 \item $\Gamma$ avoids all singularities of $g_{\mathfrak{p}, \mathfrak{p}^{(I)}}^{\e_1,\e_2}(a,\bar a, 0, 0,h)$.
\item It avoids the distinguished pole at $h=0$ from above or below.
 \end{enumerate}
In this work, we do not attempt a further analysis of the bulk contour integral. In particular, the precise definition of the limits $\delta\rightarrow 0$ and $g^2_{5d}\rightarrow 0$, as well as the contour integrals in the $h$-plane and the $a$-cylinder, remains unclear and are left for future investigation.

Based on the expectation that the zero-mode integral further localizes to the origin in the $h$-plane, it is natural to conjecture that the final result can be written as a sum of residues of the holomorphic function $g_{\mathfrak{p}, \mathfrak{p}^{(I)}}^{\e_1,\e_2}(a)$, evaluated at specific poles determined by the analytic structure of the non-holomorphic integrand. 

As we show around \eqref{eq: residue-nekrasov} in App. \ref{app:NekPartProps}, the residue of the meromorphic function $g_{\mathfrak{p}, \mathfrak{p}^{(I)}}^{\e_1,\e_2} 
(a)$ at a pole given by \eqref{a cylinder sing} depends only on the equivalence class of the fluxes $[\mathfrak{p}]$, rather than the entire $\chi$-tuple of the integers $\mathfrak{p}$. This implies that the summation over fluxes in \eqref{total D loc} must be over all inequivalent classes $[\mathfrak{p}]$. We will return to this claim in Sec.  \ref{sec:loc wall-crossing}, where we provide further justification by requiring consistency of the wall-crossing formula.

Furthermore, the function $g_{\mathfrak{p},\mathfrak{p}^{(I)}}^{\e_1,\e_2} (a,\bar a, \lambda_0, \bar\lambda_0,h)$ has the important property that its leading term in the $\e_1, \e_2 \to 0$ limit is finite. See Eq. \eqref{nekrasov conjecture holo} below. In fact, this leading term depends only on the $\chi$-tuples of integers $\mathfrak{p}$ and $\mathfrak{p}^{(I)}$ through the combinations $\boldsymbol{k}$  \eqref{kfrakpl}  and $\bfn_I$ \eqref{nfrakpIl}. While $g_{\mathfrak{p},\mathfrak{p}^{(I)}}^{\e_1,\e_2} (a,\bar a, \lambda_0, \bar\lambda_0,h)$ is metric-dependent, its non-equivariant limit at $\lambda_0=\bar\lambda_0=h=0$ is topological. 

In summary:
\begin{itemize}
\item The path integral can be expressed as a sum of contributions from the bulk and asymptotic regions of the classical Coulomb branch parametrized by $a$. 
The precise definition of the bulk contribution as an integral over the bosonic zero mode remains poorly understood, because of our ignorance of the singularity structure of $g_{\mathfrak{p}, \mathfrak{p}^{(I)}}^{\e_1,\e_2} (a,\bar a, 0, 0,h)$ for $h\not=0$. 
\item The path integral must be summed over topological sectors labeled by equivalence classes of the fluxes $[\mathfrak p]$, as defined in \eqref{equivalence relation for p}.
They can be parametrized by a $(\chi-2)$-tuple of integers $\bfk$.
\end{itemize}

\subsection{Integrand} 

In this subsection, we briefly review the K-theoretic Nekrasov partition functions.

\subsubsection{Perturbative Contributions}

The perturbative contribution of the Nekrasov partition function on $\mathbb{C}^2\times S^1$ can be written as
\be\label{Zpert equiv}
\CZ_{\text{pert}}(a, \e_1, \e_2,\Lambda,R) = \exp\left[-\widetilde\gamma_{\e_1,\e_2}(2a|R,\Lambda) -\widetilde\gamma_{\e_1,\e_2}(-2a|R,\Lambda)   \right],
\ee
where 
\bea
\widetilde\gamma_{\e_1,\e_2}(a|R,\Lambda) =& \gamma_{\e_1,\e_2}(a|R,\Lambda) + \frac{1}{\e_1\e_2}\left(\frac{\pi^2 a}{6R}-\frac{\zeta(3)}{R^2}\right) + \frac{\e_1+\e_2}{2\e_1\e_2}\left(a\log (\CR) + \frac{\pi^2}{6R}\right) \\
&+ \frac{\e_1^2 + \e_2^2 + 3\e_1\e_2}{12\e_1\e_2}\log (\CR),
\eea
with
\be
\begin{aligned}
\gamma_{\e_1,\e_2}(a|R,\Lambda) &= \frac{1}{2\e_1\e_2} \left[-\frac{R}{6}\left(a + \frac{\e_1+\e_2}{2}\right)^3 + a^2\log (\CR)\right] \\
& \quad + \sum_{n\geq 1} \frac{1}{n}\frac{e^{-n Ra}}{(e^{n R\e_1}-1)(e^{n R\e_2}-1)}.
\end{aligned}
\ee 
The series converges for ${\rm Re}(Ra)>0$ and $\e_i>0$. Under these conditions,  the last term can be written as
\cite{Gottsche:2006bm}
\be
\label{eq:PertExpLi}
\sum_{n\geq 1} \frac{1}{n}\frac{e^{-n Ra}}{(e^{nR\e_1}-1)(e^{nR\e_2}-1)} = \sum_{m\geq 0} \frac{c_m(\e_1,\e_2)}{m!} R^{m-2}\text{Li}_{3-m}(e^{-Ra}),
\ee
where $\text{Li}_{3-m}$ denotes a polylogarithm, and $c_m(\e_1,\e_2)$ is a function of $\e_1,\e_2$ which vanishes in the limit $\e_1,\e_2\rightarrow 0$ for $m>2$. 
We can then analytically continue in $a$ using the analytic properties of the polylogarithm described in App. 
\ref{app:Polylogs}. We will adopt this analytically continued expression as the definition of  $\CZ_{\text{pert}}$ in the following. Note that this definition apparently breaks gauge invariance by distinguishing the region ${\rm Re}(Ra)>0$ from ${\rm Re}(Ra)<0$. However, if the analytic continuation is performed from the opposite region, the resulting function $\CZ_{\text{pert}}$ would differ by a simple overall factor, which can be absorbed in the definition of the full partition function \eqref{partition ftn zero}.

The perturbative partition function \eqref{Zpert equiv} exhibits poles or zeros depending on the signs of $\e_1$ and $\e_2$. These poles and zeros occur at positions of the form
\begin{equation}
2a = m \e_1 + n \e_2 + \frac{2\pi i s}{R}, \quad s\in \mathbb{Z},
\end{equation}
where $m,n$ are integers whose specific ranges are determined by the signs of $\e_1$ and $\e_2$.
For more details, see App. \ref{poles-and-zeros}.  

\subsubsection{Instanton Contributions} \label{sec: instanton contribution}

The instanton partition function on $\mathbb{C}^2\times S^1$ is
\footnote{In comparing with \cite[Eq. (1.25)]{Gottsche:2006bm}, we  use the dictionary $n_I^{\rm here} = \tau^{\rm GNY}$. The relation between $\CR^4$ and $\CR^4 e^{-R(\e_1 + \e_2)}$ reflects the relation between the computation of indices of the Dirac operator and the Dolbeault operator as discussed in Eqs. \eqref{eq:EqualIndex} and \eqref{eq:ShiftIndexDensities}. Note that for $X=\IC^2$ we have the equivariant canonical bundle  $K_{\mathbb{C}^2}=-R(\eps_1+\eps_2)$.
There is an analogous shift of $\Lambda_{\rm GNY}$ in Lemma 3.4 of \cite{Gottsche:2006bm}.}
\be
\label{5d instanton sector}
\CZ_{\text{inst}} (a ,\e_1,\e_2, n_I,\Lambda,R) = \sum_{\vec{Y}} 
(\CR^{4}e^{-R (\e_1 +  \e_2) } )^{| \vec Y |}
\CZ^\text{vec}_{\vec Y}(a,\e_1,\e_2,R)\CZ^{\mathrm{U}(1)^{(I)}}_{\vec{Y}} (a,\e_1,\e_2,n_I).
\ee
Here $\vec{Y}=\{Y_1,Y_2\}$ is a pair of Young diagrams. For each Young diagram $Y=(\lambda_1\geq \lambda_2 \geq \cdots \geq 0)$, we denote its transpose by $Y^T = (\lambda_1'\geq \lambda_2' \geq \cdots\geq 0)$, and define $|Y|$ to be the total number of boxes in the Young diagram $Y$. For a box at the coordinate $s=(i,j)\in Y$, we define its arm-length and leg-length by
\be
A_Y(s)= \lambda_i-j, \quad L_Y= \lambda_j'-i. 
\ee
We also define 
\be
E(a,Y_1,Y_2,s) = a - \e_1 L_{Y_2}(s) + \e_2 (A_{Y_1}(s)+1).
\ee
The contribution from an $\CN=1$ vector multiplet is 
\be
\CZ^{\text{vec}}_{\vec{Y}} (a,\e_1,\e_2,R)= \prod_{i,j=1}^2 n_{i,j}^{(Y_i, Y_j)} (a,\e_1,\e_2,R),
\ee
where 
\be\label{definition nij} 
n_{i,j}^{(Y_i, Y_j)} = \prod_{s\in Y_j} \left( 1-e^{-R E(a_i-a_j,Y_j,Y_i,s)}\right)^{-1} \prod_{t\in Y_i}\left( 1-e^{-R \left(\e_1+\e_2 -E(a_j-a_i,Y_i,Y_j,t)\right)}\right)^{-1},
\ee
with $a_1 = -a_2=a$. 
$\CZ^{\mathrm{U}(1)^{(I)}}_{\vec Y}$ is the contribution from the background $\mathrm{U}(1)^{(I)}$ flux, which reduces to identity for vanishing $n_I$. 
We argue in App. \ref{app:Loc Hilbert Schemes} that the contribution from the background $\mathrm{U}(1)^{(I)}$ flux is 
\be\label{eq:U(1)-factor}
\CZ^{\mathrm{U}(1)^{(I)}}_{\vec{Y}} (a,\e_1,\e_2,n_I) = \exp \left[n_I \left(-|\vec Y| + \frac{a^2}{\e_1\e_2}\right)\right],
\ee
which matches \cite[Eq.  (1.24)]{Gottsche:2006bm}. The instanton partition function with background $\mathrm{U}(1)^{(I)}$ flux $n_I$ is related to that without background $\mathrm{U}(1)^{(I)}$ flux by
\be
\label{eq:ZinstnI}
\CZ_{\text{inst}} (a,\e_1,\e_2, n_I,\Lambda,R)  = \exp\left(\frac{n_I a^2}{\e_1 \e_2}\right)\CZ_{\text{inst}} (a,\e_1,\e_2, 0,\Lambda e^{-\frac{1}{4} n_I},R).
\ee

Similar to the recursion formula for $Z_{\rm inst}$ in 4d theories \cite{Poghossian:2009mk}, there exists a recursion formula for the K-theoretic instanton partition function $\CZ_{\text{inst}}$ \cite{Yanagida:2010vz},
\footnote{These recursion formulae are derived by combining the AGT relation \cite{Alday:2009aq} with Zamolodchikov's recursion relation for Virasoro conformal blocks \cite{Zamolodchikov:1984eqp}. The 5d AGT conjecture \cite{Awata:2009ur} relates the instanton partition function $\CZ_{\text{inst}}$ to the $q$-deformed Virasoro algebra \cite{Shiraishi:1995rp}.}
\begin{equation}
\label{ZinstRecursion}
\begin{split}
&  \CZ_\text{inst}(a,\e_1,\e_2,n_I =0, \Lambda, R) \\
=& 1- \sum_{m,n=1}^\infty \frac{\mathcal{R}^{4mn}\, T_{m,n}(t_1,t_2)}{(x^{-2}- t_1^m t_2^n)(1-x^2 t_1^{-m} t_2^{-n})}
\CZ_\text{inst}\left(\frac{m\e_1 - n\e_2}{2},\e_1,\e_2,n_I=0,\Lambda,R\right),
\end{split}
\end{equation}
where 
\begin{equation}
\label{eq:RecursionT}
    T_{m,n}(t_1,t_2) = \frac{(1+ t_1^{m} t_2^{n})}{(t_1 t_2)^{mn}}\underbrace{\prod_{i=-m+1}^{m} \prod_{j=-n+1}^{n}}_{(i, j) \neq\{(0,0),(m, n)\}}
    \frac{t_1^i t_2^j}{1-t_1^i t_2^j},
\end{equation}
with 
\be
x=e^{Ra},\quad t_1 = e^{ R \epsilon_1}, \quad t_2 = e^{ R \epsilon_2}.
\ee
The recursion relation simplifies the series expansion in $\CR$, and will be used later in Eqs. \eqref{Zes0} and \eqref{ZesnI}.

From the recursion relation \eqref{ZinstRecursion}, 
the instanton partition function have poles at
\begin{equation}
    2a = m \epsilon_1 + n \epsilon_2 + \frac{ 2\pi i s}{R}, 
    \quad  m n >0, \quad m,n,s\in \mathbb{Z},
\end{equation}
which correspond to the points where extra zero modes appear in the $\Omega$-deformed theory.

\subsubsection{Full Partition Function}
Combining the perturbative partition function \eqref{Zpert equiv} and the instanton partition function  \eqref{5d instanton sector}, we define the full partition function as \cite{Nakajima:2005fg,Gottsche:2006bm},
\begin{equation}\label{eq:ZnI}
\CZ\left(a, \epsilon_1, \epsilon_2, n_I, \Lambda, R\right)
:=\CZ_{\text {pert}}
\left(a, \epsilon_1, \epsilon_2, \Lambda, R\right) 
\CZ_{\text {inst}}
\left(a, \epsilon_1, \epsilon_2, n_I, \Lambda, R\right).
\end{equation}
The specialization to $n_I=0$ yields the standard Nekrasov partition function,
\begin{equation}
\label{eq: ZNkera}
\CZ(a,\e_1,\e_2, \Lambda,R) 
:= \CZ\left(a, \epsilon_1, \epsilon_2, n_I = 0, \Lambda, R\right).
\end{equation}
Notably, the full partition function \eqref{eq:ZnI} can be rewritten in terms of the Nekrasov partition function as
\begin{equation}\label{eq:Z_property}
\CZ(a, \epsilon_1, \epsilon_2, n_I, \Lambda, R) = 
\CZ(a,\e_1,\e_2, \Lambda e^{-\frac{1}{4} n_I},R) \exp\left[- \frac{n_I\left(\e_1^2 + \e_2^2 + 3\e_1 \e_2 \right)}{24 \e_1\e_2}\right].
\end{equation}
Thus, up to an elementary factor, the function $\CZ$ is essentially the $K$-theoretic Nekrasov partition function with the $n_I$-dependence absorbed into the dynamical scale $\Lambda$.

The poles or zeros of $\CZ$ are located at 
\begin{equation}
2 a=m \epsilon_1+n \epsilon_2+\frac{2 \pi i s}{R}, \quad s \in \mathbb{Z},
\end{equation}
where the integers $m,n$ belong to different sets determined by the signs of $\text{Re}(\e_1)$ and $\text{Re}(\e_2)$.
As derived in App. \ref{poles-and-zeros}, we find that
\begin{enumerate}
    \item When $\text{Re}(\epsilon_1) \text{Re}(\epsilon_2) >0$, $\CZ(a,\e_1,\e_2, n_I,\Lambda,R)$ has no poles in $a$.
    \item When $\text{Re}(\epsilon_1)\text{Re}(\epsilon_2) < 0$, $\CZ(a,\e_1,\e_2, n_I,\Lambda,R)$ has poles in $a$, coming from both perturbative and instanton partition functions. 
    Furthermore, they are all simple poles.
\end{enumerate}

\subsubsection{Non-equivariant limit}
\label{sec:non-equivlimit}
Some similar computations in this section can be also found in section 6
of \cite{Closset:2022vjj}.
In the non-equivariant limit $\e_1,\e_2 \rightarrow 0$, the Nekrasov partition function can be expanded as
\be
\begin{aligned}
\label{expansion Z nekra}
\CZ(a,\e_1,\e_2, \Lambda,R) 
= & \exp\frac{1}{\e_1\e_2}\bigg[  \mathcal{F}(a,\Lambda,R)+ (\e_1+\e_2) H(a) \\
& + \e_1\e_2 \log (A)(a,\Lambda, R) + \frac{\e_1^2+ \e_2^2}{3}\log(B)(a,\Lambda,R)+ \cdots \bigg], 
\end{aligned}
\ee
where higher-order terms in $\e_1, \e_2$ are neglected. 

By combining contributions from all fixed points, we express the function $g_{\mathfrak{p},\mathfrak{p}^{(I)}}^{\e_1,\e_2}$ as
\be
\begin{aligned}
\label{integrand}
g_{\mathfrak{p},\mathfrak{p}^{(I)}}^{\e_1,\e_2} =  &\frac{1}{\CR}\exp\left\{- \sum_{\ell=1}^\chi \frac{1}{\epsilon_1^{(\ell)}\epsilon_2^{(\ell)}} \left[\frac{R}{48}\left(\epsilon_1^{(\ell)}+\epsilon_2^{(\ell)}\right)^3 + \frac{n_I^{(\ell)}}{24}\left((\epsilon_1^{(\ell)})^2 + (\epsilon_2^{(\ell)})^2 + 3\epsilon_1^{(\ell)} \epsilon_2^{(\ell)} \right)\right]\right\} \\
&\times \prod_{\ell=1}^\chi \CZ (a^{(\ell)},\epsilon_1^{(\ell)},\epsilon_2^{(\ell)}, \Lambda e^{-\frac{1}{4}n_I^{(\ell)}},R)  \\
= & \frac{1}{\CR}\exp\left\{- \sum_{\ell=1}^\chi \frac{1}{\epsilon_1^{(\ell)}\epsilon_2^{(\ell)}} \left[\frac{R}{48}\left(\epsilon_1^{(\ell)}+\epsilon_2^{(\ell)}\right)^3 + \frac{n_I^{(\ell)}}{24}\left((\epsilon_1^{(\ell)})^2 + (\epsilon_2^{(\ell)})^2 + 3\epsilon_1^{(\ell)} \epsilon_2^{(\ell)} \right)\right]\right\} \\
& \times \exp\Bigg\{\sum_{\ell=1}^{\chi} \frac{1}{\epsilon_1^{(\ell)}\epsilon_2^{(\ell)}} \Bigg[ \mathcal{F}(a^{(\ell)},\Lambda e^{-\frac{1}{4}n_I^{(\ell)}},R)+ \left(\epsilon_1^{(\ell)}+\epsilon_2^{(\ell)}\right) H(a^{(\ell)})\\
& + \epsilon_1^{(\ell)}\epsilon_2^{(\ell)} \log(A)\left(a^{(\ell)},\Lambda e^{-\frac{1}{4}n_I^{(\ell)}}, R\right)  + \frac{(\epsilon_1^{(\ell)})^2+ (\epsilon_2^{(\ell)})^2}{3}\log(B)\left(a^{(\ell)},\Lambda e^{-\frac{1}{4}n_I^{(\ell)}},R \right)+ \cdots  \Bigg] \Bigg\}. 
\end{aligned}
\ee

Due to \eqref{epsilon relation 1}, the non-equivariant limit of $g$ is given by
\be
\begin{aligned}
\label{eq:ge120}
\lim_{\e_1,\e_2\rightarrow 0}g_{\mathfrak{p},\mathfrak{p}^{(I)}}^{\e_1,\e_2}(a) 
=&\frac{1}{\CR}\lim_{\e_1,\e_2\rightarrow 0} \exp \sum_{\ell=1}^{\chi} \frac{1}{\epsilon_1^{(\ell)}\epsilon_2^{(\ell)}} \bigg[\frac18 \frac{\partial^2 \mathcal{F}}{\partial a^2} \left(\epsilon_1^{(\ell)} {\mathfrak p}_{\ell}+ \epsilon_2^{(\ell)} {\mathfrak p}_{\ell+1} \right)^2\\
& + \frac{ R}{8}\frac{\partial^2 \mathcal{F}}{\partial a \partial (\log\mathcal{R})} \left(\epsilon_1^{(\ell)} {\mathfrak p}^{(I)}_{\ell} + \epsilon_2^{(\ell)} {\mathfrak p}^{(I)}_{\ell+1}\right)\left(\epsilon_1^{(\ell)} {\mathfrak p}_{\ell}+ \epsilon_2^{(\ell)} {\mathfrak p}_{\ell+1}\right)\\
& + \frac{ R^2}{32}\frac{\partial^2 \mathcal{F}}{\partial (\log \mathcal{R})^2}\left(\epsilon_1^{(\ell)} {\mathfrak p}^{(I)}_{\ell} + \epsilon_2^{(\ell)} {\mathfrak p}^{(I)}_{\ell+1}\right)^2 \\
& + \frac12\frac{\partial H}{\partial a}(\epsilon_1^{(\ell)} + \epsilon_2^{(\ell)})\left(\epsilon_1^{(\ell)} {\mathfrak p}_{\ell}+ \epsilon_2^{(\ell)} {\mathfrak p}_{\ell+1}\right) \\
& + \epsilon_1^{(\ell)}\epsilon_2^{(\ell)} \log(A)(a,\Lambda,R) \\
& + \frac{1}{3}\left((\epsilon_1^{(\ell)})^2+ (\epsilon_2^{(\ell)})^2\right) \log(B)(a,\Lambda,R)+ {\cal O}(\epsilon^3) \bigg].
\end{aligned}
\ee
Using relations \eqref{kfrakpl}, \eqref{nfrakpIl}, and \eqref{Bpq}, we can express the sums over fixed points in terms of the intersection form,
\be
\label{eq:epsilon-ksq}
\sum_{\ell=1}^{\chi}  \frac{\left(\epsilon_1^{(\ell)} {\mathfrak p}_{\ell}+ \epsilon_2^{(\ell)} {\mathfrak p}_{\ell+1}\right)^2}{4\epsilon_1^{(\ell)}\epsilon_2^{(\ell)}} 
= \left(\frac{1}{2}\sum_{\ell=1}^{\chi} {\mathfrak p}_{\ell} D_{\ell}\right)^2 = \bfk^2,
\ee
\be
\label{eq:epsilon-kn}
\sum_{\ell=1}^{\chi}  \frac{\left(\epsilon_1^{(\ell)} {\mathfrak p}_{\ell}+ \epsilon_2^{(\ell)} {\mathfrak p}_{\ell+1}\right) \left(\epsilon_1^{(\ell)} {\mathfrak p}^{(I)}_l + \epsilon_2^{(\ell)} {\mathfrak p}^{(I)}_{\ell+1}\right)}{2\epsilon_1^{(\ell)}\epsilon_2^{(\ell)}} 
= \left(\frac{1}{2}
\sum_{\ell=1}^{\chi} \mathfrak{p}_{\ell} D_{\ell} \right)
\left(\sum_{\ell=1}^{\chi} \mathfrak{p}^{(I)}_{\ell} D_{\ell}\right)=
B(\bfk, \bfn_I),
\ee
\be
\label{eq:epsilon-nn}
\sum_{\ell=1}^{\chi}  \frac{\left(\epsilon_1^{(\ell)} {\mathfrak p}^{(I)}_{\ell} + \epsilon_2^{(\ell)} {\mathfrak p}^{(I)}_{{\ell}+1}\right)^2}{\epsilon_1^{(\ell)}\epsilon_2^{(\ell)}} =  \left(\sum_{\ell=1}^{\chi} \mathfrak{p}^{(I)}_{\ell} D_{\ell} \right)^2
=\bfn_I^2.
\ee
Furthermore, using Eq. \eqref{Bpq} and the properties of $K_X$,
\be
K_X = - \sum_{\ell=1}^{\chi} D_{\ell}, \quad B(K_X,K_X) = 2\chi + 3 \sigma,
\ee
we obtain 
\begin{align}
&\sum_{\ell=1}^{\chi} \frac{\left(\epsilon_1^{(\ell)} + \epsilon_2^{(\ell)}\right)\left(\epsilon_1^{(\ell)} {\mathfrak p}_{\ell}+ \epsilon_2^{(\ell)} {\mathfrak p}_{\ell+1}\right)}{2\epsilon_1^{(\ell)} \epsilon_2^{(\ell)}} 
= -B(K_X,\bfk), \\ 
&\sum_{\ell=1}^{\chi} \frac{(\epsilon_1^{(\ell)})^2+ (\epsilon_2^{(\ell)})^2}{3\epsilon_1^{(\ell)} \epsilon_2^{(\ell)}}  = \sigma.
\end{align}
Eventually, the non-equivariant limit of \eqref{integrand} simplifies to
\begin{equation}
\begin{aligned}
g_{\bfmu,\bfn,\bfk}&:= 
\lim_{\epsilon_1,\epsilon_2\to 0}
g_{\mathfrak{p},\mathfrak{p}^{(I)}}^{\e_1,\e_2} 
\\ &=  \frac{1}{\CR} A^\chi B^\sigma 
\exp \left[\frac12 
\frac{\partial^2 \mathcal{F}}{\partial a^2} \bfk^2 
+\frac{ R}{4}
\frac{\partial^2 \mathcal{F}}
     {\partial a \partial (\log\mathcal{R})} B(\bfk, \bfn_I) 
     \right.\\
&\quad + \left. \frac{ R^2}{32}\frac{\partial^2 \mathcal{F}}
{\partial (\log \mathcal{R})^2} \bfn_I^2
-\frac{\partial H}{\partial a} B(K_X,\bfk)
\right].
\end{aligned}
\end{equation}
This result demonstrates that the leading term of $g_{\mathfrak{p},\mathfrak{p}^{(I)}}^{\e_1,\e_2}$ in the non-equivariant limit is finite, and depend on $\mathfrak{p}$ and $\mathfrak{p}^{(I)}$ only through their equivalence classes, namely only through the vectors $\boldsymbol{k}$ and $\boldsymbol{n}_I$. 

The explicit formula for $H(a)$ can be determined from the expansion \eqref{expansion Z nekra}. 
First, note that the instanton partition function does not contribute to $H(a)$. This follows from \cite[Lemma 4.3]{Nakajima:2005fg}, which shows that  
\begin{equation}
\begin{aligned}
\CZ_{\rm inst}
(a,\e_1, -2\e_1, n_I,\Lambda,R)
= \CZ_{\rm inst}(a,2\e_1, -\e_1, n_I,\Lambda,R).
\end{aligned}
\end{equation}
Then, through direct computation using Eqs. \eqref{Zpert equiv}, \eqref{eq:PertExpLi}, and the analytic continuation of ${\rm Li}_2$ (\ref{Liinv2}), we find
\footnote{There are different definitions of the perturbative part. Other definitions needed, for example, for the verification of the AGT conjecture, require $H=0$. See
\cite{Manschot:2019pog, Closset:2022vjj}. For our purposes, it is important to have $H=-i \pi a$. It would be good to clarify this issue.}
\begin{equation}
\label{eq:Ha}
H(a) = - i \pi a.
\end{equation}
In terms of the effective couplings $\tau, v_I$ and $\xi_{II}$ defined in \eqref{vandxi}, we have
\begin{equation}
\begin{aligned}\label{nekrasov conjecture holo}
g_{\bfmu,\bfn,\bfk}(a) 
= & \frac{1}{\CR}\exp\left[
    -\pi i \tau(a) \bfk^2 
    -2\pi i v_I(a) B(\bfk, \bfn_I) 
    -\pi i \xi_{II}(a) \bfn_I^2
    \right] \\
&\times\exp \left[\pi i B(K_X,\bfk) \right]A^\chi B^\sigma.
\end{aligned}
\end{equation}
This expression corresponds precisely to the holomorphic part of the bosonic effective action \eqref{effactionevaluated} (with $\bfn_K=0$), together with the gravitational couplings.
Moreover, the phase $\exp \left[\pi i B(K_X,\bfk) \right]$ agrees with the phase in Eq.  \eqref{eq:localmeasure} for the choice of characteristic vector $K=K_X \mod 4$.

\subsubsection*{Remark}
The first-order corrections in $\e_1, \e_2$ depend on the full $\chi$-tuples of integers $\mathfrak{p}$ and $\mathfrak{p}^{(I)}$, rather than only on their equivalence classes. 
To illustrate this, we take $X = \mathbb{CP}^2$, $n_I = 0$, and $\mathfrak{p}=(p,-p,0)$ as an example. 
\footnote{This choice in fact has $\Theta(\mathfrak{p})=0$, but it is a convenient choice to illustrate the point that, when 
$\epsilon_1,\epsilon_2$ are non-vanishing the integrand depends on the full $\chi$-tuple of integers $\mathfrak{p}$ and not just $\bfk$.}
We calculate the integrand \eqref{integrand} and expand it as a $\mathcal{R}$-series,
\begin{equation}
\begin{aligned}
g^{\epsilon_1,\epsilon_2}_{\mathfrak{p},0}
=& \left[-4 \sinh ^2(\mathfrak{a})+4 p R \sinh (2 \mathfrak{a}) \epsilon_2+\cdots\right]\CR^{-3} \\
&+ \left[\left(-\frac{9}{2} \text{csch}^2(\mathfrak{a})-2\right)+p R (25 \cosh (\mathfrak{a})+2 \cosh (3 \mathfrak{a})) \text{csch}^3(\mathfrak{a}) \epsilon_2
+\cdots\right]
 \mathcal{R} +\CO\left(\CR^5\right)
\end{aligned}
\end{equation}
Note that although $\boldsymbol{k}=0$ in this configuration, the terms linear in $\epsilon_2$ retain an explicit dependence on $p$. Similarly, one finds that the expansion of $g_{\mathfrak{p},\mathfrak{p}^{(I)}}^{\e_1,\e_2}$ contains terms linear in $\e_1,\e_2$ that depend on the full $\chi$-tuple of integers $\mathfrak{p}^{(I)}$.

\subsection{Contours And Wall-crossing}
\label{sec:loc wall-crossing}

In this subsection, we aim to examine the contribution from the asymptotic regions of the Coulomb branch, which governs the wall-crossing behavior of the partition function.

Let us return to the formal zero-mode integral discussed in Sec.  \ref{subsubsec:g-function-singularities-contours},
\be
Z_{\bfmu, \mathfrak{p}^{(I)}}^{\e_1,\e_2}(\CR) =\lim_{\substack{\delta\rightarrow 0\\ g_{5d}\rightarrow 0}}\frac{K_{\Omega}}{2}\sum_{[\mathfrak{p}]}    \int_\Gamma  \frac{\mathrm{d}h}{h}\int_{\partial\left(\frak M\backslash \Delta_\delta\right)} \mathrm{d}a\, g_{\mathfrak{p}, \mathfrak{p}^{(I)}}^{\e_1,\e_2} 
(a,\bar a, 0, 0,h),
\ee 
where the $a$-integral is taken over the union of contours around the bulk and asymptotic regions of the cylinder,
\be
\partial\left(\frak M\backslash \Delta_\delta\right) = \partial\frak M - \partial \Delta_\delta.
\ee
Under the assumption that there exists a well-defined contour $\Gamma$ as discussed in Sec.  \ref{subsubsec:g-function-singularities-contours}, the non-holomorphic integrand $g_{\mathfrak{p}, \mathfrak{p}^{(I)}}^{\e_1,\e_2} 
(a,\bar a, 0, 0,h)$ is continuous and bounded on $\Gamma$ when $a$ lies on a component of the finite contour $\partial\Delta_\delta$. 
\footnote{Since $\Gamma$ is noncompact, some justification is needed for the boundedness of $g_{\mathfrak{p}, \mathfrak{p}^{(I)}}^{\e_1,\e_2}(a,\bar a, 0, 0,h)$ as a function of $h$ on this contour. Recall that $\Gamma$ asymptotically approaches the real axis at both ends. In the non-equivariant limit, boundedness follows from Eq. \eqref{eq:g-with-nonzero-h}. When $\e_1, \e_2\not=0$, this property continues to hold for the zero-mode action, and we assume that the boundedness persists for the full function. }
Now consider the metric dependence of the contribution from the bulk contours, 
\be
\frac{K_{\Omega}}{2}\sum_{[\mathfrak p]}    \int_\Gamma  \frac{\mathrm{d}h}{h}\int_{\partial\Delta_\delta} \mathrm{d}a\, \left[g_{\mathfrak{p}, \mathfrak{p}^{(I)}}^{\e_1,\e_2, J^+} - g_{\mathfrak{p}, \mathfrak{p}^{(I)}}^{\e_1,\e_2, J^-}\right]  
(a,\bar a, 0, 0,h),
\ee
where $J^+$ and $J^-$ correspond to a pair of metrics infinitesimally close to each other. The result is expected to be infinitesimally small, which implies that the bulk contribution is invariant under the wall-crossing.
This is consistent with the claim in the $U$-plane approach that the wall-crossing of the $U$-plane integral at the strong-coupling singularities cancels against that from the SW contributions. Thus, it is a very significant claim. Furthermore, it motivates a formal equivariant wall-crossing formula
\be\label{formal equiv wc}
Z_{\bfmu, \mathfrak{p}^{(I)}}^{\e_1,\e_2, J^+} - Z_{\bfmu, \mathfrak{p}^{(I)}}^{\e_1,\e_2, J^-} = \frac{K_{\Omega}}{2}\sum_{[\mathfrak p]}    \int_\Gamma  \frac{\mathrm{d}h}{h}\int_{\partial\frak M} \mathrm{d}a \, \left[g_{\mathfrak{p}, \mathfrak{p}^{(I)}}^{\e_1,\e_2, J^+} - g_{\mathfrak{p}, \mathfrak{p}^{(I)}}^{\e_1,\e_2, J^-}\right]  
(a,\bar a, 0, 0,h),
\ee 
where the $a$-integral is a sum over contour integrals around the two asymptotic boundaries of the cylinder.

We first derive the wall-crossing formula of the path integral in the non-equivariant limit $\e_1,\e_2\rightarrow 0$, and compare the result with that obtained via the $U$-plane approach in Sec.  \ref{sec:wall-crossing}. Let us first define the non-equivariant limit of the integrand,
\be
g_{\bfmu,\bfn,\bfk} (a,\bar a, 0, 0,h) := 
\lim_{\e_1,\e_2\rightarrow 0} g_{\mathfrak{p},\mathfrak{p}^{(I)}}^{\e_1,\e_2} (a,\bar a, 0, 0,h).
\ee
This limit exists thanks to the remarkable cancellations in Eqs. 
\eqref{eq:epsilon-kn}, \eqref{eq:epsilon-ksq}, \eqref{eq:epsilon-nn},  
\eqref{epsilon relation 1}, and  \eqref{epsilon relation 2}, ultimately owing to the finite volume of $X$. 
We claim that for Eq. \eqref{formal equiv wc}, taking the non-equivariant limit $\e_1,\e_2\rightarrow 0$ commutes with the contour integrals, yielding
\be\label{eq:Z-Diff-g-function}
Z^{J^+}_{\bfmu,\bfn} - Z^{J^-}_{\bfmu,\bfn} = 
\frac{K_{\Omega}}{2}
\sum_{\bfk \in L + \bfmu} \int_\Gamma  \frac{\mathrm{d}h}{h}\int_{\partial {\frak M}} \mathrm{d}a\, \left[g_{\bfmu,\bfn,\bfk}^{+} - g_{\bfmu,\bfn,\bfk}^{-}\right](a,\bar a, 0, 0,h).
\ee
The contour $\Gamma$ for the $h$-integral is chosen as before. As demonstrated in Eq. \eqref{nekrasov conjecture holo}, the $h=0$ limit of the integrand $g_{\bfmu,\bfn,\bfk}(a,\bar a, 0,0,h)$ coincides with the bosonic Coulomb branch effective action evaluated at $h=0$. Thus, it is natural to conjecture that
\be
g_{\bfmu,\bfn,\bfk}(a,\bar a, 0, 0,h) = \exp \left[-\int_X \CL_0 \right]_{V_0 = (a,\bar a, 0, 0,h)},
\ee
where $\CL_0$ is the LEEA \eqref{LEEA bos zero} with $(\bfn_I,\bfn_K)=(\bfn,0)$. 
Turning on the non-BPS modes \eqref{def: non-bps-h} is equivalent to
\begin{equation}
    F_+ - D =  2i h \omega ,
\end{equation}
where 
\be
\omega := \frac{1}{2}I_{\mu\nu} dx^\mu \wedge dx^\nu
\ee
is the K\"ahler form of $X$, with $I_{\mu\nu}$ defined in Eq. \eqref{def: complex-structure-I}.
The LEEA with nonzero $h$ evaluates to 
\begin{equation}
\begin{aligned}
\mathcal{L}_{0,h} = &
\frac{i}{16\pi}\bar\tau F_+\wedge F_+ 
+\frac{i}{8\pi}\bar v_IF_+\wedge F_+^{(I)} 
+\frac{i}{16\pi}\bar\xi_I F_+^{(I)}\wedge F_+^{(I)}  \\
&+\frac{i}{16\pi}\tau F_-\wedge F_- 
+\frac{i}{8\pi} v_IF_-\wedge F_-^{(I)} 
+\frac{i}{16\pi}\xi_I F_-^{(I)}\wedge F_-^{(I)}
\\
& -\frac{1}{8 \pi} y 
\left(
 F_+ - 2i h \omega 
\right) \wedge 
\left( F_+ - 2i h \omega \right)
-\frac{1}{4\pi} \mathrm{Im}(v_I) 
\left(
 F_+ - 2i h \omega 
\right) 
\wedge F_+^{(I)}\\
&-\frac{1}{8\pi} \mathrm{Im}(\xi_I)
F_+^{(I)}\wedge F_+^{(I)}
\\
=&
\pi i \tau \bigg(\frac{F}{4\pi}\bigg)\wedge\bigg(\frac{F}{4\pi}\bigg)
+ \pi i v_I \bigg(\frac{F^{(I)}}{2\pi}\bigg)\wedge \bigg(\frac{F}{4\pi}\bigg)
+ \frac{1}{4}\pi i \xi_I 
\bigg(\frac{F^{(I)}}{2\pi}\bigg)\wedge\bigg(\frac{F^{(I)}}{2\pi}\bigg)\\
&+\frac{i}{2\pi} h 
\left(y F_++ \mathrm{Im}(v_I) F_+^{(I)}\right) \wedge \omega
+ \frac{1}{2\pi} y h^2 \omega \wedge \omega.
\end{aligned}
\end{equation}
We introduce the period point $J$ via 
\begin{equation}
J = \frac{\omega}{\sqrt{2\text{Vol}(X)}},
\end{equation}
such that $B(J,J)=1$.
Then the $h$-dependence of the integrand is
\begin{equation}
\begin{aligned}
\label{eq:g-with-nonzero-h}
& g_{\bfmu,\bfn,\bfk}(a,\bar a, 0, 0,h) \\
=& \exp \left[ -\pi i \tau(a) \bfk^2 
- 2\pi i v_I (a) B(\bfk, \bfn) - \pi i\xi_{II}(a) \bfn^2 \right] \\
& \times \exp \left[- \frac{y}{4\pi}\text{Vol}(X) h^2 
- iy \sqrt{2\text{Vol}(X) }
B\left(\bfk + \frac{\text{Im} (v_I)\bfn}{y},J\right) h \right] \\
& \times \exp\left[\pi i B(\bfk, K_X)\right]A^\chi B^\sigma, \\
=& g_{\bfk,\bfmu, \bfn}(a) 
\exp \left[- \frac{y}{\pi}\text{Vol}(X) h^2 
- 2iy\sqrt{2\text{Vol}(X) }B\left(\bfk 
+ \frac{\text{Im} (v_I)\bfn}{y},J\right) h \right],
\end{aligned}
\end{equation}
where $y=\text{Im}(\tau)$. Note that for $h\not=0$, the $g_{\bfmu,\bfn,\bfk}$ depends on the metric in a way not only through the period point $J$.  

Since $g_{\bfmu,\bfn,\bfk}(a,\bar a, 0, 0,h)$ is analytic in $h$ for fixed $a$, the contour $\Gamma$ satisfying the assumptions in Sec.  \ref{subsubsec:g-function-singularities-contours} can be chosen as
\be
h \in \mathbb{R}-i\eta,
\ee
where $\eta$ is a small nonzero real number avoiding the singularity at $h=0$.
The integration of $h$ along $\Gamma$ can be evaluated using
\begin{equation}\label{eq:h-integral-gives-errorfunction}
\begin{aligned}
&\lim_{\eta\rightarrow 0}\frac{1}{\pi i} \int_{\mathbb{R}-i\eta} 
e^{-t^2} e^{2\sqrt{\pi}xt i} \frac{\mathrm{d} t}{t}\\
=&\lim_{\eta \rightarrow 0}\frac{1}{\pi i} 
\left[ 
\int_{\mathbb{R}^+-i\eta}   e^{-t^2} 
  e^{2\sqrt{\pi} xt i} 
  \frac{\mathrm{d}t}{t}
-
\int_{\mathbb{R}^++i\eta}   e^{-t^2} 
  e^{-2 \sqrt{\pi} xt i} 
  \frac{\mathrm{d}t}{t}
\right]
\\ 
 =&\frac{2}{\pi} \int_0^\infty e^{-t^2}\sin(2\sqrt{\pi}xt) \frac{\mathrm{d}t}{t}  + {\rm sgn}(\eta),
\end{aligned}
\end{equation}
where the first term in the last line is an integral representation of the error function,
\be
E(x) = \frac{2}{\pi} \int_0^\infty e^{-t^2}\sin(2\sqrt{\pi}xt) \frac{\mathrm{d}t}{t}.
\ee
Note that another integral representation of the error function is given in Eq. (\ref{Eerror}).
We can therefore evaluate the $h$-integral in terms of the error function, 
\bea\label{eq:Evaluate-h-Contour-ErrorFunction}
\frac{K_{\Omega}}{2}
\int_\Gamma \frac{\mathrm{d}h}{h}\, g_{\bfmu,\bfn,\bfk}(a,\bar a, 0, 0,h) =& \frac{i\pi}{2}K_{\Omega}
E\left(\sqrt{2y} B\left(\bfk + \frac{\text{Im} (v_I)\bfn}{y},J\right)+ \text{sgn}(\eta)\right)  g_{\bfmu,\bfn,\bfk}(a),
\eea
where $g_{\bfmu,\bfn,\bfk}(a)$ is the value of $g_{\bfmu,\bfn,\bfk}(a,\bar a, 0,0,h)$ at $h=0$. 
Then, the wall-crossing formula reads
\bea
\label{eq:wc-contour}
Z^{J^+}_{\bfmu,\bfn} - Z^{J^-}_{\bfmu,\bfn} =&  -K_{\Omega} \pi^2 \sum_{\bfk \in L + \bfmu} \left(\underset{a=\infty}{\text{Res}} + \underset{a=-\infty}{\text{Res}}\right) \mathrm{d}a\, g_{\bfmu,\bfn,\bfk}(a) \\
& \times \left[E\left(\sqrt{2y} B\left(\bfk + \frac{\text{Im} (v_I)\bfn}{y},J^+\right)\right) - E \left(\sqrt{2y} B\left(\bfk + \frac{\text{Im} (v_I)\bfn}{y},J^-\right)\right)\right].
\eea
In Sec.  \ref{sec: loc to U}, we will show that this expression exactly reproduces the wall-crossing formula derived via the $U$-plane integral in Sec.  \ref{sec:wall-crossing}. Before proceeding, we briefly comment on the generalization to the equivariant wall-crossing formula. We take $\bfn=0$ for simplicity, although the extension to non-vanishing $\bfn$ is straightforward. From the bosonic action evaluated on the zero modes $V_0$ \eqref{eq:g-full-defined}, we factor out terms decoupled from the quantum fluctuations $V'$ and write the non-holomorphic integrand as 
\be\label{eq:Linear-with-tilde-g}
g_{\mathfrak{p},\mathfrak{p}^{(I)}}^{\e_1,\e_2} (a,\bar a, 0,0,h) = \exp\left[-\frac{1}{g^2_{5d}}\left( \nu_1 h^2 + i\nu_2 B(\bfk, J) h\right)\right] \tilde{g}_{\mathfrak{p},\mathfrak{p}^{(I)}}^{\e_1,\e_2} (a,\bar a, 0,0,h),
\ee
where $\nu_1 = 2\text{Vol}(X)R$ and $\nu_2 = 8\pi R/\sqrt{2\text{Vol}(X)}$ are positive constants.
This defines $\t g_{\mathfrak{p},\mathfrak{p}^{(I)}}^{\e_1,\e_2} (a,\bar a, 0,0,h)$, which satisfies 
\be\t g_{\mathfrak{p},\mathfrak{p}^{(I)}}^{\e_1,\e_2} (a,\bar a, 0,0,0) =g_{\mathfrak{p},\mathfrak{p}^{(I)}}^{\e_1,\e_2} (a,\bar a, 0,0,0) = g_{\mathfrak{p},\mathfrak{p}^{(I)}}^{\e_1,\e_2} (a).
\ee
Then the boundary contribution to the full partition function,
\be
Z_{\bfmu, \mathfrak{p}^{(I)}}^{\e_1,\e_2,\text{bdry}}(\CR) =\frac{K_{\Omega}}{2}\sum_{[\mathfrak{p}]}    \int_\Gamma  \frac{\mathrm{d}h}{h}\int_{\partial\frak M} \mathrm{d}a \, g_{\mathfrak{p}, \mathfrak{p}^{(I)}}^{\e_1,\e_2} 
(a,\bar a, 0, 0,h),
\ee 
can be written as 
\be
Z_{\bfmu, \mathfrak{p}^{(I)}}^{\e_1,\e_2,\text{bdry}}(\CR) =\frac{K_{\Omega}}{2}\sum_{[\mathfrak{p}]}    \int_\Gamma  \frac{\mathrm{d}h} {h} \exp\left[{-\frac{1}{g^2_{5d}}\left( \nu_1 h^2 + i\nu_2 B(\bfk, J) h\right)}\right]\int_{\partial\frak M} \mathrm{d}a \, \t g_{\mathfrak{p}, \mathfrak{p}^{(I)}}^{\e_1,\e_2} 
(a,\bar a, 0, 0,h).
\ee
After rescaling $h\rightarrow g^2_{5d} h$, this becomes
\be
Z_{\bfmu, \mathfrak{p}^{(I)}}^{\e_1,\e_2,\text{bdry}}(\CR) =\frac{K_{\Omega}}{2}\sum_{[\mathfrak{p}]}    \int_\Gamma  \frac{\mathrm{d}h} {h} \exp\left[{-\left( \nu_1 g^2_{5d} h^2 + i\nu_2 B(\bfk, J) h\right)}\right]\int_{\partial\frak M} \mathrm{d}a \, \t g_{\mathfrak{p}, \mathfrak{p}^{(I)}}^{\e_1,\e_2} 
(a,\bar a, 0, 0, g^2_{5d}h).
\ee
Taking the limit $g_{5d}^2\rightarrow 0$, it reduces to
\be\label{asymptotic contour loc limit}
Z_{\bfmu, \mathfrak{p}^{(I)}}^{\e_1,\e_2,\text{bdry}}(\CR) =\frac{K_{\Omega}}{2}\sum_{[\mathfrak{p}]}    \int_\Gamma  \frac{\mathrm{d}h} {h} \exp\left(-i \nu_2  B(\bfk, J) h\right)\int_{\partial\frak M} \mathrm{d}a \, g_{\mathfrak{p}, \mathfrak{p}^{(I)}}^{\e_1,\e_2} 
(a).
\ee
Now we perform the $h$-integral,
\be
\int_{\mathbb{R}-i\eta}  \frac{\mathrm{d}h} {h} \exp\left(-i \nu_2  B(\bfk, J) h\right) = \begin{cases}
2\pi iH(-B(\bfk,J)), & \eta>0\\
-2\pi iH(B(\bfk,J)), & \eta<0
\end{cases},
\ee
where $H(x)$ is the Heaviside step function,
\be
H(x)=\begin{cases}
1, & x>0\\
0, & x<0
\end{cases}.
\ee
This leads to an equivariant wall-crossing formula
\be\label{equivariant wall crossing}
\begin{aligned}
& Z^{\e_1,\e_2,J^+}_{\bfmu} - Z^{\e_1,\e_2,J^-}_{\bfmu} \\
=& -2\pi^2 K_{\Omega}\sum_{[\mathfrak{p}]} \left[H(-B(\bfk,J^+))- H(-B(\bfk, J^-))\right]  \left(\underset{a=\infty}{\text{Res}} + \underset{a=-\infty}{\text{Res}}\right)\mathrm{d}a\, g_{\mathfrak{p}, \mathfrak{p}^{(I)}}^{\e_1,\e_2}(a) \\
= & \pi^2 K_{\Omega}\sum_{[\mathfrak{p}]} \left[\text{sgn}(B(\bfk,J^+))- \text{sgn}(B(\bfk, J^-))\right]  \left(\underset{a=\infty}{\text{Res}} + \underset{a=-\infty}{\text{Res}}\right)\mathrm{d}a\, g_{\mathfrak{p}, \mathfrak{p}^{(I)}}^{\e_1,\e_2}(a).
\end{aligned}
\ee

\subsection{From Toric Localization To The $U$-plane Integral}\label{sec: loc to U}

In the weak-coupling limit, $y\to \infty$, the error function simplifies to a sign function,
\begin{equation}
    E\left(\sqrt{2y} B\left(\bfk + \frac{\text{Im}(v_I)\bfn}{y},J\right)\right) \to 
    \sgn\left( B\left(\bfk + \frac{\text{Im}(v_I)\bfn}{y},J\right)\right).
\end{equation}
Since the only poles in the non-equivariant limit $\e_1, \e_2 \to 0$ are located at $\mathfrak{a}=0,\pi i$ and $\pm \infty$, the wall-crossing formula becomes
\bea
\label{eq: wall-crossing}
Z^{J^+}_{\bfmu,\bfn} - Z^{J^-}_{\bfmu,\bfn} =& 
-\frac{\pi^2 K_{\Omega}}{R}
\sum_{\bfk \in L + \bfmu} \left(\underset{\mathfrak{a}=0}{\text{Res}} +
\underset{\mathfrak{a}=\pi i}{\text{Res}} \right)
\mathrm{d}\mathfrak{a}\, g_{\bfmu,\bfn,\bfk}(\mathfrak{a},\mathcal{R}) \\
&\quad \times \left[\sgn\left( B\left(\bfk + \frac{\text{Im} (v_I)\bfn}{y},J^+\right)\right) - \sgn \left( B\left(\bfk + \frac{\text{Im} (v_I)\bfn}{y},J^-\right)\right)\right]\\ 
=& 
-\frac{\pi^2 K_{\Omega}}{R}\sum_{\bfk \in L + \bfmu} \text{Coeff}_{\mathfrak{a}^0}\left[\Delta(\mathfrak{a}, \mathcal{R}) \mathfrak{a}\right]
+ \text{Coeff}_{(\mathfrak{a}')^0}\left[\Delta'(\mathfrak{a}', \mathcal{R}) \mathfrak{a}'\right],
\eea
where $\mathfrak{a}' = \mathfrak{a} + \pi i $, 
\bea
\label{integrand in a}
\Delta(\mathfrak{a}, \mathcal{R}) &= \exp[\pi i B(\boldsymbol{k},K_X)]A^\chi B^{\sigma} \\
&\quad\times\exp\left[-\pi i \tau \boldsymbol{k}^2-2\pi i v_I B(\boldsymbol{k},\boldsymbol{n})- \pi i \xi_{II} \boldsymbol{n}^2\right]\\
&\quad\times 
\left[\sgn\left(B\left(\boldsymbol{k}+ \frac{\text{Im}(v_I) \boldsymbol{n}}{y},J^+\right)\right)
-\sgn\left(B\left(\boldsymbol{k}+ \frac{\text{Im}(v_I) \boldsymbol{n}}{y},J^-\right)\right)
\right],
\eea
and 
\begin{equation}
    \Delta'(\mathfrak{a}', \mathcal{R})
    = \exp\left(- 4\pi i \bfk^2+2  \pi i 
    B(\bfk, \bfn) +
    \frac{\chi+\sigma}{2} \pi i 
    \right)
    \Delta(\mathfrak{a}', \mathcal{R}).
\end{equation}
For notational simplification, the dependence on $(\boldsymbol{k},\boldsymbol{n},\boldsymbol{\mu},J)$ is suppressed here.
Specializing to the case $\chi+\sigma = 4$, the wall-crossing formula simplies to 
\begin{equation}
    Z^{J+}_{\bfmu, \bfn}  - Z^{J-}_{\bfmu,\bfn} 
    = -\frac{\pi^2 K_{\Omega}}{R} \left[1+ \exp\left( 2 \pi i B(\bfmu, \bfn+ K_X\right)\right] 
    \sum_{\bfk \in L + \bfmu} \text{Coeff}_{\mathfrak{a}^0}[\Delta(\mathfrak{a}, \mathcal{R}) \mathfrak{a}].
\end{equation}
To compare this result with the wall-crossing formula derived from the $U$-plane integral, we must express the residues in $\mathfrak{a}$ as residues in the modular parameter $q$. We perform this change of variables following \cite[Corollary 4.11 and Lemma 4.12]{Gottsche:2006bm}. 
From Eqs. \eqref{prepotential} and \eqref{prepotential2}, we can express the second line in Eq. \eqref{integrand in a} in terms of the prepotential,
\begin{align}
\exp \left(-\pi i \tau \boldsymbol{k}^2\right) & =\exp \left(\frac{\boldsymbol{k}^2}{2} \frac{\partial^2 \mathcal{F}}{\partial a^2}\right) \nonumber \\
& =\exp \left(4 \mathfrak{a} \boldsymbol{k}^2\right)\left(\frac{1-e^{-2 \mathfrak{a}}}{\mathcal{R}}\right)^{4 \boldsymbol{k}^2} \exp \left(\frac{\boldsymbol{k}^2}{2} \frac{\partial^2 \mathcal{F}_{\text {inst }}}{\partial a^2}\right), \\
\exp \left(-2 \pi i v_I B(\boldsymbol{k}, \boldsymbol{n})\right) & =\exp \left(\frac{ R}{4} \frac{\partial^2 \mathcal{F}}{\partial a \partial \log (\mathcal{R})} B(\boldsymbol{k}, \boldsymbol{n})\right) \nonumber \\
& =\exp (-2  \mathfrak{a} B(\boldsymbol{k}, \boldsymbol{n})) \exp \left(\frac{ R}{4} \frac{\partial^2 \mathcal{F}_{\text {inst }}}{\partial a \partial \log (\mathcal{R})} B(\boldsymbol{k}, \boldsymbol{n})\right), \\
\exp \left(-\pi i \xi_{II} \boldsymbol{n}^2\right) & =\exp \left(\frac{ R^2}{32}  \frac{\partial^2 \mathcal{F}}{\partial \log (\mathcal{R}) ^2} \boldsymbol{n}^2\right) \nonumber \\
& =\exp \left(\frac{ R^2}{32} \frac{\partial^2 \mathcal{F}_{\text {inst }}}{\partial \log (\mathcal{R}) ^2}\boldsymbol{n}^2\right).
\end{align}
Expanding these expressions first in $\mathcal{R}$ and then in $\mathfrak{a}$ around zero yields series of the form 
\be 
\label{expansion in a & R}
\begin{aligned}
\text{Ser}_{\mathfrak{a}=0}\text{Ser}_{\mathcal{R}=0} \exp(-\pi i \tau \boldsymbol{k}^2) & = 
\sum_{n\geq 0} f^{(1)}_{n}(\mathfrak{a})\left(\frac{\mathcal{R}}{\mathfrak{a}}\right)^{4n-4\boldsymbol{k}^2}, \\
\text{Ser}_{\mathfrak{a}=0} \text{Ser}_{\mathcal{R}=0}\exp(-2\pi i v_I B(\boldsymbol{k},\boldsymbol{n}))&= \sum_{n\geq 0} f^{(2)}_{n}(\mathfrak{a})\left(\frac{\mathcal{R}}{\mathfrak{a}}\right)^{4n},\\
\text{Ser}_{\mathfrak{a}=0} \text{Ser}_{\mathcal{R}=0}\exp(-\pi i \xi_{II} \boldsymbol{n}^2)&= \sum_{n\geq 0}f^{(3)}_{n}(\mathfrak{a}) \left(\frac{\mathcal{R}}{\mathfrak{a}}\right)^{4n+4},
\end{aligned}
\ee
where the dependence of the coefficients $f^{(1)}_{n}(\mathfrak{a})$, $f^{(2)}_{n}(\mathfrak{a})$, $f^{(3)}_{n}(\mathfrak{a})$ on $\boldsymbol{k}$ and $\boldsymbol{n}$ is suppressed. 
The structure of the instanton part of the prepotential \eqref{eq:StructureInstantonExp} implies that these coefficients are formal power series in $\mathfrak{a}$ with rational coefficients. In particular, there are no negative powers of $\mathfrak{a}$. 
Using the expansion \eqref{expansion Z nekra} of the partition function, we express the gravitational couplings as
\begin{equation}
\label{gravitational couplings}
\begin{aligned}
\log (A)= & -\frac{1}{2} \log (\mathcal{R})+\frac{1}{2} \log \left(1-e^{2 \mathfrak{a}}\right)+\frac{1}{4}(-2 \mathfrak{a}+\pi i)+\sum_{n=1}^{\infty} A_n\left(\sinh ^2(\mathfrak{a})\right)\left(\frac{\mathcal{R}}{\sinh (\mathfrak{a})}\right)^{4 n} \\
= & -\frac{1}{2} \log (\mathcal{R})+\frac{1}{2} \log \left(1-e^{2 \mathfrak{a}}\right)+\frac{1}{4}(-2 \mathfrak{a}+\pi i) \\
& -\frac{2+\sinh ^2(\mathfrak{a})}{8\sinh^4 (\mathfrak{a})}\mathcal{R}^4 -\frac{38+33 \sinh ^2(\mathfrak{a})+2 \sinh ^4(\mathfrak{a})}{128\sinh^8 (\mathfrak{a})}\mathcal{R}^8+\mathcal{O}\left(\mathcal{R}^{12}\right), \\
\log (B)= & -\frac{1}{2} \log (\mathcal{R})+\frac{1}{2} \log \left(1-e^{2 \mathfrak{a}}\right)+\frac{1}{4}(-2 \mathfrak{a}+\pi i)
+\sum_{n=1}^{\infty} B_n\left(\sinh ^2(\mathfrak{a})\right)\left(\frac{\mathcal{R}}{\sinh (\mathfrak{a})}\right)^{4 n} \\
= & -\frac{1}{2} \log (\mathcal{R})+\frac{1}{2} \log \left(1-e^{2 \mathfrak{a}}\right)+\frac{1}{4}(-2 \mathfrak{a}+\pi i) \\
& -\frac{3+\sinh ^2(\mathfrak{a})}{8\sinh^4 (\mathfrak{a})}\mathcal{R}^4 -\frac{63+49 \sinh ^2(\mathfrak{a})+2 \sinh ^4(\mathfrak{a})}{128\sinh^8 (\mathfrak{a})}\mathcal{R}^8+\mathcal{O}\left(\mathcal{R}^{12}\right).
\end{aligned}
\end{equation}
Here $A_n (x), B_n (x) \in \mathbb{Q}[x]$. Consequently, the factors $A^\chi$ and $B^\sigma$ in the integrand \eqref{integrand in a} are of the form
\be
\begin{aligned}
& \text{Ser}_{\mathfrak{a}=0} \text{Ser}_{\mathcal{R}=0} A^\chi=\sum_{n \geq 0} f_n^{(4)}(\mathfrak{a} )\left(\frac{\mathcal{R}}{\mathfrak{a}}\right)^{4 n- \frac12 \chi}, \\
& \text{Ser}_{\mathfrak{a}=0} \text{Ser}_{\mathcal{R}=0} B^\sigma=\sum_{n \geq 0} f_n^{(5)}(\mathfrak{a} )\left(\frac{\mathcal{R}}{\mathfrak{a}}\right)^{4 n- \frac12 \sigma},
\end{aligned}
\ee
where $f_n^{(4)}(\mathfrak{a}), f_n^{(5)}(\mathfrak{a}) \in \mathbb{C}[\mathfrak{a}]$ implicitly depend on $\chi$ and $\sigma$, respectively. Combining all these components, we obtain
\be
\label{integrand expansion}
\text{Ser}_{\mathfrak{a}=0} \text{Ser}_{\mathcal{R}=0}\Delta(\mathfrak{a},\mathcal{R})=  \sum_{n\geq 0} f_n^{(6)}(\mathfrak{a})\left(\frac{\mathcal{R}}{\mathfrak{a}}\right)^{4n+2 - 4 \boldsymbol{k}^2},
\ee
where $f^{(6)}_n(\mathfrak{a})\in \mathbb{Q}[\mathfrak{a}]$ and depends on $(\boldsymbol{k},\boldsymbol{n},\boldsymbol{\mu},J,\chi,\sigma)$.
Here we use again $\chi + \sigma = 4$. 

We now introduce the following lemma to reorganize a double series: Consider a formal series
\be
F(\mathfrak{a}, \mathcal{R})=\sum_{n \geq 0} \sum_{l \geq 0} F_{n l} \mathfrak{a}^l\left(\frac{\mathcal{R}}{\mathfrak{a}}\right)^{4 n-m},
\ee
where $F_{n l}$ and $m$ are some constants. This can be rewritten by regrouping powers of $\CR$,
\be
\label{lemma 1}
F(\mathfrak{a}, \mathcal{R})=\sum_{l \geq 0} \sum_{n \geq 0} F_{n l}\left(\frac{\mathcal{R}}{\mathfrak{a}}\right)^{4 n-m-l} \mathcal{R}^l :=\tilde{F}\left(\frac{\mathcal{R}}{\mathfrak{a}}, \mathcal{R}\right),
\ee
where $\tilde{F}(x, \mathcal{R}) \in \mathbb{Q}(x)[\mathcal{R}]$.
Applying this lemma to Eq. \eqref{integrand expansion} with $m=-2+4 \boldsymbol{k}^2$, we get
\be
\Delta(\mathfrak{a}, \mathcal{R})=\tilde\Delta\left(\frac{\mathcal{R}}{\mathfrak{a}}, \mathcal{R}\right) \in \mathbb{Q}\left(\frac{\mathcal{R}}{\mathfrak{a}}\right)[\mathcal{R}].
\ee
The residue we need to compute transforms under this regrouping as
\be
\text{Coeff}_{\mathfrak{a}^0} \Delta(\mathfrak{a}, \mathcal{R}) \mathfrak{a}=\text{Coeff}_{\left(\frac{\mathcal{R}}{\mathfrak{a}}\right)^0} \tilde\Delta\left(\frac{\mathcal{R}}{\mathfrak{a}}, \mathcal{R}\right)\left(\frac{\mathfrak{a}}{\mathcal{R}}\right) \mathcal{R}.
\ee

We now change the variable from $a$ to $q$. The expansion of $q^{1/8}$ is given by 
\be 
q^{\frac18}(\mathfrak{a},\mathcal{R})=\text{Ser}_{a=0}\text{Ser}_{\mathcal{R}=0} \exp\left(\frac{\pi i \tau}{4}\right)  = 
\sum_{n\geq 0} f^{(7)}_n(\mathfrak{a})\left(\frac{\mathcal{R}}{\mathfrak{a}}\right)^{4n+1},
\ee
where $f^{(7)}_n(\mathfrak{a}) \in \mathbb{Q}[\mathfrak{a}]$.  From Eq. \eqref{lemma 1}, we can rewrite $q^{1 / 8}(\mathfrak{a}, \mathcal{R})$ as $q^{1 / 8}(\mathcal{R} / \mathfrak{a}, \mathcal{R}) \in \mathbb{Q}\left(\mathcal{R}/{\mathfrak{a}}\right)[\mathcal{R}]$. Explicitly, to low orders,
\be
q^{\frac18}\left(\frac{\mathcal{R}}{\mathfrak{a}}, \mathcal{R}\right)=\left(-\frac{i}{2} \frac{\mathcal{R}}{\mathfrak{a}}-\frac{3 i}{16}\left(\frac{\mathcal{R}}{\mathfrak{a}}\right)^5
-\frac{123 i}{512} 
\left(
\frac{\mathcal{R}}{\mathfrak{a}}
\right)^9
+\cdots\right)+\left(\frac{i\mathfrak{a}}{12\mathcal{R}} +\cdots\right) \mathcal{R}^2+\cdots.
\ee
The inverse relation is
\be
\label{eq:aq}
-\frac{i}{2} \frac{\mathcal{R}}{\mathfrak{a}}\left(q^{\frac18}, \mathcal{R}\right)=\left(q^{\frac18}-6 q^{\frac58}+\ldots\right)+\left(
-\frac{1}{24} q^{-\frac18} 
+ \frac{3}{4} q^{\frac38}
+\cdots\right) \mathcal{R}^2+\cdots.
\ee
The invariance of the residue under this change of variable is ensured by \cite[Lemma 4.12]{Gottsche:2006bm}: For a change of variable from $x$ to $y$ given by 
\be
y=y(x, \mathcal{R})=y_0(x)+y_1(x) \mathcal{R}+\cdots \in \mathbb{C}(x)[\mathcal{R}],
\ee
with $y_0(x)=x+a_2 x^2+\cdots$, and a function $f(y, \mathcal{R}) \in \mathbb{C}(y)[\mathcal{R}]$, we have
\be
\label{lemma 4.12}
\text{Coeff}_{y^0}\left[y f(y, \mathcal{R})\right]=\text{Coeff}_{x^0}\left[x f(y(x, \mathcal{R}), \mathcal{R}) \frac{\mathrm{d} y}{\mathrm{d} x}\right].
\ee
Applying this formula to our case with $y= -i\mathcal{R}/(2\mathfrak{a})$, 
$x = q^{1/8}$, and $f(y,\mathcal{R}) \to \tilde\Delta(\mathcal{R}/\mathfrak{a},\mathcal{R})$,
the residue formula becomes 
\bea \label{eq:residuetransform}
\text{Coeff}_{\left(\frac{\mathcal{R}}{\mathfrak{a}}\right)^0}\left[\tilde\Delta\left(\frac{\mathcal{R}}{\mathfrak{a}},\mathcal{R}\right) \left(\frac{\mathfrak{a}}{\mathcal{R}}\right)\mathcal{R}\right] = 
\text{Coeff}_{q^0}\left[q^{\frac18}\Delta(q^{\frac18},\mathcal{R}) \left(\frac{\mathfrak{a}}{\mathcal{R}}\right)^2\mathcal{R} \frac{\mathrm{d}(\mathcal{R}/{\mathfrak{a}}) }{\mathrm{d}(q^{\frac18})}\right].
\eea
The computation of the right-hand side of Eq. \eqref{eq:residuetransform} relies on the following identities:
\begin{align}
\left.\exp\left[-\pi i \tilde\tau\left(\frac{\mathcal{R}}{\mathfrak{a}},\mathcal{R}\right) \boldsymbol{k}^2\right]\right|_{\frac{\mathcal{R}}{\mathfrak{a}}=\frac{\mathcal{R}}{\mathfrak{a}}(q^{\frac18},\mathcal{R})}&=  q^{-\frac{1}{2} \boldsymbol{k}^2}, \label{eq:identity1}\\
\left.\exp\left[-2\pi i \tilde v_I\left(\frac{\mathcal{R}}{\mathfrak{a}},\mathcal{R}\right)B(\boldsymbol{k},\boldsymbol{n})\right]\right|_{\frac{\mathcal{R}}{\mathfrak{a}}=\frac{\mathcal{R}}{\mathfrak{a}}(q^{\frac18},\mathcal{R})} &= \text{Ser}_{q}\text{Ser}_{\mathcal{R}}\left[ \exp(-2\pi i v_I B(\boldsymbol{k},\boldsymbol{n}))\right], \label{eq:identity2}\\
\left.\exp\left[- \pi i \tilde\xi_{II} \left(\frac{\mathcal{R}}{\mathfrak{a}},\mathcal{R}\right)
\boldsymbol{n}^2\right]\right|_{\frac{\mathcal{R}}{\mathfrak{a}}=\frac{\mathcal{R}}{\mathfrak{a}}(q^{\frac18},\mathcal{R})} &=\text{Ser}_{q}\text{Ser}_{\mathcal{R}} \left[C_{II}(\tau,\mathcal{R})^{\boldsymbol{n}^2}\right]. \label{eq:identity3}
\end{align}
The left-hand side of these identities are evaluated by first applying \eqref{lemma 1} to \eqref{expansion in a & R} and then substituting the formal expansion $\frac{\mathcal{R}}{\mathfrak{a}}\left(q^{1 / 8}, \mathcal{R}\right)$, while the right-hand side of Eq. \eqref{eq:identity2} and Eq. \eqref{eq:identity3} are calculated directly using Eq. \eqref{vR} and Eq. \eqref{CRexp}, respectively. 
Although not \emph{a priori} obvious, Eqs \eqref{eq:identity1}, \eqref{eq:identity2} and \eqref{eq:identity3} are proven to all orders in \cite{Gottsche:2006bm}.

The other components of the measure can be expressed in a similar way \cite[P. 41 and 44]{Gottsche:2006bm},
\begin{equation}
\left.A\left(\frac{\mathcal{R}}{\mathfrak{a}}, \mathcal{R}\right)\right|_{\frac{\mathcal{R}}{\mathfrak{a}} =
\frac{\mathcal{R}}{\mathfrak{a}}(q^{\frac18},\mathcal{R})} 
= i \left(
\frac{2}{\vartheta_2(\tau)\vartheta_3(\tau)}
\right)^{\frac12},
\end{equation}
\begin{equation}
\left.B\left(\frac{\mathcal{R}}{\mathfrak{a}}, \mathcal{R}\right)\right|_{\frac{\mathcal{R}}{\mathfrak{a}} =
\frac{\mathcal{R}}{\mathfrak{a}}(q^{\frac18},\mathcal{R})} 
= 
\frac{\sqrt{2}i\vartheta_4(\tau)}{(\vartheta_2(\tau)\vartheta_3(\tau))^{\frac12}},
\end{equation}
\begin{equation}
\left. q^{\frac18} \left(
    \frac{\mathfrak{a}}{\mathcal{R}}
    \right)^2
    \mathcal{R}
    \frac{\mathrm{d}(\mathcal{R}/\mathfrak{a})}{\mathrm{d}q^{\frac18}}
    \right|_{\frac{\mathcal{R}}{\mathfrak{a}} =
\frac{\mathcal{R}}{\mathfrak{a}}(q^{\frac18},\mathcal{R})} 
= \text{Ser}_q \text{Ser}_\mathcal{R}
\left[
\frac{-i \mathcal{R}}{\sqrt{-8 \mathcal{R}^2 \ttu+ 4 \mathcal{R}^4 +4}} 
\frac{\vartheta_4(\tau)^9}{\eta(\tau)^3}
    \right]
\end{equation} 
Combining all these ingredients, using the condition $\sigma+\chi =4$, and choosing the constant 
$K_{\Omega} = -{R}/{4\pi^2}$, the wall-crossing formula becomes
\be\label{eq:WCLocalization}
Z^{J}_{\bfmu,\bfn} - Z^{J'}_{\bfmu,\bfn} =
4 K_{U,\boldsymbol{\mu},\boldsymbol{n}} \text{Coeff}_{q^0} \text{Ser}_{\mathcal{R}}\left[\nu_R(\tau)\, C_{II}^{\boldsymbol{n}^2}\, \Theta_{\boldsymbol{\mu}}^{J J^{\prime}}(\tau, \boldsymbol{n} v)\right],
\ee
with $\Theta^{JJ'}_\bfmu$ defined in Eq.  \eqref{TJJ'} with $K=K_X$. This successfully reproduces the wall-crossing formula \eqref{eq:Physical-WC}. 

\subsection{Puzzle} \label{sec: puzzle}

As discussed in Sec.  \ref{subsubsec:g-function-singularities-contours}, the full partition function in a fixed chamber is decomposed into the contributions from the bulk and the boundary contours in the $a$-cylinder,
\be
Z_{\bfmu, \bfn}^{\e_1,\e_2}(\CR) = Z_{\bfmu, \bfn}^{\e_1,\e_2,\text{bulk}}(\CR) + Z_{\bfmu, \bfn}^{\e_1,\e_2,\text{bdry}}(\CR).
\ee
Although the precise contours for the zero-mode integrals, especially for the bulk contribution, remain undetermined, we can formulate a few natural conjectures for the final result, motivated by analogous computations in lower-dimensional settings, which we briefly explore in this section.

First, from the structure of the formal bosonic zero-mode integral \eqref{total D loc} which universally appears in computations of the twisted indices defined on $X_{d-1}\times S^1$ preserving the 1d $\CN=(0,2)$ subalgebra \cite{Benini:2013nda,Hori:2014tda,Closset:2015rna,Benini:2015noa}, a natural expectation is that the bosonic contour integral reduces to a Jeffrey-Kirwan (JK) residue prescription \cite{Jeffrey:1993cun,brion1999arrangement} applied to the holomorphic integrand. This approach has also been anticipated in recent studies of 5d twisted indices \cite{Hosseini:2018uzp}.

The JK residue prescription requires a real auxiliary parameter as input, which can be identified with the Fayet-Iliopoulos parameter of the effective $\CN=(0,2)$ SQM. This parameter can be extracted from the $h$-linear term of the 5d action evaluated on the zero modes $V_0$ in the asymptotic regions of the $a$-cylinder. See the term proportional to $\nu_2$ in Eq. \eqref{eq:Linear-with-tilde-g}. From Eqs.   \eqref{eq:Linear-with-tilde-g} and \eqref{asymptotic contour loc limit}, we conjecture that for $\bfn=0$, this parameter corresponds to $B(\bfk,J)$. Following closely the procedures of  \cite{Benini:2013nda,Hori:2014tda,Closset:2015rna,Benini:2015noa}, we are led to the expression
\be\label{JK contour}
Z_{\bfmu, 0}^{\e_1,\e_2}(\CR) = -2\pi^2 K_{\Omega} \left[\sum_{B(\bfk,J)>0}  \sum_{a_*\in S_+} - \sum_{B(\bfk,J)<0}\sum_{a_*\in S_-}\right] \underset{a = a_*}{\text{Res}} \mathrm{d}a \, g_{\mathfrak{p}, 0}^{\e_1,\e_2} 
(a),
\ee
where $S_\pm$ denote the sets of singularities of the meromorphic function $g_{\mathfrak{p}, 0}^{\e_1,\e_2}(a)$ \eqref{a cylinder sing} corresponding to $\alpha=\pm 2$, respectively. A crucial requirement for this prescription is that the singularity structure of $g_{\mathfrak p, \mathfrak p^{(I)}}^{\e_1,\e_2}(a)$ must be projective, namely the poles in $S_+$ and $S_-$ never coincide. This condition is satisfied for $X=\mathbb{CP}^2$. See, e.g., \cite{Benini:2013nda} for more detail. Unfortunately, as we will soon see, Eq. 
\eqref{JK contour} cannot be correct. 

An alternative approach, motivated by analogous computations of 4d equivariant twisted partition functions in \cite{Bonelli:2020xps,Gottsche:2006tn}, suggests a somewhat different contour prescription,
\be\label{stability residues}
Z_{\bfmu, 0}^{\e_1,\e_2}(\CR) = -\frac{R}{2\pi i } \sum_{{\frak p}} \Theta_\mu({\frak p}) \oint_{2a=0} \mathrm{d}a \,g^{\e_1,\e_2}_{{\frak p},0}(a),
\ee
where the sum is over $\chi$-tuples of equivariant fluxes $\mathfrak{p}$ satisfying a set of inequalities determined by a characteristic function $\Theta(\mathfrak{p})$, and the integral is taken along a contour encircling the pole at $2a=0$.
The function $\Theta(\mathfrak{p})=\Theta_\bfmu(\mathfrak{p};J)$, which depends on the period point $J$ and the 't Hooft flux $\bfmu$, is mathematically related to stability conditions on holomorphic bundles. This relation is natural, since the contour integral around $2a=0$ localizes onto instantons corresponding to semi-stable bundles under the Donaldson-Uhlenbeck-Yau theorem. The support of $\Theta$ can be determined using toric geometry techniques, in particular for $\mathbb{CP}^2$ in \cite{Klyachko1991}, and for Hirzebruch surfaces $\mathbb{F}_n$ including the wall-crossing dependence on $J$ in  \cite{Kool_2014}. 

We will evaluate Eq. \eqref{stability residues}, or equivalently Eq. \eqref{ZesP2} below, for $X=\mathbb{CP}^2$ in Sec.  \ref{sec:equivariantKtheoreticP2}, and show that it satisfies several nontrivial consistency checks:
\begin{enumerate} 
\item Its non-equivariant limit reduces to the index computed via direct evaluation of the $U$-plane integral in Sec.  \ref{Sec:DEvalU}.
\item The series expansion contains only nonnegative powers of $\CR$, with coefficients given by Laurent polynomials in $t_i= e^{R \epsilon_i}$.
\item With finite exceptions, the coefficients in these Laurent polynomials are integers.
\end{enumerate} 
These properties align with the expected formulae \eqref{eq:Zee-QM-Equiv} and \eqref{eq:dChar}. Moreover, a 4d analog of Eq. \eqref{ZesP2} was rigorously derived in \cite{Gottsche:2006tn}.
\footnote{A path integral derivation of the result of \cite{Gottsche:2006tn} has been explored in \cite{Bonelli:2020xps}.}

Unfortunately, as shown in App. \ref{app:NekPartProps}, the expressions \eqref{JK contour} and \eqref{stability residues} do not agree. 
For $X=\mathbb{CP}^2$ and $J= D_1 + D_2 + D_3$, we obtain from Eq. \eqref{eq: residue-pro1} that
\begin{equation}\label{eq:SumAllBundles}
    Z_{\mu, 0}^{\epsilon_1, \epsilon_2}(\mathcal{R}) = 
    4\pi^2 K_{\Omega} 
    \sum_{c=0, \frac{\pi i}{R}}\left(\sum_{\{\mathfrak{p}\} \in S_{\text {unstable}}}
    +\sum_{\{\mathfrak{p}\} \in S_{\text {semi-stable}}}
    +2 \sum_{\{\mathfrak{p}\} \in S_{\text {stable}}}\right) 
    \underset{a=c}{\text{Res}} ~
    \mathrm{d} a \, g_{\mathfrak{p}, \mathfrak{p}^{(I)}=0}^{\epsilon_1, \epsilon_2}(a),
\end{equation}
while for $J = - D_1 - D_2-D_3$, Eq. \eqref{eq: residue-pro2} yields
we get 
\begin{equation}\label{eq:SumUnstableBundles}
    Z_{\mu, 0}^{\epsilon_1, \epsilon_2}(\mathcal{R}) = 
    -4\pi^2 K_{\Omega} \sum_{c=0,\frac{\pi i}{R}} \sum_{\{\mathfrak{p}\} \in S_{\text{unstable }}} 
    \underset{a=c}{\text{Res}} 
    ~\mathrm{d} a\, g_{\mathfrak{p}, \mathfrak{p}^{(I)}=0}^{\epsilon_1, \epsilon_2}(a).
\end{equation}
Hence, the result depends on the choice of $J$, and for both cases, it contains contributions from unstable fluxes, leading to a formal series in $\mathcal R$ with arbitrary negative powers. 
This suggests that the conventional localization arguments developed in the literature for lower-dimensional cases, which lead to the expression \eqref{JK contour}, must be carefully revisited in our setting.

The derivation of Eq. \eqref{JK contour} relies on several assumptions inspired by computations in lower-dimensional analogues, including:
\begin{enumerate}
\item  The specific choice of the $\bar\CQ$-exact action \eqref{eq:5d-Action-Toric-With-Epsilon} is different from alternatives such as that in \cite{Bonelli:2020xps}. Due to the BRST anomaly underlying the wall-crossing, different choices of the $\bar\CQ$-exact terms in the action may lead to different BPS equations, which correspond directly to different stability conditions in the algebraic description of the moduli space, and finally lead to different results for the partition function in a fixed chamber.
\item  The analytic structure of the non-holomorphic integrand $g_{\mathfrak{p},\mathfrak{p}^{(I)}}^{\e_1,\e_2} (a,\bar a, \lambda_0, \bar\lambda_0,h)$, which is central to the JK residue prescription in lower-dimensional computations \cite{Benini:2013nda,Hori:2014tda,Closset:2015rna,Benini:2015noa}, remains to be clarified in our case.
Related to this is the subtle definition of the zero-mode integral \eqref{partition ftn zero}. The path integral in the vicinity of the singularities of the integrand requires careful treatment of various limits. We have assumed that the correct result is obtained by taking various potentially delicate limits ($g^2\rightarrow 0, \delta\rightarrow 0, \eta\rightarrow 0)$ in a particular order, motivated by lower-dimensional analogues. However, the validity of this approach here needs justification.
\item The assumption that the solutions to the BPS equations involve only connections pulled back from $X$ (above Eq. \eqref{eq:iota-v-F4}) was not derived from the BPS equations. It is conceivable that the true BPS locus is larger than what we have considered. 
\end{enumerate}

We leave a complete path integral derivation of the expression \eqref{stability residues} for future work. We now proceed to explicit evaluations of this expression.

\subsection{Equivariant Localization For $\mathbb{CP}^2$}
\label{sec:equivariantKtheoreticP2}
In this subsection, we perform the explicit evaluation of the expression \eqref{stability residues} for the specific case of $X= \mathbb{CP}^2$ and gauge group $\mathrm{SU}(2)$. The methodology can, in principle, be extended to other toric surfaces and gauge groups. This work extends previous evaluations of partition functions for other supersymmetric theories \cite{nekrasov2006localizing, Gasparim:2009sns, Bershtein:2016mxz, Bershtein:2015xfa, Bonelli:2020xps} and computations of equivariant Donaldson invariants \cite{Gottsche:2006tn}. 
We will first discuss the evaluation for a vanishing background flux $n_I$, and later include this coupling.

The toric surface $\mathbb{CP}^2$ has three canonical patches, which we label by $\ell=1,2,3$. The equivariant parameters $\e_1^{(\ell)},\e_2^{(\ell)}$ in each patch are given in Table \ref{P2Eps}. Further aspects of the toric geometry of $\mathbb{CP}^2$ are described in App. \ref{sec:Toric Geometry}, starting from Eq. \eqref{eq:CP2fan}. 

\begin{table}[t]
\centering
\renewcommand{\arraystretch}{2.2}
\begin{tabular}{|c|c|c|c|}
\hline 
$\ell$ & 1 & 2 & 3 \\
\hline 
$\epsilon_1^{(\ell)}$ & $\e_1$ & $\e_2-\e_1$ & $-\e_2$\\
\hline 
$\epsilon_2^{(\ell)}$ & $\e_2$ & $-\e_1$ & $\e_1-\e_2$  \\
\hline 
\end{tabular}
\caption{Equivariant parameters in three patches of $\mathbb{CP}^2$.}
\label{P2Eps}
\end{table} 

Motivated by the discussions in the previous section, we consider
\be 
\label{ZesP2}
\begin{split}
Z_{\mu, 0}^{\mathbb{CP}^2\times S^1}(\e_1,\e_2,\CR)
&=- R\sum_{{\frak p}} \Theta_\mu({\frak p}) \underset{a=0,\frac{\pi i}{R}}{\text{Res}}\left[\mathrm{d}a\,g^{\e_1,\e_2}_{{\frak p},0}(a)\right]\\
&=-\frac{1}{\Lambda}\sum_{{\frak p}} \Theta_\mu({\frak p})\,\underset{a=0,\frac{\pi i}{R}}{\text{Res}}\left[ \mathrm{d}a \prod_{\ell=1,2,3} \CZ(a^{(\ell)},\epsilon_1^{(\ell)},\epsilon_2^{(\ell)},\Lambda,R)\right],
\end{split}
\ee
where the factor $-1/\Lambda$ is determined by comparison with the prefactor of the wall-crossing formula (\ref{eq:wc-contour}), and $\mathfrak{p}$ is a triplet of integers satisfying $\sum_{\ell} {\mathfrak{p}}_{\ell}=2\bfmu \mod H^2(\mathbb{CP}^2,2\mathbb{Z})$. 
The contour consists of two small circles around $a=0$ and $a=\pi i/R$ with radius $r\ll |m\epsilon_1+n\epsilon_2|$ for $m,n\in \mathbb{Z}$. Here $|m|$, $|n|$ are bounded for each given instanton charge $k$, such that $r$ can always be chosen sufficiently small assuming $|\eps_1|\neq |\eps_2|$. 
 
The set $\{{\frak p}\mid \Theta_\mu({\frak p}) \neq 0\}$ corresponds to the fixed points of the moduli space of semi-stable vector bundles under the toric action \cite{Gottsche:2006tn}. These are characterized by all ${\frak p}_{\ell}$ being positive and satisfying the triangle inequalities \cite{Klyachko1991}. Following \cite{Bershtein:2015xfa} \cite[Sec.   5.2]{Bonelli:2020xps}, we take $\Theta_\mu(\mathfrak{p})$ for $\mathfrak{p}=\{\mathfrak{p}_1,\mathfrak{p}_2,\mathfrak{p}_3\}$ to be
\be
\Theta_\mu({\frak p})=\left\{ \begin{array}{ll} 1, & \qquad \forall \ell,  \quad \mathfrak{p}_{\ell}>0,\quad \mathfrak{p}_{\ell}+\mathfrak{p}_{\ell+1} > \mathfrak{p}_{\ell+2}, \\
& \qquad \text{and } \sum_{\ell} {\frak p}_{\ell}=2\mu \mod 2,\\
\frac{1}{2}, & \qquad  \forall \ell, \quad  \mathfrak{p}_{\ell}>0,\quad \mathfrak{p}_{\ell}+\mathfrak{p}_{\ell+1} \geq  \mathfrak{p}_{\ell+2}, \\
& \qquad \exists \ell,\quad \mathfrak{p}_{\ell}+\mathfrak{p}_{\ell+1}=\mathfrak{p}_{\ell+2}, 
 \,\, \text{and}\,\,\sum_{\ell} \mathfrak{p}_{\ell}=2\mu \mod 2,\\
0, & \qquad \text{otherwise},\end{array}   \right.
\ee 
where the indices $\ell$ of $\mathfrak{p}$ are taken cyclically. Three-tuples $\mathfrak{p}$ with $\Theta_\mu(\mathfrak{p})=1$ correspond to slope stable bundles, while those with $\Theta_\mu(\mathfrak{p})=1/2$ correspond to strictly semi-stable bundles. The factor $1/2$ accounts for their larger automorphism group and can also be derived from a contour analysis \cite{Bonelli:2020xps}. Note that strictly semi-stable bundles necessarily have $\mu=0 \mod 1$.

The product of the perturbative part of Nekrasov partition function then simplifies to
\be 
\label{eq: perturbative of Nekrasov}
\begin{split}
& Z_{\mu, 0,{\rm pert}}^{\mathbb{CP}^2\times S^1}(a,\e_1,\e_2,\Lambda,R,{\frak p}) \\
=&\prod_{\ell=1,2,3} \exp\left[-\tilde \gamma_{\epsilon_1^{(\ell)},\epsilon_2^{(\ell)}}(2a^{(\ell)}|R,\Lambda)-\tilde \gamma_{\epsilon_1^{(\ell)},\epsilon_2^{(\ell)}}(-2a^{(\ell)}|R,\Lambda)\right]\\
=&\CR^{-2-(\sum_{\ell} {\frak p}_{\ell})^2} \exp\left\{\frac{1}{2}({\frak p}_1+{\frak p}_2+{\frak p}_3)\left[6a+(2\e_1-\e_2){\frak p}_1+(2\e_2-\e_1){\frak p}_2-(\e_1+\e_2){\frak p}_3\right]R\right\}\\
& \times \exp\left( -\sum_{\ell=1,2,3} \sum_{n=1}^\infty \frac{1}{n} \frac{e^{-2R n a^{(\ell)}}+e^{2R n a^{(\ell)}}}{(1-e^{Rn \epsilon_1^{(\ell)}})(1-e^{Rn \epsilon_2^{(\ell)}})} \right).
\end{split}
\ee 

For 4d SYM theory, the product of the perturbative contribution over the patches of a compact toric four-manifold is known to evaluate to a product with a finite number of terms \cite{Bershtein:2015xfa}. We find that this property also holds in the K-theoretic setting. For ${\frak p}_{\ell} >0$, we evaluate the last line of Eq. \eqref{eq: perturbative of Nekrasov} as
\be 
\label{Z1loopP2}
\begin{aligned} 
&\exp\left( -\sum_{\ell=1,2,3} \sum_{n=1}^\infty \frac{1}{n} \frac{e^{-2Rna^{(\ell)}}+e^{2Rna^{(\ell)}}}{(1-e^{Rn \epsilon_1^{(\ell)}})(1-e^{Rn \epsilon_2^{(\ell)}})} \right)\\
=& \prod_{(i,j)\in \IN^2 \atop i+j\leq \sum_{\ell} {\frak p}_{\ell}}
 \left\{1 - \exp\left[-R(2a+({\frak p}_1-j)\e_1+ ({\frak p}_2-i)\e_2 \right]\right\}\\ 
& \times \prod_{(i,j)\in \IN^2 \atop i+j\leq \sum_{\ell} {\frak p}_{\ell}-3}
\left\{1 - \exp\left[R(2a+({\frak p}_1-1-j)\e_1+ ({\frak p}_2-1-i)\e_2 )\right]\right\}.
\end{aligned}
\ee 
This reduces to the 4d case \cite{Bershtein:2015xfa} in the limit $R\to 0$. We note that Eq.  (\ref{Z1loopP2}) has a double zero at $a=0$, since the first product contains $1-e^{-2Ra}$ and the second contains $1-e^{2Ra}$. We define the product $P^{\mathbb{CP}^2\times S^1}$ with these two terms omitted and evaluated at $a=0$,
\be 
\begin{split}
P^{\mathbb{CP}^2\times S^1}(\e_1,\e_2,R,{\frak p})&=\prod_{(i,j)\in \IN^2 \setminus \{({\frak p}_2,{\frak p}_1)\} \atop i+j\leq \sum_{\ell} {\frak p}_{\ell}} 
 \left\{1 - \exp\left[-R(({\frak p}_1-j)\e_1+ ({\frak p}_2-i)\e_2 \right]\right\}
\\ 
&\quad \times 
\prod_{(i,j)\in \IN^2 \setminus \{({\frak p}_2-1,{\frak p}_1-1)\} \atop i+j\leq \sum_{\ell} {\frak p}_{\ell}-3}
\left\{1 - \exp\left[R(({\frak p}_1-1-j)\e_1+ ({\frak p}_2-1-i)\e_2 )\right]\right\}.
\end{split}
\ee 

We then consider the product of the instanton partition function, $Z_{\mu, 0,{\rm inst}}^{\mathbb{CP}^2\times S^1}$, in the absence of the background flux $n_I$. From the recursion formula (\ref{ZinstRecursion}), it is clear that $Z_{\mu, 0,{\rm inst}}^{\mathbb{CP}^2\times S^1}(a^{(\ell)}, \e_1^{(\ell)}, \e_2^{(\ell)}, \Lambda,R)$ has a single pole at $a=0, \pi i/R \mod 2\pi i/R$ for ${\mathfrak{p}}_{\ell}, {\mathfrak{p}}_{\ell+1}>0$.
Since $Z^{\mathbb{CP}^2\times S^1}_{\rm pert}$ has a double zero at $a=0 , \pi i/R \mod 2\pi i/R$, the non-vanishing contribution to Eq.  (\ref{ZesP2}) comes from terms with the pole in the instanton part for each patch. Consequently, we arrive at
\be 
\label{Zes0}
\begin{split}
&Z_{\mu, 0}^{\mathbb{CP}^2\times S^1}(\e_1,\e_2,\CR)  \\
=&\frac{1}{\Lambda}\left( 1 + (-1)^{6\mu} \right)
\sum_{{\frak p}} \Theta_\mu({\frak p}) \CR^{-2+Q(\mathfrak{p})} \exp\left(R\sum_{\ell,\ell'} \mathfrak{p}_{\ell}  a^{(\ell')}_+\right) \\   
&\times P^{\mathbb{CP}^2\times S^1}(\e_1, \e_2, R,\mathfrak{p}) \prod_{\ell=1,2,3} \frac{T_{\mathfrak{p}_{\ell},\mathfrak{p}_{\ell+1}}(t_1^{(\ell)},t_2^{(\ell)})}{e^{-2a^{(\ell)}_+}-e^{2a^{(\ell)}_+}} 
\CZ_{{\rm inst}}(a^{(\ell)}_-, \e_1^{(\ell)},\e_2^{(\ell)}, \Lambda,R),
\end{split}
\ee  
where we have introduced the indefinite quadratic form,
\be
Q(\mathfrak{p})=2{\frak p}_1{\frak p}_2+2{\frak p}_2{\frak p}_3+2{\frak p}_1{\frak p}_3-{\frak p}_1^2-{\frak p}_2^2-{\frak p}_3^2,
\ee
and defined
\be
a^{(\ell)}_{\pm}=\frac{1}{2}({\mathfrak{p}}_{\ell} \epsilon_1^{(\ell)} \pm {\mathfrak{p}}_{\ell+1}\epsilon_2^{(\ell)}).
\ee

Finally, we incorporate the background flux $n_I^{(\ell)}$ (\ref{eq:nIl}) for each patch by including the factor \eqref{eq:U(1)-factor}. This is implemented by substituting (\ref{eq:ZinstnI}) for $\CZ_{\rm inst}$ in \eqref{Zes0}, yielding
\be 
\label{ZesnI}
\begin{split}
&Z^{\mathbb{CP}^2\times S^1}_{\mu,\mathfrak{p}^{(I)}}(\e_1,\e_2,\CR) \\
=&\frac{1+ (-1)^{2\mu(n_I +3)}}
{\Lambda} 
\sum_{{\frak p}} \Theta_\mu({\frak p})\, \CR^{-2+Q({\frak p})}\,\exp\left(R\sum_{\ell,\ell'} \mathfrak{p}_{\ell}  a^{(\ell')}_+\right)\,P^{\mathbb{CP}^2\times S^1}(\e_1,\e_2,R,{\frak p})  \\
& \times \prod_{\ell=1,2,3} 
\frac{T_{{\mathfrak{p}}_{\ell},{\mathfrak{p}}_{\ell+1}}(t_1^{(\ell)},t_2^{(\ell)})}{e^{-2a^{(\ell)}_+}-e^{2a^{(\ell)}_+}}  \exp\left(\frac{n_I^{(\ell)}(a^{(\ell)}_{-})^2}{\epsilon_1^{(\ell)} \epsilon_2^{(\ell)}} \right) \CZ_{\rm inst}(a^{(\ell)}_{-},\epsilon_1^{(\ell)},\epsilon_2^{(\ell)},\Lambda\, e^{-\frac{1}{4}n_I^{(\ell)}},R),
\end{split}
\ee 
where $n_I=\sum_{\ell} \mathfrak{p}_{\ell}^{(I)}$.

We now proceed to the explicit evaluation of the first few terms in the $\CR$-expansion of Eq. (\ref{ZesnI}) for specific values of $n_I^{(\ell)}$, or equivalently $\mathfrak{p}_{\ell}^{(I)}$. We first consider the case $\mu=1/2$, for which there are no strictly semi-stable bundles. The results are presented in Table \ref{P2eqKhalf}. In agreement with the general form of the partition function, Eqs. (\ref{eq:Zee-QM-Equiv}) and (\ref{eq:dChar}), the coefficients in the $\CR$-expansion are Laurent polynomials in $t_1=e^{R\eps_1}$ and $t_2=e^{R\eps_2}$, up to an overall fractional power. This is highly nontrivial from the perspective of the explicit expression in Eq.  (\ref{ZesnI}), which involves many terms in the denominators. Moreover, in the non-equivariant limit, the terms in Table \ref{P2eqKhalf} reduce to the corresponding terms of $\Phi_{1/2,n_I}$ \eqref{PhiU12}, up to an overall sign, 
\footnote{We attribute the overall sign difference to the use of a different characteristic vector $K$ in the two calculations. As discussed in Sec.  \ref{sec:non-equivlimit}, equivariant localization gives rise to the characteristic vector $K=K_X$, while $K=-K_{\mathbb{CP}^2}$ is used in Sec.   \ref{sec:evaluate}. This results in a relative sign $(-1)^{2\mu}$ between the non-equivariant limit of $Z_{\mu,{\frak p}^{(I)}}^{\mathbb{CP}^2\times S^1}$ and $\Phi_{\mu,n_I}$.}
as verified up to $\CO(\CR^{16})$ for $|n_I| \leq 9$. Moreover, we find experimentally that the $\CR^0$ term is given by 
\be
t_1^{\frac{1}{4}\left(2{\frak p}^{(I)}_1-{\frak  p}^{(I)}_2-{\frak p}^{(I)}_3\right)} t_2^{\frac{1}{4}\left(-{\frak p}^{(I)}_1+2{\frak  p}^{(I)}_2-{\frak p}^{(I)}_3\right)}.
\ee

The results for $\mu = 0$ are presented in Table \ref{P2eqK0}. As mentioned above, the enumeration of fixed points is more subtle for $\mu=0$ due to the presence of strictly semi-stable bundles, which are weighted by $\Theta_0(\mathfrak{p})=1/2$. We have checked that the results in Table \ref{P2eqK0} agree in the non-equivariant limit with the terms in $\Phi_{0,n_I}$ (\ref{PhiP20}) up to $\mathcal{O}(\mathcal{R}^{17})$ for $|n_I|\leq 6$. 

The Tables \ref{P2eqKhalf} and \ref{P2eqK0} demonstrate that the $\e_j$-dependence is sensitive to the choice of $\{\mathfrak{p}^{(I)}_l\}$ for a fixed $\sum_{\ell} \mathfrak{p}^{(I)}_{\ell}$. Furthermore, we conjecture that the partition function satisfies the following ``reflection'' property, which is checked up to $\mathcal{O}(\mathcal{R}^{17})$ for $|n_I| \leq 9$, 
\begin{equation}
    Z_{\mu, \mathfrak{p}^{(I)}}^{\mathbb{C P}^2 \times S^1}\left(\epsilon_1, \epsilon_2, \mathcal{R}\right)
    = (-1)^{2\mu + 1}
    Z_{\mu, -\mathfrak{p}^{(I)}}^{\mathbb{C P}^2 \times S^1}\left(-\epsilon_1, -\epsilon_2, \mathcal{R}\right).
\end{equation}
This relation is the equivariant version of Eq.  \eqref{Phin-n} for $X=\mathbb{CP}^2$, and is not easily deduced from \eqref{ZesnI}.

The explicit expressions also suggest an equivariant version of ``strange duality'', which is roughly a relation among K-theoretic invariants under exchange of $n_I$ and the Chern character of the gauge bundle. See for example \cite{LePotier2005, danila2000, danila2002, Gottsche2015, Gottsche:2016}. To state the relation, we introduce coefficients $C^{\mu = 0}_{ \mathfrak{p}^{(I)}, l}(t_1,t_2)$ through the expansion
\begin{equation}
    Z^{\mathbb{CP}^2\times S^1}_{\mu = 0,\mathfrak{p}^{(I)}} = \sum_{l=0}^\infty C^{\mu = 0}_{\mathfrak{p}^{(I)}, l}(t_1,t_2) \mathcal{R}^{1+4l}.
\end{equation}
For all terms checked up to $\mathcal{O}(\mathcal{R}^{17})$, we observe
\begin{equation}\label{eq:StrangeDuality}
    C^{\mu = 0}_{\mathfrak{p}^{(I)}, |n_I'|-4}(t_1,t_2)= t_1^{f_1\left(\mathfrak{p}^{(I)},\mathfrak{p}^{(I)'}\right)}t_2^{f_2\left(\mathfrak{p}^{(I)},\mathfrak{p}^{(I)'}\right)}C^{\mu=0}_{\mathfrak{p}^{(I)'}, |n_I|-4}(t_1,t_2),
\end{equation}
where $n_I = \sum_{\ell} \mathfrak{p}^{(I)}_{\ell}$, $n_I'=\sum_{\ell} \mathfrak{p}_{\ell}^{(I)'}$, and $|n_I|,|n’_I|\geq 5$.
Here $f_1$ and $f_2$ are integers depending on $\mathfrak{p}_{\ell}^{(I)}$ and $\mathfrak{p}_{\ell}^{(I)'}$. Further exploration of these intriguing relations is left for future work.

\begin{table}[t]
\centering
\renewcommand{\arraystretch}{2.2}
\begin{tabular}{|c|l|}
\hline 
$\mathfrak{p}^{(I)}$ & $(-1)\times Z^{\mathbb{CP}^2\times S^1}_{\frac12,\mathfrak{p}^{I}}(\e_1,\e_2,\CR)$ \\
\hline 
$\{1,0,0\}$ & $t_1^{\frac12}t_2^{-\frac14}+\CO(\CR^{16})$ \\
\hline 
\{0,1,0\} & $t_1^{-\frac14}t_2^{\frac12} + \CO(\CR^{16})$ \\ 
\hline 
\{0,0,1\} & $t_1^{-\frac14}t_2^{-\frac14}+\CO(\CR^{16})$\\
\hline 
\{2,$-1$,0\} & $t_1^{\frac54}t_2^{-1}+\CO(\CR^{16})$\\
\hline
\{3,0,0\} & $t_1^{\frac32}t_2^{-\frac34}+t_1^{\frac72}t_2^{-\frac74}\CR^4
+ t_1^{\frac{11}{2}} t_2^{-\frac{11}{4}} \CR^8
+ t_1^{\frac{15}{2}} t_2^{-\frac{15}{4}}\CR^{12}
+\CO(\CR^{16})$   \\
\hline 
\{1,1,1\} & $1+ \CR^4 + \CR^8
+\CR^{12}+\CO(\CR^{16})$ \\
\hline
\{5,0,0\} & $t_1^{\frac52}t_2^{-\frac54}+t_1^{\frac92}t_2^{-\frac{17}{4}}(t_1^2+t_2^2+t_1t_2+t_1^2t_2^2+t_1^2t_2+t_1t_2^2)\CR^4+\cdots +\CO(\CR^{16})$ \\
\hline 
$\{-1, 0, 0\}$ & 
$t_1^{-\frac12} t_2^{\frac14} + \CO(\CR^{16})$ \\
\hline
$\{0, -1, 0\}$ & 
$t_1^{\frac14} t_2^{-\frac12} + \CO(\CR^{16})$ \\
\hline
$\{0, 0, -1\}$ & 
$t_1^{\frac14} t_2^{\frac14} + \CO(\CR^{16})$ \\
\hline
$\{-2, 1, 0\}$ & 
$t_1^{-\frac54} t_2 + \CO(\CR^{16})$ \\
\hline
$\{-3, 0, 0\}$ & 
$t_1^{-\frac32} t_2^{\frac34} +
t_1^{-\frac72}t_2^{\frac74} \CR^4 
+ t_1^{-\frac{11}{2}}t_2^{\frac{11}{4}}\CR^8
+ t_1^{-\frac{15}{2}}t_2^{-\frac{15}{4}}\CR^{12} +\CO(\CR^{16})$ \\
\hline
$\{-1,-1,-1\}$ & $1+ \CR^4 + \CR^8
+\CR^{12}+\CO(\CR^{16})$ \\
\hline
$\{-5,0,0\}$ & $t_1^{-\frac52}
t_2^{\frac54} + t_1^{-\frac{13}{2}} 
t_2^{\frac{9}{4}}(1 + t_1 + t_1^2 + t_2 + t_1 t_2 + t_2^2) \CR^4
+\CO(\CR^{16})$ \\
\hline
\end{tabular}
\caption{First terms of equivariant K-theoretic partition functions $Z^{\mathbb{CP}^2\times S^1}_{1/2,\mathfrak{p}^{I}}$ as function of $\mathfrak{p}^{(I)}=\{\mathfrak{p}^{(I)}_1,\mathfrak{p}^{(I)}_2,\mathfrak{p}^{(I)}_3\}$ for $\mu=1/2$, with $t_1=e^{R\eps_1}$ and $t_2=e^{R\eps_2}$.}
\label{P2eqKhalf}
\end{table}

\begin{table}[t]
\centering
\renewcommand{\arraystretch}{2.2}
\begin{tabular}{|c|l|}
\hline 
$\mathfrak{p}^{(I)}$ & $Z^{\mathbb{CP}^2\times S^1}_{0,\mathfrak{p}^{I}}(\e_1,\e_2,\CR)$ \\
\hline 
\{0,0,0\} & $0+\CO(\CR^{17})$ \\
\hline
\{1,0,0\} & $\frac{1}{2}(1+t_1+t_1t_2^{-1})\CR + \CO(\CR^{17})$ \\
\hline 
$\{0,1,0\}$ & $\frac{1}{2}(1+t_2+t_2t_1^{-1})\CR + \CO(\CR^{17})$ \\
\hline
$\{0,0,1\}$ & $\frac{1}{2}(1+t_1^{-1}+t_2^{-1})\CR + \CO(\CR^{17})$ \\
\hline
$\{1,1,0\}$ & $\frac{1}{2}t_1^{-1}t_2^{-1}(t_1^2+t_2^2+t_1t_2+t_1^2t_2^2+t_1^2t_2+t_1t_2^2)\CR+\CO(\CR^{17})$ \\
\hline
$\{2,0,0\}$ & $\frac{1}{2}t_2^{-2}(t_1^2+t_2^2+t_1t_2+t_1^2t_2^2+t_1^2t_2+t_1t_2^2)\CR+\CO(\CR^{17})$ \\
\hline  
$\{3,0,0\}$ & $\frac{1}{2}t_2^{-3}(t_2^3+t_1t_2^3+t_1t_2^2+t_1^2t_2^3+t_1^2 t_2+t_1^3t_2^3+t_1^3+t_1^3t_2+t_1^3t_2^2)\CR$\vspace{-.2cm} \\
& $+ t_1^4t_2^{-2}\CR^5+t_1^6t_2^{-3}\CR^9+\CO(\CR^{13})$\\
\hline
$\{-1,0,0\}$ & $-\frac{1}{2}(1+t_1^{-1}+t_1^{-1}t_2)\CR + \CO(\CR^{17})$ \\
\hline 
$\{0,-1,0\}$ & $-\frac{1}{2}(1+t_2^{-1}+t_1t_2^{-1})\CR + \CO(\CR^{17})$ \\
\hline 
$\{0,0,-1\}$ & $-\frac{1}{2}(1+t_1+t_2)\CR + \CO(\CR^{17})$ \\
\hline 
$\{-1,-1,0\}$ & $-\frac{1}{2}t_1^{-1}t_2^{-1}(1+t_1+t_1^2 + t_2 + t_1 t_2 + t_2^2)\CR + \CO(\CR^{17})$ \\
\hline 
$\{-2,0,0\}$ & $-\frac{1}{2}t_1^{-2}(1+t_1+t_1^2 + t_2 + t_1 t_2 + t_2^2)\CR + \CO(\CR^{17})$ \\
\hline 
\end{tabular}
\caption{First terms of equivariant K-theoretic partition functions $Z^{\mathbb{CP}^2\times S^1}_{0,\mathfrak{p}^{I}}$ as function of $\mathfrak{p}^{(I)}=\{\mathfrak{p}^{(I)}_1,\mathfrak{p}^{(I)}_2,\mathfrak{p}^{(I)}_3\}$ for $\mu=0$, with $t_1=e^{R\eps_1}$ and $t_2=e^{R\eps_2}$.}
\label{P2eqK0}
\end{table}

\section{Moving Forward}
\label{Sec:ConOUt}

In this section we highlight some other directions which seem promising and might prove fruitful.

\subsection{Other Five-dimensional Super Yang-Mills Theories}

In Sec.  \ref{Sec:E1Uplane}, we found an intriguing order-of-limits issue in taking the limit $\CR^4 \to 1$. 
We believe this is a signature of the topologically twisted superconformal fixed point theory $E_1$, although a complete understanding remains an open problem for future investigation. Analogous computations could in principle be extended to the $\mathrm{SU}(2)_{\pi}$ theory and 5d SYM with higher-rank gauge group, the low-energy effective field theories arising from relevant deformations of the $E_n$ superconformal field theories discussed in \cite{Seiberg:1996bd, Morrison:1996xf, Douglas:1996xp}. Although technically much more challenging, such studies may lead to further insight into the structure of these theories.
 
\subsection{Five-dimensional Gauge Theories With Matter}

It would also be desirable to extend the framework of the present paper to other 5d gauge theories with hypermultiplet matter, provided they admit a field-theoretic UV completion. Some aspects of the Coulomb branches of these theories are explored in \cite{Closset:2023pmc, Furrer:2024zzu}, and demonstrate a similar double cover structure. A particularly interesting example is the $\CN=1^*$ theory, obtained by adding a single massive adjoint hypermultiplet with mass $m$ to the vector multiplet.
Equivariant localization of the relevant integrals with respect to the $\mathrm{U}(1)^{(B)}$-symmetry that rotates the monopole fields leads to an integral over two types of fixed points: one corresponds to the standard moduli space of instantons where the monopole field vanishes, and the other to the ``monopole branch'', where the monopole field is nonzero, but a $\mathrm{U}(1)$ rotation is equivalent to a gauge transformation. It is natural to conjecture that, with the inclusion of a hypermultiplet with mass $m$, the partition function becomes a generating function of the character-valued Dirac indices, 
\be 
Z^{\rm 5d}_{\bfmu,\bfn}(\CR, m)=\sum_k  \CR^{d_k/2} {\rm Tr}_{\CH_k} (-1)^F e^{- \beta \slashed{D}^2  + m Q_V },
\ee 
where $\CH_k$ is the SQM with target space $M_k$, the component of nonabelian monopole moduli space containing instantons of instanton charge $k$, and $Q_V$ is the operator generating the action of $\mathrm{U}(1)^{(B)}$ on $\CH_k$. 
For $X$ a complex surface, the moduli space is also complex, and we expect this character-valued Dirac index to be related to the $\chi_y$ genus of the moduli space \cite{Hollowood:2003vt}. In this case, it would be valuable to compare with the results and conjectures of \cite{Gottsche:2017vxs}. 

The generalization to higher-rank 5d theories with matter is conceptually feasible but might present significant technical challenges. 

\subsection{Six-dimensional Gauge Theories}

Another very promising direction for future research is to generalize the results to 6d theories. For example, one could study 6d $\CN=(0,1)$ theories which are anomaly free and admit a UV completion. Such theories have been tabulated in \cite{Bhardwaj:2015xxa}. A simple example is the case of a 6d $(0,1)$ vector multiplet coupled to an adjoint hypermultiplet with mass $m$. The instanton density $J = {\rm Tr}(F\wedge F)$ now couples to a background  gerbe connection in a $(0,1)$ tensor multiplet via the  electric coupling 
\be 
\int_{X_6} B^{(I)} \wedge {\rm Tr}(F\wedge F), 
\ee
whose supersymmetric completion should give the coupling of the 6d vector multiplet with the background tensor multiplet in analogy with the result \eqref{mixed CS action}.

The 6d theory can be considered on spacetimes of the form $X \times E$, where $X$ is a four-manifold, as in this paper, and $E$ is an elliptic curve. We expect that one can again define a partial topological twist such that the resulting theory is topological along $X$ and a nontopological and spin theory on $E$. This allows us to reduce along $X$ to produce a $\CN=(0,1)$ sigma model with target space the moduli space $M$ of non-Abelian monopole equations. (A related idea was pursued in \cite{Gukov:2018iiq}.) The background tensor multiplet is expected to induce a gerbe connection over $M$, most transparently described using differential cohomology.

The natural generalization of the K-theoretic partition function would be a generating function of elliptic genera of the moduli spaces of non-Abelian monopoles, graded by the instanton charge $k$, 
\be\label{eq:EllGenGenSer}
Z^{\rm 6d} \sim \sum_k \CR^{d_k/2}_{6d} \CE(M_k; m, \tau_E),
\ee 
with $\tau_E$ the complex structure of $E$,  $\CE(M_k; m, \tau_E)$ an equivariant elliptic genus, and
\be 
\CR_{6d} \sim \exp\left( -  \frac{A}{g_{6d}^2} + i \theta \right),
\ee
where $A$ is the area of $E$, and  $\theta \sim \int_{E} B^{(I)}$ is the holonomy of the background gerbe. 

On the other hand, parallel to the 4d and 5d cases, the series \eqref{eq:EllGenGenSer} should also be computable from the LEET on $X$, derivable via SW curves. Indeed, for $X_6=X \times E$, the SW geometry of the effective 4d theory obtained by compactification along $E$ has been discussed in \cite{Ganor:1996pc,Braden:2003gv,Closset:2023pmc}.
In particular, the analog of the $U$-plane and the double-cover structure central to this work should be spelled out. 

We expect once again that the most efficient computational approach to the invariants is to start with a $U$-plane integral whose value is a modular form, akin to the case of 4d  $\CN=2^*$ theory studied in \cite{Manschot:2021qqe}. A study of the wall-crossing behavior of the integral should then indicate what other invariants are needed to cancel the metric dependence of the $U$-plane integral arising from the strong-coupling singularities. 
In this way, we hope to make contact with the work of G\"ottsche and Kool on elliptic genera of the moduli space of instantons  \cite{Gottsche:2018epz}. Given our experience in writing the present paper, we would anticipate that working out the above speculations in precise detail will uncover many interesting and nontrivial issues. 

\subsection{Mathematical Correspondences And Physical Dualities}
There are a number of relations between K-theoretic invariants known in the mathematics literature. The line bundle $L^{(I)}$ with $\overline{c_1(L^{(I)})}=\bfn_I$ is lifted to a general element of the K-theory group $K(X)$, with the rank of the vector bundle corresponding to the level of the 5d Chern-Simons term. There are two notable mathematical correspondences:
\begin{itemize}
\item Strange duality: Roughly speaking, this duality exchanges the K-theory class of the gauge bundle with the K-theory class determining the determinant line bundle \cite{Gottsche2015, Gottsche:2016}. See Eq. \eqref{eq:StrangeDuality} above. A formulation for virtual invariants is work in progress and conjectured in \cite{GottscheTalk:2022}. 
\item Segre/Verlinde correspondence: This relates K-theoretic invariants to intersection numbers of Segre classes \cite{Gottsche:2020ass, Gottsche:2021ihz}. Segre invariants are known to arise as correlation functions in twisted SYM coupled to fundamental matter multiplets \cite{Aspman:2023ate}. 
\end{itemize}

Physically, these relations can be seen as 5d analogues of level-rank duality in two-dimensional conformal field theory and three-dimensional Chern-Simons theory \cite{Naculich:1990pa, Nakanishi:1990hj, Witten:1993xi, Xu:1998nxa, Hsin:2016blu}. The notion of level in three dimensions is enhanced in five dimensions to the $\mathrm{U}(1)^{(I)}$ flux $\bfn_I$, or more general K-theory class on $X$ giving the determinant line bundle over $M_k$, while the Chern character of the gauge bundle plays the role of the rank. Moreover, it is known that the Verlinde formula in two dimensions \cite{Verlinde:1988sn} is related to gauge theory with fundamental matter \cite{Witten:1993xi}. It would be interesting to establish these equivalences more rigorously and more conceptually within the 5d setting.

\appendix

\section{Spinor Conventions} \label{app:conventions}

We take the
$\sigma$-matrices to be
\begin{align}
(\sigma^\mu)_{\alpha\dot\alpha}&:=
(\vec{\tau},i),\\
(\bar{\sigma}^\mu)^{\dot\alpha\alpha}
&:= (-\vec{\tau},i),\\
{(\sigma^{\mu\nu})_\alpha}^\beta 
&:= 
\frac{1}{4}
{(\sigma^\mu \bar\sigma^\nu - \sigma^\nu \bar\sigma^\mu)_\alpha}^\beta, \\
{(\bar\sigma^{\mu\nu})^{\dot\alpha}}_{\dot\beta} 
&:= 
\frac{1}{4}
{(\bar\sigma^\mu \sigma^\nu - \bar \sigma^\nu \sigma^\mu)^{\dot\alpha}}_{\dot\beta},
\end{align}
where $\alpha,\dot\alpha=1,2$, and $\vec{\tau}=(\tau^1,\tau^2,\tau^3)$ are the usual Pauli matrices,
\begin{equation}
\tau^{1}=\left(\begin{array}{ll}
0 & 1 \\
1 & 0
\end{array}\right), \quad \tau^{2}=\left(\begin{array}{cc}
0 & -i \\
i  & 0
\end{array}\right), \quad \tau^{3}=\left(\begin{array}{cc}
1 & 0 \\
0 & -1
\end{array}\right).
\end{equation}

The 5d $\Gamma$-matrices $\Gamma^m= (\Gamma^{\mu},\Gamma^5)$ are 
\begin{equation}
\Gamma^\mu=\left(\begin{array}{cc}
0 & \sigma^\mu \\
-\bar{\sigma}^\mu & 0
\end{array}\right), \quad \quad \Gamma^5=\Gamma^1 \Gamma^2 \Gamma^3 \Gamma^4=\left(\begin{array}{cc}
\mathbf{1} & 0 \\
0 & -\mathbf{1}
\end{array}\right),
\end{equation}
which satisfy
\begin{equation}
\left\{\Gamma^{m}, \Gamma^{n}\right\}=2 \delta^{m n}.
\end{equation}
We further define
\begin{equation}
\Gamma^{m_1 m_2 \ldots m_p} :=  \frac{1}{p !} \sum_{\sigma \in S_p} \text{sgn}(\sigma) \prod_{i=1}^p \Gamma^{m_{\sigma(i)}}
\end{equation}

The 5d Dirac spinor is denoted by $\Psi_{\mathbf{a}}$, where $\boldsymbol{a}=1,\cdots,4$ are $\mathrm{Spin}(5)$ indices. The indices are raised and lowered by 
\begin{equation}
C^{\mathbf{a b}}=\left(\begin{array}{cc}
\epsilon^{\alpha \beta} & 0 \\
0 & \epsilon_{\dot{\alpha} \dot{\beta}}
\end{array}\right),
\quad 
C_{\mathbf{a b}} = 
    \left(\begin{array}{cc}
\epsilon_{\alpha \beta} & 0 \\
0 & \epsilon^{\dot{\alpha} \dot{\beta}}
\end{array}\right),
\end{equation}
such that
$\Psi^{\boldsymbol{a}} = C^{\boldsymbol{a}\boldsymbol{b}} \Psi_{\boldsymbol{b}}$.
$\Psi_{\mathbf{a}}$ can be expressed in terms of Weyl spinors as
\be
\Psi_{\boldsymbol{a}} = \left(\begin{array}{c} \psi_\alpha \\ \bar\chi^{\dot\alpha}\end{array}\right).
\ee

The eight supercharges of 5d $\CN=1$ theory on $\mathbb{R}^5$ can be written as
\be
\mathbf{Q}_{\mathbf a}^A = \left(\begin{array}{c} Q^A_\alpha \\ \e^{AB} \bar Q_B^{\dot\alpha}\end{array}\right).
\ee
Here $A,B=1,2$ are $\mathrm{SU}(2)_R$ indices, which are raised and lowered according to 
\begin{equation}
\Psi_A=\epsilon_{A B} \Psi^B, \quad \Psi^A=\epsilon^{A B} \Psi_B
\end{equation}
where $\epsilon_{12} = \epsilon^{21} = 1$. 

The 5d $\mathcal{N}=1$ supersymmetry transformations for the vector multiplet are given in Eq. \eqref{eq:5dSUSY-TMNS}, and the supercharges satisfy the 5d $\mathcal{N}=1$ supersymmetry algebra (restricted to vector multiplets), 
\begin{equation}
\left\{\mathbf{Q}_{\mathbf{a}}^A, \mathbf{Q}_{\mathbf{b}}^B\right\}=2i \epsilon^{A B}\left(\Gamma_{\mathbf{a b}}^m D_m +  C_{\mathbf{a b}} [\sigma, \bullet]\right). 
\end{equation}

Upon dimensional reduction, $Q_\alpha^A, \bar{Q}_B^{\dot\alpha}$ can be interpreted as the supercharges of the 4d $\CN=2$ theory. Correspondingly, the 5d supersymmetry variation can be written
\begin{equation}
    \delta = i \xi_A Q^A + i \bar\xi^A \bar Q_A,
\end{equation}
where the 4d supersymmetry transformation parameters are Weyl spinors, and are obtained from the 5d supersymmetry transformation parameters by
\begin{equation}
\label{eq: 4d-weyl-supcharge}
    \xi_{\mathbf{a}A} = \binom{-\xi_{\alpha A}}{{\bar{\xi}^{\dot{\alpha}}}_A}, 
\quad 
\xi^{\mathbf{a}A} = \left(- \xi^{\alpha A}, {\bar\xi_{\dot{\alpha}}}^A\right).
\end{equation}
Here
\be
\xi_{\alpha}^{A} \in \Gamma[S_- \otimes E_R],\quad  \bar\xi_{B}^{\dot \alpha} \in \Gamma[S_+ \otimes E_R],
\ee 
with $E_R$ the associated vector bundle for the principal $\mathrm{SU}(2)_R$-bundle in the fundamental representation. Note that they are well-defined sections if
\be
w_2(E_R) \cong w_2(X).
\ee 

We perform a partial topological twist in Sec.  \ref{sec:TopTwist}.
The scalar supercharge $\mathcal{\bar Q}$ is defined by 
\begin{equation}
    \mathcal{\bar Q} := \delta_{\dot\alpha}{}^A \bar Q^{\dot \alpha}{}_A,
\end{equation}
which is equivalently to setting commutative supersymmetry parameters by
\begin{equation}
    \bar\xi= \bar \zeta, \quad \xi = 0,
\end{equation}
such that
\begin{equation}
    \bar\xi_{\dot\alpha}{}^A = \delta_{\dot\alpha}{}^A.
\end{equation}
The vector supercharge is defined by 
\begin{equation}
    \mathcal{K}_\mu := \frac{i}{4}\bar\zeta_A \bar\sigma_\mu Q^A,
\end{equation}
satisfying
\begin{equation}
    \left\{\bar Q, \mathcal{K}_{\mu}\right\} = D_{\mu}.
\end{equation}
The twisted variables are then defined as follows
\begin{align}
    \psi_\mu &:= 
    \bar\zeta_A 
    \bar\sigma_{\mu}
    \lambda^A,\\
    \eta &:= 
    \bar\zeta^A 
    \bar\lambda_A,\\
    \chi_{\mu\nu}
    &:= 
    \bar\zeta^A 
    (\bar\sigma_{\mu\nu}) \bar\lambda_A,\\
    D_{\mu\nu} & :=  
    -i 
    \bar\zeta^A 
    (\bar\sigma_{\mu\nu})\bar\zeta^B D_{AB},
\end{align}

In Sec.  \ref{Sec:Uplane}, when we dimensionally reduce the theory from 5d to 4d, the twisted fields adopted the following change of conventions 
\begin{align}
        \sigma+i A_5 &\to -2\sqrt{2} a, \\
        \sigma-i A_5 
        &\to -2\sqrt{2} \bar a, \\
        \psi_{5d} &\to -\psi_{4d}, \\
        \chi_{5d} &\to -2 \chi_{4d}, \\
        \eta_{5d} &\to 2 \eta_{4d} \\
        D_{5d} &\to -2 D_{4d}.
\end{align}

In Sec.  \ref{Sec:SusyLoc}, when $X$ is a K\"ahler surface we can decompose the positive 
chirality spinor bundle as
\be
S_+ \cong K_X^{\frac12} \oplus  K_X^{-\frac12}.
\ee 
Roughly speaking, the topological twist identifies  $E_R \cong S_+ $. The supersymmetry transformation parameters become sections of the bundle 
\begin{equation}
    S_+\otimes E_R \cong 
    \mathcal{K }\oplus \mathcal{K}^{-1}
    \oplus \mathcal{O}_{(1)}
    \oplus 
    \mathcal{O}_{(2)},
\end{equation}
and
\begin{equation}
    S_-\otimes E_R \cong
    (\mathcal{K}^{\frac{1}{2}}\otimes \Omega^{0,1})\otimes 
    (\mathcal{K}^{\frac{1}{2}}\oplus 
    \mathcal{K}^{-\frac{1}{2}})\cong 
    \Omega^{2,1}\oplus
    \Omega^{0,1} \cong 
    \Omega^{1,0}\oplus 
    \Omega^{0,1}.
\end{equation}
We can choose 
\begin{equation}
        {(\bar\zeta^{(1)})^{\dot\alpha}}_{A} 
        =\left(\begin{array}{cc}
           1  &0  \\
             0& 0
        \end{array}
        \right),\quad
        {(\bar\zeta^{(2)})^{\dot\alpha}}_{A} 
        =\left(\begin{array}{cc}
           0  &0  \\
             0& 1
        \end{array}
        \right).
\end{equation}
Then two preserved supercharges are constructed as
\begin{equation}
    \mathcal{\bar Q}_{(1)} = {(\bar\zeta_{(1)})_{\dot\alpha}}^A 
    {\bar Q^{\dot\alpha}}_{A},
    \quad 
      \mathcal{\bar Q}_{(2)} = {(\bar\zeta_{(2)})_{\dot\alpha}}^{A} 
    {\bar Q^{\dot\alpha}}_{A}.
\end{equation} 
The $\sigma^{\mu}$ matrices in the
local complex basis defined in \eqref{eq:metric-with-v} are 
\begin{equation}
    \begin{aligned}
    \sigma^1 &= 2\sigma_{\bar  1}=-\bar\sigma^1=-2 \bar\sigma_{\bar 1} =
    \left(
    \begin{array}{cc}
         0&2  \\
         0&0 
    \end{array}
    \right),\\
    \sigma^{\bar 1} &= 2\sigma_{  1}=-\bar\sigma^{\bar 1}=-2 \bar\sigma_{1}=
    \left(
    \begin{array}{cc}
         0&0  \\
         2&0 
    \end{array}
    \right), 
    \\
    \sigma^{2} &= 2\sigma_{  \bar 2}=-\bar\sigma^{\bar 2} =-2\bar\sigma_2=
    \left(
    \begin{array}{cc}
         0&0  \\
         0&-2 
    \end{array}
    \right),
    \\
    \sigma^{\bar 2} &= 2\sigma_{  2}=-\bar\sigma^{2} = 
    -2\bar\sigma_{\bar 2} =
    \left(
    \begin{array}{cc}
         2&0  \\
         0&0 
    \end{array}
    \right), 
    \end{aligned}
\end{equation}
and $\sigma^{\mu\nu}$ matrices are
\begin{equation}
    \begin{aligned}
    \sigma^{1 \bar 2} 
    &= 4 \sigma_{\bar 1 2}= 
    \left(
    \begin{array}{cc}
         0&2  \\
         0&0 
    \end{array}
    \right) ,\\
    \sigma^{2 \bar 1} 
    &= 4 \sigma_{\bar 2 1}= 
    \left(
    \begin{array}{cc}
         0&0  \\
         2&0 
    \end{array}
    \right) ,\\
    \sigma^{1 \bar 1} 
    &= 4 \sigma_{\bar 1 1}= 
    \left(
    \begin{array}{cc}
         -1&0  \\
         0&1 
    \end{array}
    \right) ,\\
    \sigma^{2 \bar 2} 
    &= 4 \sigma_{\bar 2 2}= 
    \left(
    \begin{array}{cc}
         1&0  \\
         0&-1 
    \end{array}
    \right) ,\\
    \bar\sigma^{1 \bar 1} 
    &= 4 \bar\sigma_{\bar 1 1}= 
    \left(
    \begin{array}{cc}
         -1&0  \\
         0&1
    \end{array}
    \right) ,\\
    \bar\sigma^{2 \bar 2} 
    &= 4 \bar\sigma_{\bar 2 2}= 
    \left(
    \begin{array}{cc}
         -1&0  \\
         0&1
    \end{array}
    \right) ,\\
    \bar\sigma^{\bar 1 \bar 2} 
    &= 4 \bar\sigma_{ 1  2}= 
    \left(
    \begin{array}{cc}
         0&0  \\
         -2&0
    \end{array}
    \right),
    \\
    \bar\sigma^{1 2} 
    &= 4 \bar\sigma_{\bar 1 \bar 2}= 
    \left(
    \begin{array}{cc}
         0&2  \\
         0&0
    \end{array}
    \right).
    \end{aligned}
\end{equation}

\section{Modular Forms} 
\label{App:MForms}
This appendix lists elementary functions used in the main text from the theory of modular and mock modular forms. We refer for further details to textbooks such as \cite{Apostol, Eichler:1985, Bruinier08, Bringmann:2309148}.
\subsection{Jacobi Theta Series}
The Dedekind $\eta$-function is defined as
\be
\label{DefEta}
\eta(\tau)=q^{\frac{1}{24}}\prod_{n=1}^\infty (1-q^n),\qquad q=e^{2\pi i \tau}.
\ee
It transforms under a modular transformation $\gamma \in \mathrm{SL}(2,\mathbb{Z})$ as
\be 
\label{vareps}
\eta\left(
  \frac{a\tau+b}{c\tau+d}\right)=\varepsilon(\gamma)\,(c\tau+d)^{\frac12}\,\eta(\tau),\qquad
\gamma=\left(\begin{array}{cc} a & b \\ c & d \end{array}\right),
\ee
with $\varepsilon(\gamma)$ a 24'th root of unity dependent on $\gamma$ \cite{Apostol}. 

The four Jacobi theta functions $\vartheta_j:\BH\times \BC\to \BC$ are defined by
\be
\label{Jacobitheta}
\begin{split}
&\vartheta_1(\tau,z)=i \sum_{n\in
  \mathbb{Z}+\frac12}(-1)^{n-\frac12}q^{\frac{n^2}{2}}e^{2\pi i
  nz}, \\
&\vartheta_2(\tau,z)= \sum_{n\in
  \mathbb{Z}+\frac12}q^{\frac{n^2}{2}}e^{2\pi i
  nz},\\
&\vartheta_3(\tau,z)= \sum_{n\in
  \mathbb{Z}}q^{\frac{n^2}{2}}e^{2\pi i
  n z}, \\
&\vartheta_4(\tau,z)= \sum_{n\in 
  \mathbb{Z}} (-1)^nq^{\frac{n^2}{2}}e^{2\pi i
  n z}. 
  \end{split}
\ee
$\vartheta_j(\tau,z)$ is an odd function of $z$ for $j=1$, and an even function of $z$ for $j=2,3,4$. 

The Dedekind $\eta$-function is related to the Jacobi theta function via
\be
\left.\frac{\partial}{\partial z}\vartheta_1(\tau,z)\right\vert_{z=0}=-2\pi \eta(\tau)^3.
\ee

In this paper, we use
\be
\vartheta_j(\tau) := \vartheta_j(\tau,0).
\ee
They are related to the notation $\theta_{ij}$ used in \cite{Nakajima:2003uh, Gottsche:2006bm} by
\be
\vartheta_1(\tau) = \theta_{11}(\tau),\quad \vartheta_2(\tau) = \theta_{10}(\tau),\quad 
\vartheta_{3}(\tau) = \theta_{00}(\tau),\quad \vartheta_4(\tau) =  \theta_{01}(\tau).
\ee

These functions
satisfy various identities. For their sum, we have
\be
\label{t3=t2pt4}
\begin{split}
&\vartheta_3(\tau)^4=\vartheta_2(\tau)^4 + \vartheta_4(\tau)^4,\\
\end{split}
\ee
and 
\be
\label{sumstheta}
\begin{split}
\vartheta_3(\tau/2,z/2)=\vartheta_3(2\tau,z)+\vartheta_2(2\tau,z),\\
\vartheta_4(\tau/2,z/2)=\vartheta_3(2\tau,z)-\vartheta_2(2\tau,z).
\end{split}
\ee
For their product, we have
\be
\label{tsproduct}
\begin{split}
&\vartheta_2(\tau)\,\vartheta_3(\tau)\,\vartheta_4(\tau)=2\,\eta(\tau)^3,\\
&\vartheta_1(\tau,2z)\,\vartheta_2(\tau)\,\vartheta_3(\tau)\,\vartheta_4(\tau)=2\,\vartheta_1(\tau,z)\,\vartheta_2(\tau,z)\,\vartheta_3(\tau,z)\,\vartheta_4(\tau,z).
\end{split}
\ee

The function $\vartheta_1$ transforms for an element $\gamma=\left( \begin{array}{cc}
    a & b \\ c & d \end{array} \right) \in \mathrm{SL}(2,\mathbb{Z})$ as
\be
\vartheta_1\!\left(\frac{a\tau+b}{c\tau+d},\frac{z}{c\tau+d}\right)=
\varepsilon(\gamma)^3\, (c\tau+d)^{\frac12}\exp\left(\frac{\pi i cz^2}{c\tau+d}\right)
\vartheta_1(\tau,z), 
\ee
where $\varepsilon(\gamma)$ is the multiplier system of the
$\eta$-function (\ref{vareps}).
For the two generators $S$ and $T$ of $\mathrm{SL}(2,\mathbb{Z})$, we have
\be
\begin{split}
&S:\qquad \vartheta_1\!\left(-1/\tau,z/\tau \right)=-i\sqrt{-i\tau}\, e^{\pi
  i z^2/\tau}\,\vartheta_1(\tau,z),\\
&T:\qquad \vartheta_1\!\left(\tau+1,z\right)=e^{2\pi i/8}\, \vartheta_1(\tau,z).
\end{split}
\ee
We have furthermore the following quasi-periodicity relation
\be 
\vartheta_1(\tau,z+l\tau)=(-1)^l q^{-l^2/2}e^{-2\pi i l z}\vartheta_1(\tau,z).
\ee 

\subsection{The Appell-Lerch Sum}
\label{subsecZwmu}
We recall the definition and properties of the Appell-Lerch sum
following Zwegers \cite{ZwegersThesis}. See also \cite{MR2605321, Bringmann:2309148}. We
define  
\be
\label{Mtuv}
M(\tau,u,v):=\frac{e^{\pi i u}}{\vartheta_1(\tau,v)}\sum_{n\in
  \mathbb{Z}} \frac{(-1)^n q^{n(n+1)/2}e^{2\pi i nv}}{1-e^{2\pi i u}q^n}.
\ee
With $y={\rm Im}(\tau)$ and $a=\mathrm{Im}(u)/y$, we introduce also 
\be
\label{muR}
R(\tau,\bar \tau, u,\bar u):=\sum_{n\in \mathbb{Z}+\frac{1}{2}} \left( \sgn(n) -
  E\!\left( (n+a)\sqrt{2y}
     \right) \right) (-1)^{n-\frac{1}{2}}\,e^{-2\pi i un}q^{-n^2/2},
\ee
where $\sgn:\mathbb{R}\to \{-1,0,1\}$ is defined as,
\be 
\label{def: sgn}
\sgn(x)=\left\{ \begin{array}{rr} 1, &\qquad  x>0, \\  0, & \qquad x=0, \\ -1, & \qquad x<0, \end{array}\right.
\ee 
and $E(t)$ is the error function with rescaled argument,
\be
\label{Eerror}
E(t)=2 \int_0^t e^{-\pi u^2}du=\mathrm{Erf}(\sqrt{\pi}t).
\ee
This function is anti-symmetric, $E(-t)=-E(t)$. We stress that $E$ is not a specialization of the exponential integral $E_l$ defined in Eq. \eqref{defEellz}. For $t\in \mathbb{R}-\{ 0\} $, the functions $E$ and $E_{1/2}$ are related as
\be
E(t)=\sgn(t)-t\,E_{\frac12}(\pi t^2).
\ee

The non-holomorphic completion of $M(\tau,u,v)$ is then given by
\be
\label{mucomplete}
\widehat M(\tau,\bar \tau, u,\bar u,v,\bar v) =M(\tau,u,v)+\frac{i}{2} R(\tau,\bar \tau,u-v,\bar u-\bar v). 
\ee
For an element $\gamma\in \mathrm{SL}(2,\mathbb{Z})$, it transforms as
\be
\label{whMtrafo}
\begin{split}
&\widehat M\!\left(\frac{a\tau+b}{c\tau+d},\frac{a\bar \tau+b}{c\bar \tau+d},\frac{u}{c\tau+d},\frac{\bar u}{c\bar \tau+d},\frac{v}{c\tau+d},\frac{\bar v}{c\bar \tau+d}\right)\\
=& \varepsilon(\gamma)^{-3}
(c\tau+d)^{\frac{1}{2}} \exp\left(-\frac{\pi i c(u-v)^2}{c\tau+d}\right)\widehat M(\tau,\bar \tau,u,\bar u, v,\bar v),
\end{split}
\ee
where $\varepsilon(\gamma)$ is the multiplier of the $\eta$-function (\ref{vareps}).
The anti-holomorphic derivative of $\widehat M$ is given by
\be
\begin{split}
\partial_{\bar \tau} \widehat M(\tau, \bar \tau, u,\bar u, v,\bar v)&=-i \left(\partial_{\bar \tau}
  \sqrt{2y}\right) e^{-2\pi (a-b)^2} \\
&\quad \times \sum_{n\in \mathbb{Z}+\frac{1}{2}} (n+a-b)
(-1)^{n-\frac{1}{2}} \bar q^{n^2/2}e^{-2\pi i (\bar u-\bar v) n}.
\end{split}
\ee
A number of other useful properties are \cite{ZwegersThesis}:
\begin{enumerate} 
\item $M$ is anti-periodic under shifts by 1:
\be  
M(\tau,u+1,v)=M(\tau,u,v+1)=-M(\tau,u,v).
\ee
\item Inversion of the elliptic arguments leaves $M$ invariant:
\be 
\label{M-u-v}  
M(\tau,-u,-v)=M(\tau,u,v).
\ee
\item Simultaneous shifts of $u$ and $v$ by $z$
lead to a simple transformation of $M(\tau,u,v)$. For $u,v,u+z,v+z\neq
\mathbb{Z}\tau+\mathbb{Z}$, one has 
\be
\label{mushifts} 
M(\tau,u+z,v+z)-M(\tau,u,v)=\frac{i\,\eta^3(\tau)\,\vartheta_1(\tau,u+v+z)\,\vartheta_1(\tau,z)}{\vartheta_1(\tau,u)\,\vartheta_1(\tau,v)\,\vartheta_1(\tau,u+z)\,\vartheta_1(\tau,v+z)}.
\ee 
\item Quasi-periodicity property in $u$ and $v$:
\be
\label{Mqutr}
\begin{split}
&\widehat M(\tau,\bar \tau,u+k\tau+l,\bar u+k\bar \tau+l, v+m\tau+n,\bar v+m\bar \tau+n)\\
=& (-1)^{k+l+m+n} q^{(k-m)^2/2}e^{2\pi i (k-m)(u-v)}\widehat M(\tau,\bar\tau,u,\bar u,v,\bar v).
\end{split}
\ee   
\item Symmetric under exchange of $u$ and $v$:
\be 
M(\tau,v,u)=M(\tau,u,v).
\ee 
\end{enumerate}

\section{Seiberg-Witten Solution Of Four-dimensional $\CN=2$ $\mathrm{SU}(2)$ Super Yang-Mills Theory}
\label{app4dN=2}
We review some aspects of $\CN=2$ $\mathrm{SU}(2)$ SYM in four dimensions. There are multiple conventions for the SW curve and related observables \cite{Seiberg:1994aj, Seiberg:1994rs, Klemm:1994qs, Argyres:1994xh}. The relevant curve for the main text is the elliptic curve 
\be 
\label{4dcurve}
y^2=(x^2-u)^2-4\Lambda^4,
\ee
where $u$ is the vev $\frac{1}{2}\left<{\rm Tr}\,\phi^2\right>_{\mathbb{R}^4}$ with $\phi$ the scalar field in the vector multiplet.
As discussed in Sec.  \ref{sec:4dLimitSW}, the 5d curve \eqref{5dSWcurve} reduces to this curve in the 4d limit. The singularities of the curve are given by the vanishing loci of the (physical) discriminant $\Delta_{4,{\rm phys}}$ of the curve, where
\be
\begin{split}
  \Delta_{4,{\rm phys}}=u^2-4\Lambda^4.
  \end{split}
\ee

\begin{figure}[t]\centering
	\includegraphics[width=0.8\textwidth]{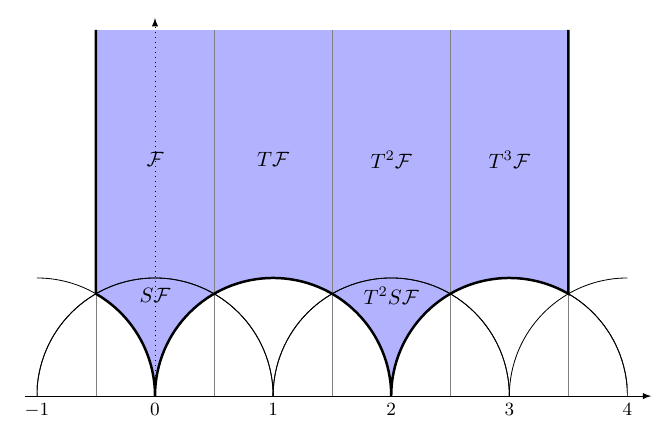} 
	\caption{Fundamental domain $\Gamma^0(4)\backslash \mathbb{H}$
          for the effective coupling $\tau$ of the 4d SW theory. The
          boundary vertical lines are identified by $\tau \to \tau
          +4$. The boundary arcs of the $S$ and $T^2S$ cusps are
          identified. And the bottom arc of the $T$ cusp is identified
        with the bottom arc of the $T^3$ cusp.}\label{fig:fund_dom_gamma04}
\end{figure}

Using the relation to the modular $j$-invariant of this curve, we have the explicit expression\footnote{Note the different sign compared to some literature.}
\be
\label{uSW}
u= -2\Lambda^2\,\left( 2\frac{\vartheta_3(\tilde \tau)^4}{\vartheta_4(\tilde \tau)^4}-1 \right)=-\Lambda^2(2+64\,\tilde q^{\frac12}+\CO(\tilde q)),
\ee
where $\tilde \tau$ is the complex structure of the curve, and we choose a duality frame that gives the $A$- and $B$-cycles discussed below Eq.  \eqref{hellipticcurve}. $u$ is left invariant by $\Gamma(2)$ transformations of $\tilde \tau$. The duality frame  for $\tilde \tau$ is not the familiar weak-coupling duality frame. We therefore make the change of variable $\tilde \tau=-2/\tau$ as below Eq.  \eqref{hellipticcurve}. The coupling $u$ can then be expressed as
\be 
\label{eq:uttu}
u(\tau)=-2\Lambda^2\,\ttu(\tau),
\ee 
where $\ttu$ is the Hauptmodul for $\Gamma^0(4)$ defined in Eq.  \eqref{ttu}. A fundamental domain for $\mathbb{H}/\Gamma_0(4)$ is plotted in Fig. \ref{fig:fund_dom_gamma04}.
This is the default duality frame in the following and in the main text. We summarize a few special points of the $u$-plane:
\begin{itemize}
\item Monopole singularity: for $\tau\to 0$, $u\to -2\Lambda^2$ and the monopole becomes massless;
\item Dyon singularity: for $\tau\to 2$, $u\to +2\Lambda^2$ and a
  dyon becomes massles;
\item For $\tau=1+i \mod 2$, $u(\tau)=0$;  
\item For $\tau=e^{\pm 2\pi i/3}$;
  $u(\tau)=-\sqrt{3}\,\Lambda^2$;
\item For $\tau=i$,
  $u(\tau)=-\frac{3\sqrt{2}}{2}\,\Lambda^2<-2\Lambda^2$.
\end{itemize}

The 4d local coordinate $a_4$ satisfies
\be\label{eq:da4du}
\frac{\mathrm{d}a_4}{\mathrm{d}u}=\frac{i}{2\Lambda}\vartheta_2(\tau)\vartheta_3(\tau),
\ee
which is one of the ingredients for the topological couplings in the $u$-plane integral. The quantities satisfies Matone's relation \cite{Matone:1995rx},
\be\label{eq:Matone}
\frac{\mathrm{d}u}{\mathrm{d}\tau}=-2\pi i\, (u^2-4\Lambda^4) \left( \frac{\mathrm{d}a_4}{\mathrm{d}u}\right)^2.
\ee 
We have furthermore
\be
\label{eq:da4dtau}
\frac{\mathrm{d}a_4}{\mathrm{d}\tau}=-\frac{\pi \Lambda}{8}\frac{\vartheta_4(\tau)^9}{\eta(\tau)^3},
\ee
from which one can derive
\be 
\begin{split}
a_4(\tau)&=-i\Lambda\,\frac{2\,E_2(\tau)+\vartheta_2(\tau)^4+\vartheta_3(\tau)^4}{3\,\theta_2(\tau)\,\theta_3(\tau)}\\
&= -\frac{i\Lambda}{2}\,q^{-\frac18}+\CO\left(q^{3/8}\right).
\end{split}
\ee

\section{Exact Results For The Order Parameter $U$}
\label{app:ExactU}

Using the instanton counting method in the $\Omega$-background, one can compute exactly the partition function and correlation functions of supersymmetry-protected operators in 5d SYM theory on $\mathbb{C}^2_{\epsilon_1,\epsilon_2}\times S^1$ 
\cite{Kim:2016qqs,Tong:2014cha,Assel:2018rcw}. In particular, the vev of the Wilson loop in the fundamental representation is given by \cite[Eq. (3.3)]{Assel:2018rcw} 
\footnote{In general, the localization formulae for vevs of supersymmetric loop defects in $\CN=2$ theories can lead to incorrect results 
\cite{Brennan:2018rcn,Assel:2019iae}. However, the monopole bubbling subtleties that lead to such problems will not appear in the present computation. }
\be 
\label{expval F wilson}
\left\langle W_{F}\right\rangle=-\oint \frac{\mathrm{d} x}{2 \pi i x}
\frac{\CZ_\text{inst}^\text{5d-1d}}{\CZ_\text{inst}},
\ee
where $x = e^m$ is the flavor symmetry fugacity of the defect theory.

$\CZ_\text{inst}$ is the 5d instanton partition function.
For $\mathrm{U}(2)$ case, the $k$ instanton sector for the 5d vector multiplet reads
\footnote{We adopt the convention for $k$ instanton sector as $\CZ_{\rm inst} = \sum_{k\geq 0} \mathcal{R}^{4k} \CZ^{(k)}_{\rm inst}$. There will be an instanton sum for both the numerator and denominator of \eqref{expval F wilson}. 
}
\be 
\label{instanton 5d}
\CZ_{\mathrm{inst}}^{(k)}=\frac{1}{k !} \oint\left[\prod_{s=1}^k \frac{R\mathrm{d}\phi_s}{2 \pi i}\right] \CZ_{\mathrm{vec}}^{(k)}, 
\ee 
where the contour choice is given by the Jeffrey-Kirwan prescription and 
\be 
\begin{aligned}
\CZ_{\mathrm{vec}}^{(k)}= & \left(-\frac{\text{sh}\left(2 \epsilon_{+}\right)}{\text{sh}\left(\epsilon_1\right) \text{sh}\left(\epsilon_2\right)}\right)^k \prod_{s \neq t}^k \frac{\text{sh}\left(\phi_s-\phi_t\right) \text{sh}\left(\phi_s-\phi_t+2 \epsilon_{+}\right)}{\text{sh}\left(\phi_s-\phi_t+\epsilon_1\right) \text{sh}\left(\phi_s-\phi_t+\epsilon_2\right)} \\
&\quad \times \prod_{s=1}^k \prod_{r=1}^2 \frac{1}{\text{sh}\left(\phi_s-a_r+\epsilon_{+}\right) \text{sh}\left(-\phi_s+a_r+\epsilon_{+}\right)}.
\end{aligned}
\ee
Here we define 
\be
\text{sh}(x) :=  2 \sinh \left(\frac{Rx}{2}\right),
\ee
and 
\be
\epsilon_{\pm} :=  \frac{\epsilon_1 \pm \epsilon_2}{2},
\ee 
where $\e_{1}, \e_2$ are the $\Omega$-deformation parameters.
The result matches with \eqref{5d instanton sector} when $\bfn_I =0$.

$\CZ^\text{5d-1d}_\text{inst}$ is the instanton part of the partition function of a 5d-1d coupled system that realizes the effect of the presence of a Wilson loop. The contribution from the $k$ instanton sector is 
\be 
\label{instanton 5d1d}
\CZ^\text{5d-1d,(k)}_{\mathrm{inst}}=\frac{1}{k !} \oint\left[\prod_{s=1}^k \frac{R\mathrm{d}\phi_s}{2 \pi i}\right] \CZ_{\mathrm{vec}}^{(k)} \CZ_{\mathrm{SQM}}^{(k)},
\ee
where the choice of integral contour is still selected by the 
Jeffrey-Kirwan prescription, and 
\be 
\CZ_{\mathrm{SQM}}^{(k)}=\prod_{r=1}^2  \text{sh}\left(m-a_r\right) \prod_{s=1}^k \frac{\text{sh}\left(\phi_s-m+\epsilon_{-}\right) \text{sh}\left(-\phi_s+m+\epsilon_{-}\right)}{\text{sh}\left(\phi_s-m+\epsilon_{+}\right) \text{sh}\left(-\phi_s+m+\epsilon_{+}\right)}.
\ee
Notice that \eqref{instanton 5d} and 
\eqref{instanton 5d1d} are both symmetric under the exchange of $a_1$ and $a_2$. Therefore,  
\be 
\left\langle W_F^{(k)}
\right\rangle
(a_1,a_2) = 
\left\langle W_F^{(k)}
\right\rangle
(a_2,a_1)
\ee
We compute $\left\langle W_F\right\rangle$ up to $\mathcal{O}(\mathcal{R}^4)$, and set $a = a_1 = -a_2$ in the end to get the result for gauge group $\mathrm{SU}(2)$, 
\be\label{eq:WF-exp-firstterms}
\langle W_F\rangle = e^{Ra} + e^{-Ra} - \CR^4 \frac{t_1t_2 (e^{Ra} + e^{-Ra})}{(1-e^{Ra}t_1t_2)( 1-e^{-Ra}t_1t_2)} + \CO(\CR^8) + \cdots,
\ee
where $t_{1,2} :=  e^{R\e_{1,2}}$. Clearly the final expression is invariant under the Weyl transformation $a\to - a$.

\section{The Polylogarithm}\label{app:Polylogs}

We review a few aspects of the polylogarithm.
For more details on the special case of $n=2$, the dilogarithm ${\rm Li}_2$, see for
example \cite{Kirillov:1994en, Zagier:2007knq, Freed:2020kzf}. For general polylogarithms see \cite{lewin1981polylogarithms,lewin1991structural}.

For integer   $n$ and $\vert z \vert < 1$ one can define the convergent series
\be
\label{defLis}
{\rm Li}_n(z)=\sum_{k=1}^\infty \frac{z^k}{k^n}
\ee  
as an analytic function in the open unit disk around $z=0$. Note that for $n=1$ 
the function is simply $-\log(1-z)$, which has a branch point at $z=1$. 
Polylogarithms are related by
\be
z \frac{\mathrm{d}}{\mathrm{d}z} {\rm Li}_{n+1}(z) ={\rm Li}_{n}(z),
\ee
and hence one can iteratively define an analytic continuation of ${\rm Li}_{n}(z)$ using the integral expression
\be 
\label{eq:Lint}
{\rm Li}_{n+1}(z) = \int_0^z  {\rm Li}_n(t) \frac{\mathrm{d}t}{t}
\ee
to define a multi-valued function on $ \IP^1 - \{0,1,\infty\}$.   
The choice of the principal branch for $-\log(1-z)$, namely,  $-\pi < \arg(1-z) \leq \pi$, defines, by iterated integration, the principal branch of the polylogarithm.  
For $x>1$  the discontinuity across the cut $[1,\infty)$ is given by 
\be 
{\rm Li}_n(x+i\epsilon)={\rm Li}_n(x-i\epsilon)+\frac{2\pi i\, {\rm log}(x)^{n-1}}{\Gamma(n)},
\ee 
where we use the principal branch of both  ${\rm Li}_n$ and $\log(x)$. 

The polylogarithm satisfies a remarkable identity \cite[Sec.  7.2]{lewin1981polylogarithms}
\be
\label{Liinv2}
{\rm Li}_n(z)+(-1)^n\,{\rm Li}_{n}(z^{-1})=\left\{ \begin{array}{ll}-\frac{(2\pi
  i)^n}{n!}B_n\left(\frac{1}{2}+\frac{\log(-z)}{2\pi i} \right), & \quad z\notin (0,1],\\ 
  -\frac{(2\pi
  i)^n}{n!}B_n\left(\frac{1}{2}-\frac{\log(-z^{-1})}{2\pi i} \right), & \quad z\notin [1,\infty). \end{array} \right.
\ee
Here $B_n$ are the Bernoulli polynomials, with the first few terms
\be
\begin{split}  
&B_0(x)=1,\\
&B_1(x)=x-\frac{1}{2},\\
&B_2(x)=x^2-x+\frac{1}{6},\\
&B_3(x)=x^3-\frac{3}{2}x^2+\frac{1}{2}x.
\end{split} 
\ee
These polynomials satisfy
\be
\label{eq:BnInt}
\int_a^b B_n(t)\mathrm{d}t=\frac{B_{n+1}(b)-B_{n+1}(a)}{n+1},
\ee 
and
\be
\label{eq:B1-x}
B_n(1-x)=(-1)^nB_n(x).
\ee 
The identity (\ref{Liinv2}) can be proven by induction as follows. The identity holds for $n=1$:
\be 
-\log(1-z)+\log(1-z^{-1})=\left\{ \begin{array}{rr} -\log(-z), & \qquad z\notin (0,1], \\ \log(-z^{-1}), & \qquad z\notin [1,\infty).\end{array} \right.
\ee 
For $n\geq 1$, we start from the definition (\ref{eq:Lint}), and substitute (\ref{Liinv2}) in the integrand on the right-hand side,
\be 
\begin{split} 
&{\rm Li}_{n+1}(z^{-1})-\zeta(n+1)=\int_1^{z^{-1}} \frac{\mathrm{d}t}{t}{\rm Li}_{n}(t) \\
=&  \int_1^{z^{-1}} \frac{\mathrm{d}t}{t}(-1)^{n+1} {\rm Li}_{n}(t^{-1}) -\frac{(2\pi i)^n}{n!} \times \left\{\begin{array}{ll} B_n\left(\frac{1}{2}+\frac{\log(-t)}{2\pi i} \right), &\quad t\notin (0,1],\\ B_n\left(\frac{1}{2}-\frac{\log(-t^{-1})}{2\pi i} \right), &\quad t\notin [1,\infty).  \end{array} \right.
\end{split}
\ee 
Evaluating the first term in terms of ${\rm Li}_{n+1}$, we arrive at
\be 
\begin{split} 
&{\rm Li}_{n+1}(z^{-1})-\zeta(n+1)\\
=&(-1)^{n} ({\rm Li}_{n+1}(z)-\zeta(n+1)) \\
&-\frac{(2\pi i)^n}{n!} \int_1^{z^{-1}} \frac{\mathrm{d}t}{t} \left\{\begin{array}{ll} B_n\left(\frac{1}{2}+\frac{\log(-t)}{2\pi i} \right), &\quad z\notin [1,\infty),\\ B_n\left(\frac{1}{2}-\frac{\log(-t^{-1})}{2\pi i} \right), &\quad z\notin (0,1], \end{array} \right.
\end{split}
\ee 
where we rewrite the condition on $t$ in terms of the condition on $z$. Manipulation of the remaining integral using Eqs. (\ref{eq:BnInt}) and (\ref{eq:B1-x}) demonstrates that the identity (\ref{Liinv2}) is maintained for $n+1$. Hence, the identity is proved.

\section{K-theoretic Donaldson Invariants Of $\mathbb{CP}^2$: From Results Of G\"ottsche-Nakajima-Yoshioka To The $U$-plane Evaluation}
\label{ResultsGNY}
The purpose of this appendix is to explicitly match the results of \cite{Gottsche:2006bm} to the result derived from the $U$-plane integral. Let us first recall that the (virtual) holomorphic Euler characteristic of a proper scheme is defined by G\"ottsche and Fantechi in \cite[Corollary 3.4 (arXiv version)]{Fantechi_2010}
in terms of the Riemann-Roch theorem. The K-theoretic Donaldson invariant is defined as \cite[Footnote 1]{Gottsche:2019vbi},
\be 
\chi^{\rm vir}(M,\mu(L))=\int_{[M]^{{\rm vir}}} e^{\mu_D(c_1(L))}\,{\rm Td}(T^{\rm vir}_M).
\ee 
This definition also covers the cases where the Donaldson line bundle $\mu(L)$ does not exist, and the right-hand side is not necessarily integral. In such cases, the right-hand side can still be evaluated using the techniques of \cite{Gottsche:2006bm, Gottsche:2019vbi}. 

We present a few explicit results in the following. For the gauge group $\mathrm{U}(1)$, or the rank-one sheaves, the moduli space of instantons corresponds to symmetric products of $X$, $M_k\simeq X^{[k]}$. We let $E={\rm det}(\CO_{X^{[k]}})$. Then the generating series for a projective surface reads \cite{Ellingsrud:1999iv}
\be 
G=\mathrm{U}(1):\qquad\Phi_{r,n}(\CR)=\sum_k \chi(M_k,\mu_D(n)\otimes E^{\otimes \kappa_{\text{CS}}})\CR^d,
\ee
with $\chi(n)$ the holomorphic Euler characteristic of the line bundle with $c_1=n$. We have for $\kappa_{\text{CS}}=0,\pm 1$,
\be 
\Phi_{0,n}(\CR)=\left(\frac{1}{1-\CR^2}\right)^{\chi(n)},\quad
\Phi_{\pm 1,n}(\CR)=\left(1+\CR^2\right)^{\chi(n)}.
\ee 
Closed expressions for other values of $\kappa_{\text{CS}}$ are unknown to us. 

For $X$ a minimal surface of general type with $b_1(X)>0$ and $p_g>0$, Conjecture 1.6 of \cite{Gottsche:2020ass} (see also Conjecture 1.2 of \cite{Gottsche:2021ihz}) gives an expression for arbitrary rank. Note that for these four-manifolds only two SW basic classes are non-vanishing, namely $c=K_X$ and $c=-K_X$. The K-theoretic Donaldson invariants for gauge group $G=\mathrm{SU}(2)$ was determined in \cite{Gottsche:2006bm}. These extend the results of Danila
\cite{danila2000} \cite{danila2002}. To arrive at this result, Ref. \cite{Gottsche:2006bm} considers the blow-up $\mathbb{F}_1\to \mathbb{CP}^2$ of $\mathbb{CP}^2$. For specific choices of the first Chern class, $\mathbb{F}_1$ gives rise to an ``empty'' chamber. Then application of the wall-crossing and blow-down formulae gives the result for $\mathbb{CP}^2$.

To state the results, we let $d=4c_2-c_1^2-3$ be the complex
dimension of the moduli space. With a few minor changes to facilitate comparison with our notation, the result by \cite{Gottsche:2006bm} reads\footnote{This equation applies to the two series presented for $c_1=0$ and $H$ in \cite[Sec.   4.5]{Gottsche:2006bm}. In the series for $c_1=H$, $2n$ is replaced by $n$. }
\be
\label{GNYseriesDef}
\Phi^{{\rm GNY}}_{\mu,n}(\CR)=\sum_{d\geq 0}\chi(M_H^{\mathbb{CP}^2}(2\mu\,H,d),\CO(\mu_D(H^{\otimes n})))\,\CR^d.
\ee
For $\mathbb{CP}^2$, $b_2(M)=2$ for the moduli space of instantons. $\mu(H)$ generates a 1d subspace of $H^2(M)$. Moreover, $\mu_D(K_X)=c_1(K_M)/2$ \cite{Gottsche:2006bm}\cite[Proposition 8.3.1]{HuyLen10} and $c_1(K_{\mathbb{CP}^2})=-3H$. 
With $w=e^{2\pi i v}$ and $v$ as in Eq.  (\ref{eq:Defv}), the result of \cite[P. 46]{Gottsche:2006bm} reads
\be
\label{GNYseries}
\begin{split}
\Phi^{{\rm GNY}}_{0,n}(\CR)&= {\rm Coeff}_{q^0}{\rm Ser}_\CR\left[\sum_{l\geq m>0} (-1)^{l+m+1}
  q^{\frac{1}{2}((l+\frac{1}{2})^2-m^2)}\,
  w^{m(n+3)-l-\frac{1}{2}} \right.\\
&\quad \times \left. \left(-\frac{\vartheta_1(\tau,v)}{\CR \vartheta_4(\tau)}
  \right)^{(n+3)^2-1} \frac{8\vartheta_4(\tau)^8}{\CR\,\vartheta_3(\tau)^3\vartheta_2(\tau)^3}\frac{1}{\sqrt{1-2\CR^2\mathtt{u}+\CR^4}}\right],\\  
\Phi^{{\rm GNY}}_{\frac12,n}(\CR)&= {\rm Coeff}_{q^0}{\rm Ser}_\CR\left[\sum_{l\geq m>0} (-1)^{l+m}
  q^{\frac{1}{2}(l^2-(m-\frac{1}{2})^2)}\,
  w^{(m-\frac{1}{2})(n+3)-l}
\right.\\
&\quad \times \left. \left(-\frac{\vartheta_1(\tau,v)}{\CR \vartheta_4(\tau)}
  \right)^{(n+3)^2-1} \frac{8\vartheta_4(\tau)^8}{\vartheta_3(\tau)^3\vartheta_2(\tau)^3}\frac{1}{\sqrt{1-2\CR^2\mathtt{u}+\CR^4}}\right]. 
\end{split}
\ee

We evaluate the $q^0$ term after making a small $\CR$ expansion. To this end, we substitute $C^{(n+3)^2-1}$ for $(\cdots)^{(n+3)^2-1}$, with $C$ as in \eqref{Ctauv} and \eqref{Cqexp}. For $v$, we substitute the expansion for small $\CR$, Eq.  (\ref{vR}). This gives for $\mu=0$ \cite[Sec.   4.5]{Gottsche:2006bm}\footnote{These data is presented in \cite{Gottsche:2006bm}  somewhat differently in terms of generating functions $P_d(t)$ (and $Q_d(t)$ for $\mu=1/2$) that run over $n$ and with $d$ the dimension of the moduli space. In addition to the series $P_d(t)$ given for $d\geq 5$ in \cite{Gottsche:2006bm}, there is also $P_1(t)=(1-t)^{-1}$ for $d=1$}
\be
\label{PhiGNYP20}
\Phi^{{\rm GNY}}_{0,n}(\CR)=\left\{\begin{array}{ll}
                               21\CR-21\CR^5-56\CR^9+\cdots, & n=-8,\\ 
                               15\CR-6\CR^5-10\CR^9+\cdots, & n=-7,\\ 
                               10\CR-\CR^5-\CR^9+\cdots, & n=-6,\\ 
                               6\CR+\CO(\CR^{13}) , & n=-5,\\ 
                               3\CR+\CO(\CR^{13}) , & n=-4,\\ 
 \CR+\CO(\CR^{13})\cdots, & n=-3,\\ 
\CO(\CR^{13})\text{ or less},& n=-2,-1, \\
                               \CR+\CR^5+\CR^9+\cdots, & n=0,\\ 3
                               \CR+6\CR^{5}+10\CR^{9}+\cdots, & n=1, \\ 6
                               \CR+21\CR^{5}+56\CR^{9}+\cdots, & n=2,\\ 10
                               \CR+56\CR^{5}+230\CR^{9}+\cdots, & n=3,  \end{array}\right.
\ee 
To compare with the $U$-plane evaluation (\ref{PhiP20}), we recall the identification \eqref{eq:c1-Lgny-bfnI},
\be 
\label{eq:nInIdent}
n_I=-n+c_1(K_{\mathbb{CP}^2})=-n-3.
\ee 
The coefficients of $\CO(\CR^5)$ and $\CO(\CR^9)$ agree with Eq. (\ref{PhiP20}). However, the coefficients of $\CO(\CR)$ do not match. We will comment on this in more detail below.

For $\mu=1/2$, determination of $\Phi^{{\rm GNY}}_{1/2,n}(\CR)$ gives for $n$ even,
\be
\label{PhiGNY12}
\Phi^{{\rm GNY}}_{\frac12,n}(\CR)=\left\{\begin{array}{ll} 
1+\CO(\CR^{13}),& n=-4,-2, \\
1+\CR^4+\CR^8+\CR^{12}+\cdots, & n=-6,0,\\ 
1+6\CR^{4}+21\CR^{8}+56\CR^{12}+\cdots, & n=-8,2, \\
1+21\CR^{4}+210\CR^{8}+1401\CR^{12}+\cdots, & n=-10,4,\\ 
1+55\CR^{4}+1310\CR^{8}+19432\CR^{12}\cdots, & n=-12,6.\\  
\end{array}\right.
\ee 
This agrees exactly with Eq.  (\ref{PhiU12}) with the identification \eqref{eq:nInIdent}.

G\"ottsche \cite{Gottsche:2016} derives that the right-hand side is in fact series expansions of rational functions in $\CR$. We list a few here,
\be 
\label{PhiG0rational}
\Phi^{{\rm GNY}}_{0,n}(\CR)=\left\{
\begin{array}{ll} 
\CR^{-3}\left(1-\CR^4\right)^{-3}-\CR^{-3},& n=1, \\
\CR^{-3}\left(1-\CR^4\right)^{-6}-\CR^{-3}, & n=2,\\
\CR^{-3}(1+\CR^8)\left(1-\CR^4\right)^{-10}-\CR^{-3}, & n=3,
\end{array}
\right.
\ee 
and
\be 
\begin{aligned}
\label{PhiG12rational}
&\Phi^{{\rm GNY}}_{\frac12,n}(\CR) \\
=&\left\{\begin{array}{ll} 
1,& n=-4,-2, \\
\left(1-\CR^4\right)^{-1} , & n=-6,0,\\
\left(1-\CR^4\right)^{-6} , & n=-8,2, \\
\left(1+6\CR^4+\CR^8\right)\left(1-\CR^4\right)^{-15}, & n=-10,4,\\ 
\left(1 + 27\CR^4 + 148\CR^{8} + 266\CR^{12} + 378\CR^{16} + 147\CR^{20} + 
  56\CR^{24} + \CR^{32}\right)\left(1-\CR^4\right)^{-28}, & n=-12,6.
\end{array}\right.
\end{aligned}
\ee 

To make contact with the $U$-plane integral, we aim to write Eq.  (\ref{GNYseries}) in terms of modular forms and mock Jacobi forms in what follows. To simplify the analysis, we replace $n+3$ by $n$ on the right-hand side of Eq.  (\ref{GNYseries}). To write the sum over $l$ and $m$ in terms of the Appell-Lerch sum \eqref{Mtuv}, note that the $q$-expansion of $v$ (\ref{vsmallR}) has exponents in $-1/8 \mod 1/4$. Since the combination of all other terms within the straight brackets in (\ref{GNYseries}) have an expansion in $q^{1/4}$, the $q^0$ term only depends on even powers of $v$ in the Taylor expansion of $w$. For $\Phi^{\rm GNY}_{0,n-3}$, we can therefore replace the series,
\be 
\label{GNYindeftheta}
\sum_{l\geq m>0} (-1)^{l+m+1}
  q^{\frac{1}{2}((l+\frac{1}{2})^2-m^2)} w^{mn-l-\frac{1}{2}},
\ee 
by\footnote{This expression demonstrates that, with the substitution described below Eq.  \eqref{GNYseries} and considering $C$ an even function of $v$, the function in straight brackets in \eqref{GNYseries} is an even function of $v$. Consequently, the explicit results of GNY are insensitive to the sign error discussed in Footnote \ref{footnote:sign}. This hinges at the fact that in \eqref{GNYseries} the exponent of $C$ is $(n+3)^2-1$ rather than $(n+3)^2$, and is contrary to the function within the straight brackets in Eq.  (\ref{eq:Phi-zero-CP2n}).}
\be 
\label{eq:GNYsymw}
\frac{1}{2}\sum_{l\geq m>0} (-1)^{l+m+1}
  q^{\frac{1}{2}((l+\frac{1}{2})^2-m^2)} (w^{mn-l-\frac{1}{2}}+w^{-(mn-l-\frac{1}{2})}).
\ee 
We proceed by rewriting the first term. After shifting $l\to l+m$, this equals
\be 
\begin{split}
&\sum_{l\geq 0, m>0} (-1)^{l+1}
  q^{\frac{1}{2}((l+m+\frac{1}{2})^2-m^2)} w^{mn-l-m-\frac{1}{2}}\\
=& \sum_{l\geq 0, m>0} (-1)^{l+1}
  q^{\frac{1}{2}(l+\frac{1}{2})^2+m(l+\frac{1}{2})} w^{m(n-1)-l-\frac{1}{2}}\\
=&\sum_{l\in \mathbb{N} +\frac{1}{2}, m>0} (-1)^{l+\frac{1}{2}}
  q^{\frac{1}{2}l^2+ml} w^{m(n-1)-l}.
\end{split}
\ee 
Now we can carry out the sum over $m$ as a geometric series,
\be 
\begin{split}
& \sum_{l\geq 0} (-1)^{l+1}
  \frac{q^{\frac{1}{2}(l+\frac{1}{2})^2+(l+\frac{1}{2})} w^{n-1-l-\frac{1}{2}}}{1-q^{l+\frac{1}{2}}w^{n-1}}\\
=& \sum_{l\in \mathbb{N} +\frac{1}{2}} (-1)^{l+\frac{1}{2}}
  \frac{q^{\frac{1}{2}l^2+l} w^{(n-1)-l}}{1-q^{l}w^{n-1}}\\
=& \sum_{l\in \mathbb{N} +\frac{1}{2}} (-1)^{l-\frac{1}{2}}
  \frac{q^{\frac{1}{2}l^2} w^{-l}}{1-q^{-l}w^{-(n-1)}},
\end{split}
\ee 
or equivalently
\be 
\sum_{l\in \mathbb{N} +\frac{1}{2}} (-1)^{l+\frac{1}{2}}
  \frac{q^{\frac{1}{2}l^2} w^{-l}}{1-q^{l}w^{n-1}}-\sum_{l\in \mathbb{N} +\frac{1}{2}} (-1)^{l+\frac{1}{2}}
  q^{\frac{1}{2}l^2} w^{-l}.
\ee 
Combining this with the second term in Eq.  \eqref{eq:GNYsymw}, we find that Eq.  (\ref{GNYindeftheta}) can be replaced with
\be 
\label{mu0AppellGNY}
\frac{1}{2}\sum_{l\in \mathbb{Z} +\frac{1}{2}} (-1)^{l+\frac{1}{2}}
  \frac{q^{\frac{1}{2}l^2} w^{-l}}{1-q^{l}w^{n-1}}-\frac{1}{2}\sum_{l\in \mathbb{N} +\frac{1}{2}} (-1)^{l+\frac{1}{2}}
  q^{\frac{1}{2}l^2} w^{-l}.
\ee 
The first term reads in terms of the Appell-Lerch sum $M$ (\ref{Mtuv}),
\be 
\frac{1}{2} q^{-\frac18} w^{-\frac{n}2}\vartheta_1(\tau,v)\, M\left(\tau,\frac{\tau}{2}+(n-1)v,-v\right).
\ee 
The sum over $l$ in the second term of Eq.  (\ref{mu0AppellGNY}) runs only over the positive half-integers. It is therefore an example of a ``partial theta function''. The modular properties of such functions, and the closely related ``false theta functions'', are much more intricate than for familiar modular forms \cite{Bringmann:2019vyd}. It may seem surprising that such functions occur here, although precisely such functions do also occur in the case of generating functions of Poincar\'e polynomials of instanton moduli spaces. See for example \cite[Proposition 6.2]{Manschot:2011ym}.
We define the partial theta function,
\be 
\vartheta^+(\tau,z)=\sum_{l\in \mathbb{N} +\frac{1}{2}} (-1)^{l+\frac{1}{2}}
  q^{\frac{1}{2}l^2} e^{2\pi i l z},
\ee 
such that we can write $\Phi^{{\rm GNY}}_{0,n-3}$ (\ref{GNYseries}) as
\be 
\begin{split}
\Phi^{{\rm GNY}}_{0,n-3}(\CR)&=\frac{1}{2}{\rm Coeff}_{q^0}{\rm Ser}_\CR\left[ C^{n^2} \frac{8\vartheta_4(\tau)^9}{\vartheta_3(\tau)^3\vartheta_2(\tau)^3}\frac{1}{\sqrt{1-2\CR^2\mathtt{u}+\CR^4}} \right. \\
&\quad \left.\times\left(q^{-\frac18}w^{-\frac{n}2}M\left(\tau,\frac{\tau}{2}+(n-1)v,-v\right)-\frac{\vartheta^+(\tau,-v)}{\vartheta_1(\tau,v)}\right)  \right].
\end{split}
\ee 

We will next relate this expression to the result in Eq.  \eqref{eq:Phi-zero-CP2n} obtained from the $U$-plane in terms of $G_{0,n}$ (\ref{G0n}). Using Eq.  (\ref{mushifts}) and $\vartheta_1(\tau,\tau/2+z)=iq^{-1/8}e^{-\pi i z}\vartheta_4(\tau,z)$, we find
\be 
\begin{split}
&M\left(\tau,\frac{\tau}{2}+(n-1)v,-v\right)\\
=& M\left(\tau,nv,-\frac{\tau}{2}\right)-\frac{1}{\CR}q^{\frac18}w^{\frac{n}2}\frac{\eta(\tau)^3\,\vartheta_1(\tau,(n-1)v)}{\vartheta_4(\tau)\,\vartheta_1(\tau,nv)\,\vartheta_4(\tau,(n-1)v)}.
\end{split}
\ee 
We furthermore have
\be 
M\left(\tau,nv,-\frac{\tau}{2}\right)=-w^{n}M\left(\tau,nv,\frac{\tau}{2}\right)-iq^{\frac{1}{8}}w^{\frac{n}{2}}.
\ee 
Then using the identity \cite[Eq. s (5.61) and (5.62)]{Korpas:2019cwg}
\be 
G_{0,n}(\tau,v)=-q^{-\frac18}w^{\frac{n}{2}}M\left(\tau,nv,\frac{\tau}{2}\right)-\frac{i}{\vartheta_4(\tau,nv)} \partial_\rho \ln\left.\left( \frac{\vartheta_1(\tau,\rho)}{\vartheta_4(\tau,\rho)}\right)\right|_{\rho=nv},
\ee 
and $i\vartheta_1(\tau,z)+\vartheta^+(\tau,-z)=\vartheta^+(\tau,z)$, we arrive at
\be 
\label{PhiGNYM0}
\begin{split}
\Phi^{{\rm GNY}}_{0,n-3}(\CR)&=\frac{1}{2}{\rm Coeff}_{q^0}{\rm Ser}_\CR\left[ C^{n^2} \frac{8\vartheta_4(\tau)^9}{\vartheta_3(\tau)^3\vartheta_2(\tau)^3}\frac{1}{\sqrt{1-2\CR^2\mathtt{u}+\CR^4}} \right. \\
&\quad\times\left(G_{0,n}(\tau,v)-\frac{1}{\CR}\frac{\eta(\tau)^3\,\vartheta_1(\tau,(n-1)v)}{\vartheta_4(\tau)\,\vartheta_1(\tau,nv)\,\vartheta_4(\tau,(n-1)v)}\right.\\
&\quad\left. \left.-\frac{i}{\vartheta_4(\tau,nv)} \partial_\rho \ln\!\left.\left( \frac{\vartheta_1(\tau,\rho)}{\vartheta_4(\tau,\rho)}\right)\right|_{\rho=nv}-\frac{\vartheta^+(\tau,v)}{\vartheta_1(\tau,v)}\right)  \right].
\end{split}
\ee 

We proceed similarly for the generating function $\Phi^{\rm GNY}_{1/2,n-3}$ in (\ref{GNYseries}). Again from the sum over $m$ and $l$, 
\be 
\sum_{l\geq m>0} (-1)^{l+m}
  q^{\frac{1}{2}(l^2-(m-\frac{1}{2})^2)}\,
  w^{(m-\frac{1}{2})n-l},
\ee
only even powers of $v$ contribute to the $q^0$ term.
We can therefore replace the sum by,
\be 
\frac{1}{2}\sum_{l\geq m>0} (-1)^{l+m}
  q^{\frac{1}{2}(l^2-(m-\frac{1}{2})^2)}\,
  (w^{(m-\frac{1}{2})n-l}+w^{-((m-\frac{1}{2})n-l)}).
\ee
Using similar manipulations as above for $\Phi^{\rm GNY}_{0,n-3}$, we find that this equals
\be  
-\frac{1}{2}\sum_{l\in \mathbb{Z}} (-1)^l \frac{q^{\frac12(l^2-\frac14)}w^{\frac{n}{2}-l}}{1-q^{l-\frac{1}{2}}w^{n-1}}, 
\ee 
which is proportional to a specialization of the Appell-Lerch sum $M$,
\be 
\label{indefM}
\frac{i}{2} \vartheta_4(\tau,v)\,M\left(\tau,-\frac{\tau}{2}+(n-1)v,-\frac{\tau}{2} -v\right).
\ee 

Using the identity (\ref{tauvid}), we can write $\Phi^{{\rm GNY}}_{1/2,n-3}$ then as 
\be 
\label{PhiGNYM} 
\begin{aligned}
\Phi^{{\rm GNY}}_{\frac{1}{2},n-3}(\CR)&= -\frac{i}{2} {\rm Coeff}_{q^0}{\rm Ser}_\CR\Bigg[ M\left(\tau,-\frac{\tau}{2}+(n-1)v,-\frac{\tau}{2} -v\right) \\
&\quad \times C^{n^2} \frac{8\vartheta_4(\tau)^9}{\vartheta_3(\tau)^3\vartheta_2(\tau)^3}\frac{1}{\sqrt{1-2\CR^2\mathtt{u}+\CR^4}}\Bigg].
\end{aligned}
\ee

It is useful to simplify $M$ with these arguments. Using (\ref{mushifts}), we have
\be 
\begin{split}
&M\left(\tau,-\frac{\tau}{2}+(n-1)v,-\frac{\tau}{2} -v\right)\\
=&M\left(\tau,-\frac{\tau}{2}+nv,-\frac{\tau}{2}\right) +\frac{i \eta(\tau)^3 \vartheta_1(\tau,(n-1)v)\,\vartheta_1(\tau,v)}{\vartheta_4(\tau,nv)\,\vartheta_4(\tau,(n-1)v)\,\vartheta_4(\tau,v)\,\vartheta_4(\tau)}.
\end{split}
\ee 
We can use (\ref{tauvid}) to further simplify
\be 
\begin{split}
&M\left(\tau,-\frac{\tau}{2}+(n-1)v,-\frac{\tau}{2} -v\right)\\
=&M\left(\tau,-\frac{\tau}{2}+nv,-\frac{\tau}{2}\right) -i\CR \frac{\eta(\tau)^3\,\vartheta_1(\tau,(n-1)v)}{\vartheta_4(\tau,nv)\,\vartheta_4(\tau,(n-1)v)\,\vartheta_4(\tau)}.
\end{split}
\ee 
Substitution of Eq.  (\ref{G12nM}), we can relate the function to $G_{1/2,n}$ used in Sec.  \ref{sec:evaluate},
\be 
\label{MG12n}
\begin{split} 
&M\left(\tau,-\frac{\tau}{2}+(n-1)v,-\frac{\tau}{2} -v\right)\\
=&  iG_{\frac{1}{2},n}(\tau,v)-i\CR \frac{\eta(\tau)^3\,\vartheta_1(\tau,(n-1)v)}{\vartheta_4(\tau,nv)\,\vartheta_4(\tau,(n-1)v)\,\vartheta_4(\tau)}.
\end{split}
\ee 
For $n=0$, we can use (\ref{tauvid}) again to obtain
\be 
M\left(\tau,-\frac{\tau}{2}-v,-\frac{\tau}{2} -v\right)=iG_{\frac12,0}(\tau) + i\CR^2 \frac{\eta(\tau)^3}{\vartheta_4(\tau)^2}.
\ee 

Finally, substituting Eq. (\ref{MG12n}) into Eq. (\ref{PhiGNYM}) gives us
\be 
\label{PhiGNYM12} 
\begin{split}
\Phi^{{\rm GNY}}_{\frac12,n}(\CR)&= \frac{1}{2} {\rm Coeff}_{q^0}{\rm Ser}_\CR\left[ C^{n^2} \frac{8\vartheta_4(\tau)^9}{\vartheta_3(\tau)^3\vartheta_2(\tau)^3}\frac{1}{\sqrt{1-2\CR^2\mathtt{u}+\CR^4}}
\right.\\
&\quad \times  \left. \left( G_{\frac12,n}(\tau,v)-\CR \frac{\eta(\tau)^3\,\vartheta_1(\tau,(n-1)v)}{\vartheta_4(\tau,nv)\,\vartheta_4(\tau,(n-1)v)\,\vartheta_4(\tau)} \right) \right].
\end{split}
\ee
The last part of Sec.  \ref{sec:evaluate} discusses in more detail the relation between Eqs. \eqref{PhiGNYM0} and \eqref{PhiGNYM12} and the evaluation of the $U$-plane integral.

\section{Expansions Near The Cusps}
\label{app:ExpCusps}
We list in this appendix various quantities near the strong-coupling cusps $U_1,\cdots,U_4$ \eqref{eq:DiscLocus}. The values of $\tau$ near the singularities are given in Eq.  (\ref{eq:tauU}).

\subsubsection*{Cusp $U_1$}
We introduce the local couplings $\tau_1$ and $v_1$ by $\tau=-1/\tau_1$ and $v=v_1/\tau_1$.
Then Eq.  (\ref{tauvid}) gives,
\be
\frac{\vartheta_1(\tau_1,v_1)}{\vartheta_2(\tau_1,v_1)}=i\,\CR.
\ee
This gives for the first terms of the $q_1$-expansion for $v_1$,
\be
\label{eq:v1app}
v_1=\frac{1}{2\pi i}\log\left(\frac{1+\CR}{1-\CR}+8(\CR+2\CR^2+5\CR^3+\CO(\CR^4))\,q_1\,+\CO\left(q_1^2\right)\right).
\ee
To define the local coupling $C_1$, we consider $C$ here as function of $\tau$ and $v$ (\ref{Ctauvv2}), and define $C_1$ as
\be
\label{eq:C1app}
C_1(\tau_1,v_1)=e^{-\pi i\, v_1^2/\tau_1}\,C\left(-\frac{1}{\tau_1},\frac{v_1}{\tau_1}\right)=\frac{\vartheta_2(\tau_1,v_1)}{\vartheta_2(\tau_1)}.
\ee
This gives for the leading behavior of $C_1$ near $U_1$,
\be
C_1(\tau_1,v_1)=\frac{1}{\sqrt{1-\CR^2}}+\CO\left(q_1\right).
\ee
The behavior of the local coordinate $a_1$ follows from Eq.  (\ref{dadtau}). Using $\mathrm{d}a_1/\mathrm{d}\tau_1=\tau_1^{-3}\mathrm{d}a/\mathrm{d}\tau$, we find
\be
\label{da1dtau1}
\frac{\mathrm{d}a_1}{\mathrm{d}\tau_1}=-\frac{2\pi i}{8\,R}\,\frac{\CR}{U}\,\frac{\vartheta_2(\tau_1)^9}{\eta(\tau_1)^3}.
\ee
As a result, we find for the leading term, 
\be
\label{lca1}
a_1=-\frac{64}{R}\,\frac{\CR}{U}\,q_1+\CO\left(q_1^2\right) =-\frac{32\,\Lambda}{1-\CR^2}\,q_1+\CO\left(q_1^2\right),
\ee
where we used that $U=U_1+\CO\left(q_1\right)=2-2\CR^2+\CO\left(q_1\right)$ for small $q_1$.

\subsubsection*{Cusp $U_2$}
We introduce the local couplings $\tau_2$ and $v_2$ through $\tau=2-1/\tau_2$ and $v=v_2/\tau_2$, or reversely
$\tau_2=-1/(\tau-2)$ and $v_2=-v/(\tau-2)$. Then  Eq.  (\ref{tauvid}) becomes near this cusp,
\be
\frac{\vartheta_1(\tau_2,v_2)}{\vartheta_2(\tau_2,v_2)}=\CR.
\ee
This gives for the leading term of $v_2$,
\be
\label{eq:v2app}
v_2=\frac{1}{2\pi i}\log\left(\frac{1-i\,\CR}{1+i\,\CR}+\CO\left(q_2\right) \right).
\ee
Then one finds for the coupling $C_2$,
\be
\label{eq:C2app}
C_2(\tau_2,v_2)= \exp\left(-\frac{\pi i v_2^2}{\tau_2}\right)\,  C\left(2-\frac{1}{\tau_2}, \frac{v_2}{\tau_2}\right)=\frac{\vartheta_2(\tau_2,v_2)}{\vartheta_2(\tau_2)}=\frac{1}{\sqrt{1+\CR^2}}+\CO\left(q_2\right).
\ee                 
We introduce the local coordinate $a_2$ using $\mathrm{d}a_2/\mathrm{d}\tau_2=\tau_2^{-3} \mathrm{d}a/\mathrm{d}\tau$. Using Eq.  (\ref{dadtau}) and $U=2+2\CR^2+\CO\left(q_2\right)$, one derives for the leading term of $a_2$,
\be
a_2=\frac{32i\,\Lambda}{1+\CR^2}\,q_2+\CO\left(q_2^2\right)
\ee

\subsubsection*{Cusp $U_3$}
We set $\tau=4-1/\tau_3$ and $v=v_3/\tau_3$, or reversely
$\tau_2=-1/(\tau-4)$ and $v_2=-v/(\tau-4)$. 
Then Eq.  (\ref{tauvid}) gives,
\be
\frac{\vartheta_1(\tau_3,v_3)}{\vartheta_2(\tau_3,v_3)}=-i\CR.
\ee
This gives for $v_3$,
\be
v_3=\frac{1}{2\pi i}\log\left(\frac{1-\CR}{1+\CR}+\CO\left(q_3\right)\,) \right).
\ee
Similarly to before, we obtain for $C_3$,
\be
C_3(\tau_3,v_3)= \exp\left(-\frac{\pi i v_3^2}{\tau_3}\right)  C\left(4-\frac{1}{\tau_3}, \frac{v_3}{\tau_3}\right)=\frac{1}{\sqrt{1-\CR^2}}+\CO\left(q_3\right).
\ee

As before, we introduce the local coordinate $a_3$ using $\mathrm{d}a_3/\mathrm{d}\tau_3=\tau_3^{-3}\mathrm{d}a/\mathrm{d}\tau$. Using Eq.  (\ref{dadtau}) and $U=U_3+\CO\left(q_3\right)=-2+2\CR^2+\CO\left(q_3\right)$, one obtains
for $a_3$, 
\be
  a_3=-\frac{32\,\Lambda}{1-\CR^2}\,q_3+\CO\left(q_3^2\right).
\ee

\paragraph{Cusp $U_4$}
We set $\tau=6-1/\tau_4$ and $v=v_4/\tau_4$, or reversely
$\tau_4=-1/(\tau-6)$ and $v_4=-v/(\tau-6)$. Then Eq.  (\ref{tauvid}) gives,
\be
\frac{\vartheta_1(\tau_4,v_4)}{\vartheta_2(\tau_4,v_4)}=-\CR.
\ee
This gives for $v_4$,
\be
v_4=\frac{1}{2\pi i}\log\left(\frac{1+i\,\CR}{1-i\,\CR}+\CO\left(q_4\right)\,) \right).
\ee

For the coupling $C_4$, we arrive at
\be
C_4(\tau_4,v_4)= \exp\left(-\frac{\pi i v_4^2}{\tau_4}\right)  C\left(6-\frac{1}{\tau_4}, \frac{v_4}{\tau_4}\right)=\frac{1}{\sqrt{1+\CR^2}}+\CO\left(q_4\right).
\ee

Using Eq.  (\ref{dadtau}) and $U=U_4+\CO\left(q_4\right)=-2-2\CR^2+\CO\left(q_4\right)$, one obtains for the local coordinate $a_4$, 
\be
\begin{split}
  a_4&=32i\,\frac{\Lambda}{1+\CR^2}\,q_4+\CO\left(q_4^2\right).
  \end{split}
\ee

\section{Monopole Equations}\label{app:MonopoleEquations}

\subsection{A Physical Derivation Of The Perturbed Seiberg-Witten Equations}\label{App:ShiftedWalls} 

We claim that the shift of the walls in \eqref{eq:tau-Walls} is related to the modification of the SW equation in the presence of the $\mathrm{U}(1)^{(I)}$ flux. This is relevant to the discussion of the shifted walls in Sec.  \ref{Sec:E1Uplane}. 

Let us consider the LEEA around one of the strong-coupling cusps, which is the $\mathrm{U}(1)$ gauge theory coupled to a spinor field $M\in \Gamma(S^+\otimes L^{1/2})$, where $L$ is the determinant line bundle of a Spin$^c$ structure. Suppose that the LEEA has a mixing between $F$ and a background flux $F^{(1)}$, where $F^{(1)}$ can be $F^{(I)}$ as in the LEEA \eqref{LEETL} in the main text, or a background flux for another global $\mathrm{U}(1)$ symmetry. The action is then a sum of two terms,
\be
S = S_{\mathrm{U}(1)} + S_{\text{hyper}},
\ee
where $S_{\mathrm{U}(1)}$ is the contribution of the abelian vector multiplet $V$ and $V_{(1)}$, while $S_{\text{hyper}}$ is the contribution of the hypermultiplet that contains the spinor field $M$. The relevant bosonic actions are
\bea
S_{\mathrm{U}(1)} = &  \frac{y}{8\pi} \int_X (F_+ \wedge F_+ - D \wedge D) + \frac{y_{1}}{2\pi} \int_X (F_+ \wedge F^{(1)}_{+} - D\wedge D^{(1)}) \\
& + \frac{i\tau}{16\pi} F\wedge F + \frac{iv}{4\pi} \int F\wedge F^{(1)} + \cdots,
\eea
where $y={\rm Im}(\tau)$ and $y_{1} = \text{Im}(v)$, and
\bea
S_{\text{hyper}} =& \int \mathrm{d}^4x \, \sqrt{g }  |D^{\alpha \dot \alpha}_{A+2A^{(1)}} M_{\dot\alpha}|^2 + c \int_X (F_+ - D) \wedge \overline M M+\cdots \\
=& \int \mathrm{d}^4x \, \sqrt{g } |D^{\alpha \dot \alpha}_{A+2A^{(1)}} M_{\dot\alpha}|^2 + c \int_X 
 \left[(F_+ - D) + \frac{2y_1}{y}(F^{(1)}_+ - D^{(1)})\right]\wedge \overline M M + \cdots,
\eea
where for the second line, we use the BPS condition $F^{(1)}_+ = D^{(1)}_{+}$ for the background fields. We also defined a two-form $(\overline M M)$, whose components are given by $ \overline M_{(\dot\alpha} M_{\dot\beta)} = \overline\sigma^{\mu\nu}_{\dot\alpha\dot\beta}(\overline M M)_{\mu\nu}$.
Note that the constant $c$ can be changed by adding a $\bar{\CQ}$-exact term 
$\sim \bar\CQ(-i\chi\wedge \bar M M)$ but that it must be nonzero and cannot be set to zero as that would change the behavior of the action at infinity in fieldspace. 

The bosonic action can be reorganized into a sum of complete squares plus topological terms:
\bea
S =& \frac{y}{8\pi} \int_X\left( F_+ +\frac{2y_{1}}{y} F^{(1)}_+ + \frac{4\pi c}{y} \overline M M\right) \wedge \left( F_+ +\frac{2y_{1}}{y} F^{(1)}_+ + \frac{4\pi c}{y} \overline M M\right) \\
& - \frac{y}{8\pi} \int_X \left(D + \frac{2y_{1}}{y} D^{(1)} + \frac{4\pi c}{y} \overline M M\right) \wedge \left(D + \frac{2y_{1}}{y} D^{(1)} + \frac{4\pi c}{y} \overline M M\right) \\
& + \int \mathrm{d}^4x \, \sqrt{g }  |D^{\alpha \dot \alpha}_{A+2A^{(1)}} M_{\dot\alpha}|^2 \\
& + \frac{i \tau}{16\pi} \int_X F\wedge F + \frac{iv}{4\pi} \int_X F\wedge F^{(1)} + \cdots.
\eea

Integrating out $D$, we arrive at
\bea
S =& \frac{y}{8\pi} \int_X\left( F_+ +\frac{2y_{1}}{y} F^{(1)}_+ + \frac{4\pi c}{y} \overline M M\right) \wedge \left( F_+ +\frac{2y_{1}}{y} F_+^{(1)} + \frac{4\pi c}{y} \overline M M\right) \\
&  + \int \mathrm{d}^4x \, \sqrt{g }  |D^{\alpha \dot \alpha}_{A+2A^{(1)}} M_{\dot\alpha}|^2 + \frac{i \tau}{16\pi} \int_X F\wedge F + \frac{iv}{4\pi} \int_X F\wedge F^{(I)} + \cdots.
\eea

The field configurations $(A,M)$ that minimize the action are the solutions to the equations:
\bea\label{eq:ModSWEq}
&F_+ +\frac{2y_{1}}{y} F^{(1)}_+ + \frac{4\pi c}{y} \overline M M =0, \\
&D^{\alpha \dot \alpha}_{A+2A^{(1)}} M_{\dot\alpha}=0.
\eea
Note that we can set the constant $c$ to $1$ by rescaling  $M$, as expected. 
Eq. \eqref{eq:ModSWEq} is the modified SW equation. The moduli space is the space of solution $(A,M)$ to these equations modulo the $\mathrm{U}(1)$ gauge transformation. At the locus where $F_+ +\frac{2y_{1}}{y} F_+^{(1)}=0$, the equations can be solved with the spinor field vanishing, such that the $\mathrm{U}(1)$ action does not act freely. This implies that the modified SW invariants are expected to jump at $B(\bfk+ \frac{\text{Im}(v)}{y}\bfn, J)=0$, with $\bfk=F/4\pi$ and $\bfn=F^{(1)}/2\pi$.

\subsubsection*{Remark}
The above discussion gives a very nice physical interpretation to the \emph{perturbed} SW equations, which are the equations usually discussed in rigorous mathematical discussions of the SW moduli space and invariants. See, for example \cite{Okonek:1996hd, Nicolaescu:2000}.

\subsection{Multi-Monopole Equations With Background Fluxes}
It is straightforward to generalize the above discussion for a single monopole field to the $N$-monopole equations with $N$ monopoles $M_j$, with $M_j$ coupling to the background flux $F^{(j)}$, 
\be
\label{eq:multiMonoEqs}
\begin{split}
&\sum_{j=1}^N F_+\!\left(A+ \frac{2N\,{\rm Im}(v_j)}{y} A^{(j)}\right) + \bar M_j M_j=0, \\
&\slashed{D}_{j}M_j=0,\qquad j=1,\cdots, N,
\end{split}
\ee 
where $\slashed{D}_{j}$ is the Spin$^c$ Dirac operator coupled to $A+2q_j A^{(j)}$.
In the limit to the singularity, $y\to\infty$, where the monopole becomes massless, $\frac{N\,{\rm Im}(v_j)}{y}$ should approach the integer charge $q_j$ of the monopole with respect to the background flux $F^{(j)}$. The normalization is such that $\tfrac{1}{2\pi}F(A)\in  \overline{w_2(X)}+2H^2(X,\mathbb{Z})$, and $\tfrac{1}{2\pi}F(A^j)\in H^2(X,\mathbb{Z})$.
The resulting equations are the generalized multi-monopole equations. 

\paragraph{Definition:} Let $A$ be a Spin$^c$ connection for a Spin$^c$ structure with characteristic class $c$. Let $(L_j, A_j)$, $j=1,\cdots, N$ be line bundles with connection.  
\footnote{The limit of Eq. \eqref{eq:multiMonoEqs} contains integers $q_j$, so we are defining $L_j$ to be the $q_j$ power of the 
line bundle for the background flux used above.}
Define $N$ corresponding   Spin$^c$ connections $A + 2 A^j$ with characteristic classes $c_j:=c+ 2 c_1(L_j)$.  Then the generalized multi-monopole equations are equations for $A$ and $N$ monopole fields 
$M_j \in \Gamma(W_j^+) $: 
\be
\label{eq:multiMonoEqs2}
\begin{split}
&\sum_{j=1}^N F_+\!\left(A+ 2   A_j\right) + \bar M_j M_j=0, \\
&\slashed{D}_{j}M_j=0,\qquad j=1,\cdots, N,
\end{split}
\ee 
where $\slashed{D}_{j}$ is the Spin$^c$ Dirac operator coupled to connection $A+ 2   A_j$. 
These are a slight generalization of the equations with $c_j=c$ for all $j=1,\cdots,N$ that appear in the literature \cite{bryan1996, LoNeSha, Dedushenko:2017tdw}.

We denote the moduli space of solutions for $(A,\{M_j\})$ modulo gauge transformations by $\CM_N(\{c_j\})$. The (virtual) complex dimension of this space is a simple adaptation of the dimension formula for equal $c_j$ \cite{bryan1996}
\be 
\label{eq:Nmondim}
m(\{c_j\})=\frac{-2\chi-2\sigma+\sum_j (c_j^2-\sigma)}{8}.
\ee 
The equations are left invariant by a  $\mathrm{U}(1)^N$ symmetry group with the $j^{th}$ $\mathrm{U}(1)$ factor 
acting only on $M_j$ and taking $M_j \mapsto e^{i\theta_j}M_j$. The diagonal $\mathrm{U}(1)$ with $\theta_j=\theta$ for all $j$ is gauged, while a complementary subgroup $\cong \mathrm{U}(1)^{N-1}$ is a global symmetry. If some of the $c_j$ coincide, the $\mathrm{U}(1)^{N-1}$ global symmetry is a subgroup of the Cartan group of an enhanced nonabelian global symmetry. 

Let $\mathbb{L}\to \CM_N(\{c_j\})\times X$ be the universal line bundle with first Chern class $c_1(\mathbb{L})$. The restriction of $c_1(\mathbb{L})$ to $\CM_N(\{c_j\})$ is the image of the local coordinate $a$ under the $\mu$-map. 
The Higgs branch integral is written naturally as an equivariant integral \cite{LoNeSha, Dedushenko:2017tdw}.
\be
  \int_{\CM_N(\{c_j\})} e^{c_1(\mathbb{L})+m_\alpha H^\alpha},
\ee 
where the equivariant parameters $m_\alpha$, $\alpha=1,\cdots,N-1$ are monopole masses.

The condition for the walls of marginal stability for these equations is 
\be 
\sum_{j=1}^N F_+(A+ 2A_j)=0,
\ee 
or
\be\label{eq:NewWalls-N}
B\left(\sum_j c_j ,J\right)=0.
\ee
In general the sum of Spin$^c$ classes is not a Spin$^c$ class, 
so these walls do not necessarily correspond to Spin$^c$ structures.
This raises many interesting questions that we leave for future investigation.

\section{Toric Geometry}\label{sec:Toric Geometry}

In this section, we review the construction of a toric variety from combinatorial data specified by a fan, focusing on the case of complex dimension $2$.  
For a comprehensive treatment, see \cite{fulton1993introduction}.

\subsection{Cones And Fans}

Let $N = \mathbb{Z}^2$ be a two-dimensional lattice, and define the associated real vector space $N_{\mathbb{R}} = N\otimes_\mathbb{Z}\mathbb{R}$. The dual lattice is $M=N^*=\text{Hom}(N,\IZ)$, with $M_{\mathbb{R}}= M \otimes_{\mathbb{Z}}\mathbb{R}$. The natural pairing between $M$ and $N$ is denoted by $\langle \cdot, \cdot \rangle:M\times N\to \IZ$. For $\vec{m}= (m_1,m_2)\in M$ and $\vec{n}= (n_1,n_2)^\text{T}\in N$,  this is given by
\begin{equation}
\langle \vec{m}, \vec{n}\rangle = m_1 n_1 + m_2 n_2.
\end{equation}

A convex rational polyhedral cone $\sigma \subset N_{\mathbb{R}}$ is a set of the form
\begin{equation}
    \sigma = \{ a_1 \vec{n}_1 + a_2 \vec{n}_2 + \cdots + 
    a_s \vec{n}_s \in N_\mathbb{R}\mid a_i \geq 0 \},
\end{equation}
where $\{\vec{n}_i\}$ is a finite set of generating vectors in $N$. A cone $\sigma$ is strongly convex if $\sigma \cap (-\sigma) = \{0\}$. 
The dual cone of $\sigma$ is defined as 
\begin{equation}
\label{def-dual-cone}
\sigma^{\vee} = \{ \vec{m} \in M_{\mathbb{R}} \mid \langle \vec{m},\vec{n} \rangle \geq 0, \forall \vec{n} \in \sigma  \}.
\end{equation}
For a given $\vec{m} \in \sigma^\vee$, we associate with it a face $\tau$ of $\sigma$, 
\begin{equation}
    \tau = \vec m^\perp= \{\vec{n}\in \sigma\mid 
    \langle \vec{m},\vec{n} \rangle = 0\}.
\end{equation}
A cone is a face of itself if $\vec{m} = 0$; all other faces are called proper faces.

A collection $\Delta$ of strongly convex rational polyhedral cones is called a fan in $N_{\mathbb{R}}$ if
\begin{enumerate}
    \item each face of a cone $\sigma \in \Delta$ is also a cone in $\Delta$;
    \item the intersection $\sigma_1\cap\sigma_2$ of any two cones $\sigma_1, \sigma_2 \in \Delta$ is a face of both.
\end{enumerate}
A compact smooth toric surface $X$ can be specified by a fan $\Delta$ in $N_{\mathbb{R}}=\mathbb{R}^2$ with the following combinatorial data:
\begin{enumerate}
\item A set of $\chi$ integer vectors $\{\vec{n}_\ell,\ell = 1, \cdots, \chi\}$ arranged in counterclockwise order in $N$;
\item Each pair of adjacent vectors $(\vec{n}_\ell, \vec{n}_{\ell+1})$ generates a cone $\sigma_\ell\in\Delta$, with cyclic identification $\vec{n}_{\chi+1} = \vec{n}_1$;
\item To ensure smoothness, each pair $\{\vec{n}_\ell,\vec{n}_{\ell+1}\}$ must form a basis for the whole lattice $N$.
\end{enumerate}

In the following, we list some properties of $\Delta$ defined above. 

\subsubsection*{Property Of The Generating Vectors} 
The vectors $\vec{n}_\ell \in N$ satisfy the following relations for all $\ell = 1, \cdots,\chi$,
\begin{equation}
\label{fan-relations}
    \vec{n}_{\ell-1} - h_s \vec{n}_\ell + \vec{n}_{\ell+1} = 0,\quad h_\ell \in \IZ.
\end{equation}

\paragraph{Proof:} 
Consider $\vec{n}_1,\vec{n}_2, \vec{n}_3$ as an example. Since they are in counterclockwise order, and $\{\vec{n}_1, \vec{n}_2\}$ and $\{\vec{n}_2,\vec{n}_3\}$ are both bases for $N$, we have 
\begin{equation}
\label{deter1}
    \det(\vec{n}_1, \vec{n}_2) = 1, \quad 
    \det(\vec{n}_2, \vec{n}_3) = 1.
\end{equation}
We write $\vec{n}_3$ as an integer linear combination of $\vec{n}_1$ and $\vec{n}_2$, where $\vec{n}_1 = (x_1, x_2)^\text{T}$ and $\vec{n}_2 = (y_1, y_2)^\text{T}$,
\be
\vec{n}_3 = (a x_1 + b y_1, a x_2 + b y_2)^\text{T}, \quad a,b \in \mathbb{Z}.
\ee
Substituting into Eq. \eqref{deter1} gives 
\begin{equation}
    x_1 y_2 - x_2 y_1 = 1,
    \quad 
    y_1(a x_2 + b y_2) - y_2 
    (a x_1 +  b y_1) =1,
\end{equation}
implying $a = -1$. Thus, we find a relation between $\{\vec{n}_1, \vec{n}_2,\vec{n}_3\}$, 
\begin{equation}
    \vec{n}_3 = - \vec{n}_1 + b \vec{n}_2,
\end{equation}
which is of the form \eqref{fan-relations} with $h_2=b$. This argument extends straightforwardly to any $\ell$.
Note that there are $\chi-2$ independent relations in Eq. \eqref{fan-relations}. 

\subsubsection*{Property of Dual Cones}

Denote the $\ell$-th vector in $N$ as $\vec{n}_\ell = (n_\ell^1, n_\ell^2)^\text{T}$. We can rotate it $90^\circ$ clockwise to get a new vector $\vec{n}_{\ell}^* = (n^2_\ell, -n^1_\ell)^\text{T}$.
Then the dual cone $\sigma^\vee_\ell$ is generated by $(\vec{n}^*_{\ell+1}, - \vec{n}_\ell^*)$.

\paragraph{Proof:}
From the definition \eqref{def-dual-cone}, for any $\vec{m} \in \sigma_{\ell}^\vee$, we have $\langle\vec{m}, \vec{n}\rangle \geq 0$ for all $n \in \sigma_{\ell}$. This implies that
\be
\langle \vec{m}, \vec{n}_{\ell} \rangle \geq 0, \quad \langle \vec{m}, \vec{n}_{\ell+1} \rangle \geq 0.
\ee
The first inequality implies that $\vec{m}$ lies in the cone generated by $\{\vec{n}_{\ell}, \vec{n}_{\ell}^* , -\vec{n}_{\ell}^*\}$, and the second implies that it lies in the cone generated by $\{\vec{n}_{\ell+1}, \vec{n}_{\ell+1}^* , -\vec{n}_{\ell+1}^*\}$.
Hence, 
\begin{equation}
    \sigma^\vee_\ell  = 
    \sigma(\{\vec{n}_{\ell}, \vec{n}_{\ell}^* , -\vec{n}_{\ell}^*\}) \cap 
    \sigma(\{\vec{n}_{\ell+1}, \vec{n}_{\ell+1}^* , -\vec{n}_{\ell+1}^*\}).
\end{equation}
Since $\sigma_{\ell}$ is strongly convex, the angle between $\vec{n}_{\ell +1}$ and $\vec{n}_{\ell}$ is less than $\pi$. Thus, the angle between $\vec{n}_{\ell+1}^*$ and $\vec{n}_{\ell}$ is less than $\pi/2$, and the angle between $n_\ell^*$ and $\vec{n}_{\ell+1}$ is strictly between $\pi/2$ and $\pi$, leading to the inequalities
\begin{equation}
\label{inequalities for vec n}
    \langle \vec{n}^*_{\ell+1}, 
    \vec{n}_{\ell} \rangle >0, \quad 
    \langle \vec{n}^*_{\ell}, 
    \vec{n}_{\ell+1} \rangle <0.
\end{equation}
Therefore, 
\begin{equation}
    \sigma^\vee_\ell  = 
    \sigma(\{\vec{n}^*_{\ell+1}, \vec{n}_{\ell}^* , -\vec{n}_{\ell}^*\}) \cap 
    \sigma(\{-\vec{n}^*_{\ell}, \vec{n}_{\ell+1}^* , -\vec{n}_{\ell+1}^*\})
    = \sigma(\{\vec{n}_{\ell +1}^*, -\vec{n}_{\ell}^*\}).
\end{equation}

\subsection{Construction Of Toric Surface}

The toric surface $X$ can be constructed from a fan $\Delta$ via homogeneous coordinates. 
Let $(y_1,\cdots,y_\chi)$ be homogeneous coordinates on $\mathbb{C}^\chi$. Let $v_\ell\in N$ be the one-dimensional cone generated by $\vec{n}_{\ell}$. 
For any $S\subset \{v_\ell\}$ that does not generate a cone in $\Delta$, let $V(S)\subset \mathbb{C}^\chi$ be the linear subspace defined by setting $y_\ell=0$ for all $v_\ell\in S$. Let $Z(\Delta)\subset \mathbb{C}^\chi$ be the union of all such $V(S)$. Then the toric variety $X$ can be defined as a quotient of an open subset in $\mathbb{C}^\chi$ by a group $G$,
\begin{equation}
   X = \left(\mathbb{C}^\chi -Z(\Delta)  \right)/ G.
\end{equation}
where $G$ consists of the equivalence relations
\begin{equation}
\label{equiv-relation}
    (y_1,\cdots,y_\chi) \sim 
    (\lambda^{C_{s,1}} y_1, 
    \lambda^{C_{s,2}} y_2, \cdots,
    \lambda^{C_{s,\chi}} y_\chi),\quad  s = 1, \cdots, \chi,\quad \lambda\in \mathbb{C}^*.
\end{equation}
Here the coefficients $C_{s,\ell}$ are obtained by rewriting Eq. \eqref{fan-relations} in the following form,
\begin{equation}
\label{relation with C}
    \sum_{\ell =1}^\chi C_{s,\ell} \vec{n}_{\ell} = 0,
    \quad s = 1, \cdots, \chi.
\end{equation}
Non-vanishing coefficients are 
\begin{equation}
\label{intersection number}
    C_{\ell,\ell} = -h_\ell, \quad 
    C_{\ell,\ell+1} = C_{\ell+1,\ell} =1.
\end{equation}
Notice that we have $\chi-2$ independent equivalent relations. 

The local patch $U_\ell \cong \mathbb{C}^2$ corresponding to the cone $\sigma^{(\ell)}$ is defined by $y_s \neq 0$ except for $s= \ell , \ell +1$. 
The local coordinates $(z_1^{(\ell)},z_2^{(\ell)})$ on $U_{\ell}$ are
\cite{Bonelli:2020xps}
\begin{equation}
    \label{local coordinate}
    z_1^{(\ell)} = \prod_{m=1}^\chi 
    y_m^{\langle \vec{n}^*_{\ell+1}, \vec{n}_m\rangle}, \quad 
    z_2^{(\ell)} = \prod_{m=1}^\chi 
    y_m^{-\langle \vec{n}^*_{\ell},  \vec{n}_m\rangle }.
\end{equation}
These local coordinates are well defined, since they are invariant under \eqref{equiv-relation}, and due to \eqref{inequalities for vec n}, $y_\ell$ and $y_{\ell+1}$ appear in the numerators.
On the intersection $U_\ell \cap U_{\ell+1} \cong \mathbb{C}^* \times \mathbb{C}$, the coordinate transformations are
\begin{equation}
\label{coordinate trans}
    z_1^{(\ell)} 
    = (z_2^{(\ell+1)})^{-1}, 
    \quad 
    z_2^{(\ell)} 
    = z_1^{(\ell+1)}
    (z_2^{(\ell+1)})^{h_{\ell+1}}.
\end{equation}

\subsection{Torus Action}

The homogeneous coordinates admit a $(\mathbb{C}^*)^2$ torus action defined by
\begin{equation}
\label{T2 action}
    y_1 \to  e^{i\epsilon_1} y_1, \quad 
    y_2 \to  e^{i\epsilon_2} y_2, 
    \quad 
    y_{\ell>2} \to y_\ell,
\end{equation}
with $\epsilon_{1,2} \in \mathbb{C}$.
The local coordinates on $U_\ell$ transform as 
\begin{equation}
\begin{aligned}
    z_1^{(\ell)} &\to
    \exp\left(i\epsilon_1\langle \vec{n}^*_{\ell+1}, \vec{n}_1\rangle + i\epsilon_2\langle \vec{n}^*_{\ell+1}, \vec{n}_2\rangle\right) z_1^{(\ell)}, \\
    z_2^{(\ell)} &\to
    \exp\left(-i\epsilon_1\langle \vec{n}^*_{\ell}, \vec{n}_1\rangle -i\epsilon_2\langle\vec{n}^*_{\ell}, \vec{n}_2\rangle\right) z_2^{(\ell)}.
\end{aligned}
\end{equation}
Choosing $\vec{n}_1 = (1,0)^\text{T}$ and $\vec{n}_2 = (0,1)^\text{T}$, the action on $U_\ell$ simplifies,
\begin{equation}
    z_1^{(\ell)} \to
    e^{i\epsilon_1^{(\ell)}} z_1^{(\ell)},   \quad
    z_2^{(\ell)} \to
    e^{i\epsilon_2^{(\ell)}} z_2^{(\ell)},
\end{equation}
where 
\begin{equation}
\label{def: epsilonl}
    \epsilon_1^{(\ell)}  := \langle \vec{n}_{\ell+1}^*, \vec \epsilon \rangle, 
    \quad 
    \epsilon_2^{(\ell)}  :=  
    -\langle \vec{n}_{\ell}^*, \vec \epsilon\rangle ,
\end{equation}
with $\vec{\epsilon} = (\epsilon_1, \epsilon_2)^\text{T}$.
In particular, on $U_1$, we have $\epsilon_1^{(1)} = \epsilon_1$ and $\epsilon_2^{(1)} = \epsilon_2$. Using \eqref{fan-relations}, the parameters on the other patches can be determined recursively,
\begin{equation}
\label{recursion epsilon}
    \epsilon_1^{(\ell+1)} 
    =  h_{\ell+1} \epsilon_1^{(\ell)} + 
    \epsilon_2^{(\ell)}, \quad 
    \epsilon_2^{(\ell+1)} = 
    - \epsilon_1^{(\ell)}.
\end{equation}
This leads to the identities
\begin{equation}
\label{epsilon relation 1}
   \begin{aligned}
   \sum_{\ell=1}^{\chi} 
   \frac{1}{\epsilon_1^{(\ell)} \epsilon_2^{(\ell)}} &= \sum_{\ell =1}^\chi  
   \frac{\epsilon_1^{(\ell)}+ \epsilon_2^{(\ell)}}{\epsilon_1^{(\ell)} \epsilon_2^{(\ell)}}
   = 0,
   \end{aligned}
\end{equation}
and
\begin{equation}
\label{epsilon relation 2}
\frac{\epsilon_1^{(\ell)}}{\epsilon_2^{(\ell)}} + 
   \frac{\epsilon_2^{(\ell-1)}}{\epsilon_1^{(\ell-1)}} = -h_\ell.
\end{equation}
Let $\vec{\epsilon}^{(\ell)} = (\epsilon_1^{(\ell)},\epsilon_2^{(\ell)})^\text{T}$. Then $\vec{\epsilon}^{(\ell)}$ and $\vec{\epsilon}^{(\ell')}$ are related by a linear transformation, 
\begin{equation}
\label{eq: A-trans-epsilon}
    \vec{\epsilon}^{(\ell)} =  A^{\ell,\ell'} \vec{\epsilon}^{(\ell')}, 
\end{equation}
where $A^{\ell,\ell'}$ is defined recursively by \eqref{recursion epsilon}.
From \eqref{recursion epsilon}, the transformation between adjacent patches is
\begin{equation}
\label{eq: Allp1}
    A^{\ell +1, \ell} = \begin{pmatrix}
        h_{\ell+1} & 1\\
        -1 & 0
    \end{pmatrix},\quad
    A^{\ell, \ell+1} = \begin{pmatrix}
        0 & -1 \\
        1 & h_{\ell + 1}
    \end{pmatrix}.
\end{equation}
From \eqref{eq: A-trans-epsilon} and \eqref{eq: Allp1}, we have
\begin{equation}
\label{eq: A-property-1}
    A^{\ell, \ell'+1} = A^{\ell,\ell'}A^{\ell', \ell'+1} = 
    \begin{pmatrix}
        (A^{\ell,\ell'})_{12}
        &
        -(A^{\ell,\ell'})_{11}+ h_{\ell'+1}(A^{\ell,\ell'})_{12}\\
        (A^{\ell,\ell'})_{22}
        &
        -(A^{\ell,\ell'})_{21}+ h_{\ell'+1}(A^{\ell,\ell'})_{22}
    \end{pmatrix},
\end{equation}
and
\begin{equation}
\label{eq: A-property-2}
    A^{\ell+1,\ell'} = A^{\ell+1, \ell} A^{\ell,\ell'} = \begin{pmatrix}
        h_{\ell+1} (A^{\ell,\ell'})_{11} + (A^{\ell,\ell'})_{21}
        & 
        h_{\ell+1} (A^{\ell,\ell'})_{12} + (A^{\ell,\ell'})_{22}\\
        -(A^{\ell,\ell'})_{11}
        & - (A^{\ell,\ell'})_{12}
    \end{pmatrix}.
\end{equation}
From the definition \eqref{def: epsilonl}, we can express $A^{\ell,1}$ and $A^{1,\ell}$ in terms of the generating vectors $\vec{n}_{\ell}$,
\begin{equation}
\label{eq: Al1-A1l}
    A^{\ell,1} = \begin{pmatrix}
        n^2_{\ell + 1} & - n^1_{\ell+1}\\
        -n^2_\ell      & n^1_\ell
    \end{pmatrix}, 
    \quad 
    A^{1,l} = \begin{pmatrix}
        n^1_\ell & - n^1_{\ell + 1 }\\
        -n^2_\ell & n^2_{\ell+1}
    \end{pmatrix}.
\end{equation}

\subsection{Divisors}

To each $\vec{n}_{\ell}$, we associate a Weil divisor $D_{\ell} \cong \mathbb{CP}^1$, defined by $y_\ell =0$. 
The number of independent divisors is $b_2=\chi-2$. For a symplectic toric four-manifold, $b_1(X)=0$ and $b_2^{+}(X)=1$, so $\chi$ is the Euler characteristic of $X$.

The divisor $D_{\ell}$ is preserved by the torus action, making it an equivariant divisor supported on $U_{\ell-1} \cap U_{\ell}$.
A general $(\mathbb{C}^*)$-invariant Weil divisor can be written as
\begin{equation}\label{eq:WeilDiv-1}
    \boldsymbol{p} = \sum_{\ell=1}^\chi 
        \mathfrak{p}_{\ell} D_{\ell}, \quad \mathfrak{p}_{\ell} \in \mathbb{Z},
\end{equation}
which defines a co-dimensional one subvariety. On each local patch, it is given by
\begin{equation}
\label{eq:WeilDiv-2}
    \boldsymbol{p} 
    \cap U_{\ell} = 
    \{
    f^{(\ell)}_{\boldsymbol{p}} = 
    (z_1^{(\ell)})^{\mathfrak{p}_{\ell}} 
    (z_2^{(\ell)})^{\mathfrak{p}_{\ell+1}} =0
    \}.
\end{equation}
On the intersection $U_{\ell} \cap U_{\ell+1}$, defined by $y_s \neq 0$ for all $s\neq \ell +1$, we find using \eqref{inequalities for vec n} that
\begin{equation}
\label{divisor on intersection}
    \boldsymbol{p} 
    \cap U_{\ell} \cap U_{\ell+1} = 
    \{  y_{\ell+1}^{\mathfrak{p}_{\ell+1}} =0 
    \}.
\end{equation}
Since all homogeneous coordinates are nonzero on $U_{\ell} \cap U_s$ for $s \neq \ell-1, \ell+1$, we have
\begin{equation}
    \boldsymbol{p} 
    \cap U_{\ell} \cap U_{s} = \emptyset.
\end{equation}
Therefore, Eq. \eqref{eq:WeilDiv-2} is well defined on all intersections. 
Especially, when $\boldsymbol{p} = D_{\ell}$, it follows from \eqref{divisor on intersection} that $D_{\ell}$ supported on $U_{\ell-1} \cap U_{\ell}$ is defined by $y_\ell =0$.
The associated holomorphic line bundle can be determined from a nonzero meromorphic section $\{ U_{\ell}, f^{(\ell)}_{\boldsymbol{p}}\}$, with the transition function on $U_{\ell} \cap U_{\ell+1}$
\begin{equation}
\label{cocycle1}
\left(f^{(\ell)}_{\boldsymbol{p}}\right)
   ^{-1} f^{(\ell+1)}_{\boldsymbol{p}} 
   = (z_1^{(\ell)})^{h_{\ell+1} \mathfrak{p}_{\ell+1}- \mathfrak{p}_{\ell+2}- 
   \mathfrak{p}_{\ell}}
   = (z_2^{(\ell+1)})^{-h_{\ell+1} \mathfrak{p}_{\ell+1}+ \mathfrak{p}_{\ell+2}+
   \mathfrak{p}_{\ell}}.
\end{equation}
The transition function for the canonical class of $X$ on $U_{\ell}\cap U_{\ell+1}$  can be computed from \eqref{coordinate trans},
\begin{equation}
    \left(\frac{\mathrm{d} z_1^{(\ell+1)}}
    {\mathrm{d} z_1^{(\ell)}} 
    \frac{\mathrm{d} z_2^{(\ell+1)}}
    {\mathrm{d} z_2^{(\ell)}}-
    \frac{\mathrm{d} z_1^{(\ell+1)}}
    {\mathrm{d} z_2^{(\ell)}} 
    \frac{\mathrm{d} z_2^{(\ell+1)}}
    {\mathrm{d} z_1^{(\ell)}}\right)^{-1} 
    =(z_2^{(\ell+1)})^{-2+ h_{\ell+1}}.
\end{equation}
Matching this with Eq. \eqref{cocycle1} requires setting $\mathfrak{p}_{\ell} =-1$ for all $\ell$. Therefore, the canonical class of $X$ satisfies
\be 
K_X = -\sum_{\ell} D_{\ell}.
\ee

The intersection numbers are given by 
\begin{equation}
B(D_s, D_{\ell}) = C_{s,\ell}, 
\end{equation}
with $C_{s,\ell}$ defined in \eqref{relation with C}. Explicitly, the non-vanishing intersection numbers are
\begin{equation}
\label{intersectionnumber}
    B(D_{\ell}, D_{\ell}) = - h_\ell, \quad 
    B(D_{\ell + 1}, D_{\ell}) = 
    B(D_{\ell}, D_{\ell+1}) = 1.
\end{equation}
The divisors also satisfy two linear relations,
\begin{equation}
\label{relation-divisor}
\sum_{\ell=1}^{\chi}  n_{\ell}^i D_{\ell}=0, \quad  i=1,2,
\end{equation}
which can be proved by showing that the intersection number between $\sum_{\ell}  n_\ell^i D_{\ell}$ and any 2-cycle $\boldsymbol{p}=\sum_{\ell}  \mathfrak{p}_{\ell} D_{\ell}$ vanishes, 
\begin{equation}
   \begin{aligned}
    B\left(\sum_{\ell=1}^{\chi} \mathfrak{p}_{\ell} D_{\ell}, \sum_{\ell=1}^{\chi} 
    n_{\ell}^i D_{\ell} \right) &= \sum_{\ell=1}^{\chi}
    \left(- \mathfrak{p}_{\ell} n_{\ell}^i h_\ell + \mathfrak{p}_{\ell} n^i_{\ell+1} 
    + \mathfrak{p}_{\ell+1} n_{\ell}^i\right) \\
    & = 
    \sum_{\ell=1}^{\chi} \mathfrak{p}_{\ell} 
    \left(
    - n_{\ell}^i h_\ell 
    + n^i_{\ell-1} 
    + n^i_{\ell+1}
    \right)=0,
    \end{aligned}
\end{equation}
where the last step follows from \eqref{fan-relations}.

\subsection{Fluxes}

Given a flux $\bfk  = \left[\frac{F_4}{4\pi}\right] = \frac{1}{2} \sum_{\ell} {\mathfrak{p}}_{\ell} D_{\ell} \in L+\bfmu$, the coefficients $\mathfrak{p}_{\ell}$ are not unique due to \eqref{relation-divisor}. We define an equivalence relation: $\{\mathfrak{p}_\ell\} \sim \{\mathfrak{p}'_\ell\} $ if 
\begin{equation}\label{equivalence relation for p}
    \mathfrak{p}_\ell - \mathfrak{p}'_\ell = 
    \sum_{i=1,2} s_i n^i_\ell, \quad s_1, s_2\in \IZ, \quad \ell = 1, \cdots, \chi.
\end{equation}
The equivalence class $[\{\mathfrak{p}_{\ell}\}]$ is determined by $\boldsymbol{k}$.

We can use $\{D_{\ell}, \ell=3, \cdots, \chi \}$ as a basis for the integer lattice $L$. Using Eq. \eqref{relation-divisor}, we have 
\begin{equation}
    D_1 = - \sum_{\ell = 3}^\chi n_\ell^1 D_{\ell}, \quad
    D_2 = -\sum_{\ell=3}^\chi n_\ell^2 D_{\ell}.
\end{equation}
Then, $\boldsymbol{k}$ is given by
\begin{equation}
\label{eq: kp-rel}
    \boldsymbol{k} 
    = \sum_{\ell=3}^\chi 
    \boldsymbol{k}_\ell D_{\ell} =
    \frac{1}{2}
    \sum_{\ell=3}^\chi 
    (\mathfrak{p}_{\ell}-\mathfrak{p}_1 n_\ell^1 -\mathfrak{p}_2 n_\ell^2)D_{\ell}.
\end{equation}
We write $\boldsymbol{\mu}$ in the same basis,
\be
\boldsymbol{\mu} = \sum_{\ell=3}^\chi {\mu}_\ell D_{\ell}.
\ee
If $\boldsymbol{k}\in L + \boldsymbol{\mu}$, its representatives $\{\mathfrak{p}_{\ell}\}$ belong to the set
\begin{equation}
\label{eq: pmu-condition}
    S_{\boldsymbol{\mu}} =\{ \{\mathfrak{p}\} \mid\mathfrak{p}_{\ell} -\mathfrak{p}_1 n_\ell^1 - \mathfrak{p}_2 n_\ell^2 = 2\mu_\ell \mod 2,\quad  \ell = 3, \cdots, \chi\}.
\end{equation}
The representative of $\boldsymbol{k}$ is labeled by two arbitrary integers, $\mathfrak{p}_1$ and $\mathfrak{p}_2$, and we have the summation rule
\begin{equation}
\label{eq: pk-summation}
\sum_{\mathfrak{p} \in S_{\boldsymbol{\mu}}} = \sum_{(\mathfrak{p}_1, \mathfrak{p}_2)\in \mathbb{Z}^2}\sum_{\boldsymbol{k}}.
\end{equation}

The difference between Coulomb branch variables in adjacent patches depends only on the equivalence class $[\mathfrak{p}]$:
\bea
\label{eq: al-alprime}
a^{(\ell)} - a^{(\ell+1)} &= \frac12\left[\left(\mathfrak{p}_\ell \epsilon_1^{(\ell)} + \mathfrak{p}_{\ell+1}\epsilon_2^{(\ell)}\right) - \left(\mathfrak{p}_{\ell+1}\e_1^{(\ell+1)}+\mathfrak{p}_{\ell+2}\e_2^{(\ell+1)}\right) \right] \\
& = \frac12 \left(\mathfrak{p}_\ell +\mathfrak{p}_{\ell+2}-\mathfrak{p}_{\ell+1}h_{\ell+1}\right)\epsilon_1^{(\ell)}.
\eea
If $\mathfrak p$ and $\mathfrak p'$ are related by \eqref{equivalence relation for p}, then 
\be
(\mathfrak{p}_\ell +\mathfrak{p}_{\ell+2}-\mathfrak{p}_{\ell+1}h_{\ell+1}) - (\mathfrak{p}'_\ell +\mathfrak{p}'_{\ell+2}-\mathfrak{p}'_{\ell+1}h_{\ell+1}) = \sum_{i=1,2} s_i (n_{\ell}^i+n_{\ell+2}^i-n^i_{\ell+1}h_{\ell+1}) = 0,
\ee
which implies that $a^{(\ell)} - a^{(\ell+1)}$ depends only on the equivalence class. Therefore, $a^{(\ell)} - a^{(\ell')}$ for any pair $(\ell,\ell')$ depends only on the equivalence class determined by $H^2(X,\mathbb{Z})$.

For two fluxes $\boldsymbol{p} = \sum_{\ell} \mathfrak{p}_{\ell} D_{\ell}$ and $\boldsymbol{q} = \sum_{\ell} \mathfrak{q}_{\ell} D_{\ell}$, using Eqs. \eqref{epsilon relation 2} and \eqref{intersectionnumber}, we compute
\begin{equation}
\label{Bpq}
    \begin{aligned}
        &\sum_{\ell=1}^{\chi}  \frac{1}{\epsilon_1^{(\ell)}\epsilon_2^{(\ell)}} (\epsilon_1^{(\ell)} {\mathfrak p}_{\ell}+ \epsilon_2^{(\ell)} {\mathfrak p}_{\ell+1}) (\epsilon_1^{(\ell)} {\mathfrak q}_\ell + \epsilon_2^{(\ell)} {\mathfrak q}_{\ell+1})\\ =& 
       \sum_{\ell=1}^{\chi} \left[
        \left(\frac{\epsilon_1^{(\ell)}}{\epsilon_2^{(\ell)}} 
        +\frac{\epsilon_2^{(\ell-1)}}{\epsilon_1^{(\ell-1)}} \right)\mathfrak{p}_{\ell} 
        \mathfrak{q}_\ell 
        + \mathfrak{p}_\ell \mathfrak{q}_{\ell+1} +
        \mathfrak{q}_\ell \mathfrak{p}_{\ell+1} \right]\\
        =& \sum_{\ell=1}^{\chi}  
        B(D_{\ell}, D_{\ell} )\mathfrak{p}_{\ell} 
        \mathfrak{q}_\ell
        +B(D_{\ell}, D_{\ell +1})\mathfrak{p}_\ell \mathfrak{q}_{\ell+1} +B(D_{\ell}, D_{\ell +1})
        \mathfrak{q}_\ell \frak p_{\ell+1}\\
     =& B\left( 
\sum_{\ell=1}^{\chi}\frak p_\ell D_{\ell}, \sum_{\ell=1}^{\chi}\frak q_\ell D_{\ell} \right).
    \end{aligned}
\end{equation}

Finally, we claim that the equivariant extension of the divisor $D_s$ is given by
\begin{equation}
D_s^\text{eq} = D_s + \nu_s,
\end{equation}
where the value of $\nu_s$ at the $\ell$-th fixed point is 
\begin{equation}
    \nu_s^{(\ell)}=-\frac{i}{2 \pi}\left(\epsilon_1^{(\ell)} \delta_{s,\ell}+\epsilon_2^{(\ell)} \delta_{s-1, \ell}\right).
\end{equation}
To validate this expression, we compute the integral of the equivariantly closed form $D_s^\text{eq} \wedge D_{s'}^\text{eq}$ over the toric surface $X$ in two distinct ways and demonstrate that the results coincide. We start by performing the integral directly. Since $X$ is a 4d compact manifold, the integral only receives contributions from the four-form part of the integrand. Hence, only $D_s$ part of $D_s^\text{eq}$ contributes to the integral. The equivariant integral reduces to the classical intersection product,
\be
\label{eq:DsIntegralDirect}
\int_{X} D_s^\text{eq} \wedge D_{s'}^\text{eq} 
= \int_{X} D_s \wedge  D_{s'} = C_{s,s'} = -h_s \delta_{s,s'} + \delta_{s,s'-1} + \delta_{s,s'+1}.
\ee
On the other hand, the equivariant localization formula allows us to express the integral of an equivariantly closed form $\alpha$ as a sum over the fixed points of the torus action. When the fixed points are isolated, the formula reads
\be
\int_X \alpha = (-2\pi)^{\frac12 \dim(X)} \sum_{\ell=1}^{\chi} \frac{\alpha_0^{(\ell)}}{(i\e_1^{(\ell)}) (i\e_2^{(\ell)})},
\ee
where $\alpha_0^{(\ell)}$ is the zero-form component of $\alpha$ evaluated at the $\ell$-th fixed point. We apply this to $\alpha = D_s^\text{eq} \wedge D_{s'}^\text{eq}$. At each fixed point, only the zero-form part, namely $\nu_s$, contributes to the localization formula. Therefore, 
\be
\begin{aligned}
\int_{X} D_s^\text{eq} \wedge D_{s'}^\text{eq}  
&= -(-2\pi)^2\sum_{\ell=1}^{\chi} \frac{\nu_s^{(\ell)} \nu_{s'}^{(\ell)}}{\e_1^{(\ell)} \e_2^{(\ell)}} \\
&= -4\pi^2\sum_{\ell=1}^{\chi} \frac{1}{\epsilon_1^{(\ell)} \epsilon_2^{(\ell)}} \left( -\frac{i}{2\pi} \right)^2 \left( \epsilon_1^{(\ell)} \delta_{s,\ell} + \epsilon_2^{(\ell)} \delta_{s-1, \ell} \right) \left( \epsilon_1^{(\ell)} \delta_{s',\ell} + \epsilon_2^{(\ell)} \delta_{s'-1, \ell} \right) \\
&=  \frac{\epsilon_1^{(s)}}{\epsilon_2^{(s)}} \delta_{s,s'} + \delta_{s,s'-1} + \delta_{s,s'+1} + \frac{\epsilon_2^{(s-1)}}{\epsilon_1^{(s-1)} } \delta_{s,s'}, 
\end{aligned}
\ee
which matches \eqref{eq:DsIntegralDirect} with 
\be
h_s = - \frac{\epsilon_1^{(s)}}{\epsilon_2^{(s)}} - \frac{\epsilon_2^{(s-1)}}{\epsilon_1^{(s-1)} }.
\ee

\subsubsection*{Example: $\mathbb{CP}^2$}

The fan $\Delta$ for $\mathbb{CP}^2$ can be defined by three vectors in the lattice $N$,
\begin{equation}
\label{eq:CP2fan}
\vec{n}_1 = (1,0)^{\text{T}},  \quad 
\vec{n}_2 = (0,1)^{\text{T}}, \quad 
\vec{n}_3 = (-1,-1)^{\text{T}}.
\end{equation}
They satisfy the relation 
\begin{equation}
    \vec{n}_1 + \vec{n}_2 +
    \vec{n}_3 = 0,
\end{equation}
which gives $h_1=h_2=h_3=-1$.
The homogeneous coordinates on $\mathbb{CP}^2$ are denoted by $(y_1, y_2, y_3)$, with the equivalence relation
\begin{equation}
    (y_1,y_2,y_3) \sim (\lambda y_1, \lambda y_2, \lambda y_3), \quad \forall\lambda \in \mathbb{C}^*.
\end{equation}
The dual vectors in $M$ are
\begin{equation}
    \vec{n}_1^* = (0,-1), 
    \quad 
    \vec{n}_2^* = (1,0),
    \quad 
    \vec{n}_3^* = (-1,1).
\end{equation}
For each cone $\sigma_\ell \in \Delta$, the dual cone $\sigma^\vee_\ell$ is generated by the semi-group $S_\ell = M \cap \sigma_\ell^\vee$, which consists of integer linear combinations of the vectors $(\vec{n}_{\ell+1}^*, -\vec{n}_{\ell}^*)$.
According to Eq. \eqref{local coordinate}, the local coordinates on three patches $U_{\ell}$ are written in terms of homogeneous coordinates as
\begin{equation}
\begin{aligned}
    &U_1: & (u_1,u_2) &= (y_1 y_3^{-1}, y_2 y_3^{-1}), \\
    &U_2: & (v_1,v_2) &= (y_2 y_1^{-1}, y_3 y_1^{-1}), \\ 
    &U_3: & (w_1,w_2) &= (y_3 y_2^{-1}, y_1 y_2^{-1}).
\end{aligned}
\end{equation}
The Weil divisor $D_{\ell}\cong \mathbb{CP}^1$ is the subvariety defined by $y_\ell =0$. The support of $D_{\ell}$ is contained in $U_{\ell-1}\cup U_{\ell}$. In terms of local coordinates, the defining equations are
\begin{equation}
   \begin{aligned}
    D_1 \cap U_1 &=\{ u_1 = 0\}, &
    D_1 \cap U_3 &=\{ w_2 = 0\}, \\
    D_2 \cap U_2 &=\{ v_1 = 0\}, &
    D_2 \cap U_1 &=\{ u_2 = 0\}, \\
    D_3 \cap U_3 &=\{ w_1 = 0\}, &
    D_3 \cap U_2 &=\{ v_2 = 0\}.
    \end{aligned}
\end{equation}
A general $(\mathbb{C}^*)^2$-invariant Weil divisor $\boldsymbol{p}$ on $\mathbb{CP}^2$ is of the form 
\begin{equation}
    \boldsymbol{p} = 
    \mathfrak{p}_1 D_1 
    +
    \mathfrak{p}_2 D_2
    +
    \mathfrak{p}_3 D_3,
\end{equation}
subject to the equivalence relations 
\begin{equation}
\label{P2 divisor relation}
    D_1 = D_3, \quad  D_2 = D_3.
\end{equation}
Locally, $\boldsymbol{p}$ is defined on each patch $U_{\ell}$ as
\begin{equation}
\begin{aligned}
    \boldsymbol{p}\cap U_1 &= \{f_{\boldsymbol{p}}^{(1)} = (u_1)^{\mathfrak{p}_1} (u_2)^{\mathfrak{p}_2}=0\},\\
    \boldsymbol{p}\cap U_2 &= \{f_{\boldsymbol{p}}^{(2)} = (v_1)^{\mathfrak{p}_2} (v_2)^{\mathfrak{p}_3}=0\}, \\
    \boldsymbol{p}\cap U_3 &= \{f_{\boldsymbol{p}}^{(3)} = (w_1)^{\mathfrak{p}_3} (w_2)^{\mathfrak{p}_1}=0\}.
\end{aligned}
\end{equation}
Note that there is a map between the divisor $\mathfrak{p}$ and the holomorphic line bundle with a nonzero meromorphic section given by $\{(U_{\ell}, f^{(\ell)}_{\boldsymbol{p}})\}$.
The transition functions on the overlaps $U_{\ell} \cap U_{\ell+1}$ are given by
\begin{equation}
    s_{\ell+1, \ell} = (f^{(\ell)}_{\boldsymbol{p}})^{-1}f^{(\ell+1)}_{\boldsymbol{p}},
\end{equation}
or, more explicitly,
\begin{equation}
\label{transition function P}
        s_{2,1} =
        (u_1)^{-\mathfrak{p}_1-\mathfrak{p}_2-\mathfrak{p}_3}, \quad
        s_{3,2} =
        (v_1)^{-\mathfrak{p}_1-\mathfrak{p}_2-\mathfrak{p}_3},
        \quad 
        s_{1,3} =
        (w_1)^{-\mathfrak{p}_1-\mathfrak{p}_2-\mathfrak{p}_3},
\end{equation}
which depend only on the equivalent class $[\mathfrak{p}]$.

The canonical bundle $K_{\mathbb{CP}^2}$ is the determinant bundle of the holomorphic cotangent bundle.
A global section of $K_{\mathbb{CP}^2}$ is written on each patch as
\begin{equation}
\begin{aligned}
    &U_1 : & &\sigma_1 \mathrm{d}u_1\wedge \mathrm{d}u_2, \\
    &U_2 : &  
    &\sigma_2 
    \mathrm{d}v_1\wedge \mathrm{d}v_2, \\
    &U_3 : &  
    &\sigma_3 \mathrm{d}w_1\wedge \mathrm{d}w_2.
\end{aligned}
\end{equation}
The relation between $K_{\mathbb{CP}^2}$ and $D_{\ell}$ can be derived from the transition functions. On the overlap $U_1 \cap U_2$, 
\begin{equation}
    \mathrm{d}v_1 \wedge \mathrm{d}v_2 = 
    \frac{\partial(v_1, v_2)}{\partial(u_1, u_2)} \mathrm{d} u_1 \wedge \mathrm{d} u_2 
    = u_1^{-3}
    \mathrm{d} u_1 \wedge \mathrm{d} u_2.
\end{equation}
Matching the sections $\sigma_1 \mathrm{d}u_1\wedge \mathrm{d}u_2 = \sigma_2 \mathrm{d}v_1 \wedge \mathrm{d}v_2$ yields
\begin{equation}
    \sigma_1 =u_1^{-3} \sigma_2,
\end{equation}
and therefore the transition function for the canonical bundle on the overlap $U_1 \cap U_2$ is 
\be
s_{2,1}(K_{\mathbb{CP}^2}) = u_1^3.
\ee
Similar computations on other overlaps give 
\be
s_{3,2}(K_{\mathbb{CP}^2}) = v_1^3,\quad s_{1,3}(K_{\mathbb{CP}^2}) = w_1^3.
\ee
Comparing these with the general form of the transition functions in Eq. \eqref{transition function P}, we find that the divisor $\boldsymbol{p}$ corresponding to the canonical line bundle of $K_{\mathbb{CP}^2}$ satisfies $\mathfrak{p}_1+\mathfrak{p}_2+\mathfrak{p}_3=-3$. Using the relation \eqref{P2 divisor relation}, we conclude that 
\begin{equation}
    K_{\mathbb{CP}^2}     = -D_1 -D_2 -D_3.
\end{equation}

\section{Toric Localization And The K-theoretic Instanton Partition Functions} \label{app:Loc Hilbert Schemes}

Let us assume that $X$ is an algebraic toric surface. The BPS equations \eqref{eq:BPS-eq-v1} and \eqref{eq:BPS-eq-v2} imply that for a generic value of $\sigma$,
\footnote{Here $\sigma$ is the scalar field rather than the signature.}
a rank-two sheaf $E$ with $c_1(E)=c_1$ and $c_2(E)=c_2$ splits into rank-one sheaves, $L_1 \oplus  L_2$. We denote $c_1(L_1)=m_1$, $c_1(L_2)=m_2$. Then $c_1=m_1+m_2$. For fixed $c_1,c_2$, the path integral localizes to the product of rank-one torsion-free sheaves:
\be
X_{[d]} = \coprod_{q_1+q_2=d} X^{[q_1]}\times X^{[q_2]}.
\ee
Here $X^{[n]}$ denotes the Hilbert scheme of $n$ points on $X$, which is identical to the moduli space of rank-one torsion-free sheaves with $c_2=n$, and 
\footnote{
This relation arises from the Whitney sum formula, 
\[
c(E)=c(L_1)c(L_2) = (1+c_1(L_1)+c_2(L_1)+\cdots)(1+c_1(L_2)+c_2(L_2)+\cdots),
\]
comparing degree-two terms yields $c_2= d + \frac14 c_1^2-\bfk^2$.
}
\be
d:= c_2(L_1) + c_2(L_2)=c_2-B(m_1,m_2).
\ee
These abelian fluxes are related to $\bfk$ defined in Eq. \eqref{kfrakpl} by 
\be
m_1 = \frac{1}{2} c_1 + \bfk,\quad m_2=\frac12 c_1-\bfk.
\ee
The total instanton charge is 
\be
\label{eq: kd}
k=c_2-\frac{1}{4}c_1^2=d-{\bfk^2}.
\ee
We define the projections:
\bea
&\pi_a :&  &X \times X^{[q_1]}\times X^{[q_2]} \rightarrow X\times X^{[q_a]},\quad a=1,2,\\
&\pi_0 :&  &X \times X^{[q_1]}\times X^{[q_2]} \rightarrow X,\\
&\pi:&  &X \times X^{[q_1]}\times X^{[q_2]} \rightarrow X^{[q_1]}\times X^{[q_2]}.
\eea
Let $\CI_a$ be the universal sheaf over $X \times X^{[q_1]}\times X^{[q_2]} $ such that $I_a = \pi_a^* (Z_a)$ for a universal subscheme $Z_a$. We also define $I_a(m_a) = I_a \otimes \pi_0^* \CO(L_a)$ for $a=1,2$. 

As reviewed in the previous appendix, we consider the toric action $T = \mathbb{C}^*\times \mathbb{C}^*$ on $X$, with the set of $T$-fixed points $X^T = \{p_1,\cdots, p_\chi\}$. Around each $p_\ell$, we can choose local coordinates $(z_1^{(\ell)}, z_2^{(\ell)})$ such that
\be
T \triangleright (z_1^{(\ell)},z_2^{(\ell)}) = (e^{i\epsilon_1^{(\ell)}} z_1^{(\ell)}, e^{i\epsilon_2^{(\ell)}}z_2^{(\ell)}).
\ee
This induces a $T$-action on the Hilbert scheme of points $X^{[q_a]}$ for $a=1,2$. A subscheme $Z_a \in X^{[q_a]}$ is a $T$-fixed point if and only if its support lies in $X^T$, and locally the defining ideal $I_{Z_a}$ is generated by monomials in $z_1^{(\ell)}$, $z_2^{(\ell)}$. Such ideals are in one-to-one correspondence with Young diagrams,
\be
Y_a^{(\ell)},\quad \ell=1,\cdots,\chi,\quad a = 1,2. 
\ee
We can label the $T$-fixed points on the moduli space $X_{[d]}$ by $2\chi$-tuples of Young diagrams:
\be
\label{eq: fixed points}
X_{[d]}^T \leftrightarrow \left. \left\{ \vec{Y} =  (Y_1^{(\ell)}, Y_2^{(\ell)}) \right| \sum_{\ell,a} \left|Y_a^{(\ell)}\right|=d \right\}.
\ee

We consider the path integral of the effective 1d $\CN=(0,2)$ SQM on $S^1$, obtained through twisted compactification of the 5d theory. After applying localization techniques, the degrees of freedom organize into $\CN=2$ multiplets, which correspond to terms in the Atiyah-Hitchin-Singer (AHS) complex for the virtual tangent bundle of $M_{\bfmu,k}$, 
\be 
0 \rightarrow \Omega^0(X,\mathrm{ad}(P)) \rightarrow \Omega^1(X,\mathrm{ad}(P)) \rightarrow \Omega^{2,+}(X,\mathrm{ad}(P) ) \rightarrow 0.
\ee
The various components of this complex have the following interpretations in the effective SQM: 
The $\Omega^0$ terms correspond to fluctuations of the 1d $\CN=2$ vector multiplet, $\delta V_{1d} = (i\delta A_5+\delta\sigma, \delta\bar\lambda^{(0,0)})$, which are responsible for imposing the gauge invariance; The $\Omega^1$ terms correspond to the fluctuations of 1d $\CN=2$ chiral multiplets, $\delta\Phi_{1d} = ( \delta A_\mu,\delta\lambda_\mu)$. The integration over auxiliary fields imposes the constraint $\delta\Phi_{1d} \in \ker d^+$.

After restriction to $X_{[d]}\subset M_{\bfmu,k}$, where $E \cong  L_1 \oplus L_2$, we can split $\mathrm{ad}(P) = \mathfrak{t} \oplus W$, where $W  = (L_1 \otimes L_2^{-1}) \oplus (L_1^{-1}\otimes L_2)$. 
The AHS complex becomes
\be 
0 \rightarrow \Omega^0(X,\mathfrak{t}) \oplus  \Omega^0(X, W)\rightarrow \Omega^1(X,\mathfrak{t}) \oplus  \Omega^1(X, W) \rightarrow \Omega^{2,+}(X,\mathfrak{t}) \oplus \Omega^{2,+}(X, W)\rightarrow 0,
\ee
where the sheaves $\Omega^{\bullet}(X,W)$ contribute from the fluctuations of the root part.  
Since the index of SQM depends only on the K-theory class of the complex, we can decompose the virtual tangent bundle over $X_{[d]}$ (noncanonically) as
\be
TM_{\bfmu,k}|_{X_{[d]}} = TX_{[d]} + N,
\ee
where 
\be
-TX_{[d]} = \left[\Omega^0(X,\mathfrak{t})\right] - \left[ \Omega^1(X,\mathfrak{t})\right] + \left[ \Omega^{2,+}(X,\mathfrak{t})\right]
\ee
and
\be
-[N] = \left[\Omega^0(X,W)\right] - \left[ \Omega^1(X,W)\right] + \left[ \Omega^{2,+}(X,W)\right]
\ee
are the K-theory classes of the virtual tangent bundle and the virtual normal bundle over $X_{[d]}$, respectively. We can further decompose $[N]= [W_+] + [W_-]$, where $W_\pm$ corresponds to $[\Omega^\bullet (X,L_1^\pm\otimes L_2^{\mp})]$.

At a fixed zero-mode background $a$ (with $h=0$), the path integral over the quantum fluctuations can be identified with the Dirac index of $X_{[d]}$ twisted by the symmetric algebra of $[N^\vee]$. Including the contribution from the $U(1)^{(I)}$ flux, it may be formally expressed as the intersection integral,
\bea\label{intersection 1d}
& \sum_{d}\CR^{4k-3}\int_{X_{[d]}} \hat A\left(TX_{[d]}\right)\wedge \text{ch}\left(\hat S^\bullet [N^\vee]\right) \wedge \text{ch}\left(\widetilde\CL^{(I)}\right)\\
=&\sum_{d}\CR^{4k-3}\int_{X_{[d]}} \frac{\hat A\left(TX_{[d]}\right)}{\text{ch}\left(\hat \wedge^\bullet [N^\vee]\right)} \wedge \text{ch}\left(\widetilde\CL^{(I)}\right),
\eea
where $k$ is given in terms of $d$ by Eq. \eqref{eq: kd}. We define the symmetric and exterior algebras of $V$ by
$S^\bullet V$ and $\wedge^\bullet V$, respectively. We further define 
\be
\hat S^\bullet V = (\det V)^{\frac12} S^\bullet V, \quad \hat\wedge^\bullet V = (\det V)^{-\frac12} \wedge^\bullet V,
\ee
using their symmetrized versions. 
\footnote{We use the ``symmetric quantization" of the 1d fermions such that a 1d $\CN=(0,2)$ chiral multiplet valued in a complex vector bundle $V$ contributes $\prod_{i} (e^{x_i/2}-e^{-x_i/2})^{-1}$ to the index, where $x_i$'s are Chern roots of $V$. Then $\prod_{i} (e^{x_i/2}-e^{-x_i/2})^{-1}= \prod_{i} e^{-x_i/2}\prod_{i}(1-e^{-x_i})^{-1} = \text{ch}[\hat S^\bullet V^\vee] = 1/\text{ch}[\hat\wedge^\bullet V^{\vee}]$. } 
Here, $\tilde \CL^{(I)}$ denotes the restriction of the line bundle $\CL^{(I)}\to M_{\bfmu,k}$ from Sec.  \ref{subsec:ChernSimonsLineBundle} to the localization locus. One can show that   
\be\label{c1 LI tilde}
c_1(\widetilde\CL^{(I)}) = \pi_* \left[\frac{F^{(I)}}{2\pi} \wedge \text{ch}_2 \left(\CI_1(m_1)\oplus \CI_2(m_2)\right)\right],
\ee
Notice that the expression \eqref{intersection 1d} can also be written as
\be\label{intersection symmetrized}
\sum_{d}\CR^{4k-3}\int_{X_{[d]}} \frac{\text{td}(T^{(1,0)}X_{[d]})}{\text{ch}( \wedge^\bullet [N^\vee])} \wedge \exp\left(-\frac12 c_1 \left(T^{(1,0)}X_{[d]}+ [N]\right)\right)\wedge\text{ch}(\widetilde\CL^{(I)}).
\ee

Consider the $\mathbb{T}$-action on the moduli space, where $\mathbb{T}= T\times \mathbb{C}^*$ corresponds to the complexified maximal torus of $\mathrm{U}(1)^2\times G$ for gauge group $G=\mathrm{SU}(2)$. Then the sheaves $TX_{[d]}$, $[N]$, and $\widetilde \CL^{(I)}$ can be promoted to $\mathbb{T}$-equivariant sheaves, and the intersection integrals are replaced by the equivariant pushforward to a point. 
We can write the virtual tangent bundle as
\be
\left[TX_{[d]}\right] = -\pi_* \left(\CI_1\otimes \CI_1^\vee + \CI_2\otimes \CI_2^\vee\right),
\ee
and the normal bundle as
\be
[N] = -\pi_* \left(\sum_{a\neq b} \CI_a(m_a) \otimes \CI_b(m_b)^\vee\right).
\ee
The Chern characters of these sheaves can be computed by the Grothendieck Riemann-Roch theorem:
\bea
\text{ch}(TX_{[d]}) &= \pi_* \left[ \text{td}(X)\text{ch}\left(-\left(\CI_1\otimes \CI_1^\vee + \CI_2 \otimes \CI_2^\vee\right)\right)\right],\\
\text{ch}([N]) &= \pi_* \left[ \text{td}(X) \text{ch}\left(-\left(\CI_1(m_1) \otimes \CI_2^\vee(m_2) + \CI_2(m_2)\otimes \CI_1^\vee(m_1)\right)\right)\right].
\eea
Define the Chern roots of $TX_{[d]}$ and $[N]$ by
\be
\text{ch} (TX_{[d]}):= \sum_{p_{\ell}\in X^T}\sum_{x(p_{\ell})} e^{x(p_{\ell})},\quad \text{ch}([N]) :=\sum_{p_{\ell}\in X^T}\sum_{{\tilde x}(p_{\ell})} e^{{\tilde x}(p_{\ell})}.
\ee
Performing the equivariant pushforward, we can write these characters as a sums over the fixed points over $X$,
\be
\begin{aligned}
\text{ch} (TX_{[d]}) &= \sum_{p_{\ell}\in X^T} \left[\frac{\text{td}(T_{p_{\ell}}X)}{e(T_{p_{\ell}}X)} \iota^*_{p_{\ell}} \text{ch}\left(-(\CI_1\otimes \CI_1^\vee + \CI_2 \otimes \CI_2^\vee)\right)\right], \\
\text{ch}([N]) 
&= \sum_{p_{\ell}\in X^T} \left[\frac{\text{td}(T_{p_{\ell}}X)}{e(T_{p_{\ell}}X)} \iota^*_{p_{\ell}}\text{ch} \left(-(\CI_1(m_1)\otimes \CI_2^\vee(m_2) + \CI_2(m_2) \otimes \CI_1^\vee(m_1))\right)\right].
\end{aligned}
\ee
We write
\be
\text{td}(T^{(1,0)}X_{[d]}) = \prod_{p_{\ell}\in X^T} \prod_{x(p_{\ell})} \frac{x(p_{\ell})}{1-e^{-x(p_{\ell})}},
\ee
and 
\be
\text{ch} (\wedge^\bullet [N^\vee]) =\prod_{p_{\ell}\in X^T}\prod_{{\tilde x}(p_{\ell})} \left(1- e^{-{\tilde x}(p_{\ell})}\right),
\ee
where these characters are written as products over the fixed loci $X^T$. Altogether, the equivariant integral \eqref{intersection symmetrized} becomes a sum over fixed points $X^T_{[d]}$ on the moduli space, which are labeled by $2\chi$-tuples of Young diagrams $\vec{Y}$ \eqref{eq: fixed points}, 
\bea
\label{fixed point formula}
\sum_{d}\CR^{4k-3}\sum_{\vec{Y}} \sum_{\substack{m_1+m_2=c_1 \\ q_1+q_2=d}} & \frac{1}{e(T_{\vec{Y}}X_{[d]})} \\ &\iota^*_{\vec{Y}}
 \left[\frac{\text{td}(T^{(1,0)}X_{[d]})}{\text{ch}( \wedge^\bullet [N^\vee])} \wedge \exp\left(-\frac12 c_1 \left(T^{(1,0)}X_{[d]}\oplus [N]\right)\right)\wedge\text{ch}(\widetilde\CL^{(I)})\right].
\eea
The characteristic classes evaluated at fixed points are expressed in terms of the Nekrasov partition functions. Using the formula from \cite{Losev:2003py}
\be\label{ch of uni sheaf}
\iota^*_{p_{\ell}, \vec{Y}}\text{ch}(\CI_\alpha(m_\alpha)) = e^{a^{(\ell)}_\alpha} \left[1-(1-e^{-\epsilon_1^{(\ell)}})(1-e^{-\epsilon_2^{(\ell)}})\sum_{(i,j)\in Y_\alpha^{(\ell)}} e^{-(i-1)\epsilon_1^{(\ell)}- (j-1)\epsilon_2^{(\ell)}}\right],
\ee
where $a_1^{(\ell)} = -a^{(\ell)}_2 = a^{(\ell)}$,
one can show that  \cite{Gottsche:2006tn} 
\be\label{factor 1}
\frac{\iota^*_{\vec{Y}} \text{td}(T^{(1,0)}X_{[d]})}{e(T_{\vec{Y}}X_{[d]})} = \prod_{\ell=1}^{\chi}\prod_{\alpha=1}^2 n_{\alpha,\alpha}^{(Y_{\alpha}^{(\ell)},Y_{\alpha}^{(\ell)})}(\epsilon_1^{(\ell)}, \epsilon_2^{(\ell)}, R),
\ee
with $n_{\alpha,\beta}$'s defined in \eqref{definition nij}. We also have 
\be\label{factor 2}
\CR^{4k-3}\iota_{\vec{Y}}^*\frac{1}{\text{ch} (\wedge^\bullet [N^\vee])} = K(a^{(\ell)}, \epsilon_1^{(\ell)},\epsilon_2^{(\ell)}, \Lambda, R) \prod_{\ell=1}^{\chi}\prod_{i\neq j} n_{\alpha,\beta}^{(Y_{\alpha}^{(\ell)},Y_{\beta}^{(\ell)})}(a^{(\ell)},\epsilon_1^{(\ell)}, \epsilon_2^{(\ell)}, R),
\ee
where
\be
\begin{aligned}
& K(a^{(\ell)}, \epsilon_1^{(\ell)},\epsilon_2^{(\ell)}, \Lambda, R) \\
= & \CR^{d-1} \exp\left\{ R\sum_{\ell=1}^{\chi}\left[ -\frac{\left(\epsilon_1^{(\ell)}+\epsilon_2^{(\ell)}\right)^3}{48\epsilon_1^{(\ell)}\epsilon_2^{(\ell)}} +\frac{(\epsilon_1^{(\ell)} +\epsilon_2^{(\ell)})}{24\epsilon_1^{(\ell)}\epsilon_2^{(\ell)}}\left((\epsilon_1^{(\ell)})^2+(\epsilon_2^{(\ell)})^2 + 3\epsilon_1^{(\ell)}\epsilon_2^{(\ell)}\right) \right]\right\}\\ 
&\times \prod_{\ell=1}^{\chi} \CZ_{\text{pert}}(a^{(\ell)}, \epsilon_1^{(\ell)}, \epsilon_2^{(\ell)},\Lambda e^{(\e_1+\e_2)/4},R).
\end{aligned}
\ee
The four-form part of \eqref{ch of uni sheaf} is given by
\be
\iota_{p_\ell,\vec{Y}}^*\text{ch}_2 \left(\CI_1(m_1)\oplus \CI_2(m_2)\right) = (a^{(\ell)})^2 - \epsilon_1^{(\ell)}\epsilon_2^{(\ell)}\left(|Y_1^{(\ell)}|+|Y_2^{(\ell)}|\right),
\ee
and thus we obtain from Eq. \eqref{c1 LI tilde} that
\be\label{factor 3}
\iota_{\vec{Y}}^*\text{ch}(\widetilde\CL^{(I)}) = \exp\left[\sum_{\ell=1}^{\chi} n_I^{(\ell)} \left(-|Y| + \frac{(a^{(\ell)})^2}{\e_1^{(\ell)}\e_2^{(\ell)}}\right)\right],
\ee
where $n_I^{(\ell)}$ is defined in Eq. \eqref{eq:nIl}. Combining the results \eqref{factor 1}, \eqref{factor 2}, and \eqref{factor 3}, the fixed point formula \eqref{fixed point formula} reduces to
\be
\begin{aligned}
& \frac{1}{\CR} \exp\left\{ R\sum_{\ell=1}^{\chi}\left[  -\frac{\left(\epsilon_1^{(\ell)}+\epsilon_2^{(\ell)}\right)^3}{48\epsilon_1^{(\ell)}\epsilon_2^{(\ell)}} +\frac{\epsilon_1^{(\ell)} +\epsilon_2^{(\ell)}}{24\epsilon_1^{(\ell)}\epsilon_2^{(\ell)}}((\epsilon_1^{(\ell)})^2+(\epsilon_2^{(\ell)})^2 + 3\epsilon_1^{(\ell)}\epsilon_2^{(\ell)}) \right]\right\} \\
&\times \exp\left[\sum_{\ell=1}^{\chi} (n_I^{(\ell)} +\epsilon_1^{(\ell)}+\epsilon_2^{(\ell)})\left(-|Y| + \frac{(a^{(\ell)})^2}{\e_1\e_2}\right)\right]\\
&\times\prod_{\ell=1}^{\chi} \CZ_{\text{pert}}(a^{(\ell)}, \epsilon_1^{(\ell)}, \epsilon_2^{(\ell)},\Lambda e^{\frac14(\e_1+\e_2)},R) \CZ_{\text{inst}}(a^{(\ell)}, \epsilon_1^{(\ell)}, \epsilon_2^{(\ell)},0,\Lambda e^{\frac14(\e_1+\e_2)},R).
\end{aligned}
\ee
Using the identity \eqref{eq:ZinstnI}, this becomes
\be
\begin{aligned}
& \frac{1}{\CR} \exp\left\{ -R\sum_{\ell=1}^{\chi}  \frac{\left(\epsilon_1^{(\ell)}+\epsilon_2^{(\ell)}\right)^3}{48\epsilon_1^{(\ell)}\epsilon_2^{(\ell)}} \right\} \\ 
&\times\prod_{\ell=1}^{\chi} \CZ_{\text{pert}}(a^{(\ell)}, \epsilon_1^{(\ell)}, \epsilon_2^{(\ell)},\Lambda ,R) \CZ_{\text{inst}}(a^{(\ell)}, \epsilon_1^{(\ell)}, \epsilon_2^{(\ell)},n_I^{(\ell)},\Lambda ,R),
\end{aligned}
\ee
which is precisely the function $g_{\mathfrak{p},\mathfrak{p}^{(I)}}^{\e_1,\e_2} (a)$ in \eqref{ga}.
The factor $\sum_{\ell=1}^{\chi}  \frac{\left(\epsilon_1^{(\ell)}+\epsilon_2^{(\ell)}\right)^3}{48\epsilon_1^{(\ell)}\epsilon_2^{(\ell)}}$ has a geometric interpretation as the equivariant integral of $(K_X^{\rm eq})^3$ over $X$. 

\section{Properties Of Five-dimensional $\mathrm{SU}(2)$ Nekrasov Partition Function}\label{app:NekPartProps}
\subsection{Poles And Zeros}
\label{poles-and-zeros}

\subsubsection*{On $\mathbb{C}^2 \times S^1$}

We first study the locations of poles or zeros in the perturbative part of the Nekrasov partition function \eqref{Zpert equiv}.
Note that
\begin{equation}
    \CZ_\text{pert}
    (a, \epsilon_1, \epsilon_2, \Lambda, R) \sim 
    \text{PE}
    \left[
    -\frac{x^{-2}+x^2}
    {(t_1 -1)(t_2 - 1)}
    \right],
\end{equation}
where $\text{PE}$ denotes the plethystic exponential,
\begin{equation}
    \text{PE}\left[f(x,t_1,t_2)\right]:= 
    \exp\left(
    \sum_{m\geq 1}
    \frac{1}{m} f(x^{m}, t_1^{m}, t_2^{m} )
    \right).
\end{equation}
Whether $\CZ_\text{pert}$ contains poles or zeros depends on the sign of $\text{Re}(\epsilon_1)$ and $\text{Re}(\epsilon_2)$.

\paragraph{Case 1:} 
$\text{Re}(\epsilon_1) < 0$ and
$\text{Re}(\epsilon_2) < 0$
\begin{equation}
    \CZ_\text{pert}
    (a, \epsilon_1, \epsilon_2, \Lambda, R) \sim 
    \prod_{i,j = 0}^\infty 
    \left(1 - x^{-2}t_1^i t_2^j
    \right)
    \left(1 - x^2 t_1^i t_2^j
    \right),
\end{equation}
where zeros are at
\begin{equation}
    2 a = m \epsilon_1 + n \epsilon_2 + \frac{2\pi i s}{R}, 
    \quad 
    mn \geq0, \quad m,n,s\in \mathbb{Z}.
\end{equation}

\paragraph{Case 2:} 
$\text{Re}(\epsilon_1) > 0$ and
$\text{Re}(\epsilon_2) > 0$
\begin{equation}
    \CZ_\text{pert}
    (a, \epsilon_1, \epsilon_2, \Lambda, R) \sim 
    \prod_{i,j = 0}^\infty 
    \left(1 - x^{-2}t_1^{-i-1} t_2^{-j-1}
    \right)
    \left(1 - x^2 t_1^{-i-1} t_2^{-j-1}
    \right),
\end{equation}
where zeros are at
\begin{equation}
    2 a = m \epsilon_1 + n \epsilon_2 + \frac{2\pi i s}{R}, 
    \quad 
    mn >0,\quad m,n,s\in \mathbb{Z}.
\end{equation}

\paragraph{Case 3:} 
$\text{Re}(\epsilon_1) > 0$ and
$\text{Re}(\epsilon_2) < 0$
\begin{equation}
    \CZ_\text{pert}
    (a, \epsilon_1, \epsilon_2, \Lambda, R) \sim 
    \prod_{i,j = 0}^\infty 
    \frac{1}{(1-x^{-2} t_1^{-i -1} t_2^j)}
    \frac{1}{(1- x^2 t_1^{-1-i}t_2^j)},
\end{equation}
where poles are at
\begin{equation}
    2 a = m \epsilon_1 + n \epsilon_2 + \frac{2\pi i s}{R}, 
    \quad 
    m<0\leq n\, 
    \text{or}\,
    n\leq0<m,\quad m,n,s\in \mathbb{Z}.
\end{equation}

\paragraph{Case 4:} 
$\text{Re}(\epsilon_1) < 0$ and
$\text{Re}(\epsilon_2) > 0$
\begin{equation}
    \CZ_\text{pert}
    (a, \epsilon_1, \epsilon_2, \Lambda, R) \sim 
    \prod_{i,j = 0}^\infty 
    \frac{1}{(1-x^{-2} t_1^{i} t_2^{-1-j})}
    \frac{1}{(1- x^2 t_1^{i}t_2^{-1-j})}, 
\end{equation}
where poles are at
\begin{equation}
    2 a = m \epsilon_1 + n \epsilon_2 + \frac{2\pi i s}{R}, 
    \quad 
    m\leq 0 < n\, 
    \text{or}\,
    n < 0 \leq m,\quad m,n,s\in \mathbb{Z}.
\end{equation}

From the recursion relation \eqref{ZinstRecursion}, the 5d $\mathrm{SU}(2)$ instanton partition function has poles at
\begin{equation}
    2a = m \epsilon_1 + n \epsilon_2 + \frac{2\pi i s}{R}, 
    \quad 
    mn >0,\quad m,n,s\in \mathbb{Z}.
\end{equation}
We then conclude that for the full partition function $\CZ(a,\e_1,\e_2, n_I,\Lambda,R)$ defined in \eqref{eq:ZnI},
\begin{enumerate}
    \item When $\text{Re}(\epsilon_1) \text{Re}(\epsilon_2) >0$, $\CZ(a,\e_1,\e_2, n_I,\Lambda,R)$ does not have poles.
    \item When $\text{Re}(\epsilon_1) \text{Re}(\epsilon_2) <0$, $\CZ(a,\e_1,\e_2, n_I,\Lambda,R)$  has poles coming from both perturbative and instanton partition functions. Moreover, these poles are all simple.
\end{enumerate}

\subsubsection*{On $X\times S^1$}
Recall our definition of 5d $\mathrm{SU}(2)$ Nekrasov partition function \eqref{ga}
\begin{equation}
g_{\mathfrak{p}, \mathfrak{p}^{(I)}}^{\epsilon_1, \epsilon_2}(a)=\prod_{\ell=1}^{\chi} \CZ\left(a^{(\ell)}, \epsilon_1^{(\ell)}, \epsilon_2^{(\ell)}, n_I^{(\ell)}, \Lambda, R\right).
\end{equation}
Since for any compact toric manifold $X$, there are two and only two patches with $\epsilon_1^{(\ell)}\epsilon_2^{(\ell)} > 0$, the order of poles in $g_{\mathfrak{p}, \mathfrak{p}^{(I)}}^{\epsilon_1, \epsilon_2}(a)$ should be at most $\chi - 2$ \cite{Bonelli:2020xps}. In particular, when $X$ is $\mathbb{CP}^2$, all the poles in $g_{\mathfrak{p}, \mathfrak{p}^{(I)}}^{\epsilon_1, \epsilon_2}(a)$ are simple.

\paragraph{Claim:} 
When $X = \mathbb{CP}^2$, only the instanton part contributes to the poles in \eqref{ga}. 
The positions of poles in $g_{\mathfrak{p}, \mathfrak{p}^{(I)}}^{\epsilon_1, \epsilon_2}(a)$ are solutions to
\begin{equation}
\label{eq:CP2-pole-pos}
    2 a^{(\ell)}=m \epsilon_1^{(\ell)}+n \epsilon_2^{(\ell)}+\frac{2 \pi i s}{R}, \quad  mn>0, \ell=1,\cdots,\chi, m,n,s \in \mathbb{Z}.
\end{equation}

\paragraph{Proof:} 
We need to show that the perturbative part does not contain poles. This is done by straightforward computations
\begin{equation}
\begin{aligned}
    \prod_{\ell=1}^3 
    \CZ_\text{pert} (a^{(\ell)},\epsilon_1^{(\ell)},\epsilon_2^{(\ell)},\Lambda,R)
    &\sim 
    \text{PE}
    \left[
    -\sum_{\ell=1}^3 
    \frac{x^2(t_1^{(\ell)})^{\mathfrak{p}_{\ell}} (t_2^{(\ell)})^{\mathfrak{p}_{\ell+1}} + 
    x^{-2}
    (t_1^{(\ell)})^{-\mathfrak{p}_{\ell}}
    (t_2^{(\ell)})^{-\mathfrak{p}_{\ell+1}}
    }
    {(t_1^{(\ell)}-1)(t_1^{(\ell)}-1)}
    \right]\\
    &=:
    \text{PE}\left[
    -x^2 \chi^+_{\mathfrak{p}}(t_1,t_2) - x^{-2}
    \chi^{-}_{\mathfrak{p}}(t_1,t_2) 
    \right],
\end{aligned}
\end{equation}
where $t_1^{(\ell)}:= e^{R\epsilon_1^{(\ell)}}$, $t_2^{(\ell)}:= e^{R \epsilon_2^{(\ell)}}$, and 
$\chi^{\pm}_{\mathfrak{p}}(t_1,t_2)$ are finite polynomials,
\begin{equation}
\begin{aligned}
    \chi^{+}_{\mathfrak{p}}(t_1,t_2) &= 
    \begin{cases}
        \sum_{j=0}^{-r}\sum_{i=0}^{-r-j}t_1^{i+\mathfrak{p}_1} t_2^{j+\mathfrak{p}_2}, & r \leq 0,\\
        0, & r = 1,2,\\
        \sum_{j=2-r}^{-1}\sum_{i = -r+1-j}^{-1} 
        t_1^{i+\mathfrak{p}_1} t_2^{j+\mathfrak{p}_2},
        & r \geq 3,
    \end{cases} \\
    \chi^{-}_{\mathfrak{p}}(t_1,t_2) &= 
    \begin{cases}
       \sum_{j=0}^r
       \sum_{i=0}^{k-j} t_1^{i-\mathfrak{p}_1}
       t_2^{j-\mathfrak{p}_2},
       & r \geq 0, \\
       0, & r = -1,-2,\\
       \sum_{j=2+r}^{-1}
       \sum_{i=r+1-j}^{-1} t_1^{i-\mathfrak{p}_1}
       t_2^{j-\mathfrak{p}_2}, & r \leq -3. 
    \end{cases}
\end{aligned}
\end{equation}
The relevant part in the perturbative partition function for $\mathbb{CP}^2$ is a finite product 
\begin{equation}
\label{eq:pert-cp2-plus}
    \text{PE}[-x^2 \chi^+_{\mathfrak{p}}(t_1, t_2)] = 
    \begin{cases}
        \prod_{j=0}^{-r}
        \prod_{i=0}^{-r-j}\left(
        1 - x^2 t_1^{i + \mathfrak{p}_1} t_2^{j+\mathfrak{p}_2}
        \right), & r \leq 0\\
        1, & r = 1,2,\\
        \prod_{j=2-r}^{-1} \prod_{i = -r +1 -j}^{-1}
        \left(
        1 - x^2 t_1^{i + \mathfrak{p}_1} t_2^{j+\mathfrak{p}_2}
        \right), &
        r \geq 3,
    \end{cases}
\end{equation}
and 
\begin{equation}
\label{eq:pert-cp2-minus}
    \text{PE}[-x^{-2} \chi_{\mathfrak{p}}^{-}(t_1, t_2)] =
    \begin{cases}
        \prod_{j=0}^{r}
        \prod_{i=0}^{r-j} 
        \left(
        1-x^{-2} t_1^{i - \mathfrak{p}_1
        }
        t_2^{j-\mathfrak{p}_2}\right), & r \geq 0,\\
        1, & r = -1, -2\\
        \prod_{j=2+r}^{-1}
        \prod_{i=r+1-j}^{-1}
        \left(
        1-x^{-2} t_1^{i - \mathfrak{p}_1
        }
        t_2^{j-\mathfrak{p}_2}\right), & r \leq -3.
    \end{cases}
\end{equation}
Since there are no poles in the perturbative part, we finish the proof.

In general, the perturbative part might have new poles other than the poles determined by \eqref{eq:CP2-pole-pos}.  However, the positions of poles
in $g_{\mathfrak{p}, \mathfrak{p}^{(I)}}^{\epsilon_1, \epsilon_2}(a)$ can still be written as solutions to the following equations
\begin{equation}
\label{eq:gen-pole-structure}
    2 a^{(\ell)}=m \epsilon_1^{(\ell)}+n\epsilon_2^{(\ell)}+ \frac{2 \pi i  s }{ R}, \quad \ell=1,\cdots,\chi,  s \in \mathbb{Z},
\end{equation}
where $(m,n)$ is in some subset of integer pairs $\mathbb{Z}\times \mathbb{Z}$.

\paragraph{Poles at $a= 0 \mod \pi i / R$:}

We restrict our discussion to $X = \mathbb{CP}^2$.
From \eqref{eq:pert-cp2-plus} and \eqref{eq:pert-cp2-minus},
for the perturbative part, we have 
\begin{enumerate}
    \item The perturbative part has zeros at $a= 0 \mod \pi i/R$ of order $2$, if $\mathfrak{p}_{\ell}\geq 1$ for all $\ell=1,\cdots,3$.
    \item The perturbative part has zeros at $a= 0 \mod \pi i/R$ of order $2$, if $\mathfrak{p}_{\ell}\leq -1$ for all $\ell=1,\cdots,3$.
    \item The perturbative part has zeros at $a= 0 \mod \pi i/R$ of order $1$, if $\mathfrak{p}_{\ell}\geq 0$ for all $\ell=1,\cdots,3$ and there exists at least one $\mathfrak{p}_{\ell} = 0$. 
    \item The perturbative part has zeros at $a= 0 \mod \pi i/R$ of order $1$, if $\mathfrak{p}_{\ell}\leq 0$ for all $\ell=1,\cdots,3$ and there exists at least one $\mathfrak{p}_{\ell} = 0$.
\end{enumerate}
From the instanton contribution \eqref{eq:CP2-pole-pos}, we have the following:
\begin{enumerate}
    \item The instanton part has poles at $a= 0 \mod \pi i/R$ of order $3$, if $\mathfrak{p}_{\ell}\geq 1$, for all $\ell=1,\cdots,3$.
    \item The instanton part has poles at $a= 0 \mod \pi i/R$ of order $3$, if $\mathfrak{p}_{\ell}\leq -1$, for all $\ell=1,\cdots,3$.
    \item The instanton part has poles at $a= 0 \mod \pi i/R$ of order $1$, if there exists $\ell$, such that $\mathfrak{p}_{\ell}\leq 0$ and $\mathfrak{p}_{\ell+1},\mathfrak{p}_{\ell+2}\geq 1$.
    \item 
    The instanton part has poles at $a= 0 \mod \pi i/R$ of order $1$, if there exists $\ell$, such that $\mathfrak{p}_{\ell}\geq 0$ and $\mathfrak{p}_{\ell+1},\mathfrak{p}_{\ell+2}\leq -1$.
\end{enumerate}
From the previous discussion, we concluded that the full 5d $\mathrm{SU}(2)$ Nekrasov partition function \eqref{ga} on $\mathbb{CP}^2\times S^1$ has a simple pole at $a= 0 \mod \pi i/R$, if and only if $\mathfrak{p}_{\ell}\neq 0$ for all $\ell=1,2,3$. $a= 0 \mod \pi i/R$ becomes a regular point if and only if there exists $\mathfrak{p}_{\ell} = 0$.

\subsection{5d $\mathrm{SU}(2)$ Abstruse Duality}

An ``abstruse duality'' relating the residues at the poles of the 1-loop and instanton parts of the 4d Nekrasov partition function was derived in \cite{Bonelli:2020xps}. 
In this appendix, we show how to derive the following analogous relations for the 5d $\mathrm{SU}(2)$ Nekrasov partition function defined in \eqref{eq:ZnI},
\begin{equation}
\label{eq:duality-nekrasov-e1}
    \lim_{a\to 0}
    \frac{
    \CZ\left(a + a^{(m,n)}/2, \epsilon_1, \epsilon_2,n_I,  \Lambda, R\right)
    }{
    \CZ\left(a + \hat{a}^{(m,n)}/2, \epsilon_1, \epsilon_2,   n_I, \Lambda, R\right)
    } = -\text{sgn}(\text{Re}(\epsilon_1)),
\end{equation}
and 
\begin{equation}
\label{eq:duality-nekrasov-e2}
    \lim_{a\to 0}
    \frac{
    \CZ\left(a + a^{(m,n)}/2, \epsilon_1, \epsilon_2,  n_I, \Lambda, R\right)
    }{
    \CZ\left(a - \hat{a}^{(m,n)}/2, \epsilon_1, \epsilon_2, n_I, \Lambda, R\right)
    } = -\text{sgn}(\text{Re}(\epsilon_2)),
\end{equation}
where $m,n$ are nonzero integers, and 
\begin{equation}
    a^{(m, n)}:=m \epsilon_1+n \epsilon_2, \quad \hat{a}^{(m, n)}:=m \epsilon_1-n \epsilon_2.
\end{equation}

It would be interesting to have a better physical understanding of these identities. 
We note that changing the sign of $\text{Re}(\epsilon_2)$ holding $\epsilon_1$ fixed is related to the parity symmetry of the theory. However, we must stress that this does \emph{not} explain Eq. \eqref{eq:duality-nekrasov-e1} since we do not flip the sign of $\epsilon_2$ in the third argument. 

\paragraph{Proof:} 
First, we use the relation \eqref{eq:ZnI} to rewrite the left-hand side of \eqref{eq:duality-nekrasov-e1} as
\begin{equation}
\label{eq: ratio-Z-ZN}
    \lim_{a\to 0}
    \frac{
    \CZ\left(a + a^{(m,n)}/2, \epsilon_1, \epsilon_2, n_I, \Lambda, R\right)
    }{
    \CZ\left(a + \hat{a}^{(m,n)}/2, \epsilon_1, \epsilon_2, n_I, \Lambda, R\right)
    } = 
    \lim_{a\to 0}
    \frac{
    \CZ\left(a + a^{(m,n)}/2, \epsilon_1, \epsilon_2,  \Lambda e^{-\frac{1}{4} n_I}, R\right)
    }{
    \CZ\left(a + \hat{a}^{(m,n)}/2,  \epsilon_1, \epsilon_2,  \Lambda e^{-\frac{1}{4} n_I}, R\right)
    },
\end{equation}
where $\CZ$ is given in \eqref{eq: ZNkera}.

To set the stage, we focus on the perturbative contribution $\CZ_\text{pert}$ that appears in \eqref{Zpert equiv} and evaluate the $a\to 0$ limit of its ratio,
\begin{equation}
\begin{aligned}
    &\lim_{a\to 0} 
    \frac{
    \CZ_\text{pert}
    (a+ a^{(m,n)}/2, \epsilon_1, \epsilon_2, \Lambda, R)
    }{
    \CZ_\text{pert}
    (a + \hat{a}^{(m,n)}/2, \epsilon_1, \epsilon_2, \Lambda, R)
    }\\=& 
    \mathcal{R}^{-4mn} (t_1 t_2)^{mn} \times\\
    &\times\lim_{a\to 0}\exp\left(
    -\sum_{l\geq 1}
    \frac{1}{l(t_1^{l} - 1)(t_2^{l} -1)}
    \left(
    x^{-2l} t_1^{-lm}t_2^{-ln} + 
    x^{2l} t_1^{lm}t_2^{ln}
    -x^{-2l} t_1^{-lm}t_2^{ln}
    -x^{2l} t_1^{lm}t_2^{-ln}
    \right)
    \right)\\
    =& 
    \mathcal{R}^{-4mn} (t_1 t_2)^{mn}
    \lim_{a\to 0}\text{PE}\left[
    -\frac{
    x^{-2}t_1^m t_2^{-n} 
    + x^{2}t_1^m t_2^{n}
    -x^{-2}t_1^{-m} t_2^{n}
    -x^{2}t_1^{m} t_2^{-n}
    }
    {
    (t_1 - 1)(t_2 -1)    
    }
    \right],
\end{aligned}
\end{equation}
where 
\be
x = e^{Ra}, \quad t_1 = e^{R\epsilon_1},\quad t_2 = e^{R \epsilon_2}.
\ee
We consider the following four cases separately.
\paragraph{Case 1:} $\text{Re}(\epsilon_1) < 0$ and $\text{Re}(\epsilon_2) < 0$
\begin{equation}
\begin{aligned}
    &\lim_{a\to 0} 
    \frac{
    \CZ_\text{pert}
    (a+ a^{(m,n)}/2, \epsilon_1, \epsilon_2, \Lambda, R)
    }{
    \CZ_\text{pert}
    (a + \hat{a}^{(m,n)}/2, \epsilon_1, \epsilon_2, \Lambda, R)
    }\\=&\mathcal{R}^{-4mn} (t_1 t_2)^{mn}
    \lim_{a\to 0}\text{PE}\left[
    -\sum_{i,j = 0}^\infty \left(x^{-2} t_1^{-m + i} t_2^{-n + j} + x^2 t_1^{m+i}t_2^{n+j}-x^{-2}t_1^{-m+i}t_2^{n+j}-x^2 t_1^{m+i}t_2^{-n+j}\right)
    \right]\\
    =&\mathcal{R}^{-4mn} (t_1 t_2)^{mn}
    \lim_{a\to 0}\prod_{i,j =0}^\infty
    \frac{
    (1 - x^{-2}t_1^{-m + i}t_2^{-n + j})
    (1 - x^2 t_1^{m+i}t_2^{n + j})
    }{
    (1 - x^{-2}t_1^{-m + i}t_2^{n + j})
    (1 - x^2 t_1^{m+i}t_2^{-n + j})
    },
\end{aligned}
\end{equation}
\begin{itemize}
\item $m>0$ and $n>0$:
\begin{equation}
    \lim_{a\to 0}
    \frac{1}{1-x^{-2}}
    \frac{
    \CZ_\text{pert}
    (a+ a^{(m,n)}/2, \epsilon_1, \epsilon_2, \Lambda, R)
    }{
    \CZ_\text{pert}
    (a + \hat{a}^{(m,n)}/2, \epsilon_1, \epsilon_2, \Lambda, R)
    } =\mathcal{R}^{-4 mn} (t_1 t_2)^{mn}
    \underbrace{\prod_{i=-m}^{m-1} \prod_{j=-n}^{n-1}}_{(i, j) \neq\{(0,0)\}}
    (1 - t_1^i t_2^j).
\end{equation}
\item $m<0$ and $n<0$:
\begin{equation}
    \lim_{a\to 0}
    \frac{1}{1-x^{2}}
    \frac{
    \CZ_\text{pert}
    (a+ a^{(m,n)}/2, \epsilon_1, \epsilon_2, \Lambda, R)
    }{
    \CZ_\text{pert}
    (a + \hat{a}^{(m,n)}/2, \epsilon_1, \epsilon_2, \Lambda, R)
    } =\mathcal{R}^{-4 mn} (t_1 t_2)^{mn}
    \underbrace{\prod_{i=m}^{-m-1} \prod_{j=n}^{-n-1}}_{(i, j) \neq\{(0,0)\}}
    (1 - t_1^i t_2^j).
\end{equation}
\end{itemize}

\paragraph{Case 2:} $\text{Re}(\epsilon_1) > 0$ and $\text{Re}(\epsilon_2) > 0$
\begin{equation}
\begin{aligned}
    &\lim_{a\to 0} 
    \frac{
    \CZ_\text{pert}
    (a+ a^{(m,n)}/2, \epsilon_1, \epsilon_2, \Lambda, R)
    }{
    \CZ_\text{pert}
    (a + \hat{a}^{(m,n)}/2, \epsilon_1, \epsilon_2, \Lambda, R)
    }\\
    =& 
    \mathcal{R}^{-4mn} (t_1 t_2)^{mn}
    \lim_{a\to 0}\text{PE}\left[
    -\frac{
    x^{-2} t_1^{-m-1}  t_2^{-n-1} 
    + x^{2}  t_1^{m-1}  t_2^{n-1}
    -x^{-2}  t_1^{-m-1}  t_2^{n-1}
    -x^{2} t_1^{m-1}  t_2^{-n-1}
    }
    {
    (1- t_1^{-1})(1 - t_2^{-1})    
    }
    \right]
    \\
    =&\mathcal{R}^{-4mn} (t_1 t_2)^{mn}
    \lim_{a\to 0}\prod_{i,j =0}^\infty
    \frac{
    (1 - x^{-2}t_1^{-m -1 - i}t_2^{-n -1- j})
    (1 - x^2 t_1^{m-1-i}t_2^{n -1 - j})
    }{
    (1 - x^{-2}t_1^{-m -1 - i}t_2^{n -1 - j})
    (1 - x^2 t_1^{m -1 -i}t_2^{-n -1 - j})
    },
\end{aligned}
\end{equation}
\begin{itemize}
\item $m>0$ and $n>0$:
\begin{equation}
    \lim_{a\to 0}
    \frac{1}{1-x^{2}}
    \frac{
    \CZ_\text{pert}
    (a+ a^{(m,n)}/2, \epsilon_1, \epsilon_2, \Lambda, R)
    }{
    \CZ_\text{pert}
    (a + \hat{a}^{(m,n)}/2, \epsilon_1, \epsilon_2, \Lambda, R)
    } =\mathcal{R}^{-4 mn} (t_1 t_2)^{mn}
    \underbrace{\prod_{i=-m}^{m-1} \prod_{j=-n}^{n-1}}_{(i, j) \neq\{(0,0)\}}
    (1 - t_1^i t_2^j).
\end{equation}
\item $m<0$ and $n<0$:
\begin{equation}
    \lim_{a\to 0}
    \frac{1}{1-x^{-2}}
    \frac{
    \CZ_\text{pert}
    (a+ a^{(m,n)}/2, \epsilon_1, \epsilon_2, \Lambda, R)
    }{
    \CZ_\text{pert}
    (a + \hat{a}^{(m,n)}/2, \epsilon_1, \epsilon_2, \Lambda, R)
    } =\mathcal{R}^{-4 mn} (t_1 t_2)^{mn}
    \underbrace{\prod_{i=m}^{-m-1} \prod_{j=n}^{-n-1}}_{(i, j) \neq\{(0,0)\}}
    (1 - t_1^i t_2^j).
\end{equation}
\end{itemize}

\paragraph{Case 3:} $\text{Re}(\epsilon_1) > 0$ and $\text{Re}(\epsilon_2) < 0$
\begin{equation}
\begin{aligned}
    &\lim_{a\to 0} 
    \frac{
    \CZ_\text{pert}
    (a+ a^{(m,n)}/2, \epsilon_1, \epsilon_2, \Lambda, R)
    }{
    \CZ_\text{pert}
    (a + \hat{a}^{(m,n)}/2, \epsilon_1, \epsilon_2, \Lambda, R)
    }\\
    =& 
    \mathcal{R}^{-4mn} (t_1 t_2)^{mn}
    \lim_{a\to 0}\text{PE}\left[
    -\frac{
    x^{-2} t_1^{-m-1}  t_2^{-n} 
    + x^{2}  t_1^{m-1}  t_2^{n}
    -x^{-2}  t_1^{-m-1}  t_2^{n}
    -x^{2} t_1^{m-1}  t_2^{-n}
    }
    {
    (1- t_1^{-1})(t_2 -1)    
    }
    \right]
    \\
    =&\mathcal{R}^{-4mn} (t_1 t_2)^{mn}
    \lim_{a\to 0}\prod_{i,j =0}^\infty
    \frac{
    (1 - x^{-2}t_1^{-m -1 - i}t_2^{n + j})
    (1 - x^2 t_1^{m-1-i}t_2^{-n + j})
    }{
    (1 - x^{-2}t_1^{-m -1 - i}t_2^{-n + j})
    (1 - x^2 t_1^{m -1 -i}t_2^{n + j})
    },
\end{aligned}
\end{equation}
\begin{itemize}
\item $m>0$ and $n>0$:
\begin{equation}
    \lim_{a\to 0}
    \frac{1}{1-x^{2}}
    \frac{
    \CZ_\text{pert}
    (a+ a^{(m,n)}/2, \epsilon_1, \epsilon_2, \Lambda, R)
    }{
    \CZ_\text{pert}
    (a + \hat{a}^{(m,n)}/2, \epsilon_1, \epsilon_2, \Lambda, R)
    } =\mathcal{R}^{-4 mn} (t_1 t_2)^{mn}
    \underbrace{\prod_{i=-m}^{m-1} \prod_{j=-n}^{n-1}}_{(i, j) \neq\{(0,0)\}}
    (1 - t_1^i t_2^j).
\end{equation}
\item $m<0$ and $n<0$:
\begin{equation}
    \lim_{a\to 0}
    \frac{1}{1-x^{-2}}
    \frac{
    \CZ_\text{pert}
    (a+ a^{(m,n)}/2, \epsilon_1, \epsilon_2, \Lambda, R)
    }{
    \CZ_\text{pert}
    (a + \hat{a}^{(m,n)}/2, \epsilon_1, \epsilon_2, \Lambda, R)
    } =\mathcal{R}^{-4 mn} (t_1 t_2)^{mn}
    \underbrace{\prod_{i=m}^{-m-1} \prod_{j=n}^{-n-1}}_{(i, j) \neq\{(0,0)\}}
    (1 - t_1^i t_2^j).
\end{equation}
\end{itemize}

\paragraph{Case 4:} $\text{Re}(\epsilon_1) < 0$ and $\text{Re}(\epsilon_2) > 0$
\begin{equation}
\begin{aligned}
    &\lim_{a\to 0} 
    \frac{
    \CZ_\text{pert}
    (a+ a^{(m,n)}/2, \epsilon_1, \epsilon_2, \Lambda, R)
    }{
    \CZ_\text{pert}
    (a + \hat{a}^{(m,n)}/2, \epsilon_1, \epsilon_2, \Lambda, R)
    }\\
    =& 
    \mathcal{R}^{-4mn} (t_1 t_2)^{mn}
    \lim_{a\to 0}\text{PE}\left[
    -\frac{
    x^{-2} t_1^{-m}  t_2^{-n-1} 
    + x^{2}  t_1^{m}  t_2^{n-1}
    -x^{-2}  t_1^{-m}  t_2^{n-1}
    -x^{2} t_1^{m}  t_2^{-n-1}
    }
    {
    (t_1 -1)(1- t_2^{-1})    
    }
    \right]
    \\
    =&\mathcal{R}^{-4mn} (t_1 t_2)^{mn}
    \lim_{a\to 0}\prod_{i,j =0}^\infty
    \frac{
    (1 - x^{-2}t_1^{-m + i}t_2^{n-1 - j})
    (1 - x^2 t_1^{m+i}t_2^{-n -1 -j})
    }{
    (1 - x^{-2}t_1^{-m + i}t_2^{-n -1 - j})
    (1 - x^2 t_1^{m  +i}t_2^{n-1 - j})
    },
\end{aligned}
\end{equation}
\begin{itemize}
\item $m>0$ and $n>0$:
\begin{equation}
    \lim_{a\to 0}
    \frac{1}{1-x^{-2}}
    \frac{
    \CZ_\text{pert}
    (a+ a^{(m,n)}/2, \epsilon_1, \epsilon_2, \Lambda, R)
    }{
    \CZ_\text{pert}
    (a + \hat{a}^{(m,n)}/2, \epsilon_1, \epsilon_2, \Lambda, R)
    } =\mathcal{R}^{-4 mn} (t_1 t_2)^{mn}
    \underbrace{\prod_{i=-m}^{m-1} \prod_{j=-n}^{n-1}}_{(i, j) \neq\{(0,0)\}}
    (1 - t_1^i t_2^j).
\end{equation}
\item $m<0$ and $n<0$:
\begin{equation}
    \lim_{a\to 0}
    \frac{1}{1-x^{2}}
    \frac{
    \CZ_\text{pert}
    (a+ a^{(m,n)}/2, \epsilon_1, \epsilon_2, \Lambda, R)
    }{
    \CZ_\text{pert}
    (a + \hat{a}^{(m,n)}/2, \epsilon_1, \epsilon_2, \Lambda, R)
    } =\mathcal{R}^{-4 mn} (t_1 t_2)^{mn}
    \underbrace{\prod_{i=m}^{-m-1} \prod_{j=n}^{-n-1}}_{(i, j) \neq\{(0,0)\}}
    (1 - t_1^i t_2^j).
\end{equation}
\end{itemize}
We can summarize the above results for $\CZ_\text{pert}$ as follows:
\begin{itemize}
    \item  $m>0$ and $n>0$:
\begin{equation}
\label{eq:duality-pert-pos}
    \lim_{a\to 0}
    \frac{1}{1-x^{2\,\text{sgn}(\text{Re}(\epsilon_1))}}
    \frac{
    \CZ_\text{pert}
    (a+ a^{(m,n)}/2, \epsilon_1, \epsilon_2, \Lambda, R)
    }{
    \CZ_\text{pert}
    (a + \hat{a}^{(m,n)}/2, \epsilon_1, \epsilon_2, \Lambda, R)
    } =(\mathcal{R}^{-4}t_1 t_2)^{mn}
    \underbrace{\prod_{i=-m}^{m-1} \prod_{j=-n}^{n-1}}_{(i, j) \neq\{(0,0)\}}
    (1 - t_1^i t_2^j).
\end{equation}
\item $m<0$ and $n<0$:
\begin{equation}
\label{eq:duality-pert-neg}
    \lim_{a\to 0}
    \frac{1}{1-x^{-2\, \text{sgn}(\text{Re}(\epsilon_1))}}
    \frac{
    \CZ_\text{pert}
    (a+ a^{(m,n)}/2, \epsilon_1, \epsilon_2, \Lambda, R)
    }{
    \CZ_\text{pert}
    (a + \hat{a}^{(m,n)}/2, \epsilon_1, \epsilon_2, \Lambda, R)
    } =(\mathcal{R}^{-4}t_1 t_2)^{mn}
    \underbrace{\prod_{i=m}^{-m-1} \prod_{j=n}^{-n-1}}_{(i, j) \neq\{(0,0)\}}
    (1 - t_1^i t_2^j).
\end{equation}
\end{itemize}
From the recursion relation for the $\mathrm{SU}(2)$ instanton partition function \eqref{ZinstRecursion}, we get
\begin{itemize}
    \item $m > 0$ and $n > 0 $:
    \begin{equation}
    \label{eq:duality-inst-pos}
    \lim_{a\to 0} (1-x^2)
    \frac{
    \CZ_\text{inst}
    (a+ a^{(m,n)}/2, \epsilon_1, \epsilon_2, n_I = 0, \Lambda, R)
    }
    {
    \CZ_\text{inst}
    (a+ \hat a^{(m,n)}/2, \epsilon_1, \epsilon_2, n_I = 0, \Lambda, R)
    }
    =-
    \frac{
    \mathcal{R}^{4mn} T_{m,n}(t_1,t_2)
    }
    {
    t_1^{-m}t_2^{-n}-t_1^m t_2^n
    },
    \end{equation}
    \item $m < 0$ and $n < 0 $:
    \begin{equation}
    \label{eq:duality-inst-neg}
    \lim_{a\to 0} (1-x^{-2})
    \frac{
    \CZ_\text{inst}
    (a+ a^{(m,n)}/2, \epsilon_1, \epsilon_2, n_I = 0, \Lambda, R)
    }
    {
    \CZ_\text{inst}
    (a+ \hat a^{(m,n)}/2, \epsilon_1, \epsilon_2, n_I = 0, \Lambda, R)
    }
    =-
    \frac{
    \mathcal{R}^{4mn} T_{-m,-n}(t_1,t_2)
    }
    {
    t_1^{-m}t_2^{-n}-t_1^m t_2^n
    },
    \end{equation}
\end{itemize}
where $T_{m,n}$ can be rewritten as 
\begin{equation}
    T_{m,n}(t_1, t_2) = 
    -\frac{
    (1 + t_1^m t_2^n)
    (1- t_1^{-m} t_2^{-n})
    }{
    (t_1 t_2)^{mn}
    }
    \underbrace{\prod_{i=-m}^{m-1} \prod_{j=-n}^{n-1}}_{(i, j) \neq\{(0,0)\}}
    \frac{1}{(1 - t_1^i t_2^j)}.
\end{equation}
Combining \eqref{eq:duality-pert-pos}, 
\eqref{eq:duality-pert-neg}, 
\eqref{eq:duality-inst-pos} and
\eqref{eq:duality-inst-neg}, we can get 
\begin{equation}
    \lim_{a\to 0}
    \frac{
    \CZ\left(a + a^{(m,n)}/2, \epsilon_1, \epsilon_2,  \Lambda , R\right)
    }{
    \CZ\left(a + \hat{a}^{(m,n)}/2,  \epsilon_1, \epsilon_2,  \Lambda , R\right)
    } = - \text{sgn}(\text{Re}(\epsilon_1)).
\end{equation}
Shifting $\Lambda$ to $\Lambda e^{-\frac{1}{4} n_I}$ will not change the ratio, then \eqref{eq:duality-nekrasov-e1} follows from \eqref{eq: ratio-Z-ZN}. By exchanging $\epsilon_1$ and $\epsilon_2$, we can get \eqref{eq:duality-nekrasov-e2}.

We can apply the abstruse duality \eqref{eq:duality-nekrasov-e1} and \eqref{eq:duality-nekrasov-e2} to the $g$ function \eqref{ga} when $X = \mathbb{CP}^2$. 
If we introduce the operator $\mathcal{P}_{\ell}$ that flips the sign of the $\ell$-th component in 
$\mathfrak{p}$, and due to the fact that $\text{Re}(\epsilon_1^{(\ell)}) \text{Re}(\epsilon_2^{(\ell+1)}) <0$ and the pole at $a=0$ is simple, we have that
\begin{equation}
\label{eq:Pl-ga}
    \mathop{\mathrm{Res}}_{a = 0} \,\mathcal{P}_{\ell} \,g^{\epsilon_1, \epsilon_2}_{\mathfrak{p},\mathfrak{p}^{[I]}}(a) = -\mathop{\mathrm{Res}}_{a = 0}g^{\epsilon_1, \epsilon_2}_{\mathfrak{p},\mathfrak{p}^{[I]}}(a).
\end{equation}
We can extend the result to the poles
at $a = s \pi i /R$ with $s\in \mathbb{Z}$
\begin{equation}
\label{eq:Pl-ga2}
    \mathop{\mathrm{Res}}_{a = s \pi i /R} \,\mathcal{P}_{\ell} \,g^{\epsilon_1, \epsilon_2}_{\mathfrak{p},\mathfrak{p}^{[I]}}(a) = -\mathop{\mathrm{Res}}_{a = s \pi i /R}g^{\epsilon_1, \epsilon_2}_{\mathfrak{p},\mathfrak{p}^{[I]}}(a).
\end{equation}
\subsection{Residue Reduction}
\subsubsection*{A residue property}

\paragraph{Claim:}
The residue 
\begin{equation}\label{eq: residue-nekrasov}
\mathop{\mathrm{Res}}_{a = a^*}\mathrm{d}a\,
g_{\mathfrak{p}, \mathfrak{p}^{(I)}}^{\epsilon_1, \epsilon_2}(a)
\end{equation}
depends only on the equivalence class $[\{\mathfrak{p}\}]$. Here $a^*$ is any singularity of $g_{\mathfrak{p}, \mathfrak{p}^{(I)}}^{\epsilon_1, \epsilon_2}(a)$, which is a solution to the equation
\begin{equation}\label{eq: a*}
2 a^{(\ell)}=m \epsilon_1^{(\ell)}+n \epsilon_2^{(\ell)}
+2 \pi i s / R, \quad s \in \mathbb{Z},
\end{equation}
where $(m,n)$ is in some subset of integer pairs $\mathbb{Z}\times \mathbb{Z}$, and $\ell\in \{1, \ldots \chi\}$.

\paragraph{Proof:}
Let us denote the shift of a as
\begin{equation}
    2a^{(\ell)} = 2a + f_{\ell}(\{\mathfrak{p}\}).
\end{equation}
We have shown that $f_{\ell}(\{\mathfrak{p}\}) - f_{\ell'}(\{\mathfrak{p}\})$
depends only on the equivalence class $[\{\mathfrak{p}\}]$.
We can rewrite the Eq. \eqref{eq: a*} as
\begin{equation}
\begin{aligned}
2 a &= -f_1(\{\mathfrak{p}\}) + (f_1(\{\mathfrak{p}\}) 
- f_{\ell}(\{\mathfrak{p}\})) + m \epsilon_1^{(\ell)}+n \epsilon_2^{(\ell)}\\
& =: -f_1(\{\mathfrak{p}\}) + n_1 \epsilon_1 + n_2 \epsilon_2.
\end{aligned}
\end{equation}
The integer pair $(n_1, n_2)$ labels the position of the pole given the equivalence class $[\{\mathfrak{p}\}]$.
We can then rewrite the residue formula \eqref{eq: residue-nekrasov} as
\begin{equation}
\begin{aligned}
&\mathop{\mathrm{Res}}_{2a=-f_1(\{\mathfrak{p}\}) + n_1 \epsilon_1 + n_2 \epsilon_2}\mathrm{d}a\,
g^{\epsilon_1, \epsilon_2}_{\mathfrak{p}, \mathfrak{p}^{(I)}}(a) \\
=&
\mathop{\mathrm{Res}}_{2a=n_1 \epsilon_1 + n_2 \epsilon_2} \mathrm{d} a\,
g^{\epsilon_1, \epsilon_2}_{\mathfrak{p}, \mathfrak{p}^{(I)}}
\left(a - f_1(\{\mathfrak{p}\})\right)\\
=& \mathop{\mathrm{Res}}_{2a=n_1 \epsilon_1 + n_2 \epsilon_2} \mathrm{d} a\,
\prod_{\ell=1}^{\chi} Z(a + f_{\ell}(\{\mathfrak{p}\}) - f_1(\{\mathfrak{p}\})
, \epsilon_1^{(\ell)}, \epsilon_2^{(\ell)}, n_I^{(\ell)}, \Lambda, R).
\end{aligned}
\end{equation}
The right-hand side on the second line is independent of the choice of the representative
in $\{\mathfrak{p}\}$.

\subsubsection*{Summing over all the poles}

Next, we establish the following residue identity for the integrand $g_{\mathfrak{p},\mathfrak{p}^{(I)}}^{\e_1,\e_2} (a)$ in \eqref{ga}
\begin{equation}
\label{eq:reduce-pole-sum}
\sum_{\boldsymbol{k}\in L + \boldsymbol{\mu}}
    \sum_{a^*\in S}\mathop{\mathrm{Res}}_{a = a^*}\, \mathrm{d}a
    f(\boldsymbol{k}) \,g_{\mathfrak{p},\mathfrak{p}^{(I)}}^{\e_1,\e_2} (a) = \sum_{\{\mathfrak{p}\}\in S_{\boldsymbol{\mu}}} 
    \left(\mathop{\mathrm{Res}}_{a =0}+\mathop{\mathrm{Res}}_{a =\pi i/R}\right)
    da
    f([\{\mathfrak{p}\}]) \,g_{\mathfrak{p},\mathfrak{p}^{(I)}}^{\e_1,\e_2} (a),
\end{equation}
From the discussion below \eqref{eq: residue-nekrasov}, the summand on the left-hand side does not depend on the choice of representative of $\boldsymbol{k}$. Thus, we can pick a representative $\{\mathfrak{p}\}$ from each equivalence class
$[\{\mathfrak{p}\}]$ in the $g$ function; 
On the right-hand side the sum runs over all $\chi$-tuples of integers subjected to the condition \eqref{eq: pmu-condition}. 
$f([\{\mathfrak{p}\}])$ is an arbitrary function of the equivalence class $[\{\mathfrak{p}\}]$.
$S$ is the set of all poles in $g_{\mathfrak{p},\mathfrak{p}^{(I)}}^{\e_1,\e_2} (a)$. 

\paragraph{Proof:}

From \eqref{eq:gen-pole-structure}, we extend $S$ to 
\begin{equation}
\label{eq: pole equation}
    \bar S := \{ a \mid 2a^{(\ell)} = \epsilon_1^{(\ell)} m + \epsilon_2^{(\ell)} n + 2c, m, n \in \mathbb{Z} ; \, \ell =1,\cdots,\chi;\, c= 0, \pi i/R \}.
\end{equation}
All additional points in $\bar S\backslash S $ are regular. Hence, replacing $S$ with $\bar S$ leaves the left-hand side of \eqref{eq:reduce-pole-sum} unchanged.
Let the solutions to \eqref{eq: pole equation} arising from the $\ell$-th local patch be labeled by
\begin{equation}
    2 a^{(\ell,m,n)}+2c := 
    (m- \mathfrak{p}_{\ell}) \epsilon_1^{(\ell)} + 
    (n- \mathfrak{p}_{\ell+1})\epsilon_2^{(\ell)} + 2c, 
\end{equation}
where $c\in \{0, \pi i / R\}$ and $a^{(\ell,m,n)}$ depend implicitly on $\mathfrak{p}$. We first choose some fixed $\mathfrak{p}$ for the discussion below.
A particular solution may arise in more than one patch.
To avoid overcounting, we introduce an equivalence relation on the triples, 
\begin{equation}
(\ell, m, n) \sim\left(\ell^{\prime}, m^{\prime}, n^{\prime}\right) \Longleftrightarrow a^{(\ell, m, n)}=a^{\left(\ell^{\prime}, m^{\prime}, n^{\prime}\right)}.
\end{equation}
In vector form, this is
\begin{equation}
    (m - \mathfrak{p}_{\ell}, n - \mathfrak{p}_{\ell+1} ) \vec\epsilon^{(\ell)} = 
    (m' - \mathfrak{p}_{\ell'}, n' - \mathfrak{p}_{\ell'+1} ) \vec\epsilon^{(\ell')},
\end{equation}
where we denote $\vec{\epsilon}^{(\ell)} = (\epsilon_1^{(\ell)}, \epsilon_2^{(\ell)})^\text{T}$. 
Equivalently, using \eqref{eq: A-trans-epsilon},
\begin{equation}
\label{eq: equiv-rela-overlap}
    (m - \mathfrak{p}_{\ell}, n - \mathfrak{p}_{\ell+1} ) A^{\ell,\ell'} = 
    (m' - \mathfrak{p}_{\ell'}, n' - \mathfrak{p}_{\ell'+1} ).
\end{equation}
The sum over poles on the left-hand side of \eqref{eq:reduce-pole-sum} (for a fixed $\mathfrak{p}$) can be organized as
\begin{equation}
    \sum_{a^*\in S} \mathop{\mathrm{Res}}_{a = a^*}\, \mathrm{d}a= 
    \sum_c\sum_{[(\ell,m,n)]} \mathop{\mathrm{Res}}_{a = a^{(\ell,m,n)}+c}\, da
    = 
    \sum_c\sum_{\ell=1}^{\chi}
    \sum_{(m,n)\in \mathbb{Z}^2}
    \frac{1}{\Delta^{[(\ell,m,n )]}}
    \mathop{\mathrm{Res}}_{a = a^{(\ell,m,n)}+c}\, \mathrm{d}a,
\end{equation}
where $\Delta^{[(\ell,m,n)]}$ is the cardinality of the corresponding equivalence class, inserted to cancel the overcounting caused by patch overlaps.
However, since $A^{\ell,\ell'}\in \mathrm{SL}(2,\mathbb{Z})$, from \eqref{eq: equiv-rela-overlap}, for a given triple $(\ell,m,n)$, there exists a unique triple $(\ell', m', n')$ for all $\ell'=1,\cdots,\chi$, 
such that
\begin{equation}
    (\ell,m,n)\sim (\ell',m',n').
\end{equation}
Therefore, $\Delta^{[(\ell,m,n)]}=\chi$. We have
\begin{equation}
    \sum_{a^*\in S} \mathop{\mathrm{Res}}_{a = a^*}\, \mathrm{d}a = \frac{1}{\chi}
    \sum_c\sum_{\ell=1}^{\chi}
    \sum_{(m,n)\in \mathbb{Z}^2}
    \mathop{\mathrm{Res}}_{a = a^{(\ell,m,n)}+c}\, \mathrm{d}a.
\end{equation}
Let us examine the residue that appears on the left–hand side of \eqref{eq:reduce-pole-sum},
\begin{equation}
\label{eq: change-variable-a}
    \begin{aligned}
        &\mathop{\mathrm{Res}}_{a = a^{(\ell,m,n)}+c}
        \mathrm{d}a \, f(\boldsymbol{k}) 
        \prod_{\ell'=1}^\chi 
        \CZ\left(a + \frac{1}{2}(\mathfrak{p}_{\ell'}, \mathfrak{p}_{\ell'+1})\vec{\epsilon}^{l'},
        \epsilon_1^{(\ell')}, \epsilon_2^{(\ell')}, 
        n_I^{(\ell')}, \Lambda, R
        \right)\\
        =& \mathop{\mathrm{Res}}_{a =c}\mathrm{d}a \, f(\boldsymbol{k})
        \prod_{\ell'=1}^\chi 
        \CZ\left(
        a + \frac{1}{2}
        \left(
        (m- \mathfrak{p}_{\ell},n- \mathfrak{p}_{\ell+1})
        A^{\ell,\ell'}+(\mathfrak{p}_{\ell'}, \mathfrak{p}_{\ell'+1})
        \right)\vec{\epsilon}^{(\ell')},
        \epsilon_1^{(\ell')}, \epsilon_2^{(\ell')}, 
        n_I^{(\ell')}, \Lambda, R
        \right)
    \end{aligned}
\end{equation}

\paragraph{Claim:}
\begin{enumerate}
    \item The effective shift of the Coulomb parameter that appears in the second line of \eqref{eq: change-variable-a} can be written as
    \begin{equation}
    \label{def: bar-p}
             (m- \mathfrak{p}_{\ell},n- \mathfrak{p}_{\ell+1})
        A^{\ell,\ell'}+(\mathfrak{p}_{\ell'}, \mathfrak{p}_{\ell'+1})
        =: 
        \left(\bar{\mathfrak{p}}^{[\ell]}_{\ell'}, \bar{ \mathfrak{p}}^{[\ell]}_{\ell' +1} \right). 
    \end{equation}
    \item $\{\bar{\mathfrak{p}}^{[\ell]}\}$ lies in the same 
    equivalence class as the original 
    $\{\mathfrak{p}\}$,
    thus is a representative of $\boldsymbol{k}$
    \begin{equation}
        [\{\mathfrak{p}\}] = [\{\bar{\mathfrak{p}}^{[\ell]}\}], 
    \end{equation}
    where the equivalent relation is introduced in
    \eqref{equivalence relation for p}.
\end{enumerate}

\paragraph{Proof:}

We first verify that the definition of \eqref{def: bar-p} is self-consistent, namely the definition of $\bar{\mathfrak{p}}_{\ell'}$ read off from the $\ell'$-th patch and that from the $(\ell'+1)$-th patch coincide.
From the $\ell'$-th patch, we obtain
\begin{equation}
    \bar{\mathfrak{p}}^{[\ell]}_{\ell'+1} = 
    (m - \mathfrak{p}_{\ell}) (A^{\ell,\ell'})_{12}
    + (n - \mathfrak{p}_{\ell+1})(A^{\ell,\ell'})_{22} + 
    \mathfrak{p}_{\ell'+1},
\end{equation}
whereas the $(\ell'+1)$-th patch yields 
\begin{equation}
    \bar{\mathfrak{p}}^{[\ell]}_{\ell'+1} = 
    (m - \mathfrak{p}_{\ell}) (A^{\ell,\ell'+1})_{11}
    + (n - \mathfrak{p}_{\ell+1})(A^{\ell,\ell'+1})_{21} + 
    \mathfrak{p}_{\ell'+1}.
\end{equation}
From \eqref{eq: A-property-1}, we have
\begin{equation}
    (A^{\ell,\ell'+1})_{11} = (A^{\ell,\ell'})_{12}, \quad
    (A^{\ell,\ell'+1})_{21} = (A^{\ell,\ell'})_{22}.
\end{equation}
Thus, these two expressions for $\bar{\mathfrak{p}}^{[\ell]}_{\ell'+1}$ coincide, establishing the
self-consistency of $\{\bar{\mathfrak{p}}^{[\ell]}\}$.

To show $\{\bar{\mathfrak{p}}^{[\ell]}\} \sim \{{\mathfrak{p}}\}$, we compute the difference
\begin{equation}
\begin{aligned}
    \bar{\mathfrak{p}}^{[\ell]}_{\ell'} - \mathfrak{p}_{\ell'}
    &= (m - \mathfrak{p}_{\ell}) (A^{\ell,\ell'})_{11}
    + (n - \mathfrak{p}_{\ell+1})(A^{\ell,\ell'})_{21}\\
    &= 
    (m - \mathfrak{p}_{\ell})\sum_{i=1}^2 
    (A^{\ell,1})_{1i}(A^{1,\ell'})_{i1}+
    (n - \mathfrak{p}_{\ell+1})
    \sum_{i=1}^2 
    (A^{\ell,1})_{2i}(A^{1,\ell'})_{i1}.
\end{aligned}
\end{equation}
From \eqref{eq: Al1-A1l}, we can write the difference in terms of $\vec{n}_{\ell}$ as
\begin{equation}
\begin{aligned}
    \bar{\mathfrak{p}}^{[\ell]}_{\ell'} - \mathfrak{p}_{\ell'} 
    =& \left((m - \mathfrak{p}_{\ell}) (A^{\ell,1})_{11} + 
    (n - \mathfrak{p}_{\ell+1})(A^{\ell,1})_{21}\right)
    n_{\ell'}^1\\ &-
    \left((m - \mathfrak{p}_{\ell}) (A^{\ell,1})_{12} + 
    (n - \mathfrak{p}_{\ell+1})(A^{\ell,1})_{22}\right)
    n_{\ell'}^2.
\end{aligned}
\end{equation}
The right-hand side is an integral linear combination of $n_{\ell'}^1$ and $n_{\ell'}^2$. From \eqref{equivalence relation for p}, we conclude $\{\bar{\mathfrak{p}}^{[\ell]}\} \sim \{{\mathfrak{p}}\}$ as claimed.

Rewriting the left–hand side of \eqref{eq:reduce-pole-sum} with the change of variables introduced above, we obtain
\begin{equation}
\label{eq: reduction-pole-2}
    \frac{1}{\chi}\sum_{
    \boldsymbol{k}\in L+\boldsymbol{\mu}}\sum_c\sum_{\ell=1}^{\chi}\sum_{(m,n)\in \mathbb{Z}^2}
    \mathop{\mathrm{Res}}_{a = c}\, \mathrm{d}a\,
f(\boldsymbol{k})\,g^{\epsilon_1,\epsilon_2}_{\bar{\mathfrak{p}}^{[\ell]}, \mathfrak{p}^{(I)}}(a).
\end{equation}
From the definition of $\bar{\mathfrak{p}}^{[\ell]}$, \eqref{def: bar-p}, we obtain
\begin{equation}
\label{eq:pl-m-n}
    \bar{\mathfrak{p}}_l^{[\ell]} =m, \quad 
    \bar{\mathfrak{p}}_{\ell+1}^{[\ell]} =n.
\end{equation}
\eqref{eq: reduction-pole-2} becomes 
\begin{equation}
\label{eq: reduction-pole-3}
    \frac{1}{\chi}\sum_{
    \boldsymbol{k}\in L+\boldsymbol{\mu}}\sum_c\sum_{\ell=1}^{\chi}\sum_{(\bar{\mathfrak{p}}_l^{[\ell]},\bar{\mathfrak{p}}_{\ell+1}^{[\ell]})\in \mathbb{Z}^2}
    \mathop{\mathrm{Res}}_{a = c}\, \mathrm{d}a\,
    f(\boldsymbol{k})\,g^{\epsilon_1,\epsilon_2}_{\bar{\mathfrak{p}}^{[\ell]}, \mathfrak{p}^{(I)}}(a).
\end{equation}
From a similar discussion around \eqref{eq: pk-summation}, each representative in $\boldsymbol{k}$ can be fixed by two integers $(\bar{\mathfrak{p}}^{[\ell]}_\ell, bar{\mathfrak{p}}^{[\ell]}_{\ell+1})$.
The following summation rule holds:
\begin{equation}
    \sum_{\boldsymbol{k}\in L + \boldsymbol{\mu}}\sum_{(\bar{\mathfrak{p}}_\ell^{[\ell]},\bar{\mathfrak{p}}_{\ell+1}^{[\ell]})\in \mathbb{Z}^2} =\sum_{\{\mathfrak{p}\} \in S_{\boldsymbol{\mu}}}.
\end{equation}
Using the summation rule in \eqref{eq: reduction-pole-3}, we get the right-hand side of \eqref{eq:reduce-pole-sum}.

\subsubsection*{Summing over half the poles}

When $X = \mathbb{CP}^2$, the positions of the poles in Eq. \eqref{eq:CP2-pole-pos} can be divided into two sets.
\begin{equation}
    S^\pm := \{ a \mid 2a^{(\ell)} = \epsilon_1^{(\ell)} m + \epsilon_2^{(\ell)} n + 2c, m, n \in \mathbb{Z}^\pm ; \, \ell =1,\cdots,\chi;\, c= 0, \pi i/R \}.
\end{equation}
We can write residue identities similar to \eqref{eq:reduce-pole-sum},
\begin{equation}
\label{eq:reduce-pole-sum-half}
\sum_{\boldsymbol{k}\in L + \boldsymbol{\mu}}
    \sum_{a^*\in S^\pm}\mathop{\mathrm{Res}}_{a = a^*}\, \mathrm{d}a\,
    f(\boldsymbol{k}) g_{\mathfrak{p},\mathfrak{p}^{(I)}}^{\e_1,\e_2} (a) = \sum_{\{\mathfrak{p}\}\in S_{\boldsymbol{\mu}}^\pm} 
    \left(\mathop{\mathrm{Res}}_{a =0}+\mathop{\mathrm{Res}}_{a =\pi i/R}\right)
    \mathrm{d}a\,
    f([\{\mathfrak{p}\}]) g_{\mathfrak{p},\mathfrak{p}^{(I)}}^{\e_1,\e_2} (a),
\end{equation}
where $S_{\boldsymbol{\mu}}^\pm$ is defined by 
\begin{equation}
    S_{\boldsymbol{\mu}}^\pm := 
    \{ \{\mathfrak{p}\}\in S_{\boldsymbol{\mu}} | \exists  \ell,  \text{ s.t. } \mathfrak{p}_{\ell}, \mathfrak{p}_{\ell+1}\in\mathbb{Z}^\pm \}.
\end{equation}

\paragraph{Proof:}
Let us focus on the sum over residues in $S^+$. 
We can rewrite the sum over residues on the left-hand side of \eqref{eq:reduce-pole-sum-half}
as
\begin{equation}
    \sum_{a^*\in S^+} \mathop{\mathrm{Res}}_{a = a^*}\, \mathrm{d}a = 
    \sum_c\sum_{[(\ell,m,n)]} \mathop{\mathrm{Res}}_{a = a^{(\ell,m,n)}+c}\, \mathrm{d}a
    = 
    \sum_c\sum_{\ell=1}^{\chi}
    \sum_{m,n\in \mathbb{Z}^+}
    \frac{1}{\Delta^{[(\ell,m,n)]}}
    \mathop{\mathrm{Res}}_{a = a^{(\ell,m,n)}+c}\, \mathrm{d}a,
\end{equation}
where $\Delta^{[(\ell,m,n)]}$ is the cardinality of the equivalence class defined by \eqref{eq: equiv-rela-overlap} but with $m,n$ all positive integers. Following a similar procedure, we can change the variables and rewrite the left-hand side of \eqref{eq:reduce-pole-sum-half} as
\begin{equation}
\label{eq:pole-reduction-half-2}
\begin{aligned}
    &\sum_{
    \boldsymbol{k}\in L+\boldsymbol{\mu}}\sum_c\sum_{\ell=1}^{\chi}\sum_{\bar{\mathfrak{p}}_\ell^{[\ell]},\bar{\mathfrak{p}}_{\ell+1}^{[\ell]}\in \mathbb{Z}^+}
    \frac{1}{\Delta^{[(\ell,\bar{\mathfrak{p}}_\ell^{[\ell]},\bar{\mathfrak{p}}_{\ell+1}^{[\ell]})]}}
    \mathop{\mathrm{Res}}_{a = c}\, \mathrm{d}a\,
f(\boldsymbol{k})\,g^{\epsilon_1,\epsilon_2}_{\bar{\mathfrak{p}}^{[\ell]}, \mathfrak{p}^{(I)}}(a)\\
    =&\sum_c \sum_{\ell=1}^{\chi} 
    \sum_{\{\bar{\mathfrak{p}}^{[\ell]}\}\in S_{\boldsymbol{\mu}}^{[\ell]}}
    \frac{1}{\Delta^{[(\ell,\bar{\mathfrak{p}}_\ell^{[\ell]},\bar{\mathfrak{p}}_{\ell+1}^{[\ell]})]}}
    \mathop{\mathrm{Res}}_{a = c}\, \mathrm{d}a\,
f(\boldsymbol{k})\,g^{\epsilon_1,\epsilon_2}_{\bar{\mathfrak{p}}^{[\ell]}, \mathfrak{p}^{(I)}}(a),
\end{aligned}
\end{equation}
where $S_{\boldsymbol{\mu}}^{[\ell]}$ is defined by
\begin{equation}
    S_{\boldsymbol{\mu}}^{[\ell]} := 
    \{ \{\mathfrak{p}\}\in S_{\boldsymbol{\mu}} |  \mathfrak{p}_{\ell}, \mathfrak{p}_{\ell+1}\in\mathbb{Z}^+ \}.
\end{equation}

\paragraph{Claim:}
The equivalent relation $(\ell,m,n)\sim (\ell',m',n')$ is equivalent to the degeneracy condition
\begin{equation}
\label{eq:pl-pl'}
\bar{\mathfrak{p}}^{[\ell]}=\bar{\mathfrak{p}}^{[\ell']},
\end{equation}
where $\bar{\mathfrak{p}}^{[\ell]}$ 
is defined in Eq. \eqref{def: bar-p}. $(m,n)$,$(m',n')$ are related with $\bar{\mathfrak{p}}^{[\ell]}$ and $\bar{\mathfrak{p}}^{[\ell']}$ respectively via \eqref{eq:pl-m-n}

\paragraph{Proof:}
Note that the condition \eqref{eq:pl-pl'} is equivalent to $\bar{\mathfrak{p}}^{[\ell]}_{1,2}=\bar{\mathfrak{p}}^{[\ell']}_{1,2}$, where
\begin{equation}
\label{eq: barp12}
\begin{aligned}
    \bar{\mathfrak{p}}_1^{[\ell]} = (m - \mathfrak{p}_{\ell}) (A^{\ell,1})_{11} 
    + (n - \mathfrak{p}_{\ell+1}) (A^{\ell,1})_{21} +
    \mathfrak{p}_1, \\
    \bar{\mathfrak{p}}^{[\ell]}_2 = (m - \mathfrak{p}_{\ell}) (A^{\ell,2})_{11} 
    + (n - \mathfrak{p}_{\ell+1}) (A^{\ell,2})_{21} +
    \mathfrak{p}_2.
\end{aligned}
\end{equation}
Then \eqref{eq:pl-pl'} is equivalent to the pair of relations
\begin{equation}
\label{eq: p1p2-equality}
\begin{aligned}
    (m-\mathfrak{p}_{\ell}) (A^{\ell,1})_{11} + 
    (n - \mathfrak{p}_{\ell+1}) (A^{\ell,1})_{21}
    =
    (m'-\mathfrak{p}_{\ell'}) (A^{\ell',1})_{11} + 
    (n' - \mathfrak{p}_{\ell'+1}) (A^{\ell',1})_{21},\\
    (m-\mathfrak{p}_{\ell}) (A^{\ell,2})_{11} + 
    (n - \mathfrak{p}_{\ell+1}) (A^{\ell,2})_{21}
    =
    (m'-\mathfrak{p}_{\ell'}) (A^{\ell',2})_{11} + 
    (n' - \mathfrak{p}_{\ell'+1}) (A^{\ell',2})_{21}.   
\end{aligned}
\end{equation}
From \eqref{eq: A-property-1}, $A$-matrice satisfy
\begin{equation}
    (A^{\ell,2})_{11} = (A^{\ell,1})_{12}, \quad 
    (A^{\ell,2})_{21} = (A^{\ell,1})_{22}.
\end{equation}
The two equalities in \eqref{eq: p1p2-equality} can be combined into a single matrix identity
\begin{equation}
    (m-\mathfrak{p}_{\ell}, n - \mathfrak{p}_{\ell+1}) 
    A^{\ell,1} = 
    (m'-\mathfrak{p}_{\ell'}, n' - \mathfrak{p}_{\ell'+1}) A^{\ell', 1}.
\end{equation}
Multiplying from the right by $A^{1, \ell'}$, reproduces \eqref{eq: equiv-rela-overlap}, the defining criterion for the equivalent relation $(\ell,m,n)\sim (\ell',m', n')$. Hence, the two statements are indeed equivalent. 
Due to this one-to-one correspondence, the degeneracy of $\bar{\mathfrak{p}}^{[\ell]}$ for different $\ell$ will cancel the denominator $\Delta^{[(\ell,m,n)]}$ in \eqref{eq:pole-reduction-half-2}. Thus, we get the right-hand side of \eqref{eq:reduce-pole-sum-half}.

\subsubsection*{Inserting Heaviside step function}
From the discussion at the end of Sec.  \ref{poles-and-zeros}, $\{\mathfrak{p}\}$ with vanishing component will not contribute to the residue on the right-hand side of \eqref{eq:reduce-pole-sum-half}.
We choose $f$ to be Heaviside step function $H$ 
and prove the following identities 
\begin{equation}\label{eq: residue-pro1}
    \begin{aligned}
&\sum_{\boldsymbol{k}\in L + \boldsymbol{\mu}}
    \sum_{a^*\in S^\pm}
    \mathop{\mathrm{Res}}_{a = a^*}\, \mathrm{d}a\, H(\pm r)
    g_{\mathfrak{p},\mathfrak{p}^{(I)}}^{\e_1,\e_2} (a)\\
    =&\mp
    \sum_c\left(\sum_{\{\mathfrak{p}\} \in S_{\text {unstable }}}
    +\sum_{\{\mathfrak{p}\} \in S_{\text {semi-stable }}}
    +2 \sum_{\{\mathfrak{p}\} \in S_{\text {stable }}}\right) 
    \underset{a=c}{\text{Res}} ~
    \mathrm{d} a\, g_{\mathfrak{p}, \mathfrak{p}^{(I)}}^{\epsilon_1, \epsilon_2}(a) ,
    \end{aligned}
\end{equation}
\begin{equation}\label{eq: residue-pro2}
    \sum_{\boldsymbol{k}\in L + \boldsymbol{\mu}}
    \sum_{a^*\in S^\pm}
    \mathop{\mathrm{Res}}_{a = a^*}\, \mathrm{d}a\, H(\mp r)
    g_{\mathfrak{p},\mathfrak{p}^{(I)}}^{\e_1,\e_2} (a)
    =\mp\sum_c \sum_{\{\mathfrak{p}\} \in S_{\text {unstable }}} 
    \underset{a=c}{\text{Res}}
    ~\mathrm{d} a \,g_{\mathfrak{p}, \mathfrak{p}^{(I)}}^{\epsilon_1, \epsilon_2}(a).
\end{equation}
where $S_\text{stable}$ is the set of all stable fluxes,
 $S_\text{semi-stable}$ is the set of all semi-stable fluxes, and $S_\text{semi-stable}$ is the set of all unstable fluxes when $\mathfrak{p}>0$,
\begin{equation}
\begin{aligned}
    S_\text{stable} &:= 
    \{\{\mathfrak{p}\}\in S_{\boldsymbol{\mu}}
    \mid \mathfrak{p}_{\ell} >0, 
    \mathfrak{p}_{\ell} +
    \mathfrak{p}_{\ell+1} >
    \mathfrak{p}_{\ell+2},\,
    \forall \ell
    \},\\
    S_\text{unstable} &:= 
    \{\{\mathfrak{p}\}\in S_{\boldsymbol{\mu}}
    \mid \mathfrak{p}_{\ell} >0, \,
    \forall \ell;\,
    \exists\, \ell, \text{ s.t. }
    \mathfrak{p}_{\ell} +
    \mathfrak{p}_{\ell+1} <
    \mathfrak{p}_{\ell+2}
    \}, \\
    S_\text{semi-stable} &:= 
    \{\{\mathfrak{p}\}\in S_{\boldsymbol{\mu}}
    \mid \mathfrak{p}_{\ell} >0, \,
    \forall \ell;\,
    \exists\, \ell, \text{ s.t. }
    \mathfrak{p}_{\ell} +
    \mathfrak{p}_{\ell+1} =
    \mathfrak{p}_{\ell+2}
    \}.
\end{aligned}
\end{equation}

\paragraph{Proof:}
To prove \eqref{eq: residue-pro1}, we add the residues in $S^+$.
Using \eqref{eq:reduce-pole-sum-half}
\begin{equation}
    \begin{aligned}
\sum_{\boldsymbol{k}\in L + \boldsymbol{\mu}}
    \sum_{a^*\in S^+}
    \mathop{\mathrm{Res}}_{a = a^*}\, \mathrm{d}a\, H( r)\,
    g_{\mathfrak{p},\mathfrak{p}^{(I)}}^{\e_1,\e_2} (a)
    &=
    \sum_c \sum_{\{\mathfrak{p}\}\in S_{\boldsymbol{\mu}}^+}
    \mathop{\mathrm{Res}}_{a = c}\, 
    \mathrm{d}a \, H(r) \, g_{\mathfrak{p},\mathfrak{p}^{(I)}}^{\e_1,\e_2} (a)\\
    &= 
    \sum_c \sum_{\{\mathfrak{p}\}\in \tilde{S}^+}
    \mathop{\mathrm{Res}}_{a = c}\, 
    \mathrm{d}a \, g_{\mathfrak{p},\mathfrak{p}^{(I)}}^{\e_1,\e_2} (a),
    \end{aligned}
\end{equation}
where we define
\begin{equation}
    \tilde{S}^{+}:=\left\{\{\mathfrak{p}\} \in S_{\boldsymbol{\mu}} 
    \mid \exists\, \ell, \text { s.t. } \mathfrak{p}_{\ell}, \mathfrak{p}_{\ell+1} 
    \in \mathbb{Z}^{+},r>0\right\}.
\end{equation}
$\tilde{S}^+$ can be divided into $4$ subsets
\begin{equation}
\label{eq:split-Sp}
    \tilde{S}^{+} = 
    \tilde{S}^{+}_0 \sqcup
    \tilde{S}^{+}_1 \sqcup
    \tilde{S}^{+}_2\sqcup
    \tilde{S}^{+}_3,
\end{equation}
where the subsets are defined by
\begin{equation}
    \tilde{S}^+_0 := 
    \{\{\mathfrak{p}\} \in S_{\boldsymbol{\mu}} | 
    \mathfrak{p}_{\ell} >0, \forall \ell\},
\end{equation}
and 
\begin{equation}
    \tilde{S}^+_\ell := 
    \{\{\mathfrak{p}\} \in S_{\boldsymbol{\mu}} | 
    \mathfrak{p}_{\ell} <0,\mathfrak{p}_{\ell+1},\mathfrak{p}_{\ell+2}>0, r>0\}.
\end{equation}
We can further split $\tilde{S}^+_0$
\begin{equation}
\label{eq: sp1}
    \sum_c \sum_{\{\mathfrak{p}\}\in \tilde{S}^+_0}
    \mathop{\mathrm{Res}}_{a = c}\, \mathrm{d} \,g_{\mathfrak{p},\mathfrak{p}^{(I)}}^{\e_1,\e_2} (a) 
    = \sum_c \left(\sum_{\{\mathfrak{p}\}\in S_\text{unstable}}
    +\sum_{\{\mathfrak{p}\}\in S_\text{semi-stable}}
    +\sum_{\{\mathfrak{p}\}\in S_\text{stable}}\right)
    \mathop{\mathrm{Res}}_{a = c}\, \mathrm{d} \,g_{\mathfrak{p},\mathfrak{p}^{(I)}}^{\e_1,\e_2} (a).
\end{equation}
Moreover, we can use \eqref{eq:Pl-ga2} to flip the negative element in $\mathfrak{p}$
\begin{equation}
\label{eq: sp2}
\begin{aligned}
    &\sum_c \sum_{\{\mathfrak{p}\}\in \tilde{S}^+_l}
    \mathop{\mathrm{Res}}_{a = c}\, \mathrm{d} \,g_{\mathfrak{p},\mathfrak{p}^{(I)}}^{\e_1,\e_2} (a) 
    \\=& -\sum_c \left(\sum_{\{\mathfrak{p}\}\in S^{[\ell+1]}_\text{semi-stable}}
    +\sum_{\{\mathfrak{p}\}\in S^{[\ell+2]}_\text{semi-stable}}
    +\sum_{\{\mathfrak{p}\}\in S^{[\ell+1]}_\text{unstable}}
    +\sum_{\{\mathfrak{p}\}\in S^{[\ell+2]}_\text{unstable}}
    +\sum_{\{\mathfrak{p}\}\in S_\text{stable}}\right)
    \mathop{\mathrm{Res}}_{a = c}\, \mathrm{d} \,g_{\mathfrak{p},\mathfrak{p}^{(I)}}^{\e_1,\e_2} (a),
    \end{aligned}
\end{equation}
where we define 
\begin{equation}
    \begin{aligned}
    S^{[\ell]}_\text{semi-stable} &:= 
    \{ \{\mathfrak{p}\}\in S_\text{semi-stable} \mid 
    \mathfrak{p}_{\ell} = \mathfrak{p}_{\ell-1} + \mathfrak{p}_{\ell+1}
    \}. \\
    S^{[\ell]}_\text{unstable} &:= 
    \{ \{\mathfrak{p}\}\in S_\text{unstable} \mid 
    \mathfrak{p}_{\ell} > \mathfrak{p}_{\ell-1} + \mathfrak{p}_{\ell+1}
    \}.
    \end{aligned}
\end{equation}
Combining \eqref{eq: sp1} and \eqref{eq: sp2}, we get \eqref{eq: residue-pro1}.

To prove \eqref{eq: residue-pro2}, we add residues in $S^+$,
\begin{equation}
    \begin{aligned}
\sum_{\boldsymbol{k}\in L + \boldsymbol{\mu}}
    \sum_{a^*\in S^+}
    \mathop{\mathrm{Res}}_{a = a^*}\, \mathrm{d}a\, H(-r)\,
    g_{\mathfrak{p},\mathfrak{p}^{(I)}}^{\e_1,\e_2} (a)
    &=
    \sum_c \sum_{\{\mathfrak{p}\}\in S_{\boldsymbol{\mu}}^+}
    \mathop{\mathrm{Res}}_{a = c}\, 
    \mathrm{d}a \, H(-r) \, g_{\mathfrak{p},\mathfrak{p}^{(I)}}^{\e_1,\e_2} (a)\\
    &= 
    \sum_c \sum_{\{\mathfrak{p}\}\in \tilde{S}^-}
    \mathop{\mathrm{Res}}_{a = c}\, 
    \mathrm{d}a \, g_{\mathfrak{p},\mathfrak{p}^{(I)}}^{\e_1,\e_2} (a),
    \end{aligned}
\end{equation}
and we define 
\begin{equation}
    \tilde{S}^{-}:=\left\{\{\mathfrak{p}\} \in S_{\boldsymbol{\mu}} 
    \mid \exists \ell, \text { s.t. } \mathfrak{p}_{\ell}, \mathfrak{p}_{\ell+1} 
    \in \mathbb{Z}^{+},r<0\right\}.
\end{equation}
We can split $\tilde{S}^-$ into three subsets,
\begin{equation}
    \tilde{S}^- = \tilde{S}^-_1 \sqcup
    \tilde{S}^-_2\sqcup
    \tilde{S}^-_3,
\end{equation}
where
\begin{equation}
    \tilde{S}^-_\ell := 
    \{\{\mathfrak{p}\} \in S_{\boldsymbol{\mu}} \mid 
    \mathfrak{p}_{\ell} <0,\mathfrak{p}_{\ell+1},\mathfrak{p}_{\ell+2}>0, r<0\}.
\end{equation}
Flipping $\mathfrak{p}_{\ell}$ from negative to positive using
\eqref{eq:Pl-ga}, we have
\begin{equation}
    \sum_c \sum_{\{\mathfrak{p}\}\in \tilde{S}^-_\ell}
    \mathop{\mathrm{Res}}_{a = c}\, \mathrm{d}a \,g_{\mathfrak{p},\mathfrak{p}^{(I)}}^{\e_1,\e_2} (a) 
    =- \sum_c \sum_{\{\mathfrak{p}\}\in S^{[\ell]}_\text{unstable}}
    \mathop{\mathrm{Res}}_{a = c}\,\mathrm{d}a \,g_{\mathfrak{p},\mathfrak{p}^{(I)}}^{\e_1,\e_2} (a).
\end{equation}
Thus, we obtain \eqref{eq: residue-pro2}.

Combining Eqs. \eqref{eq: residue-pro1} and \eqref{eq: residue-pro2}, we reproduce Eq. \eqref{ZesP2},
\begin{equation}
-\frac{1}{2}\sum_{\boldsymbol{k}\in L + \boldsymbol{\mu}}
    \sum_{a^*\in S^+}
    \mathop{\mathrm{Res}}_{a = a^*}\, \mathrm{d}a\, \text{sgn}(r) \,g_{\mathfrak{p},\mathfrak{p}^{(I)}}^{\e_1,\e_2} (a)= \sum_{c\in \{0, \pi i /R\}} \sum_{\{\mathfrak{p}\}}
    \Theta_{\boldsymbol{\mu}}(\mathfrak{p})
    \mathop{\mathrm{Res}}_{a = c}\, \mathrm{d}a \,g_{\mathfrak{p},\mathfrak{p}^{(I)}}^{\e_1,\e_2} (a),
\end{equation}
\begin{equation}
-\frac{1}{2}\sum_{\boldsymbol{k}\in L + \boldsymbol{\mu}}
    \sum_{a^*\in S^-}
    \mathop{\mathrm{Res}}_{a = a^*}\, \mathrm{d}a\, \text{sgn}(r) \,g_{\mathfrak{p},\mathfrak{p}^{(I)}}^{\e_1,\e_2} (a)= \sum_{c\in \{0, \pi i /R\}} \sum_{\{\mathfrak{p}\}}
    \Theta_{\boldsymbol{\mu}}(\mathfrak{p})
    \mathop{\mathrm{Res}}_{a = c}\, \mathrm{d}a \,g_{\mathfrak{p},\mathfrak{p}^{(I)}}^{\e_1,\e_2} (a).
\end{equation}
Putting the above two equations together, we get
\begin{equation}\label{eq:AppendixeK:Final}
-\frac{1}{4}\sum_{\boldsymbol{k}\in L + \boldsymbol{\mu}}
    \sum_{a^*\in S}
    \mathop{\mathrm{Res}}_{a = a^*}\, \mathrm{d}a\, \text{sgn}(r) \,g_{\mathfrak{p},\mathfrak{p}^{(I)}}^{\e_1,\e_2} (a)= \sum_{c\in \{0, \pi i /R\}} \sum_{\{\mathfrak{p}\}}
    \Theta_{\boldsymbol{\mu}}(\mathfrak{p})
    \mathop{\mathrm{Res}}_{a = c}\, \mathrm{d} a \,g_{\mathfrak{p},\mathfrak{p}^{(I)}}^{\e_1,\e_2} (a).
\end{equation}

\section{Supersymmetric Quantum Mechanics In $\Omega$-background}\label{app:SQMinOmega}

The generalization of the 5d twisted action  (\ref{eq:Stwisted-nI=0}) when $n_I = 0$ to the $\Omega$-background is given by \cite{Nekrasov:2003rj, Nekrasov:2010ka}
\begin{equation}\label{eq:TwistedOmegaAction}
S^{\text{twisted}, \Omega}_\text{SYM} = 
\frac{1}{g^2_{5d}}\int_{S^1 \times X} 
\sqrt{g_5} 
\mathcal{L}^{\text{twisted}, \Omega}_\text{SYM} + 
\frac{\log(\mathcal{R}^4)}{8\pi^2}\int_X F\wedge F, 
\end{equation}
where $\CL^{\text{twisted}, \Omega}_\text{SYM}$ is given by $\bar{\mathcal{Q}} \mathcal{V}$ in (\ref{eq:Stwisted-nI=0}) in the $\Omega$-background 
\begin{equation}\label{eq:TwistedOmegaAction2}
    \begin{aligned}
\mathcal{L}^{\text{twisted}, \Omega}_\text{SYM} =& \text{tr}\left(  (F_{\mu\nu}^+)^2  + \left(F_{\mu 5} + v^\nu F_{\nu\mu}\right) \left(F^{\mu 5} + v_\nu F^{\nu\mu} \right) - \frac{1}{4} D_{\mu\nu}D^{\mu\nu} + D_\mu \sigma D^\mu \sigma \right.\\&
+ (D_5\sigma+\mathcal{L}^A_v\sigma) 
(D_5\sigma+\mathcal{L}^A_v\sigma) + 2i \chi^{\mu\nu}D_\mu \psi_\nu -\frac{\sqrt{2}}{2}i \chi^{\mu\nu}[D_5+\mathcal{L}_v^A-\sigma, \chi_{\mu\nu}]\\
&\left.-\frac{i}{2}\eta[D_5 +\mathcal{L}^A_v-\sigma, \eta] 
+ i \psi^\mu D_\mu \eta 
+ \frac{i}{2} \psi^\mu [D_5 + \mathcal{L}_v^A + \sigma,\psi_\mu ]\right),
\end{aligned}
\end{equation}
where $\mathcal{L}_v^A:=\mathcal{L}_v-i v^\mu A_\mu$ and $v$ is the Killing vector which generates isometries $\mathrm{U}(1) \times$ $\mathrm{U}(1)$ on a toric K\"ahler manifold $X$ with equivariant parameters $\e_1,\e_2$. 
Since $S^{\text{twisted}, \Omega}_\text{SYM}$ is $\bar{\mathcal{Q}}$-exact,
a procedure similar to that in Sec.  \ref{subsec:redu2SQM} can be done. 
Similarly to Eq. \eqref{twisted Q exact}, the relevant part of the bosonic action that will contribute to the SQM in the $\Omega$-background is  
\begin{equation}
\int_{X \times S^1} \mathrm{d}^5 x \, \sqrt{g} \text{tr}\left(\left(F_{\mu 5}+v^\nu F_{\nu \mu}\right)\left(F^{\mu 5}+v_\nu F^{\nu \mu}\right)+ D_\mu \sigma D^\mu \sigma\right).
\end{equation}
Note that
\begin{equation}
\begin{aligned}
v^\mu F_{\mu \nu}+F_{5 \nu} & =v^\mu \partial_\mu A_\nu-v^\mu D_\nu A_\mu+\partial_5 A_\nu-D_\nu A_5 \\
& =v^\mu \partial_\mu A_\nu+\partial_\nu v^\mu A_\mu-D_\nu\left(v^\mu A_\mu+A_5\right)+\partial_5 A_\nu \\
& =\left(\mathcal{L}_v A\right)_\nu-D_\nu\left(v^\mu A_\mu+A_5\right)+\partial_5 A_\nu.
\end{aligned}
\end{equation}

We now introduce collective coordinates. This is a standard procedure. For a detailed treatment of the closely related derivation of effective actions for monopoles, see \cite{Moore:2015szp}. Let $A(x, Z)$ be an ASD instanton solution with collective coordinates $Z^I$ and set
\begin{equation}
\mathcal{L}_v A=V^I(Z)\, \partial_I A,
\end{equation}
where we introduce $V^I(Z)$ as the induced Killing vector for $\mathrm{U}(1) \times \mathrm{U}(1)$ acting on the moduli space of instantons. We then get
\begin{equation}
v^\mu F_{\mu \nu}+F_{5 \nu}=\left(\dot{Z}^I+V^I\right) \delta_I A-D_\nu\left(v^\mu A_\mu+A_5-\alpha_I\left(\dot{Z}^I+V^I\right)\right).
\end{equation}
The bosonic term leads to
\begin{equation}
\begin{aligned}
& \int_{X \times S^1} \mathrm{d}^5 x \, \sqrt{g} \text{tr}\left(\left(F_{\mu 5}+v^\nu F_{\nu \mu}\right)\left(F^{\mu 5}+v_\nu F^{\nu \mu}\right)+ D_\mu \sigma D^\mu \sigma\right) \\
= & \int_{S^1} \mathrm{d} t g_{I J}\left(\dot{Z}^I+V^I\right)\left(\dot{Z}^J+V^J\right)+\int_{X \times S^1} \mathrm{d}^5 x \sqrt{g} \text{tr}\left(D_\mu \Phi_V D^\mu \bar{\Phi}_V\right),
\end{aligned}
\end{equation}
where
\begin{equation}
\begin{aligned}
& \Phi_V :=  \sigma+i A_5+i v^\mu A_\mu-i \alpha_I\left(\dot{Z}^I+V^I\right), \\
& \bar{\Phi}_V :=  \sigma-i A_5-i v^\mu A_\mu+i \alpha_I\left(\dot{Z}^I+V^I\right).
\end{aligned}
\end{equation}
The relevant fermionic terms are given by
\begin{equation}
\label{eq: fermionOmega}
\begin{aligned}
& \int_{X \times S^1} \mathrm{d}^5 x \sqrt{g} \text{tr}\left(\frac{i}{2} \psi^\mu\left[D_5+\mathcal{L}_v^A+\sigma, \psi_\mu\right]\right) \\
= & \int_{X \times S^1} \mathrm{d}^5 x \sqrt{g} \text{tr}\left(\frac{i}{2} \psi^\mu\left(\partial_5-i \alpha_I \dot{Z}^I\right) \psi_\mu+\frac{i}{2} \psi^\mu\left(\mathcal{L}_v-i \alpha_I V^I\right) \psi_\mu+\frac{i}{2} \psi^\mu\left[\bar{\Phi}_V, \psi_\mu\right]\right).
\end{aligned}
\end{equation}
Using the following relations \cite{Moore:2015szp}
\begin{equation}
\label{eq:trans_isome}
\begin{gathered}
\psi(x, t)=\zeta^I(t) \delta_I A, \\
\mathcal{L}_v \zeta^I(t)=\zeta^J(t) \partial_J V^I(Z), \\
\mathcal{L}_v \delta_I A=V^J(Z) \partial_J \delta_I A,
\end{gathered}
\end{equation}
and the definition of the Christoffel symbol, the first term in \eqref{eq: fermionOmega} gives
\begin{equation}
    \begin{aligned}
& \int_{X \times S^1} \mathrm{d}^5 x \sqrt{g} \text{tr}\left(\frac{i}{2} \psi^\mu\left(\partial_5-i \alpha_I \dot{Z}^I\right) \psi_\mu\right) \\
= & \frac{i}{2} \int_{S^1} \mathrm{d} t g_{I J} \zeta^I \partial_t \zeta^J+\int_{S^1} \mathrm{d} t \int_X \mathrm{d}^4 x \sqrt{g_4} \text{tr}\left(\frac{i}{2} \zeta^I \delta_I A^\mu\left(\partial_K \dot{Z}^K-i \alpha_K \dot{Z}^K\right) \zeta^J \delta_J A_\mu\right) \\
= & \frac{i}{2} \int_{S^1} \mathrm{d} t g_{I J} \zeta^I \partial_t \zeta^J+\frac{i}{2} \int_{S^1} \mathrm{d} t \zeta^I \Gamma_{I K, J} \dot{Z}^K \zeta^J \\
= & \frac{i}{2} \int_{S^1} \mathrm{d} t g_{I J} \zeta^I\left(\partial_t \zeta^J+\Gamma_{K L}^J \dot{Z}^K \zeta^{(\ell)}\right) \\
= & \frac{i}{2} \int_{S^1} \mathrm{d} t g_{I J} \zeta^I \nabla_t \zeta^J,
\end{aligned}
\end{equation}
while the second term in \eqref{eq: fermionOmega} reads
\begin{equation}
    \begin{aligned}
& \int_{X \times S^1} \mathrm{d}^5 x \sqrt{g} \text{tr}\left(\frac{i}{2} \psi^\mu\left(\mathcal{L}_v-i \alpha_I V^I\right) \psi_\mu\right) \\
= & \frac{i}{2} \int_{S^1} \mathrm{d} t g_{I J} \zeta^I \partial_K V^J \zeta^K+\int_{S^1} \mathrm{d} t \int_X \mathrm{d}^4 x \sqrt{g_4} \text{tr}\left(\frac{i}{2} \zeta^I \delta_I A^\mu\left(V^K \partial_K-i \alpha_K V^K\right) \zeta^J \delta_J A\right. \\
= & \frac{i}{2} \int_{S^1} \mathrm{d} t \zeta^I \partial_J V_I \zeta^J+\frac{i}{2} \int_{S^1} \mathrm{d} t \zeta^I \Gamma_{I K, J} V^K \zeta^J \\
= & \frac{i}{2} \int_{S^1} \mathrm{d} t \zeta_I\left(\partial_J V^I \zeta^J+\Gamma_{J K}^I V^K \zeta^J\right) \\
= & \frac{i}{2} \int_{S^1} \mathrm{d} t \zeta_I \nabla_J V^I \zeta^J=\frac{i}{2} \int_{S^1} \mathrm{d} t \nabla_J V_I \zeta^I \zeta^J.
\end{aligned}
\end{equation}
The fermionic part of the action \eqref{eq: fermionOmega} reduces to
\begin{equation}
\frac{i}{2} \int_{S^1} \mathrm{d} t\left(g_{I J} \zeta^I \nabla_t \zeta^J-\nabla_I V_J \zeta^I \zeta^J\right)+\int_{X \times S^1} \mathrm{d}^5 x \sqrt{g} \text{tr}\left(\frac{i}{2} \psi^\mu\left[\bar{\Phi}_V, \psi_\mu\right]\right).
\end{equation}
Combining the bosonic and fermionic contributions and integrating out $\bar{\Phi}_V$, we obtain the action for SQM with Killing vector $V^I$
\footnote{The action is slightly different from \cite[Eq. (4.20)]{Alvarez-Gaume:1983zxc}.}
\begin{equation}
\label{eq:SQM_Omega}
S_V = \frac{1}{2}\int_{S^1} \mathrm{d} t \,g_{I J}\left(\dot{Z}^I+V^I\right)\left(\dot{Z}^J+V^J\right)+\frac{i}{4} \int_{S^1} \mathrm{d} t\left(g_{I J} \zeta^I \nabla_t \zeta^J-\nabla_I V_J \zeta^I \zeta^J\right),
\end{equation}
where we scale by $1/2$ since the action is $Q$-exact.
The path integral of this action will be the character-valued Dirac index.
Since $X$ is a toric K\"ahler manifold, there exists an induced symplectic K\"ahler structure $\omega_{IJ}$ that is preserved by $V(Z)$ on the moduli space $M_{\boldsymbol{\mu},\boldsymbol{k}}$
\begin{equation}
    \mathcal{L}_V \omega = 0.
\end{equation}
The Killing vector $V(Z)$ is holomorphic with respect to the complex structure given
by
\begin{equation}
    J_I{}^{J}=\omega_{IK}g^{KJ}.
\end{equation}
There are two conserved supercharges in \eqref{eq:SQM_Omega} (need to recheck the sign carefully in \eqref{eq:SQM_Omega})
\begin{equation}
\label{eq:SQM_susy_tran_Omega}
\begin{aligned}
    \delta Z^I &=  -\frac{i\varepsilon_1}{\sqrt{2}}
    \zeta^I + \frac{\varepsilon_2}{\sqrt{2}}J^I{}_J \zeta^J,\\
    \delta\zeta^I & = 
    \sqrt{2}\varepsilon_1({\dot Z}^I - V^I) + 
    \varepsilon_2 J^{I}{}_{J}(\dot Z^J - V^J)-i\varepsilon_2 \zeta^J\zeta^K J^L{}_J 
    \Gamma^I{}_{LK}.
\end{aligned}
\end{equation}
With the canonical momentum given by 
\begin{equation}
    P_I := g_{IJ}(\dot Z^J + V^J),   
\end{equation}
and $\zeta^a = \zeta^I e^a{}_I$, where $e^a{}_I$ is a frame, we introduce the canonical commutator relations
\begin{equation}
\begin{aligned}
    \left[Z^I,P_J\right] &= i \delta^I{}_J, \\
    \left\{\zeta^a, \zeta^b\right\}& = 2 \delta^{ab}.
\end{aligned}
\end{equation}
Note that $\{\zeta^a\}$ behaves as Gamma matrices on $M_{\boldsymbol{\mu},k}$.
Moreover, by introducing the momentum operator with spin connection $\omega_\mu$
\begin{equation}
    \pi_I =P_I -i\omega_\mu.
\end{equation}
The supercharges (density) in \eqref{eq:SQM_susy_tran_Omega} take the form
\begin{equation}
    Q_1 = \frac{1}{\sqrt{2}} \zeta^I (\pi_I - V_I), \quad
    Q_2 = \frac{1}{\sqrt{2}} \zeta^I 
    J_I{}^J(\pi_J - V_J).
\end{equation}
$Q_1$ acts as a nonchiral Dirac operator. Linear combinations of $Q_1$ and $Q_2$ give the 
chiral Dirac operators. 
The algebra of supercharges is given by
\begin{equation}
    Q_1^2 = Q_2^2 = 
    \frac{1}{2\sqrt{g}} 
    \pi_I \sqrt{g} g^{IJ}\pi_J +\frac{1}{2}V_I V^I +\frac{i}{2}\zeta^I \zeta^J D_I V_J -V^I P_I.
\end{equation}
From \eqref{eq:trans_isome}, the transformation generated by the Killing vector $V(Z)$ is
\begin{equation}
    \delta_V Z^I = V^I, \quad 
    \delta _V\zeta^I = \partial_J V^I \zeta^J.
\end{equation}
The corresponding Noether charge density is \cite{Moore:2015szp} 
\begin{equation}
    \epsilon_1J_1+\epsilon_2J_2 = -V^I \pi_I +\frac{i}{4}\zeta^I\zeta^J D_IV_J,
\end{equation}
where $J_1$ and $J_2$ are the generators for $\mathrm{U}(1)\times \mathrm{U}(1)$.
The Hamiltonian density with $V$ turned on is
\begin{equation}
    \mathcal{H}_V = \frac{1}{2}\left(
    \frac{1}{\sqrt{g}}\pi_I \sqrt{g} g^{IJ}\pi_J + V_I V^J + \frac{i}{2}\zeta^I\zeta^J D_{I}D_J
    \right) = Q_1^2 -\epsilon_1 J_1 -
    \epsilon_2 J_2.
\end{equation}
Therefore, the path integral becomes the character-valued Dirac index,
\begin{equation}
    \Tr_{\mathcal{H}_{k,\boldsymbol{\mu}}}(-1)^F e^{-\int_{S^1} \mathcal{H}_V} =
    \Tr_{\mathcal{H}_{k,\boldsymbol{\mu}}} \Gamma e^{- R(\slashed{D}^2 + \epsilon_1 J_1 + \epsilon_2 J_2)},
\end{equation}
where $\mathcal{H}_{k,\boldsymbol{\mu}}$ is the space of $L^2$ integrable sections of the spinor bundle on $M_{\boldsymbol{\mu},k}$, and  
\begin{equation}
\Gamma=\left(\begin{array}{cc}
1 & 0 \\
0 & -1
\end{array}\right)
\end{equation}
is identified with $(-1)^F$ with respect to the chiral decomposition of $\mathcal{H}_{k,\boldsymbol{\mu}}$.

Taking into account the second term in Eq. \eqref{eq:TwistedOmegaAction} we see that the full partition function $Z^{\e_1,\e_2}_{\bfmu,\mathfrak{p}^{(I)}}(\CR)$ is a generating function for the character-valued Dirac indices. 

\section{The Donaldson $\mu$-map}\label{app:DonaldsonMuMap}

At several points in the text we invoke the famous Donaldson $\mu$-maps. We briefly recall the definition here. Let $\mathbb{E}\to X\times M_{\bfmu,k}$ be the universal bundle
\footnote{We ignore subtleties associated with the existence of the universal bundle since we want to avoid the use of stacks.}
associated to the adjoint representation of $\mathrm{SO}(3)$, with natural maps $p:X\times M_{\bfmu,k}\to M_{\bfmu,k}$ and $q:X\times M_{\bfmu,k}\to X$. 
The Donaldson map \cite{Donaldson90}
\be
\mu_D: H_i(X,\mathbb{Z} ) \rightarrow H^{4-i} (M_{\bfmu,k},\mathbb{Q})
\ee
is defined by the slant product
\be
\label{eq:p1Universal}
\mu_D(x) = -\frac{1}{4}p_1(\mathbb{E})/x, \quad x\in H_{\bullet} (X,\mathbb{Z}),
\ee 
where $\text{tr}$ is the trace in the fundamental representation of $\mathfrak{su}(2)$. The real $\mathrm{SO}(3)$-bundle $\mathbb{E}$ can be lifted to a $\mathrm{U}(2)$-bundle $\mathcal{E}\to X\times M_{\bfmu,k}$. 
The Pontrjagin class in Eq. \eqref{eq:p1Universal} can be expressed in terms of the Chern classes $c_j(\mathcal{E})\in H^{2j}(X\times M_{\bfmu,k})$ as
\be 
-\frac{1}{4}p_1(\mathbb{E})= c_2({\mathcal{E}})-\frac{1}{4} c_1(\mathcal{E})^2. 
\ee 
Moreover, 
\be 
w_2(\mathbb{E})=c_1(\mathcal{E}) \mod H^2(X\times M_{\bfmu,k}, 2\mathbb{Z}).
\ee

Since the first Chern class decomposes as $c_1(\mathbb{E})=p^*c_1(\mathcal{E})+q^*c$ for some $c\in H^2(M_{\bfmu,k},\mathbb{Z})$, we deduce that when evaluated on integral homology classes $S \in H_2(X,\mathbb{Z})$ we have an integral homology class $2\mu_D(S) \in  H^2(M_{\bfmu,k},\mathbb{Z})/\mathrm{Tors}$. In the case where $X$ admits a complex structure, one can use the Riemann-Roch theorem to show that \cite[Chap.  8]{HuyLen10}
\be
\mu_D(K_X)=\frac{1}{2}K_{M_{\bfmu,k}}.
\ee

%References

\bibliographystyle{ytamsalpha} 
\bibliography{K-DW} 

\end{document}